%% file: oooeee.tex
\documentclass{aa}
\usepackage{grffile}
\usepackage{graphicx}
\usepackage{txfonts}
\usepackage{mathtools}
\usepackage{lscape}

\bibpunct{(}{)}{;}{a}{}{,}

\include{common-def}

\def\hho{H$_{\rm 2}$O}
\def\co{CO}
\def\coo{CO$_{\rm 2}$}
\def\chhhh{CH$_{\rm 4}$}
\def\criresplus{CRIRES$+$}
\def\taurex{$\tau$-REx}

\begin{document}

\title{Remote sensing of exoplanetary atmospheres with ground-based high-resolution near-infrared spectroscopy}

\author{
D. Shulyak\inst{1}
\and
M. Rengel\inst{1}
\and
A. Reiners\inst{2}
\and
U. Seemann\inst{2}
\and
F. Yan\inst{2}
}

\institute{
Max-Planck Institut f\"ur Sonnensystemforschung, Justus-von-Liebig-Weg 3, D-37077, G\"ottingen, Germany\\
\email{shulyak@mps.mpg.de}
\and
Institute for Astrophysics, Georg-August University, Friedrich-Hund-Platz 1, D-37077 G\"ottingen, Germany
}

\date{Received ; accepted}

% \abstract{}{}{}{}{} 
% 5 {} token are mandatory
 
  \abstract
   {Thanks to the advances in modern instrumentation we have learned about many exoplanets 
that span a wide range of masses and composition. Studying
their atmospheres provides insight into planetary origin, evolution, dynamics, and habitability.
Present and future observing facilities will address these important topics in great detail by using
more precise observations, high-resolution spectroscopy, and improved analysis methods.}
{We  investigate the feasibility of retrieving the vertical temperature distribution
and molecular number densities from expected exoplanet spectra in the near-infrared. 
We use the test case of the \criresplus\, instrument at the Very Large Telescope 
which will operate in the near-infrared between $1$~\mum\ and $5$~\mum\ 
and resolving powers of R=$100\,000$ and R=$50\,000$. 
We also determine the optimal wavelength coverage and observational strategies for increasing accuracy in the retrievals.
}
   {We used the optimal estimation approach to  retrieve the atmospheric parameters from the simulated emission
   observations of the hot Jupiter HD~189733b. The radiative transfer forward model is calculated using a public version 
   of the \taurex\, software package.}
   {Our simulations show that we can retrieve accurate temperature distribution in a very wide range 
of atmospheric pressures between $1$~bar and $10^{-6}$~bar depending on the chosen spectral region.
Retrieving molecular mixing ratios is very challenging, but a simultaneous observations in two separate infrared regions
around $1.6$~\mum\ and $2.3$~\mum\ helps to obtain accurate estimates; the exoplanetary spectra must be of relatively
high signal-to-noise ratio S/N$\geqslant$10, while the temperature can already be derived accurately  with the lowest
value that we considered in this study (S/N=$5$). }
   {The results of our study suggest that high-resolution near-infrared spectroscopy is a powerful tool for studying
   exoplanet atmospheres because numerous lines of different molecules can be analyzed simultaneously. 
   Instruments similar to \criresplus\, will provide data 
   for detailed retrieval and will provide new important constraints on the atmospheric chemistry and physics.}


   \maketitle
%
%-------------------------------------------------------------------

\section{Introduction}

Studying exoplanetary atmospheres is a key to understanding physico-chemical
processes, origin and evolution paths, and habitability conditions of these distant worlds.
This is usually done with the help of spectroscopic and/or (spectro-)photometric
observations from space and from the ground. However, obtaining
radiation from an exoplanetary atmosphere at an accuracy sufficient for its detailed investigation
is a difficult task due to the  large distance to even nearby exoplanets, which results in a very weak
photon flux reaching the top of the Earth's atmosphere. 
Photometric observations are the more efficient than spectroscopic observations in terms of the
 time needed to achieve a required signal-to-noise ratio (S/N) for the exoplanetary
signal.
%because the ratiation is collected in wide spectral bands
%of a dozen of filters rather being dispersed 
By using photometric observations obtained in a wide wavelength range (which may cover dozens of microns
in the infrared domain) it is possible to constrain the temperature stratification and mixing ratios of most abundant
molecules in the atmospheres of exoplanets. So far, the atmospheres of hot Jupiters (HJ) have been
a subject of intense investigations 
\citep[e.g.,][]{2019AJ....157..114B,2016Natur.529...59S,2015ApJ...813...13W,2015A&A...576A.111S,2013ApJ...775..137L,2012MNRAS.420..170L}. 
These gas giants have masses between $0.5-13$~\Mjup\,  and short rotation
periods $P<10$~days \citep{2015ApJ...799..229W} so that their
atmospheres are hot due to a close distance to their parent stars \citep{2010ApJ...718L.145W}. 
This results in strong irradiation
from their atmospheres that is much more easy to detect than the atmospheres of similar exoplanets that are at  larger distances from their stars.
To date, there are about a dozen   HJs that have been investigated   using photometric observations from different space missions \citep[e.g.,][]{2016Natur.529...59S}.

Spectroscopic techniques, on the other hand, are capable of resolving individual spectral lines of atmospheric molecules,
but require much longer observing times and are usually limited to a narrow wavelength range compared to photometry.
Nevertheless, modern high-resolution spectroscopy has been successfully used to detect the presence of many molecules
in atmospheres of different types of exoplanets via cross-correlation techniques 
\citep[e.g.,][]{2010Natur.465.1049S,2014A&A...561A.150D,2014A&A...565A.124B,2015ApJ...814...66K,2016ApJ...817..106B}. 
Unfortunately, extracting
profiles of individual molecular lines from available spectroscopic observations remains very challenging due
to the high noise level, telluric removal problems, and purely instrumental effects. 
Recently, \citet{2019AJ....157..114B} showed that the cross-correlation
technique can be used to asses the temperature structure in the atmospheres of HJs via the differential
analysis  of weak and strong features in the retrieved exoplanet spectra without the need to resolve profiles of individual lines
\citep[see, e.g.,][]{2015A&A...576A.111S}.
However, only very limited information can be retrieved and it is still not possible to obtain robust results, for example
due to  the quality of the observed spectrum. 
This is why most  past and present studies utilize spectroscopy   to detect the presence 
of molecules, but they do not attempt to study atmospheric temperature stratification and accurate mixing ratios. 

The advantage of spectroscopy in providing accurate estimates of molecular mixing ratios is currently
limited by the capabilities of the available instruments. In particular, the quality of the observed spectrum 
is often not good enough to attempt studying atmospheric temperature stratification and accurate mixing ratios. 
However, with the advent of future instrumentation it will be possible to study the accurate shapes of atmospheric
lines on diverse exoplanetary atmospheres. 
This will open a way to constrain atmospheric chemistry, global circulation (winds),
and signatures of trace gases produced by possible biological activity.

Motivated by recent progresses in exoplanetary science and advances in instrumentation, 
in this work we investigate the potential
of applying very high-resolution spectroscopy to the study of HJ atmospheres.
We focus our research on the CRyogenic high-resolution IR Echelle Spectrometer (\criresplus) on the Very Large Telescope scheduled
for the end of 2019 at the European Southern Observatory (ESO)\citep{2014Msngr.156....7D,2014SPIE.9147E..19F}.
\criresplus\, is an upgrade of the well- known old CRIRES; it has  improved
efficiency thanks to new detectors,
polarimetric capability, and an order of magnitude larger wavelength coverage
that can be achieved in single-setting observation thanks to a new cross-disperser.
In particular, this improvement is essential
for exoplanetary research because it opens a way to observe many molecular
features and to study atmospheric physics in great detail. 
Similar to the previous version of the instrument, \criresplus\, 
will operate between $0.9-5.3$~\mum\, with a highest achievable resolving 
power of $\lambda/\Delta\lambda=R=100\,000$\footnote{\tt https://crir.es/}.

Our main goal was to test the near-infrared wavelength domain at very high spectral resolution 
for the retrieval of atmospheric temperature profiles and mixing ratios of molecular species
using \criresplus\ as a test case. The key difference between our research and similar projects
is that our  aim was to retrieve assumption-free temperature profiles
and that we used different near-infrared spectral regions that contain numerous lines of atmospheric
molecules expected to be observed with \criresplus.
We simulated observations at different wavelength regions and their combinations, 
spectral resolutions, and S/N  to test our ability to derive
accurate temperatures and chemical compositions of HJ atmospheres. 
We made predictions to identify observational strategies and potential targets to study exoplanet atmospheres
 with high-resolution spectroscopy.
Finally, we note that our results can also be applied to any other 
present or future instrument with capabilities similar to \criresplus.

\section{Methods}
\subsection{Estimating the state of the atmosphere}

In our retrieval approach we used an algorithm based on optimal estimation (OE) \citep{1976RvGSP..14..609R}.
The main idea of OE is to derive a maximum \textit{\emph{a posteriori}} information about parameters
that define the state of the system under investigation. 
The OE approach finds an optimal solution that minimizes the difference between the model and observations 
under an additional constraint, which is our knowledge of a priori parameter values and their errors.
Following \citet{2000imas.book.....R} we solve the OE problem for the state vector of the system

\begin{equation}
\begin{multlined}
\vec{x_{i+1}} = \vec{x_{i}} + \left[(1+\gamma)\vec{S_{a}^{-1}} + \vec{K_i^{\rm T}\vec{S_y}^{-1}\vec{K_i}}\right]^{-1}\times\\
                \left[\vec{K_i}^{\rm T}\vec{S_y}^{-1}(\vec{y}-\vec{y_i}) + \vec{S_a}^{-1}(\vec{x_a}-\vec{x_i})\right],
\end{multlined}
\end{equation}

\noindent
where $\vec{x_{i+1}}$ and $\vec{x_{i}}$ are respectively the new and old solutions for the state vector at iteration $i+1$ and $i$, 
$\vec{x_a}$ and $\vec{S_{a}}$ are the a priori vector and its co-variance matrix, $\vec{y}$ and $\vec{S_y}$ are the measurements
and measurement errors, and  $\vec{y_i}$ and $\vec{K_i}$ are respectively the model prediction for the state vector $\vec{x_i}$ and the corresponding Jacobian
matrix calculated at $\vec{x_i}$. An adjustable parameter ($\gamma$) is used  to fine-tune the balance between the measurement and the a priori constraint
\citep{2000imas.book.....R,2008JQSRT.109.1136I}. We searched for the solution that optimizes the total cost function $\phi$
that can be written in the form

\begin{equation}
\phi = \vec{(\vec{y}-\vec{y_i})}^{\rm T}\vec{S_y}^{-1}\vec{(\vec{y}-\vec{y_i})}
       +\vec{(\vec{x_i}-\vec{x_a})}^{\rm T}\vec{S_a}^{-1}\vec{(\vec{x_i}-\vec{x_a})},
\label{eq:oe}
\end{equation}

\noindent
where the first term is the usual $\chi^2$ computed as a weighted difference between the model and observations,
while the second term accounts for the deviation of the state vector from its a priori assumption.
After the solution has converged, the final errors on the state vector are

\begin{equation}
\vec{S_x} = \left(\vec{S_a}^{-1} + \vec{K_i}^{\rm T}\vec{S_y}^{-1}\vec{K_i}\right)^{-1}.
\end{equation}

It is seen that the obtained solution depends on how accurately we know our a priori.
By setting large errors on the a priori,\, the OE eventually turns to a regular nonlinear 
$\chi^2$-minimization problem. In the case of solar system planets, the a priori\, values of physical parameters
such as  temperature stratification and surface pressures  can be known from  \textit{\emph{in situ}}
measurements, if available. This helps to substantially narrow down the parameter range and the application of OE provides
robust solutions with realistic error estimates \citep{2010A&A...521L..49H,2008JQSRT.109.1136I,2008P&SS...56.1368R}. 
In the case of extrasolar planets, little to nothing
is usually known about their physical and chemical structures. Nevertheless, even in these cases
it is possible to build initial constraints upon some simplistic analytical models and to estimate realistic
error bars a priori. 
Because of its relatively fast computational performance compared to other methods, such as the Markov chain Monte Carlo  (MCMC) algorithm, the OE method was successfully used to study the atmospheres of exoplanets
\citep[see, e.g.,][]{2012MNRAS.420..170L,2012ApJ...749...93L,2013ApJ...775..137L,2013MNRAS.434.2616B,2014ApJ...789...14L,2014ApJ...786..154B,2017ApJ...834...50B}.

Like every retrieval technique, OE has its own advantages and weaknesses. An obvious limitation
is the use of a priori\ information, which penalizes the solution toward the one that provides
the best balance between the observed and a priori\ uncertainties. Because the inverse problem we are dealing with is ill posed,
there can often be a variety of solutions that provide the same good fit but very different values of fit parameters. 
This happens if, for example, some of the parameters are strongly correlated and/or the data quality is poor.
In these cases using the a priori\ constraints helps  to keep retrievals 
 from finding the solution with the lowest $\chi^2$ but  not expected from the physical point of view
(i.e., strong fluctuations in temperature profile, molecular mixing ratios that are too far from the predictions of chemical models, among others.).
Unfortunately, in the case of exoplanets the use of inappropriate a priori\ information can
lead retrievals to fall into local minima rather than finding the true solutions.
In the OE method this difficulty is easy to overcome by choosing large enough errors
on (poorly known) a priori\ parameters whose influence on the final solution is thus substantially reduced.
Usually, a family of retrievals with different sets of initial guesses and a priori\ errors
are used to ensure that the found solution is robust \citep[see, e.g.,][]{2012MNRAS.420..170L}.

Another limitation of the OE method is the assumption that the distribution of the \textit{\emph{a posteriori}} parameters are Gaussian.
This is usually the case for data that is of high spectral resolution and S/N, which is
satisfied in most of our retrievals presented below. 
When the data is of low quality the usual workaround is again to perform retrievals with multiple initial guesses
to ensure that the global minimum is found.
The relative comparison of different retrieval methods can be found in \citet{2013ApJ...775..137L}. In our work we do not
favor any of the retrieval methods specifically. The choice of the retrieval approach normally depends on the goals of the 
particular investigation, quality of the available data, number of free parameters, available computing resources, among others.
Here we chose the OE method because 1) it has  long been used for the analysis of high-resolution spectroscopic data of planets
in our solar system, which means  that the expertise can be easily shared between research fields; 
2) future instruments will eventually provide us with high-resolution data for the brightest 
exoplanets, and this data will be of better quality compared to what we have now, thus allowing the OE method 
to use all its advantages; and 
3) we carried out assumption-free retrievals of temperature distribution, which means that
the local value of temperature in each atmospheric layer is a free parameter in our approach; 
because the atmosphere is split into several tens of layers, this results in too many free parameters 
to be treated with purely Bayesian approaches.

\subsection{Forward model}

In order to use the OE method,   an appropriate radiative transfer forward model is needed that computes the outgoing radiation
from the visible planetary surface for a given set of free parameters.
We did this by employing the Tau Retrieval for Exoplanets (\taurex) software package \citet{2015ApJ...802..107W,2015ApJ...813...13W}. The
\taurex\, forward model uses up-to-date molecular cross sections based on line lists provided by the \textsc{ExoMol}\footnote{\tt www.exomol.com}
project \citep{2012MNRAS.425...21T} and \textsc{HITEMP} \citep{2010JQSRT.111.2139R}. These cross sections are pre-computed on a grid of temperature and pressure
pairs and are stored in binary opacity tables that are available for a number of spectral resolutions. The continuum opacity includes
Rayleigh scattering on molecules and collisionaly induced absorption due to H$_{\rm 2}$-H$_{\rm 2}$ and H$_{\rm 2}$-He either
after \citet{2011mss..confEFC07A} and \citet{2012JChPh.136d4319A} or \citet{2001JQSRT..68..235B,2002A&A...390..779B} and \citet{1989ApJ...341..549B}, respectively.
The original line opacity tables contain high-resolution
cross sections for each molecular species and for a set of temperature and pressure pairs.
%which, when running multimolecular retrievals, 
%require too much shared computer memory than those provided by cluster facilities available to us.
Thus, we optimized the public version
of \taurex\footnote{\tt https://github.com/ucl-exoplanets/TauREx\_public} to compute spectra in wide wavelength intervals 
and with very high spectral resolution R$=100\,000$ expected for \criresplus\, via the more efficient usage of opacity tables inside the code. 
Finally, we adapted our retrieval code for multiprocessing which is essential for the analysis of altitude 
dependent quantities (e.g., T, mixing ratios, winds) when the number of free parameters can be very high.

In order to keep solutions stable and to ensure that the retrieved altitude dependent quantities (e.g., temperature) are smooth, it is usually
assumed that these quantities are correlated with a characteristic correlation length $l_{\rm corr}$ expressed
in pressure scale heights, for example. Then, the off-diagonal elements of the a priori\, co-variance matrix can be written as

\begin{equation}
S_{ij} = \sqrt{S_{ii}S_{jj}} \exp{\left(-\frac{|\ln p_i/p_j|}{l_{\rm corr}}\right)},
\end{equation}

\noindent
where $p_i$ and $p_j$ are the pressures at atmospheric layers $i$ and $j$, respectively, and $l_{\rm corr}$ is the number of pressure scale heights
within which the correlation between layers drops by a factor of $e$. We found that a value of $l_{\rm corr}=1.5$ provides a necessary
amount of smoothing without losing the physical information that we retrieve, and thus we adopted this value in all our retrievals
(the same value of $l_{\rm corr}$ was also used in \citealt{2008JQSRT.109.1136I}).

If no correlation between atmospheric layers is used then the inverse of co-variance matrices can be analytically computed.
To the contrary, when the correlation length is non-zero the $\vec{S_a}$ matrix contains off-diagonal elements whose values
can differ by many orders of magnitude which leads  rounding-off errors in the numerical matrix
inversion.
In order to reduce numerical problems we normalize the state vector co-variance matrix $\vec{S_a}$ by its diagonal elements
and re-normalize obtained solution in Eq.~\ref{eq:oe} accordingly.

In our retrievals we make no assumptions about the altitude dependence of temperature profile. 
However we assumed constant mixing ratios of molecular species. 
We do this to be consistent with the original work by \citet{2012MNRAS.420..170L} (see next section). 
This also decreases the number of free parameters in our model and significantly
reduces calculation time.

The final set of free parameters in our retrieval model includes $50$ temperature values corresponding to different
altitude levels, four molecular species (\hho, \co, \coo, \chhhh), and a continuum scaling factor for each wavelength region that we investigated.
The scaling factors account for any possible normalization inaccuracies between observed and predicted
spectra. This adjustment will be needed because  after the reduction each \criresplus\, spectra will be normalized 
to an arbitrary chosen level (e.g., pseudo continuum).
In this work we make no attempts to simulate these very complicated reduction steps that should be a subject
of a separate investigation. 
Instead, our simulations represent an ideal case where we do not consider all instrument systematics and data
reduction inaccuracies.
Thus, we normalized the observed spectrum to some arbitrary value (mean flux level in our case) and did the same with the fluxes computed with our
forward model. The additional scaling factor is then applied to our forward model to account for any remaining mismatch
between the observed and modeled spectra. Clearly, these scaling factors should be very close to unity in the case 
of accurate retrieval because the observed and predicted spectra were computed with the same code. 
Still, these additional scaling factors will be needed in retrievals performed on real data and we include them in our model.

In this work we concentrate only on emission spectroscopy. We do this because emission spectra are formed in a very
wide range of atmospheric temperatures and pressures, while transmission spectroscopy 
(performed when a planet transits in front of the host star)   senses mainly regions of high altitudes.
In addition, the duration of transits of all potential targets are shorter than off-transit times when the day side
of the planet is visible. This  makes it easier to observe emission spectra with a required noise level.

\section{Validation of the method}

Before addressing the main goal of our work, we validated our method 
against some representative examples. To do this we 
chose a test case of a well-studied exoplanet, \object{HD~189733b}. This HJ transits a young 
main sequence star of spectral type K0 at a distance of about $19.3$~pc \citep{2005A&A...444L..15B}.
%This choice is somewhat arbitrary because our goal is to test the retrieval approach against
%some representative atmosphere of a HJ rather than to study atmospheres of each available HJ in detail.
The atmospheric structure of HD~189733b was studied by different groups and with different methods
in the past \citep[see, e.g.,][]{2012MNRAS.420..170L,2015ApJ...813...13W,2019AJ....157..114B}.

\begin{figure*}
\centerline{
\includegraphics[width=0.7\hsize]{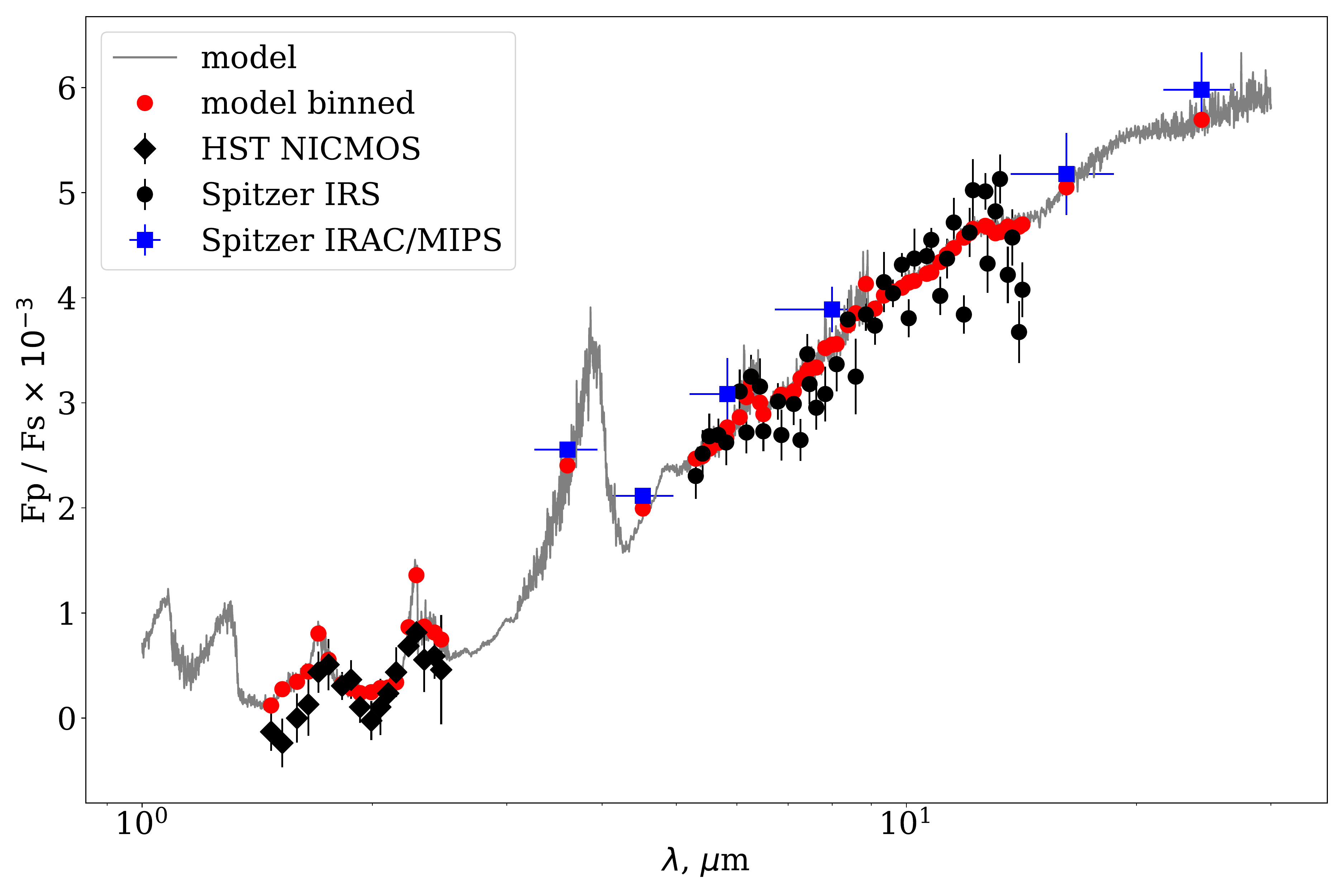}
}
\begin{minipage}{0.6\hsize}
\centerline{
\includegraphics[trim={0 0 21.5cm 0},clip,width=0.9\hsize]{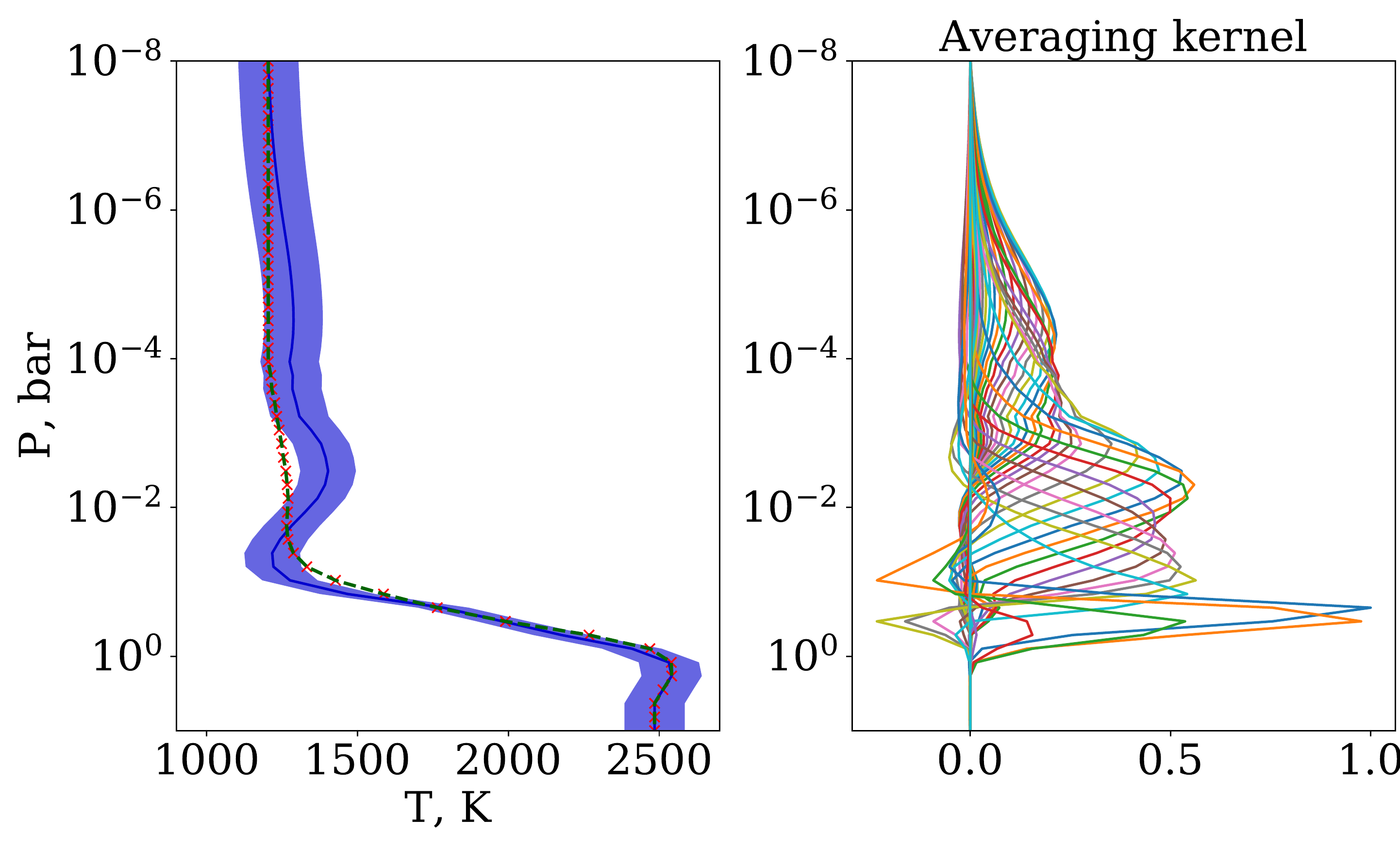}
}
\end{minipage}
\begin{minipage}{0.4\hsize}
\centerline{
\includegraphics[width=\hsize]{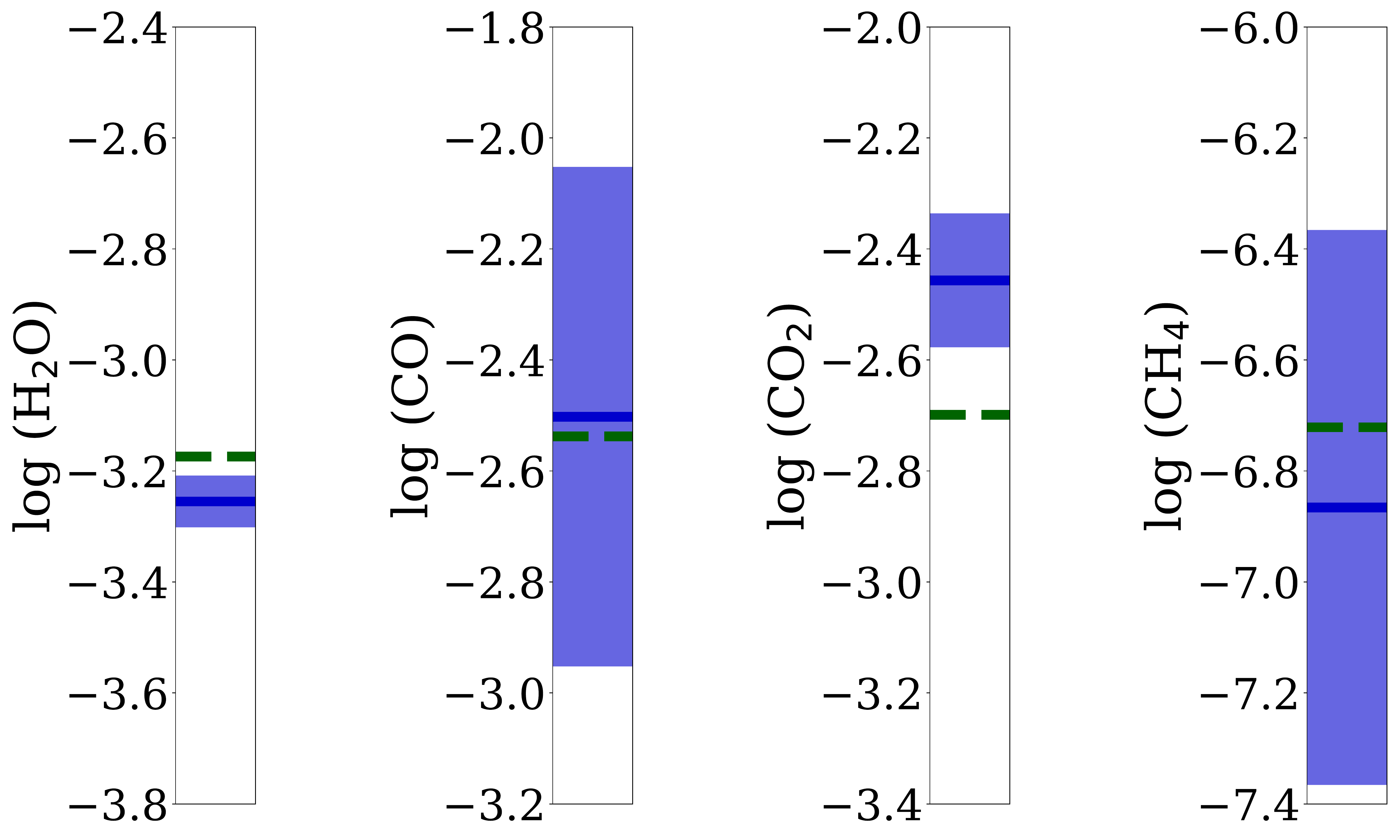}
}
\end{minipage}
\caption{\label{fig:hd18-1}
\textbf{Top panel:} Observed day side fluxes of HD~189733b and our best fit model predictions (see legend in the plot). 
\textbf{Bottom panel:} Retrieved T-P profile along with corresponding averaging kernels and mixing ratios
of molecular species.
We used the T-P profile of \cite{2012MNRAS.420..170L}  a priori\, (shown as red crosses). 
Our best fit model is shown with a solid blue line and the shaded area represents 1$\sigma$ error bars.
The assumed a priori\, are shown as a green dashed line (which coincides with the red crosses in this particular case).
The same color-coding is used for the four   side plots of mixing ratios (on the right).
}
\end{figure*}

Figure~\ref{fig:hd18-1} summarizes the retrieved temperature and mixing ratio of \hho, \co, \coo, and \chhhh\,
in the atmosphere of HD~189733b using photometric observations obtained with different space missions (see figure legend). 
We used results of \citet{2012MNRAS.420..170L} as our a priori\, and assumed $100$~K uncertainties
for our initial temperature profile. We note that here and throughout the paper our a priori\, is the same as an initial guess in our
iterative retrieval approach, although in a more general case these two can differ.
The analysis of corresponding averaging kernels for the temperature distribution (second plot in the bottom panel of Fig.~\ref{fig:hd18-1})
shows that the available observations probe temperature structures
in a wide range of atmospheric depths between $10^{-6}$~bar and $1$~bar. 
The averaging kernels describe a response of the retrieval to a small perturbation in temperature at each atmospheric depth \citep{2000imas.book.....R},
and thus characterize the contribution of each depth to the final result.
The zero values for averaging kernels correspond to the case when no information can be retrieved from corresponding depths
and they are nonzero otherwise.
It is seen that averaging kernels peak at appropriate atmospheric depths between $10^{-6}$~bar and $1$~bar, thus indicating that
our retrievals are robust at these altitudes;  they  approach zero outside that range.

Figure~\ref{fig:hd18-1} shows   that we retrieved parameters
in agreement with original work by \citet{2012MNRAS.420..170L}. 
We also confirm the high content of water in the atmosphere of HD~198733~b, 
and the general shape of the temperature distribution.
Finally, the concentrations of \coo\ and \chhhh\, are badly constrained and we could not measure their content with any
initial guess assumed, again in agreement with \citet{2012MNRAS.420..170L}.

\begin{figure*}
\begin{minipage}{0.45\hsize}
\centerline{
\includegraphics[width=\hsize]{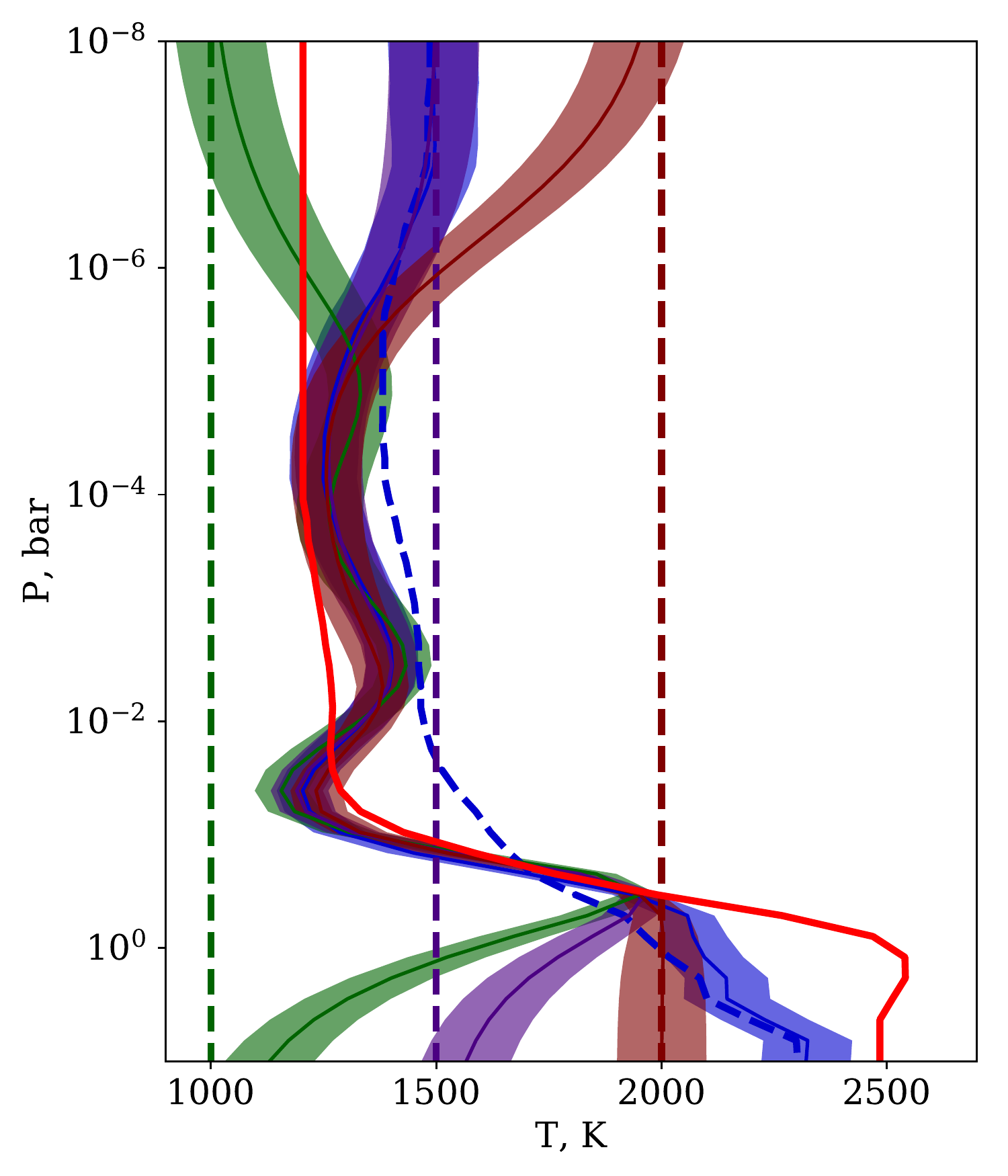}
}
\end{minipage}
\begin{minipage}{0.05\hsize}
\flushright
{\small\rotatebox{90}{P, bar}}
\end{minipage}
\begin{minipage}{0.45\hsize}
\centerline{
\includegraphics[trim={15.2cm 0 20.0cm 0},clip,width=0.5\hsize]{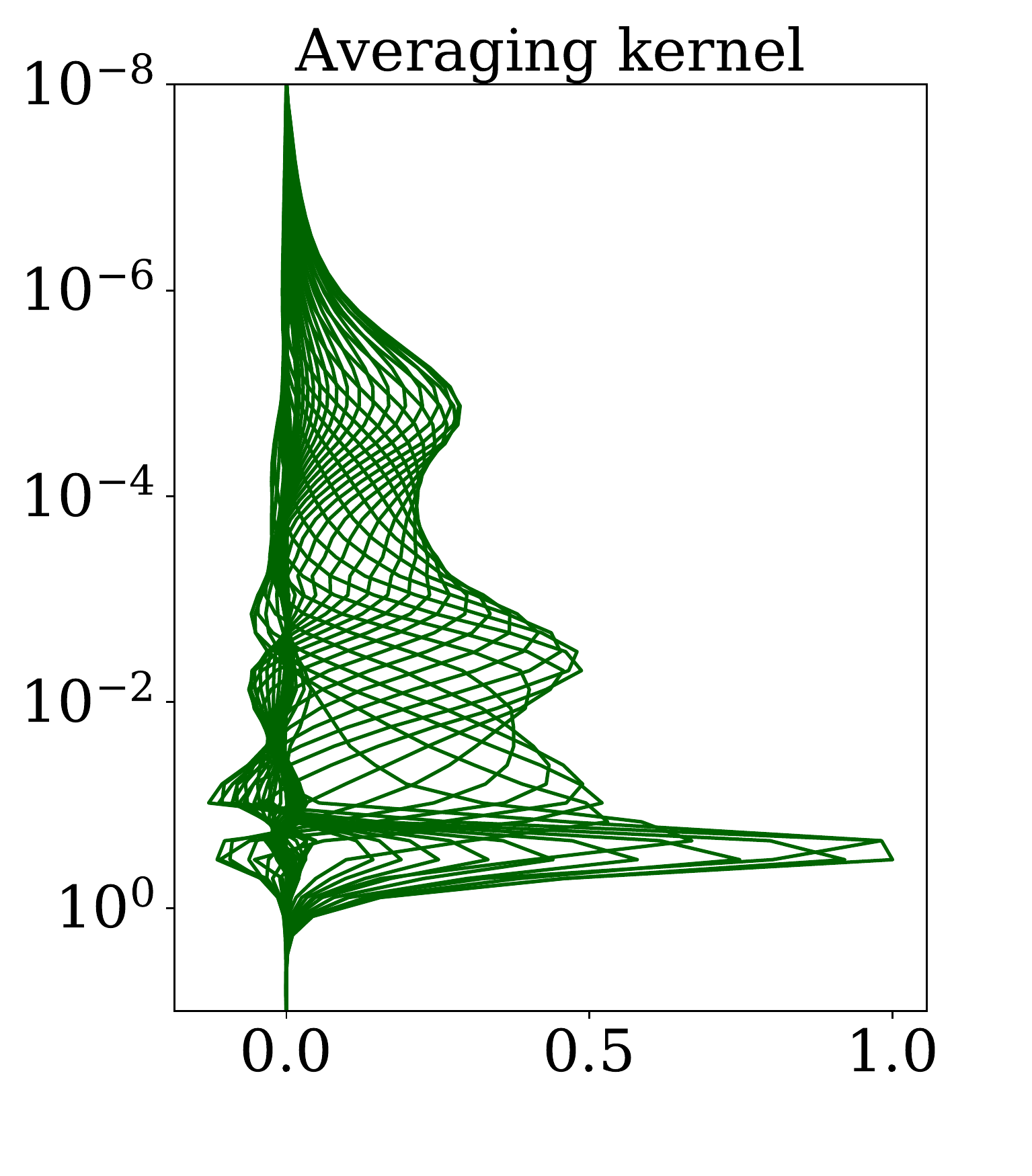}
\includegraphics[trim={15.2cm 0 20.0cm 0},clip,width=0.5\hsize]{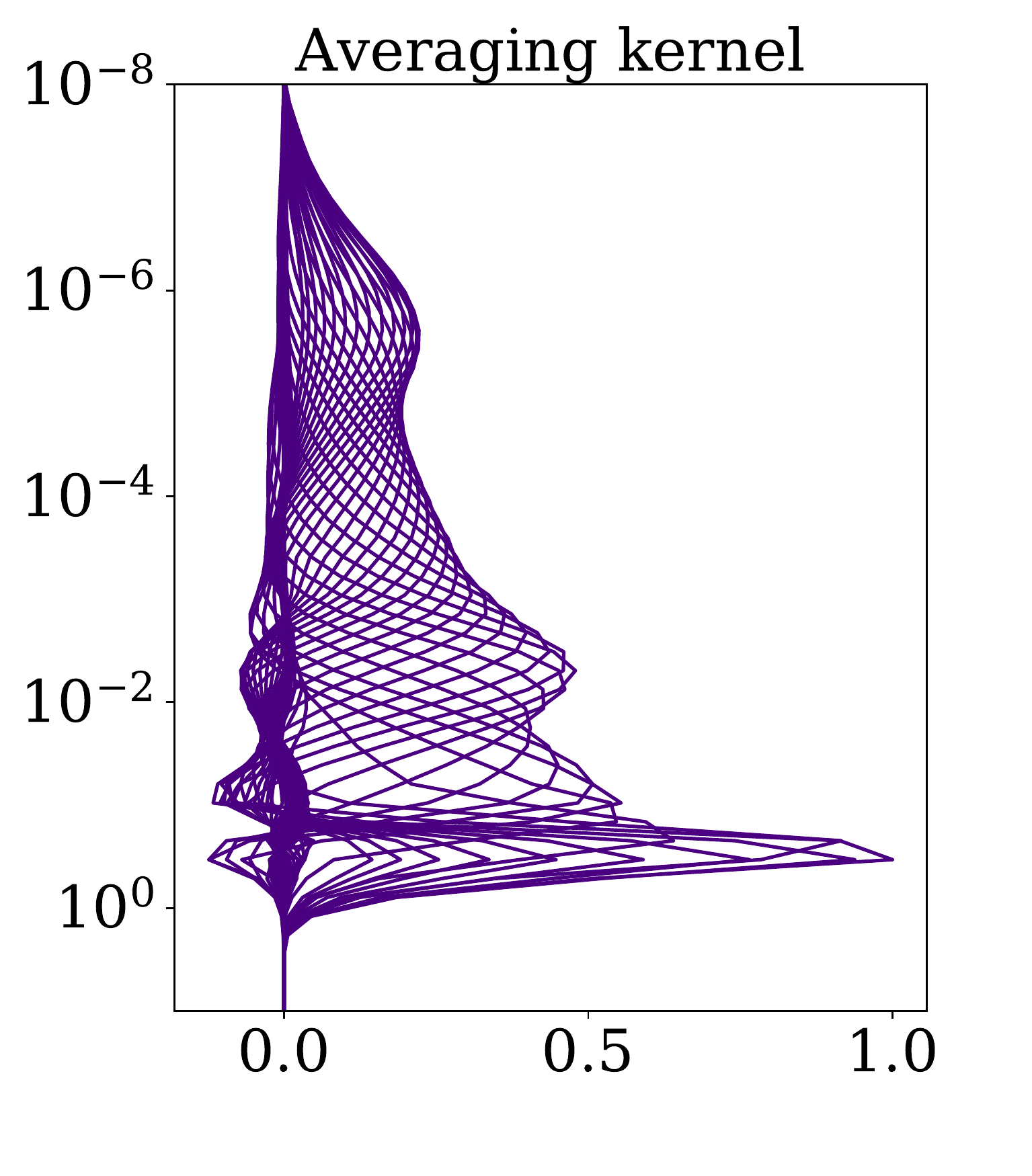}
}
\centerline{
\includegraphics[trim={15.2cm 0 20.0cm 0},clip,width=0.5\hsize]{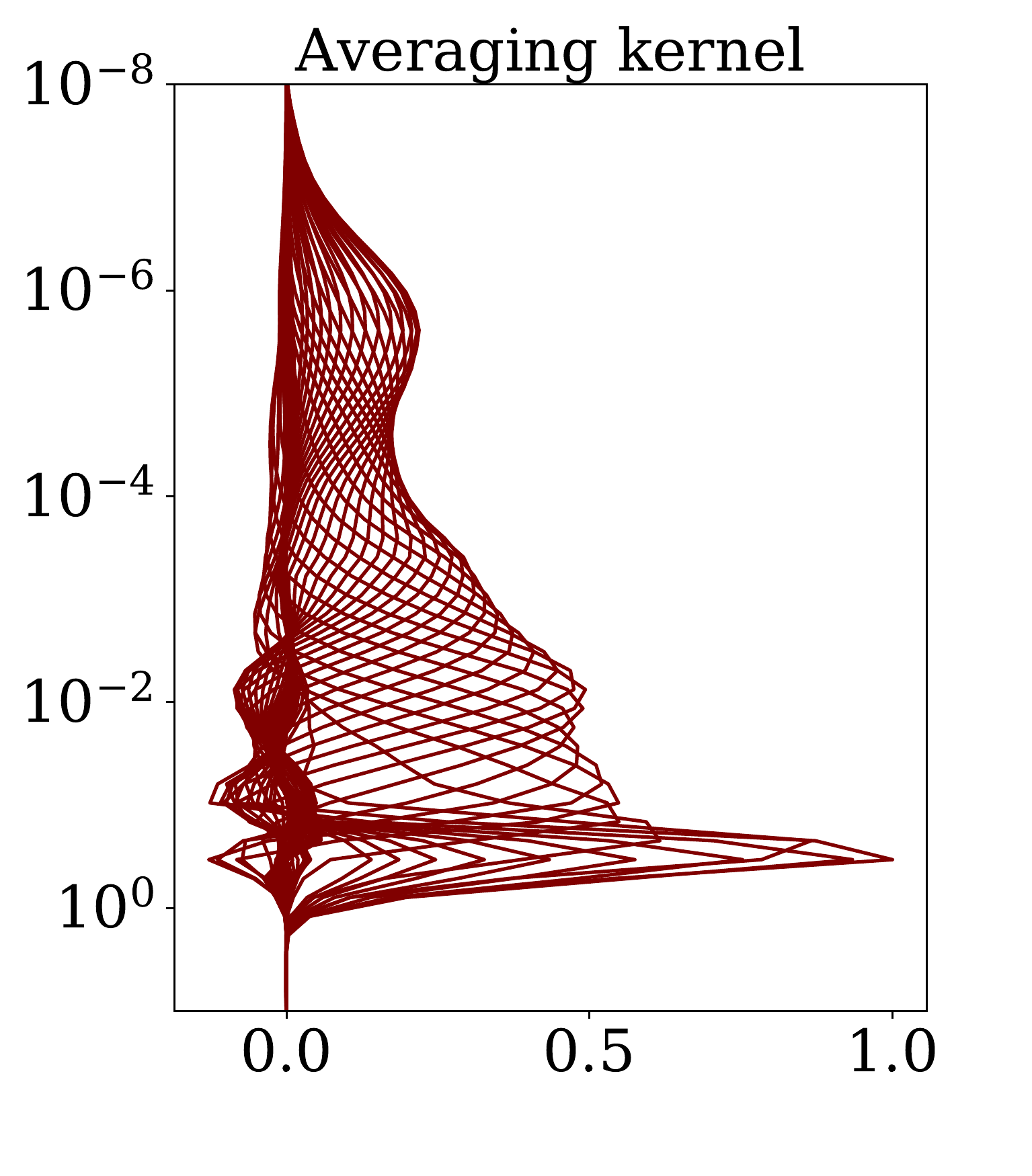}
\includegraphics[trim={15.2cm 0 20.0cm 0},clip,width=0.5\hsize]{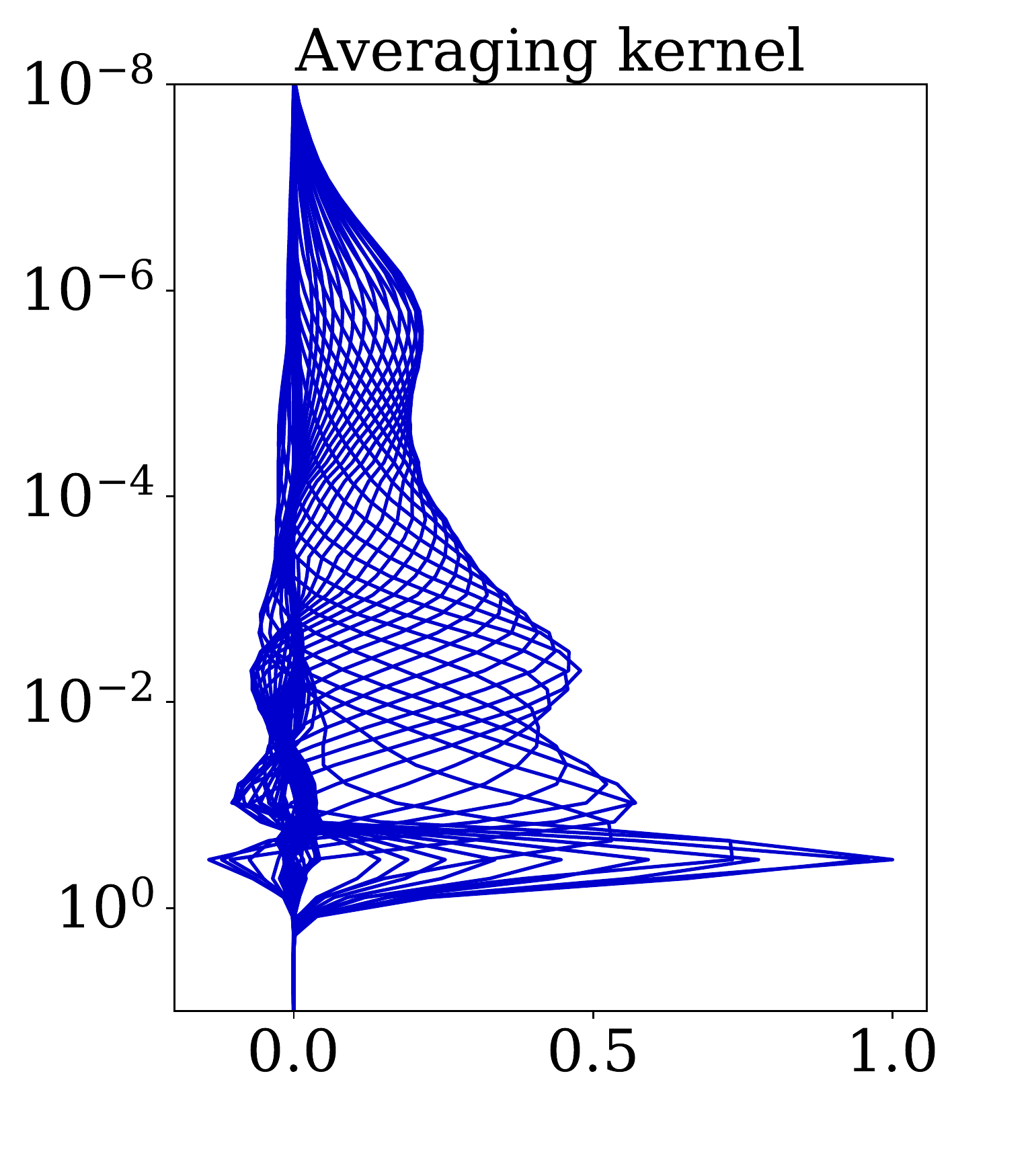}
}
\end{minipage}

\caption{\textbf{Left panel:} Retrieved temperature distribution assuming four different initial guesses: 
three homogeneous at 1000~K, 1500~K, and 2000~K (vertical dashed lines), and the one from \citet{2015ApJ...813...13W} (blue dashed line).
The retrieved best fit profiles are shown with solid lines along with corresponding error bars
as shaded areas. The profile from \cite{2012MNRAS.420..170L} is shown as a solid red line.
\textbf{Right panel:} Averaging kernels for the four retrieved temperature distributions (from left to right and from top to bottom, color-coded
accordingly).}
\label{fig:hd18-2}
\end{figure*}

To check the robustness of our results we performed four retrievals with different 
initial temperature profiles to investigate the sensitivity of our solutions to the initial guess.
The result are presented in Fig.~\ref{fig:hd18-2} where we show retrieved temperature profiles
assuming four different initial guesses.
In all cases we constrain very similar temperatures which confirms the robustness of our retrieval approach.
In the lower and upper atmosphere our solution remains unconstrained because the observations do not sense these altitudes,
as can be seen from the right panel of Fig.~\ref{fig:hd18-2} which shows corresponding averaging kernels for each of the
retrieved temperature distribution.

Finally, we note that our temperature distribution and that published in \citet{2012MNRAS.420..170L}
are noticeably different from the temperature found by \citet{2015ApJ...813...13W} (blue dashed line in Fig.~\ref{fig:hd18-2}). 
The largest deviation is found in the region of the temperature drop around $0.1$~bar. This is likely because of the different opacity tables used 
in the old version of  \taurex\, (I.~Waldmann, private communication).
We conclude that our retrieval method is accurate and robust, and that it can be used for atmospheric studies.

\section{Results}

Compared to its old version, the essential new capability of \criresplus\, is the increased wavelength coverage when
observing with a single setting. In one shot, the instrument is able to cover a region corresponding to about one spectral band (i.e., ten times more
than the old CRIRES). This gives us the possibility to study many spectral lines simultaneously with our retrieval
method and to increase the amount of information that we can learn about exo-atmospheres.
In what follows we study several cases of simulated observations at different spectral bands, their combinations,
spectral resolving powers, and S/N values. We chose to use five different spectral regions that  are free from strong
atmospheric telluric absorption and that  contain molecular bands of main molecules found in atmospheres of HJs: \hho, \co, \coo, and \chhhh.
These regions are: $1.50-1.70$~\mum, $2.10-2.28$~\mum, $2.28-2.38$~\mum, $3.80-4.00$~\mum, and $4.80-5.00$~\mum.
Each of these regions can be observed with single \criresplus\ setting; however, some gaps between spectral orders within each setting are expected.
For each of these regions we computed the spectra of HD~189733b assuming spectral resolutions of $R=100\,000$
and $R=50\,000$ (corresponding to slit widths of $40\arcsec$ and $10\arcsec$, respectively) and four S/N values
of $5$, $10$, $25$, and $50$, respectively. The S/N values are given per spectral dispersion element (1 pixel), which would be
obtained in real observations after integrating the spectrum profile along the spatial direction on the CCD. 
We note that the S/N of the planetary spectrum is obtained from the S/N of the stellar spectrum by taking into account
a characteristic planet-to-star flux ratio S/N(planet)=F$_{\rm p}$/F$_{\rm s}$$\cdot$S/N(star).
The values of S/N(star) can be directly computed using  exposure time calculators, for example, as we discuss  below.

In order to obtain accurate results, we performed retrievals with different initial guesses
and their uncertainties. We used three starting guesses for molecular volume mixing ratios (VMRs) of $\pm0.5$~dex and $-1.0$~dex from the true solution.
For each of the initial guess we tried  uncertainties of $0.5$~dex and $1.0$~dex, respectively.
By choosing such large uncertainties on the VMRs we avoid the problem of falling into local minima, 
as was discussed in the previous section.
We found   that we always converge to nearly the same solution     whether $0.5$~dex or $1.0$~dex error bars
were assumed. However, this was not always the case, especially for data of low S/N, and we discuss these cases below. 
For the temperature retrievals we tried different uncertainties between $100$~K and $500$~K
and found   that the value of $200$~K allows us to retrieve a temperature structure that is smooth and very close to the true one. 
Choosing too high temperature errors causes strong nonphysical fluctuations in the finally retrieved
temperature distribution and we do not discuss this solution  further. 
%In the retrievals that we present below we start from the homogenous temperature distribution
%close to the expected equilibrium temperature of the planet and mixing ratio uncertainties of $1.0$~dex.

\subsection{Single spectral region retrievals}

%\begin{landscape}
\begin{figure*}
\centerline{
\includegraphics[height=0.11\vsize]{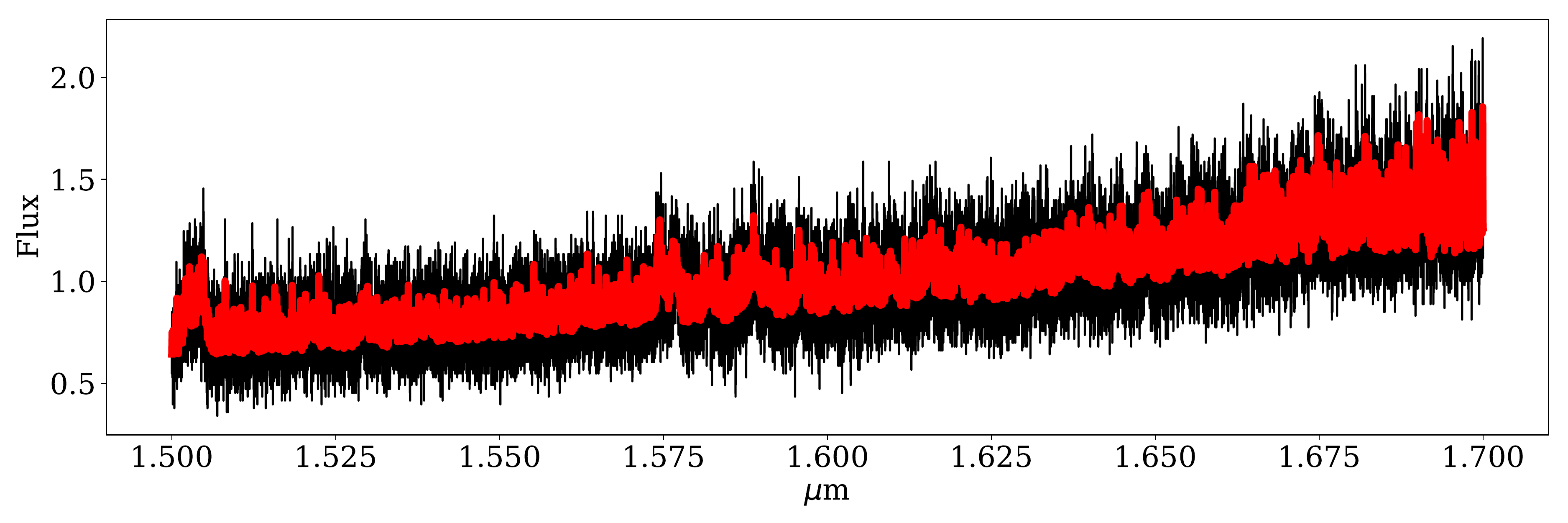}
\hspace{0.02\hsize}
\includegraphics[trim={0 0 20.0cm 0},clip,height=0.11\vsize]{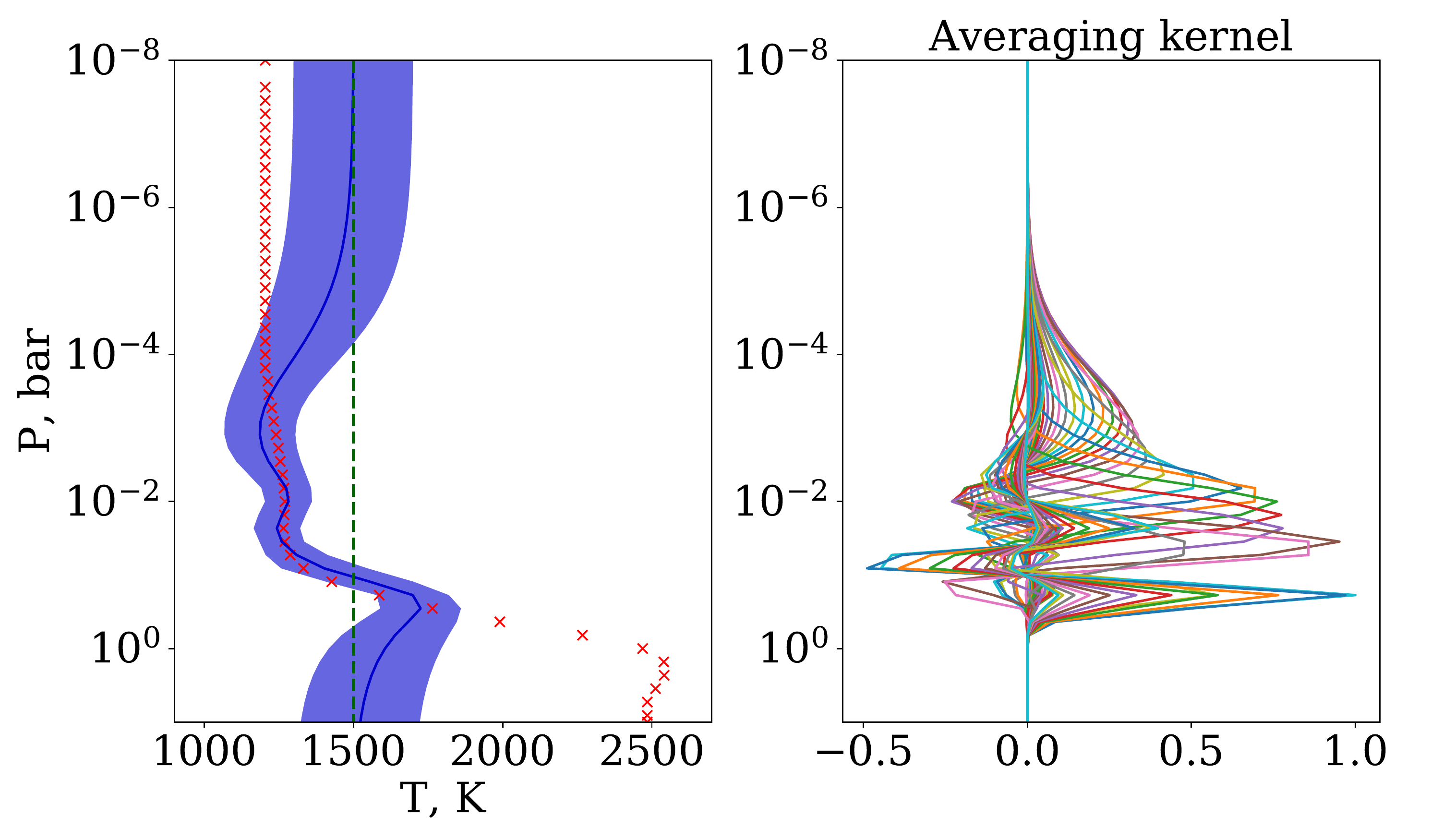}
\hspace{0.02\hsize}
\includegraphics[height=0.11\vsize]{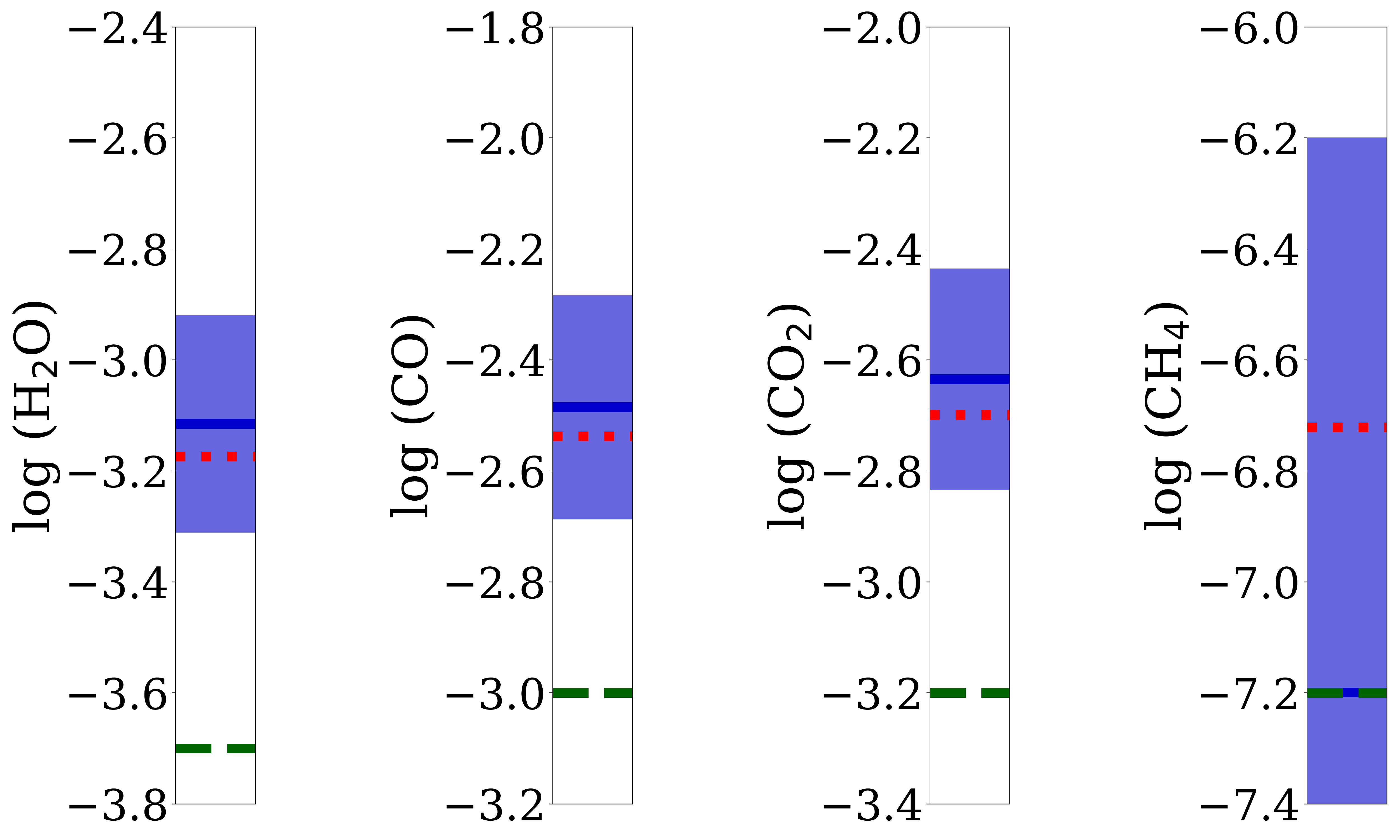}
}
\centerline{
\includegraphics[height=0.11\vsize]{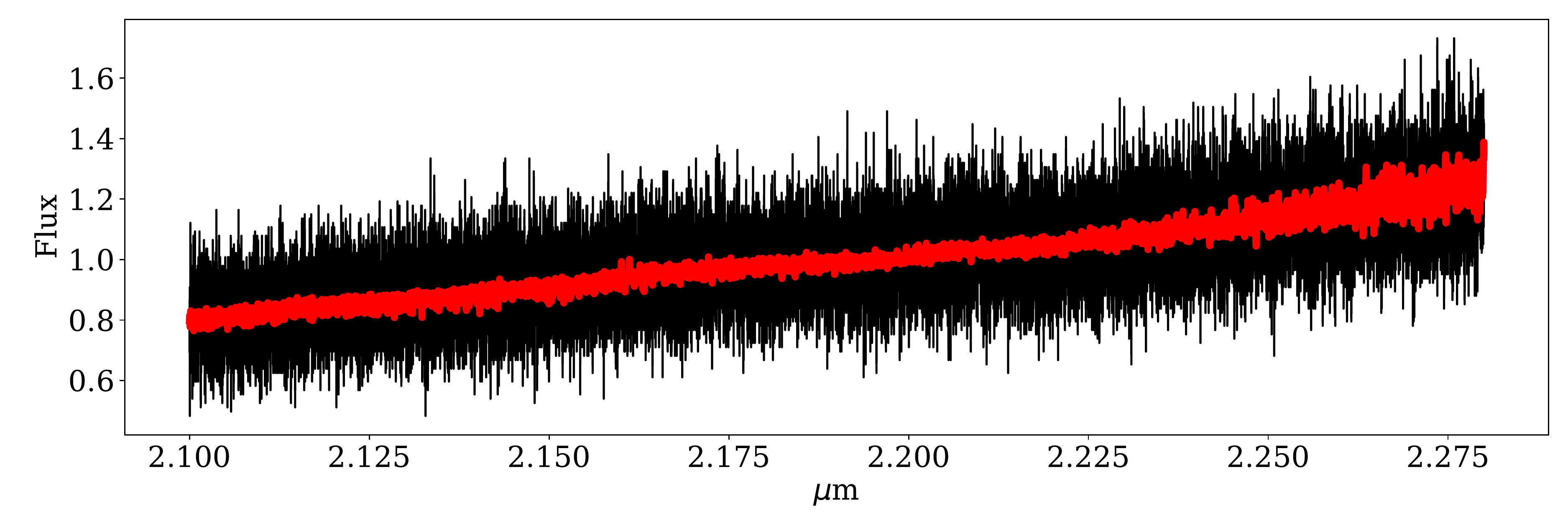}
\hspace{0.02\hsize}
\includegraphics[trim={0 0 20.0cm 0},clip,height=0.11\vsize]{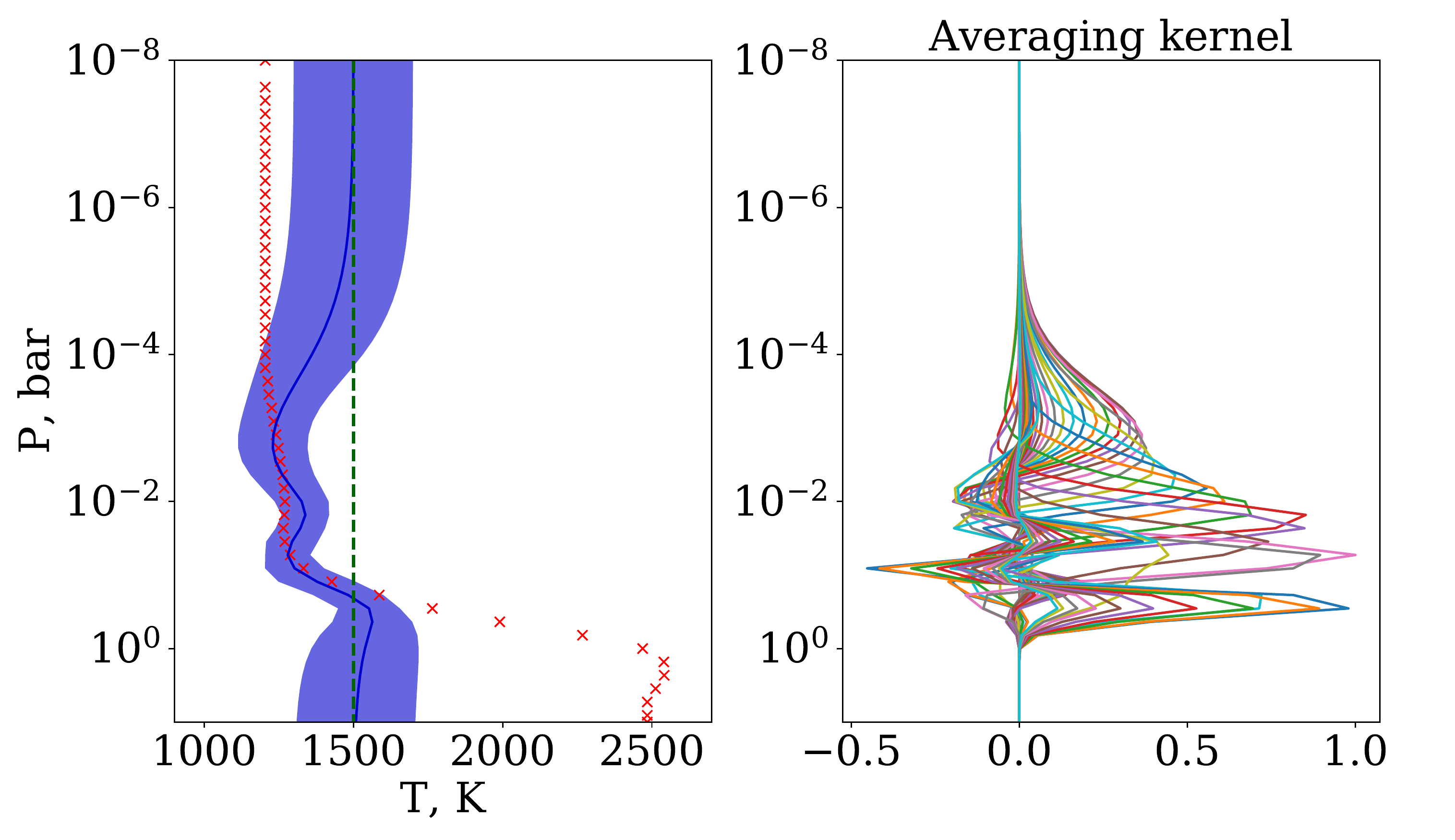}
\hspace{0.02\hsize}
\includegraphics[height=0.11\vsize]{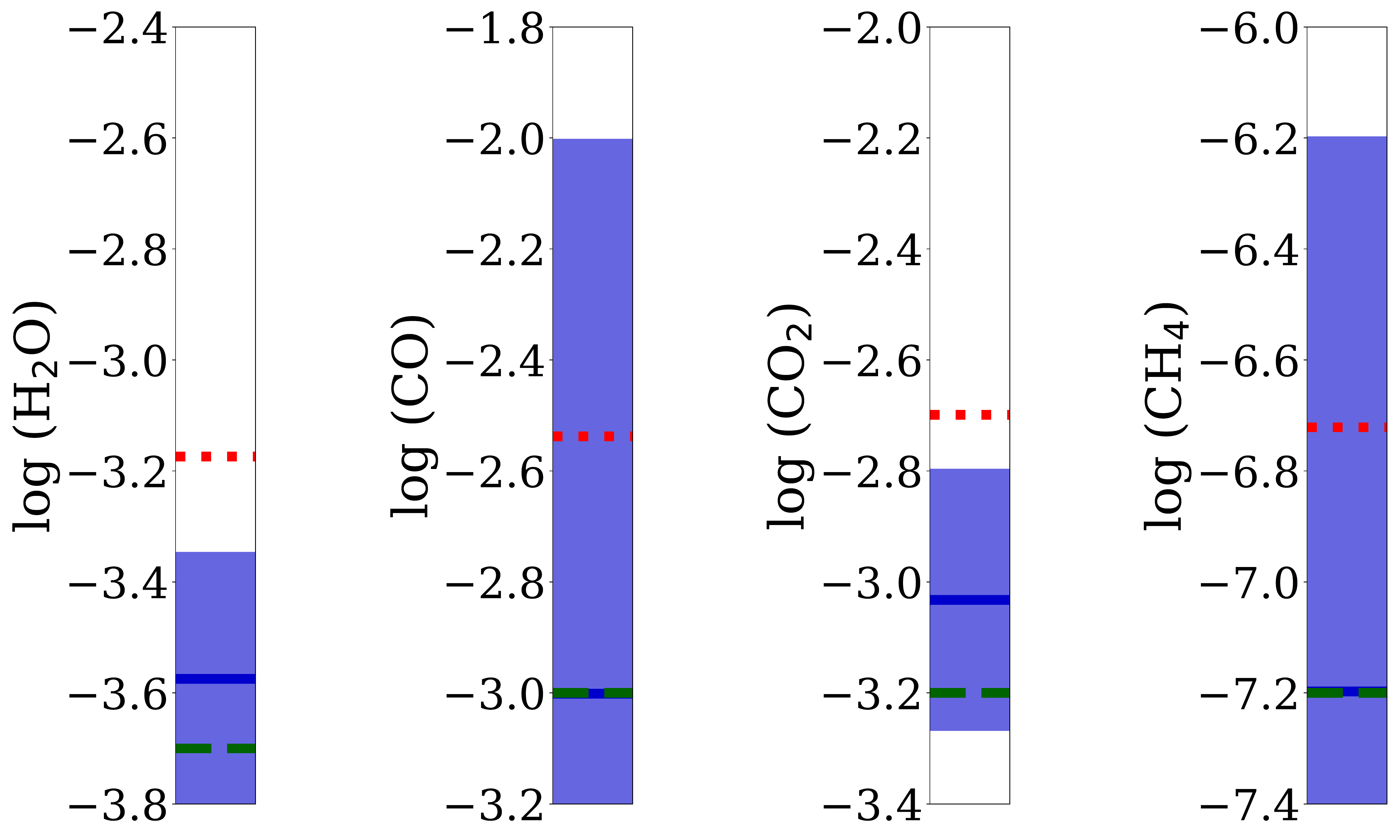}
}
\centerline{
\includegraphics[height=0.11\vsize]{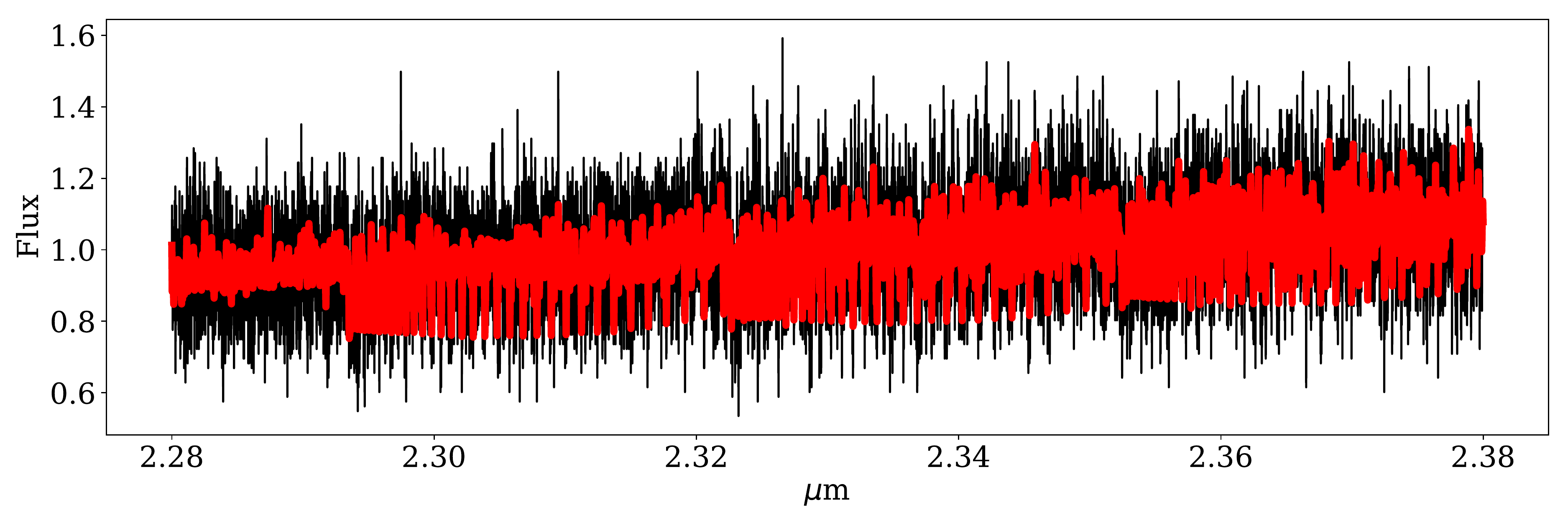}
\hspace{0.02\hsize}
\includegraphics[trim={0 0 20.0cm 0},clip,height=0.11\vsize]{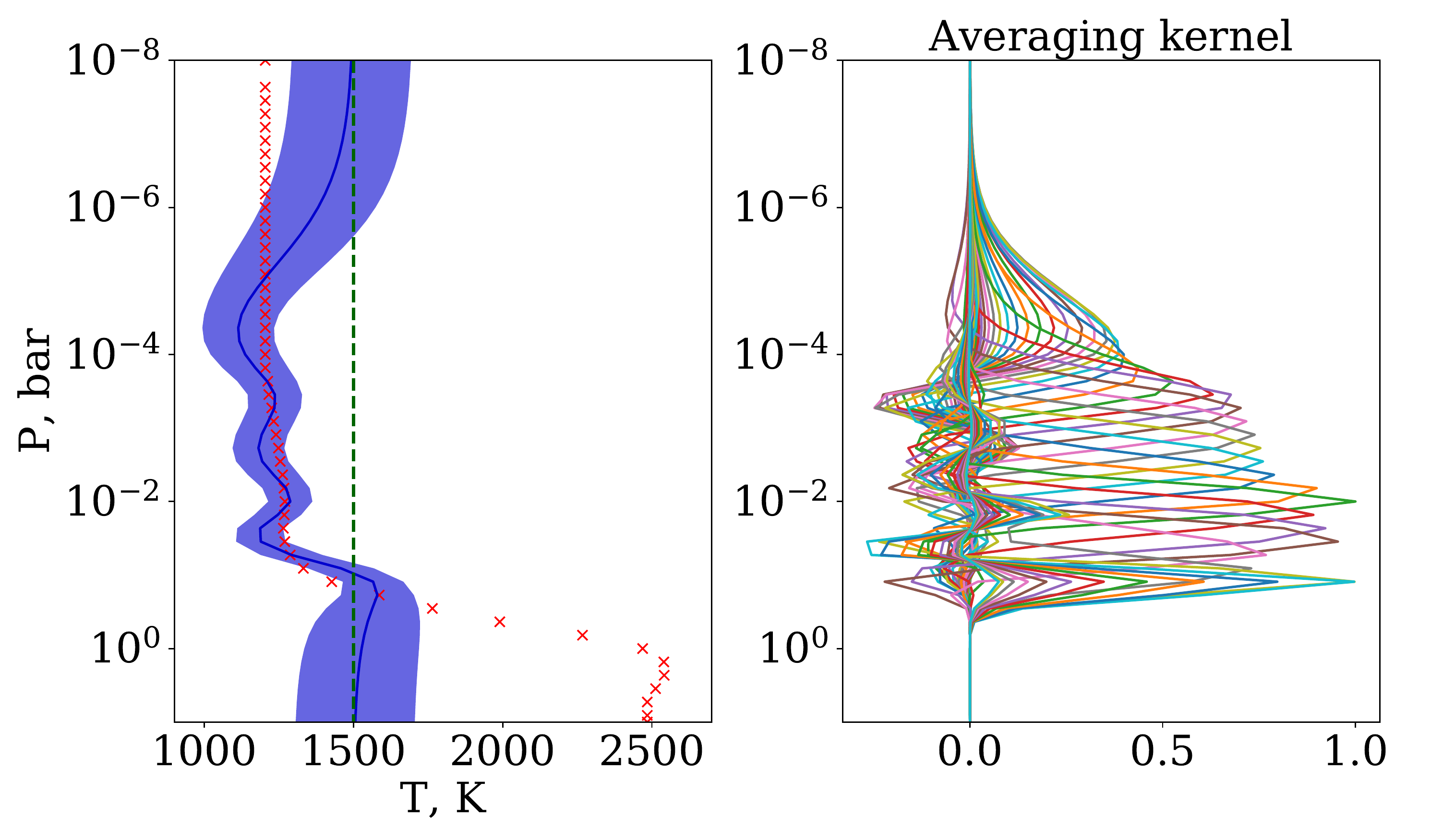}
\hspace{0.02\hsize}
\includegraphics[height=0.11\vsize]{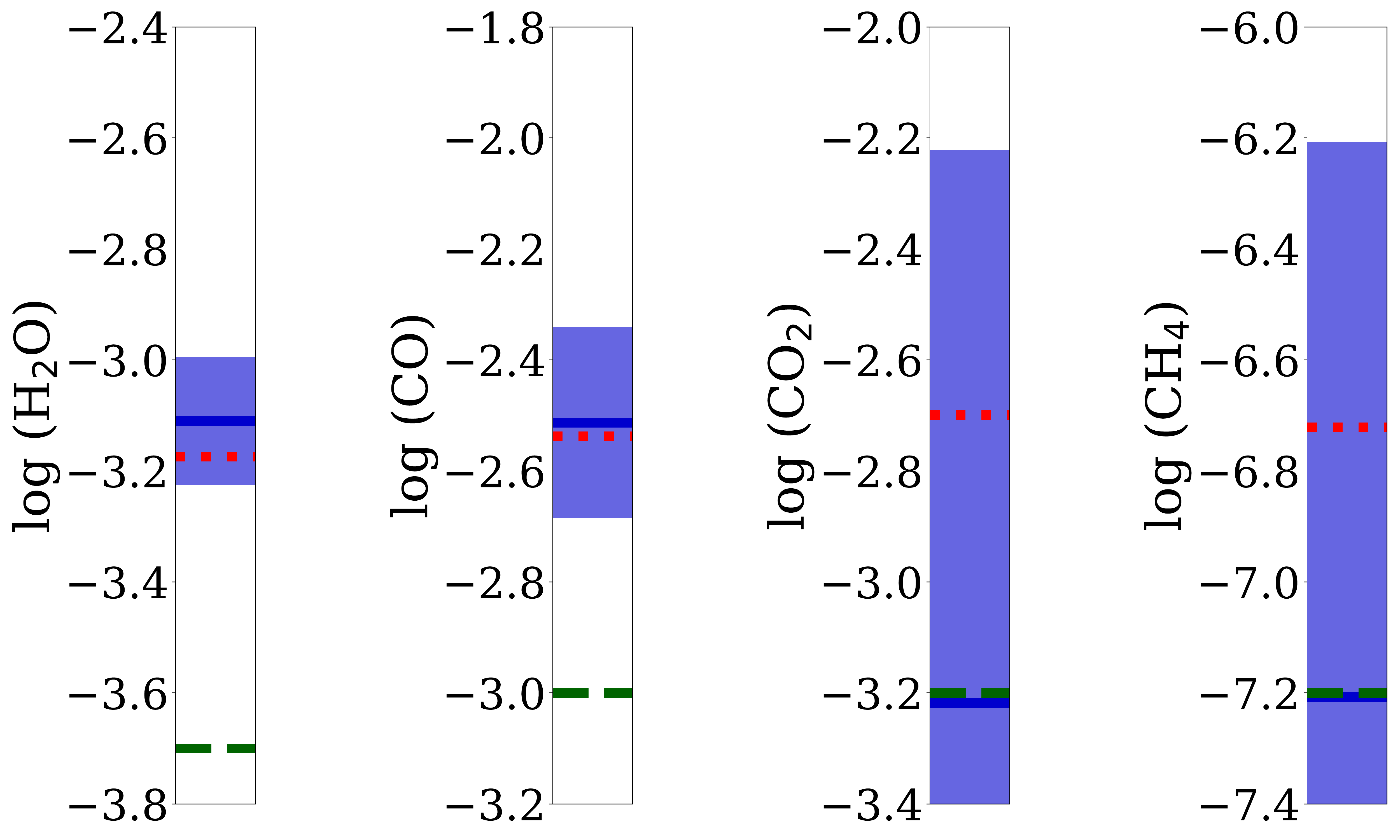}
}
\centerline{
\includegraphics[height=0.11\vsize]{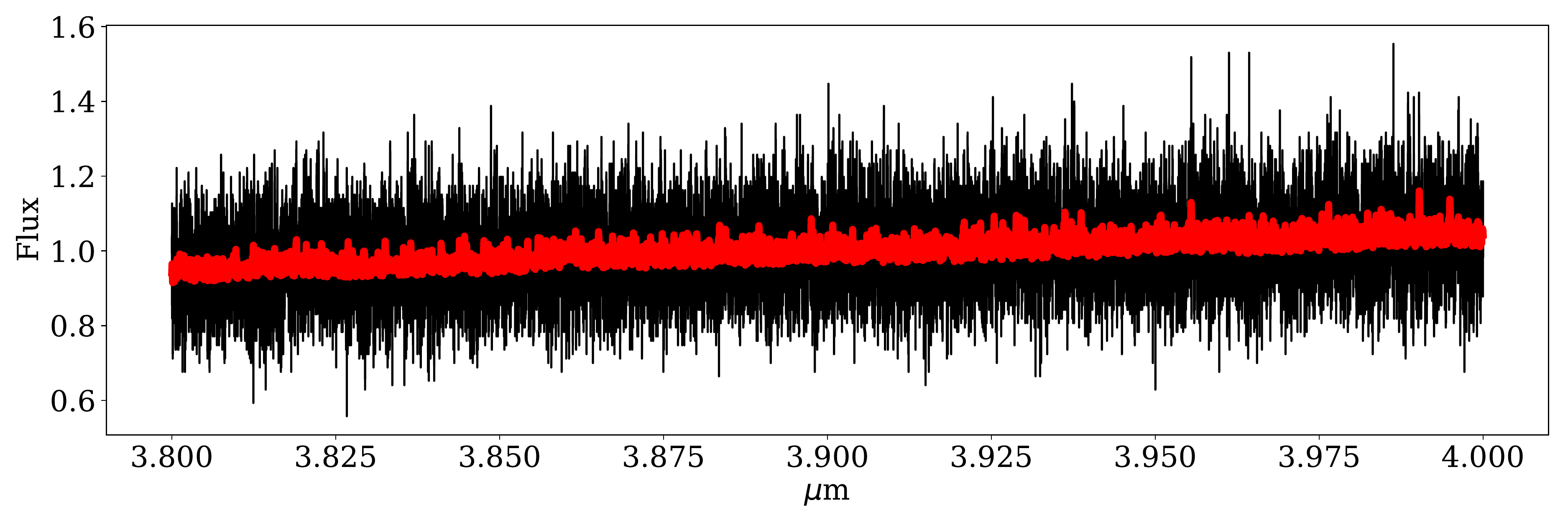}
\hspace{0.02\hsize}
\includegraphics[trim={0 0 20.0cm 0},clip,height=0.11\vsize]{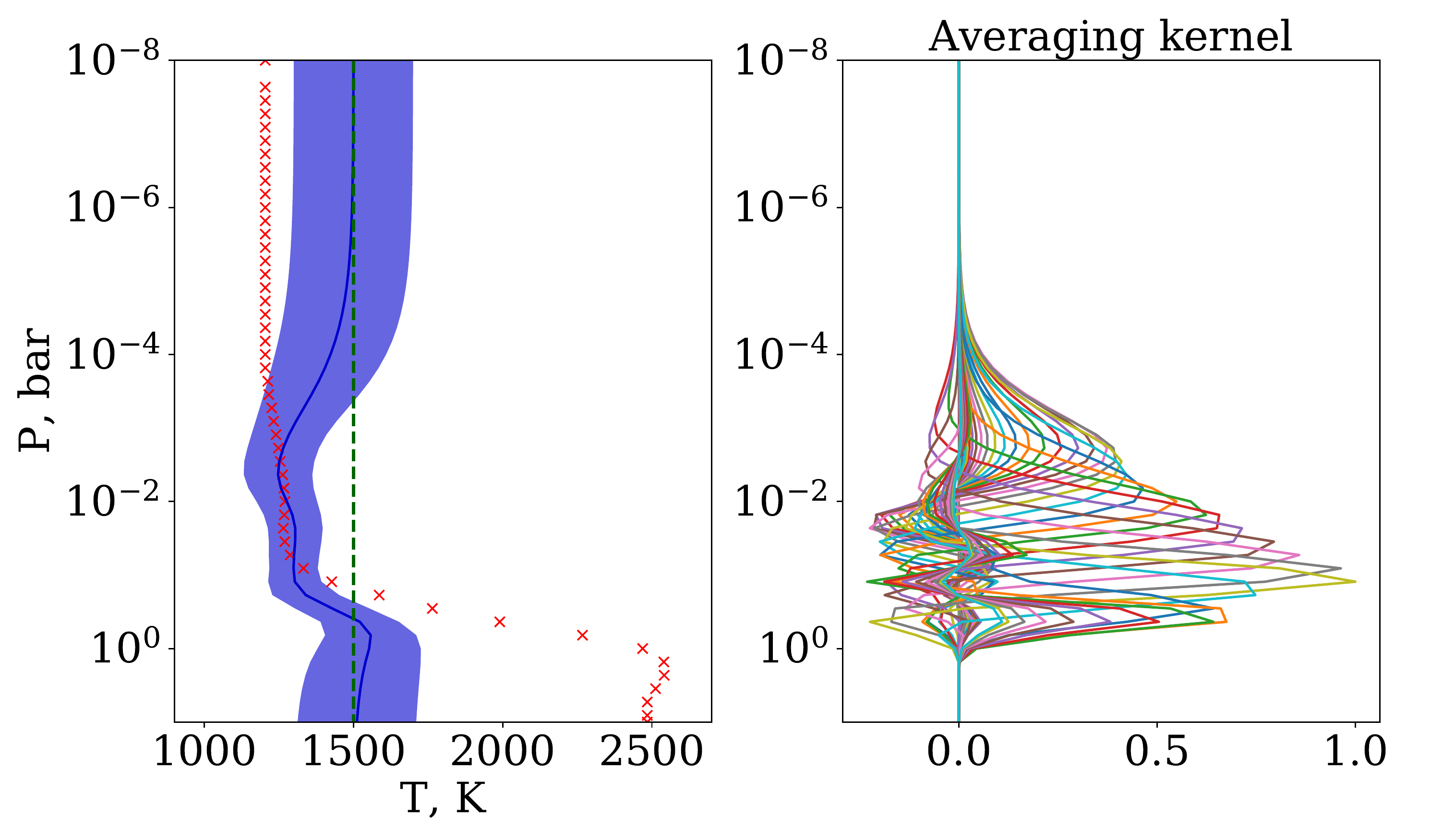}
\hspace{0.02\hsize}
\includegraphics[height=0.11\vsize]{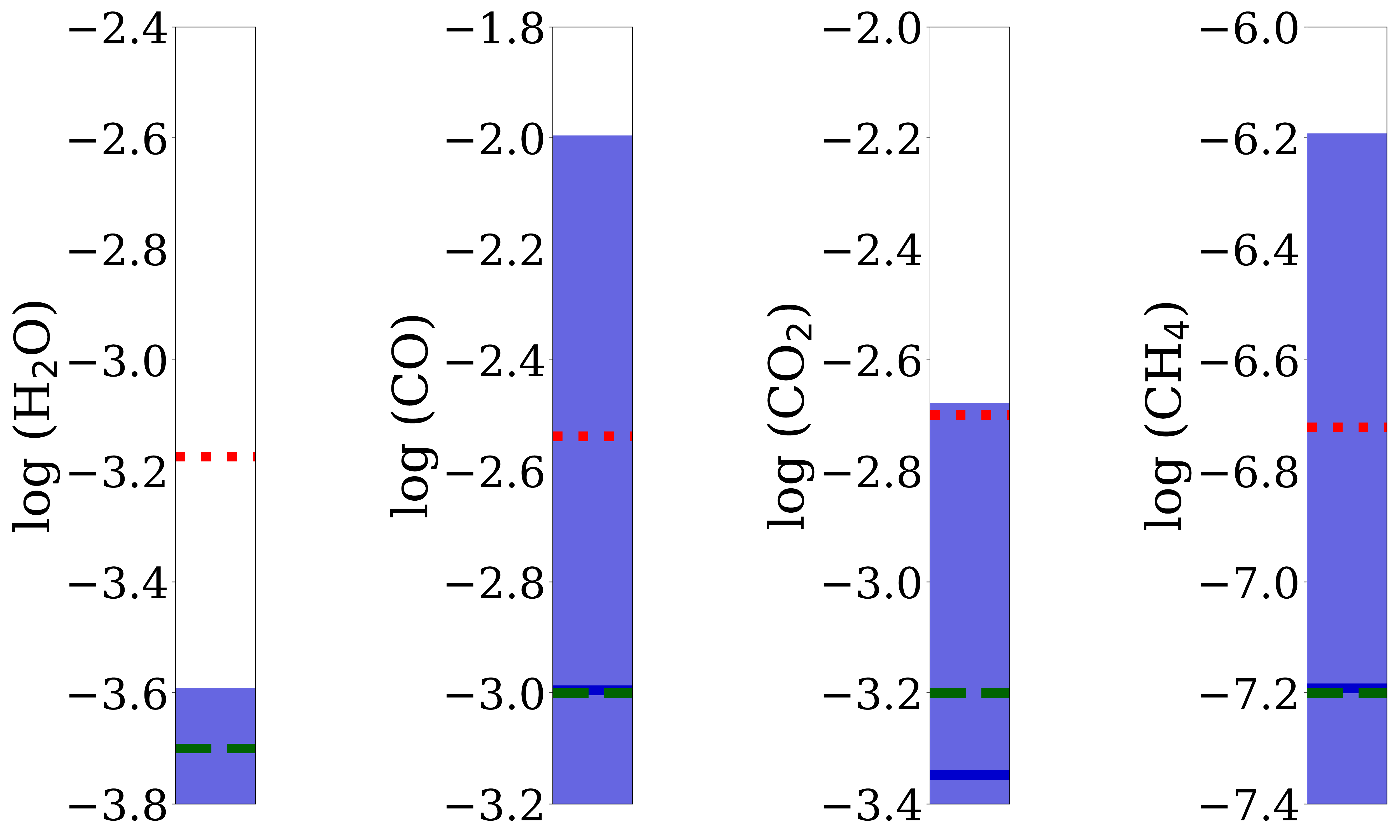}
}
\centerline{
\includegraphics[height=0.11\vsize]{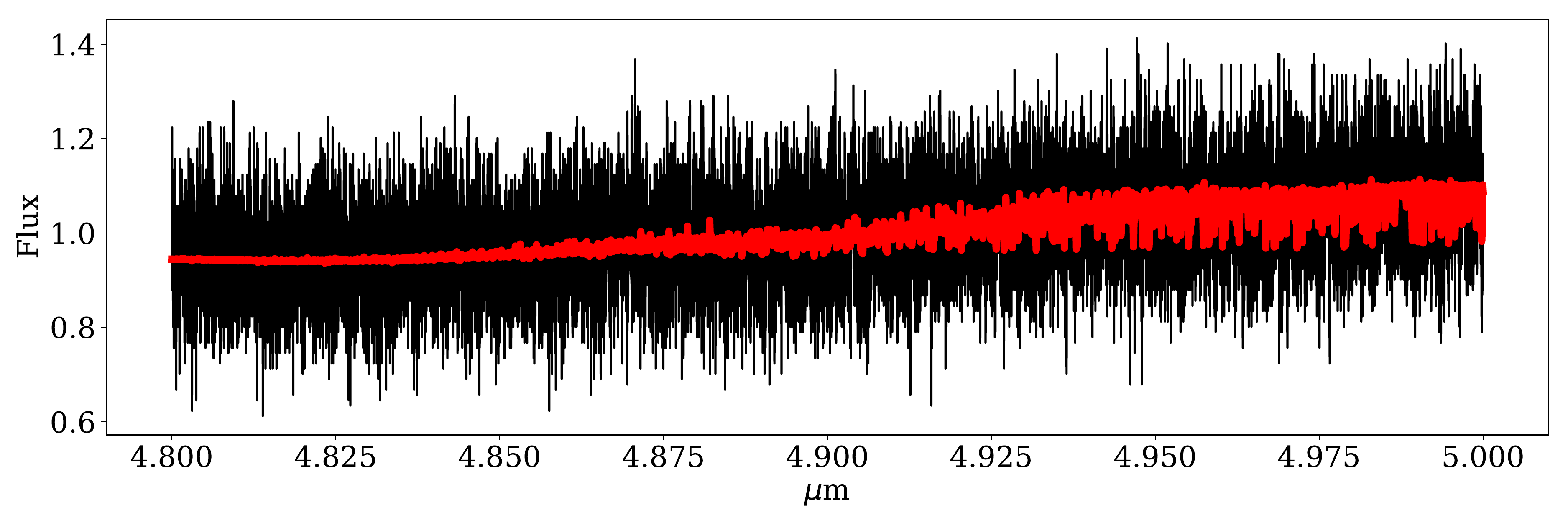}
\hspace{0.02\hsize}
\includegraphics[trim={0 0 20.0cm 0},clip,height=0.11\vsize]{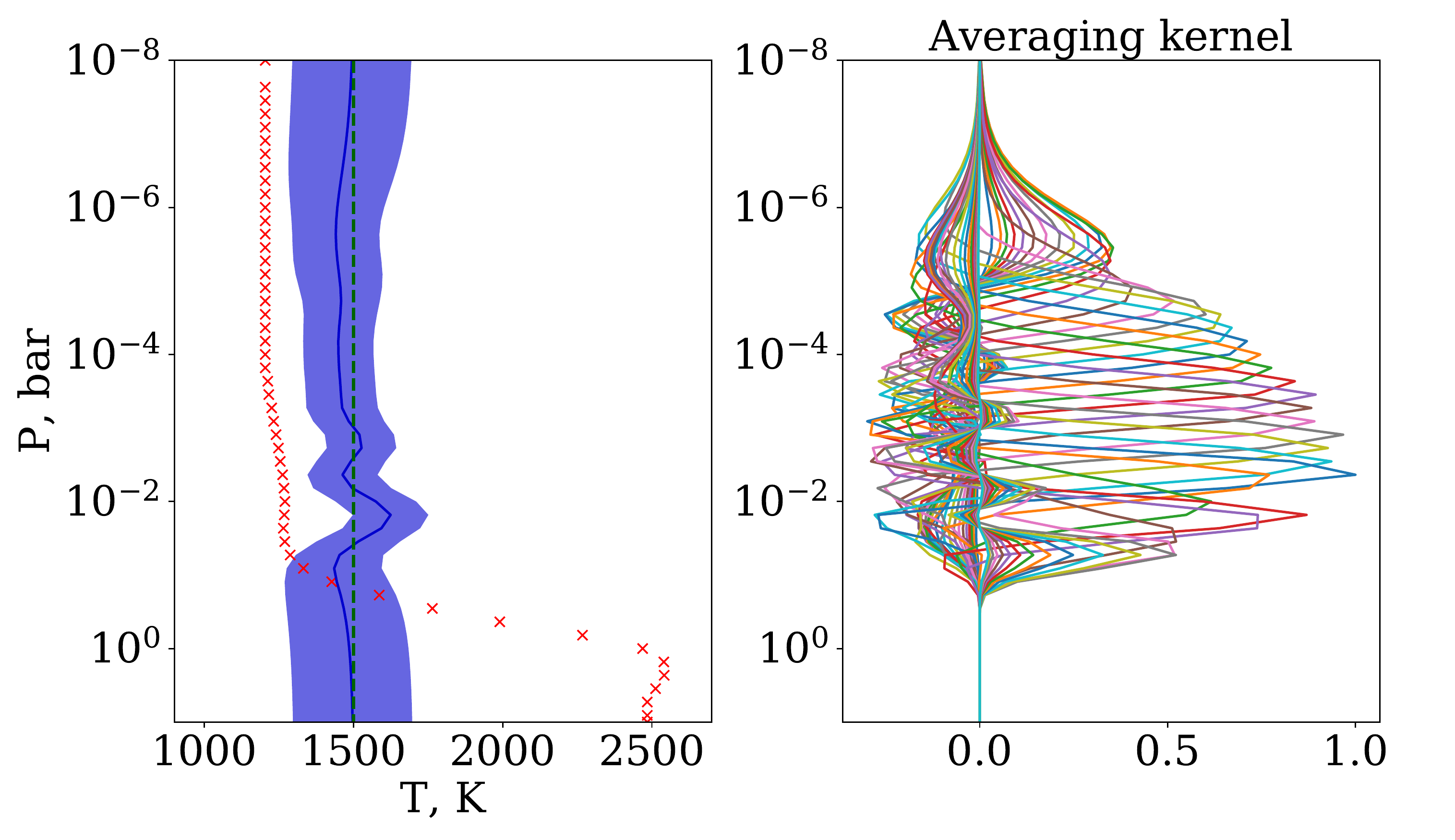}
\hspace{0.02\hsize}
\includegraphics[height=0.11\vsize]{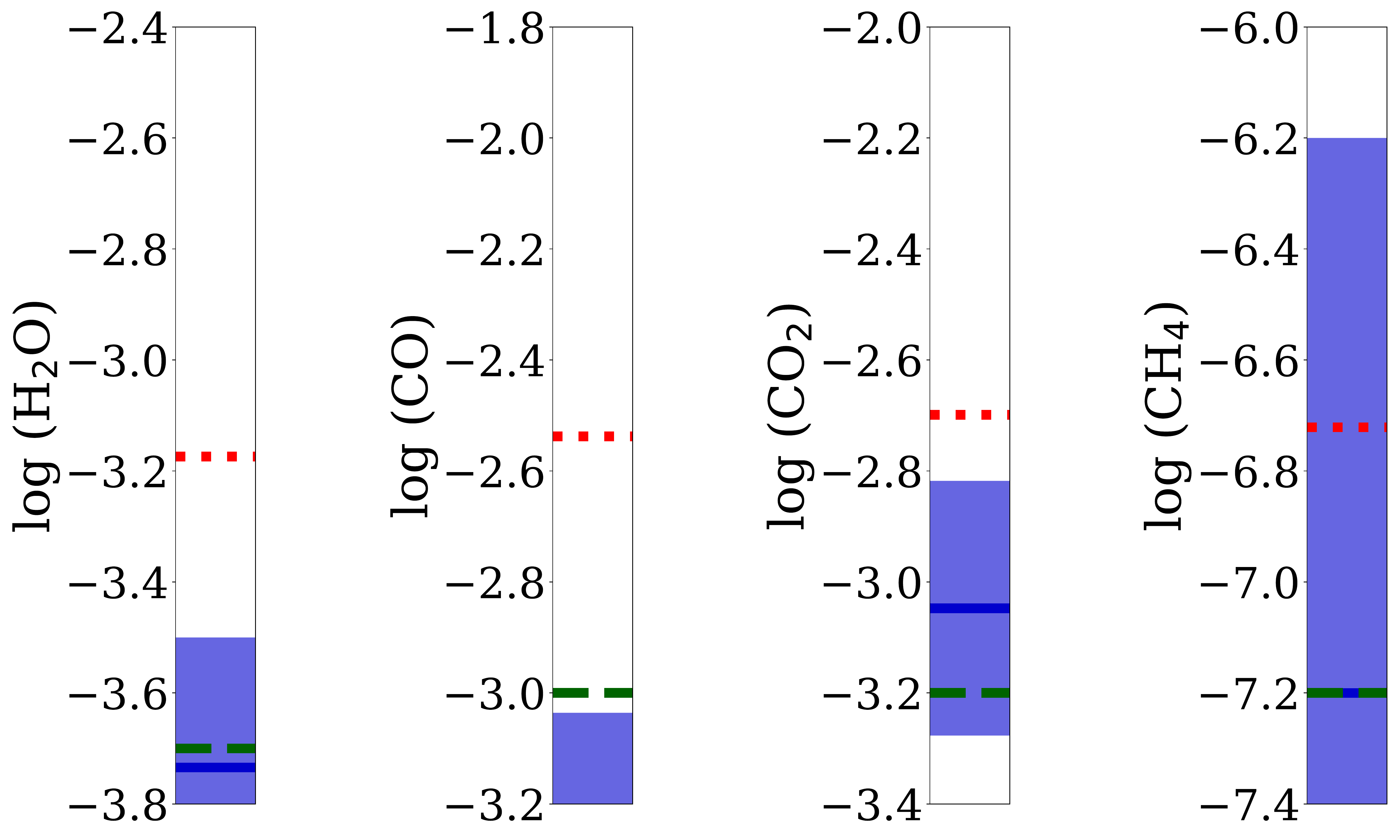}
}
\caption{\label{fig:snr10-r100k}
Retrieved temperature and mixing ratios from five spectral regions with R=$100\,000$ and S/N=$10$. In each panel, 
the first plot compares the best fit predicted spectrum (solid red  line) with the simulated observations (black).
The second plot is the temperature distribution as a function of atmospheric pressure (solid blue line) and with error bars shown as shaded areas.
The red crosses and green dashed line are the true solution and initial guess, respectively. The third plot shows
averaging kernels derived for the temperature distribution. Here averaging kernels for different atmospheric depths are randomly
color-coded for clarity? The next four plots are the values of the retrieved mixing ratios of four
molecular species (color-coded   as in the first plot). The mixing ratios were assumed to be constant with 
atmospheric depth, and we show their values on the $y$-axis.
}
\end{figure*}
%\end{landscape}

\begin{figure*}
\hspace{0.1\hsize} \textbf{S/N=5} \hspace{0.18\hsize} \textbf{S/N=10} \hspace{0.18\hsize} \textbf{S/N=25} \hspace{0.18\hsize} \textbf{S/N=50}
\par\vspace{0.005\vsize}
\centerline{
\includegraphics[trim={0 0 35.5cm 0},clip,height=0.13\hsize]{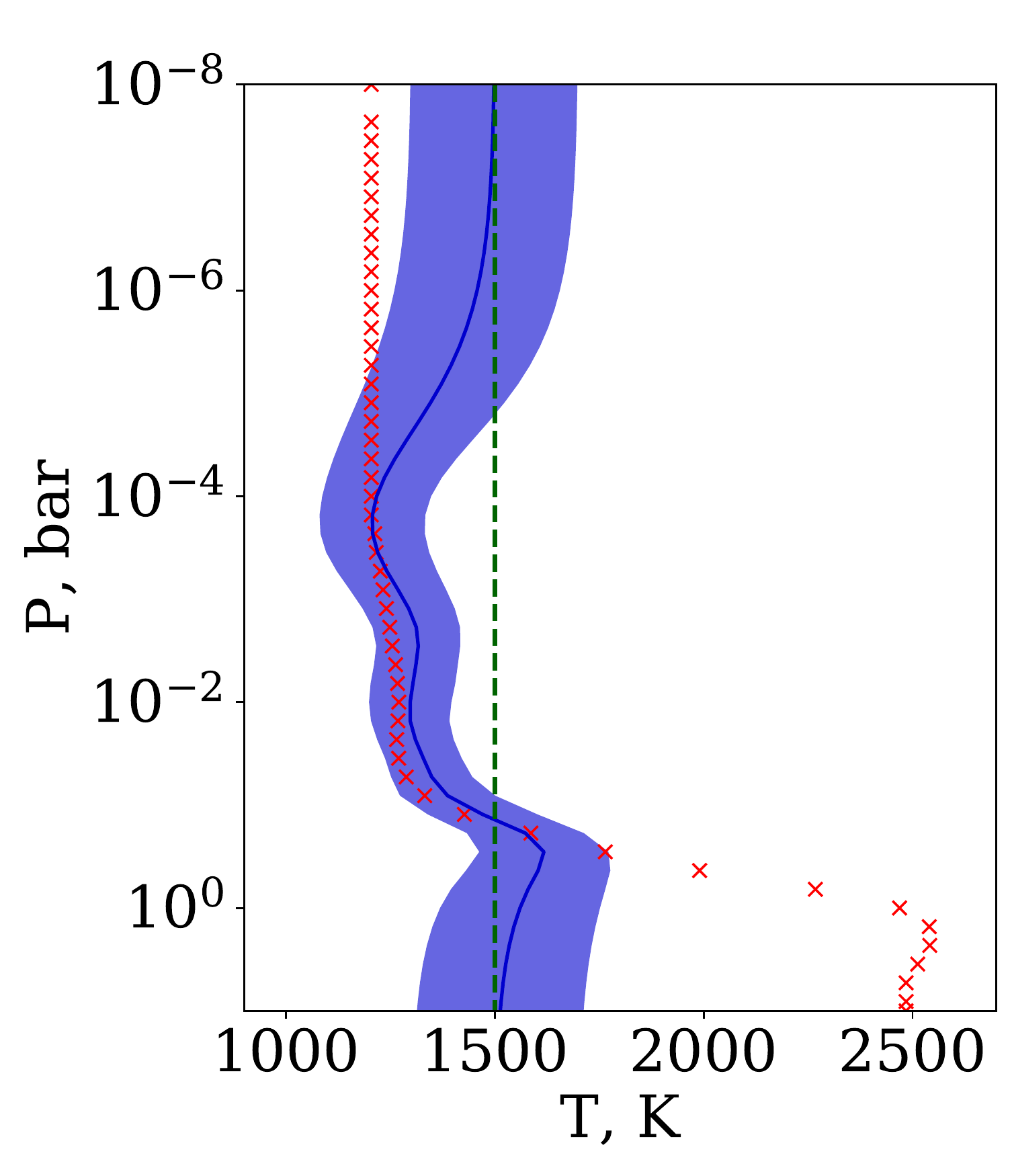}
\includegraphics[trim={0 0 25.0cm 0},clip,height=0.13\hsize]{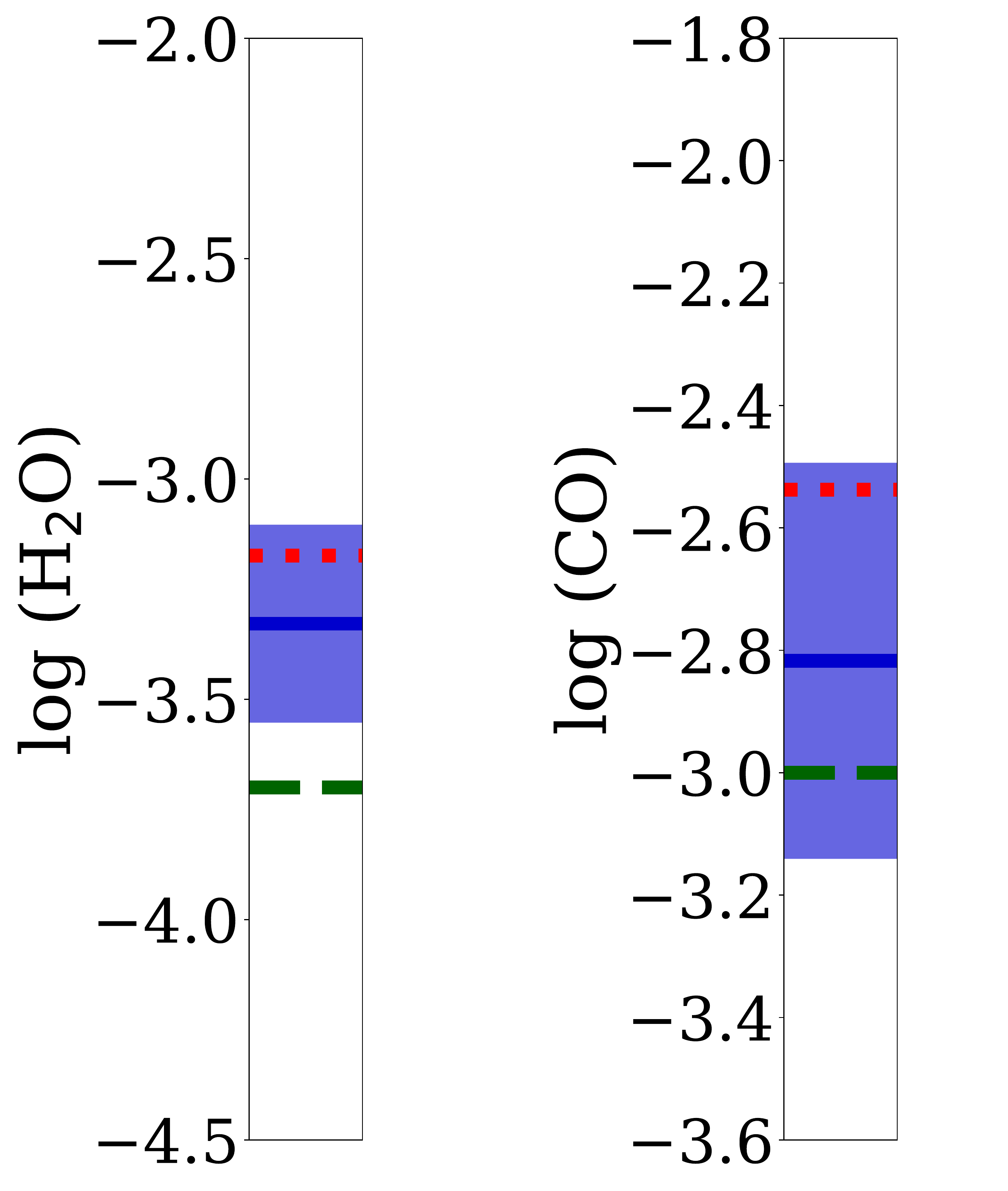}
\hspace{0.02\hsize}
\includegraphics[trim={0 0 35.5cm 0},clip,height=0.13\hsize]{figures/r100k/snr10/h2o+co+ch4+co2/ll-2.28-2.38/out-vmr_init_m0.5dex-vmr_err_1.0dex.bwd_param1.pdf}
\includegraphics[trim={0 0 25.0cm 0},clip,height=0.13\hsize]{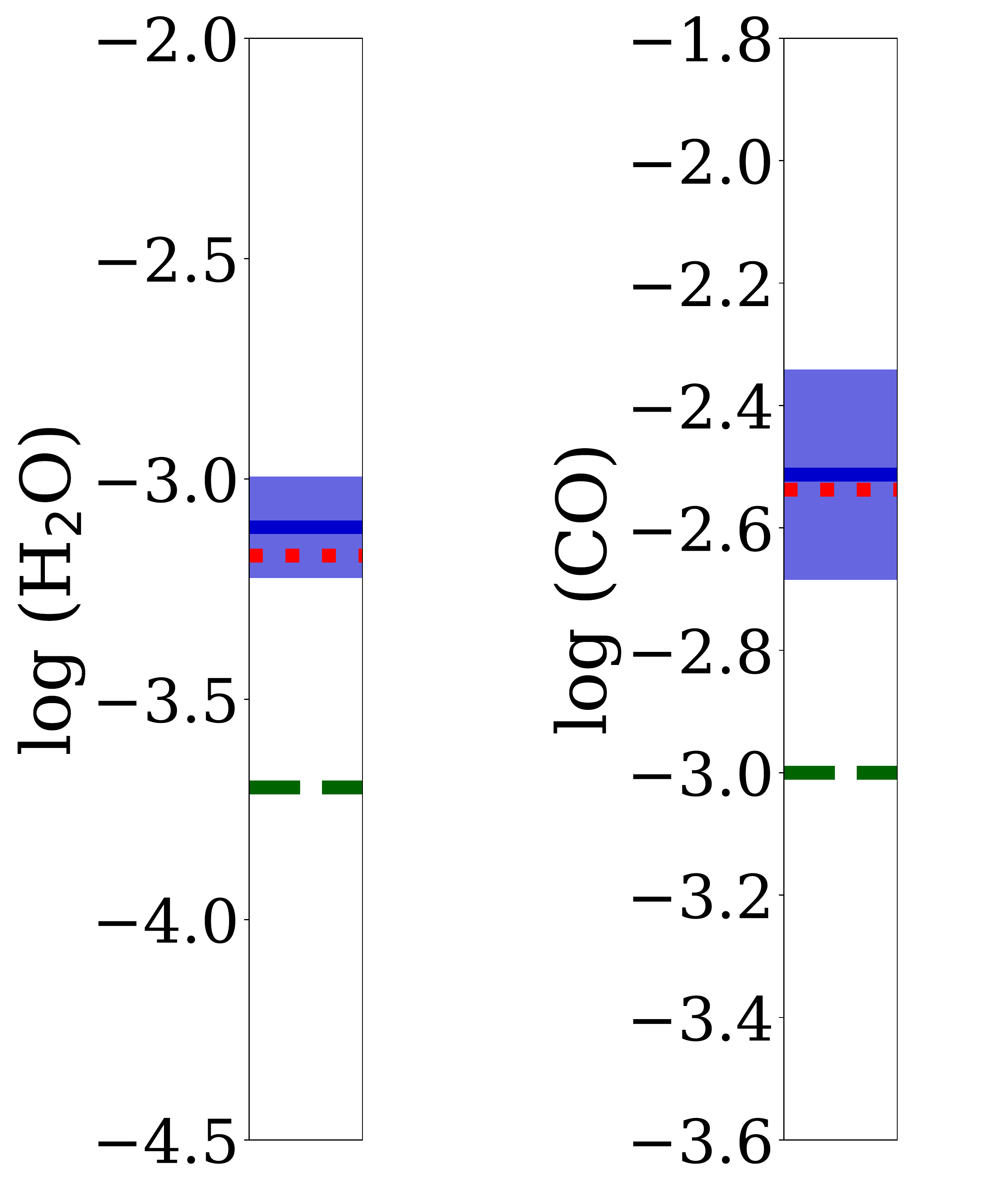}
\hspace{0.02\hsize}
\includegraphics[trim={0 0 35.5cm 0},clip,height=0.13\hsize]{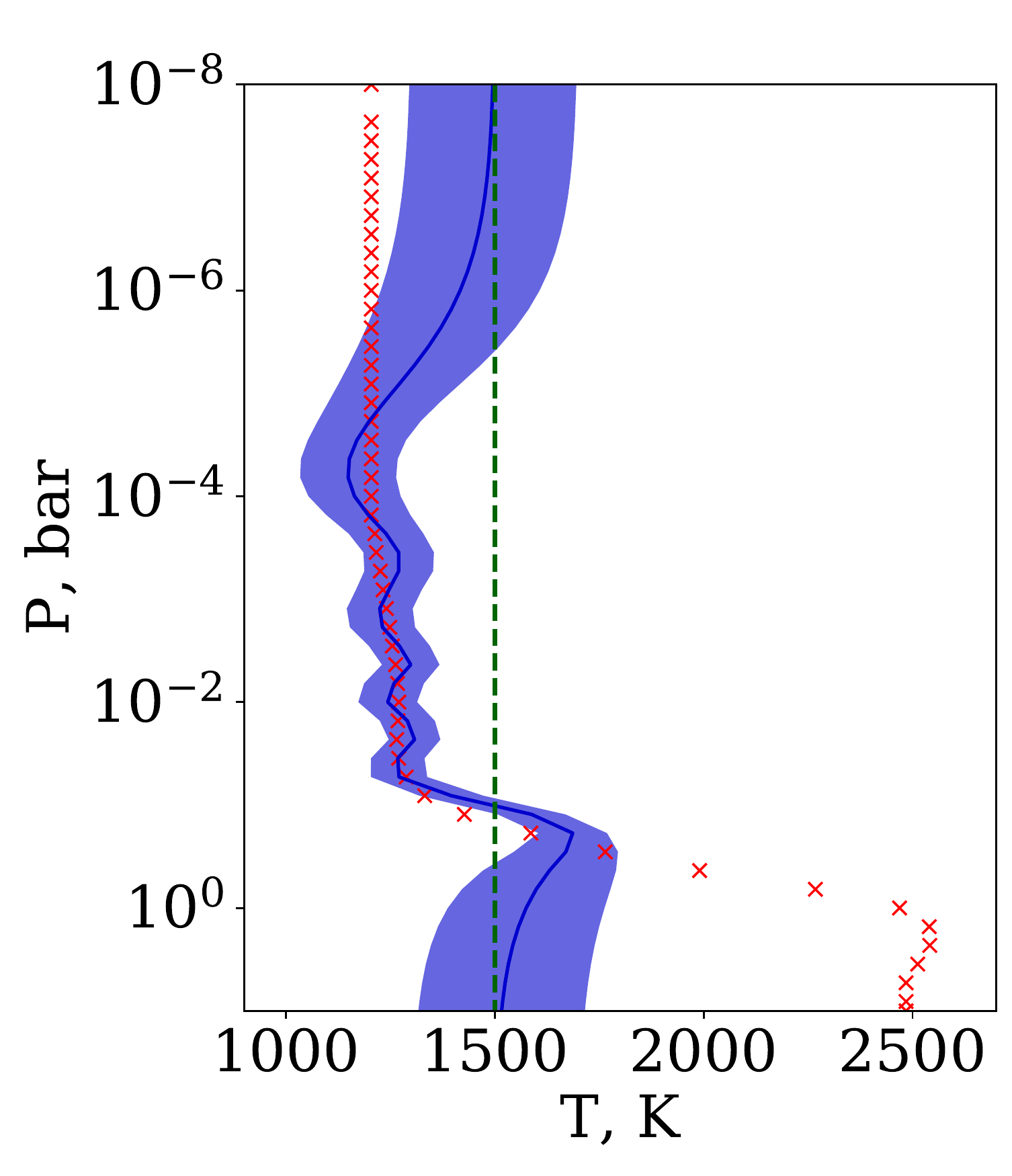}
\includegraphics[trim={0 0 25.0cm 0},clip,height=0.13\hsize]{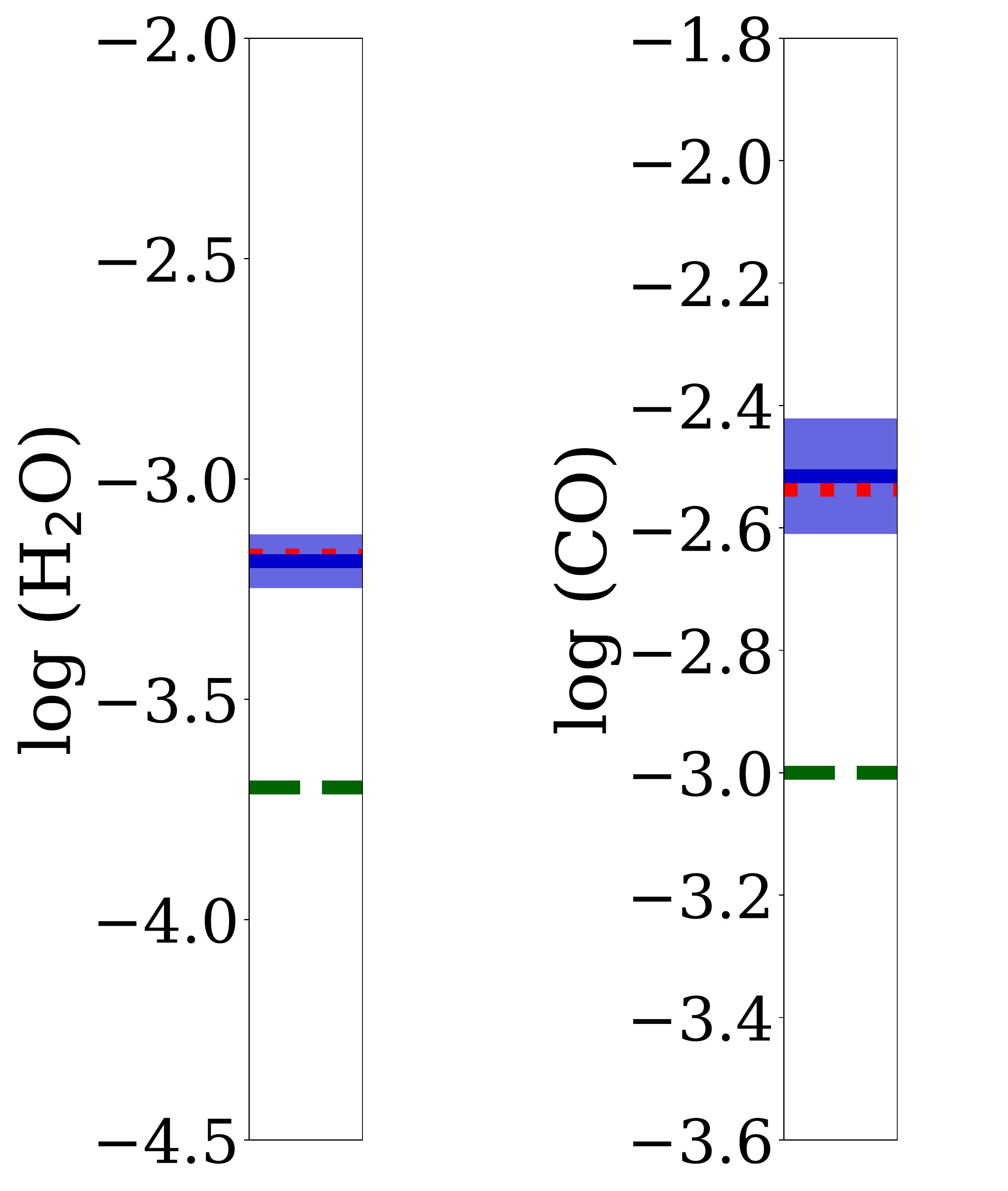}
\hspace{0.02\hsize}
\includegraphics[trim={0 0 35.5cm 0},clip,height=0.13\hsize]{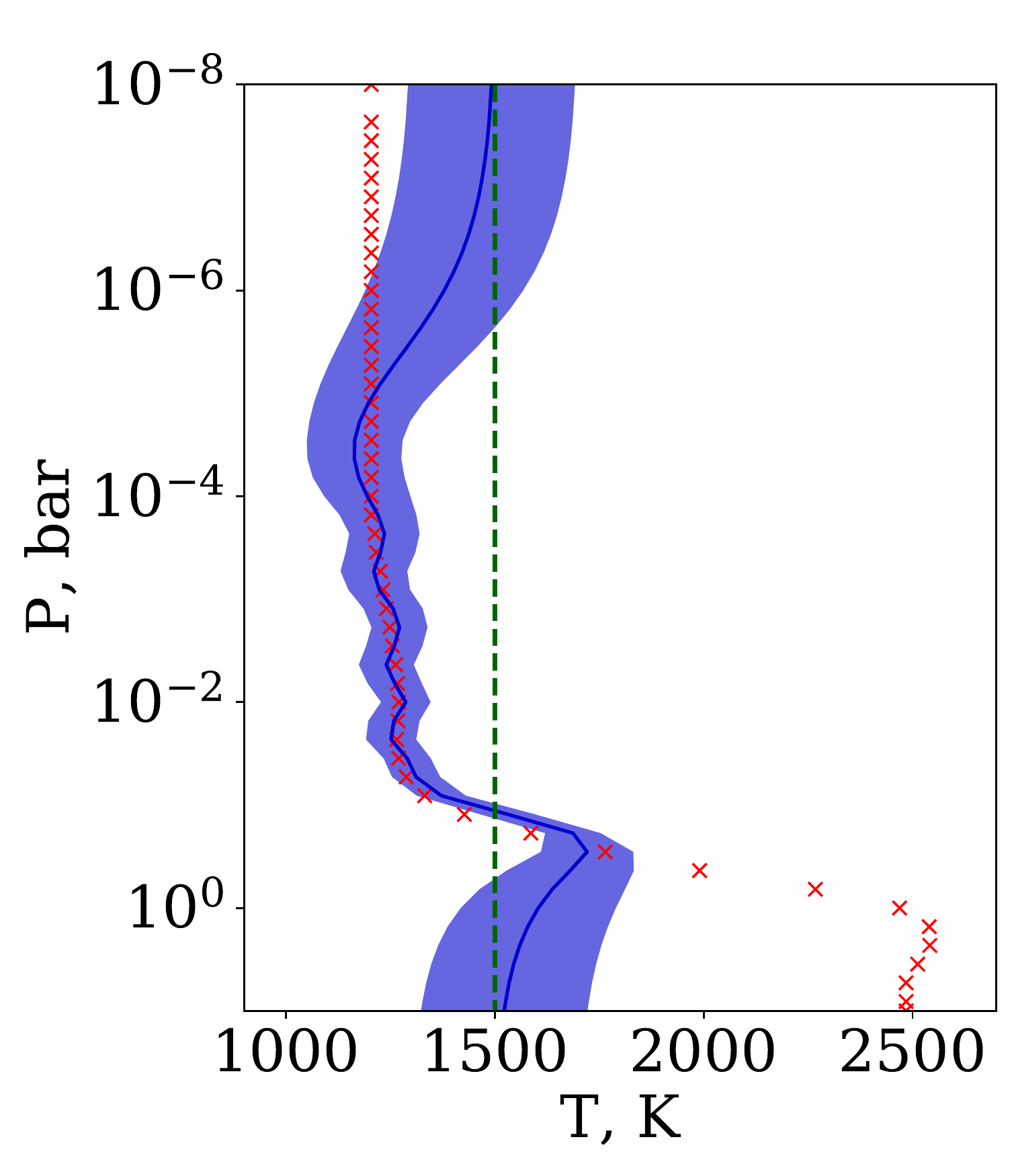}
\includegraphics[trim={0 0 25.0cm 0},clip,height=0.13\hsize]{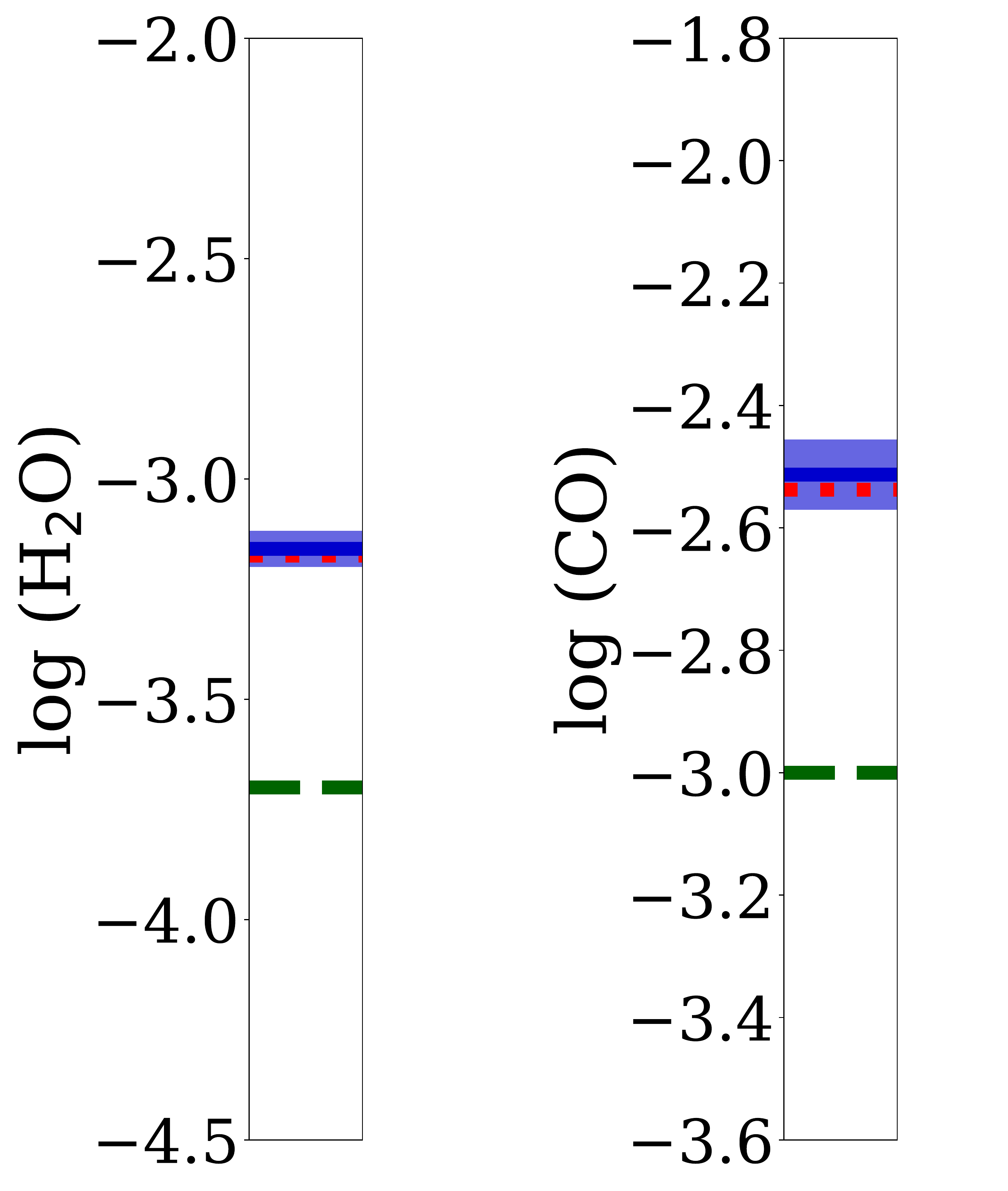}
}
\centerline{
\includegraphics[trim={0 0 35.5cm 0},clip,height=0.13\hsize]{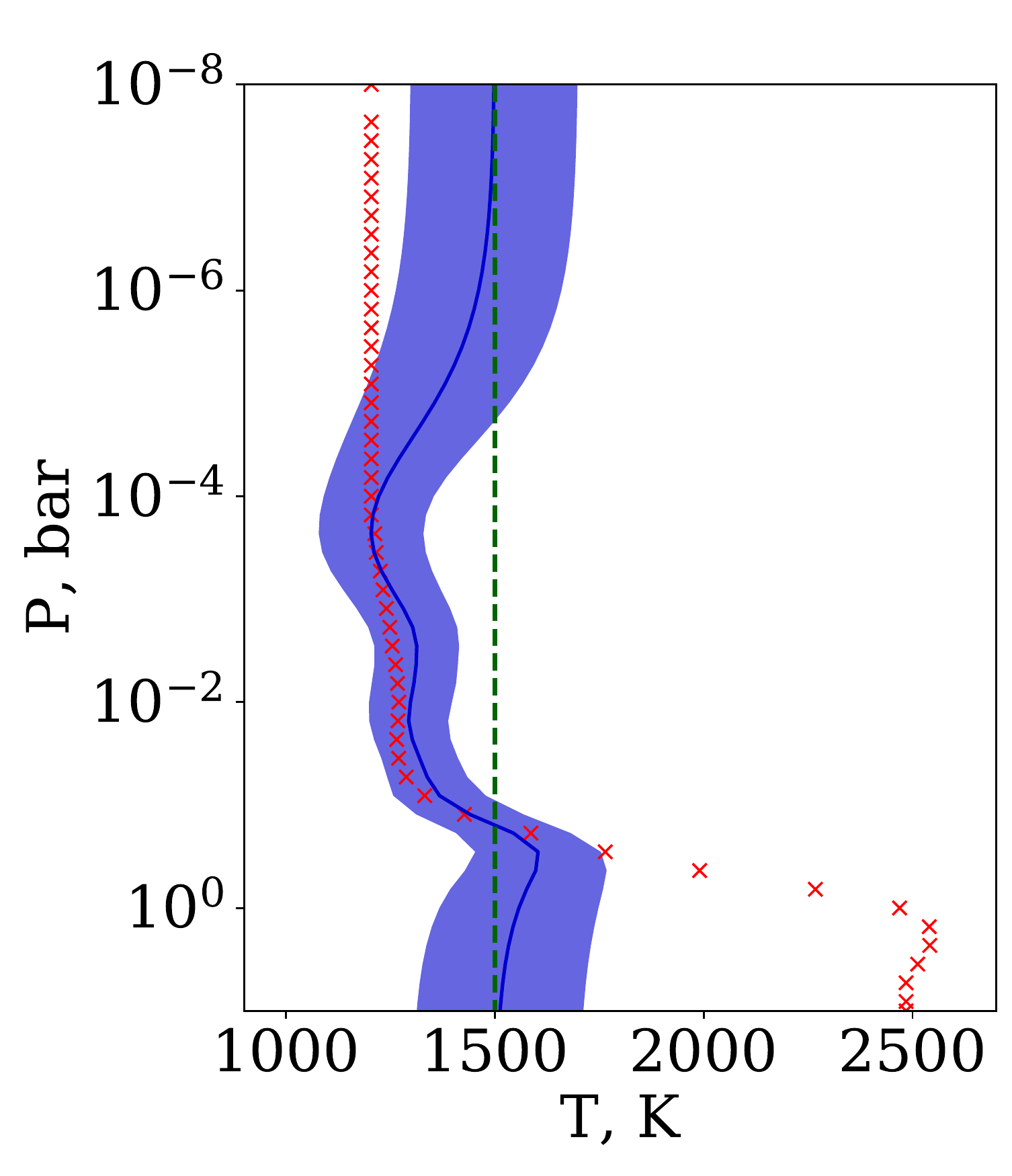}
\includegraphics[trim={0 0 25.0cm 0},clip,height=0.13\hsize]{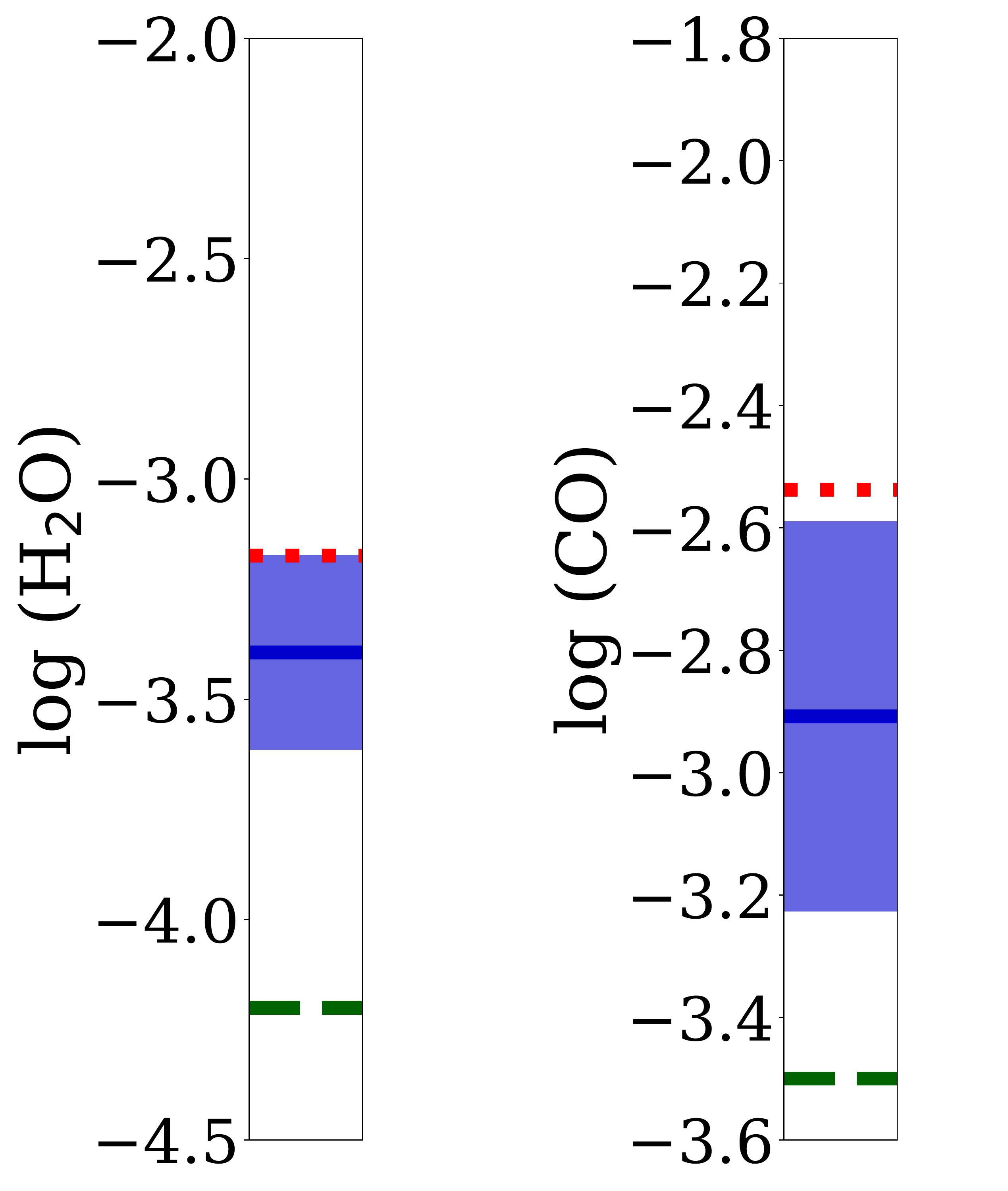}
\hspace{0.02\hsize}
\includegraphics[trim={0 0 35.5cm 0},clip,height=0.13\hsize]{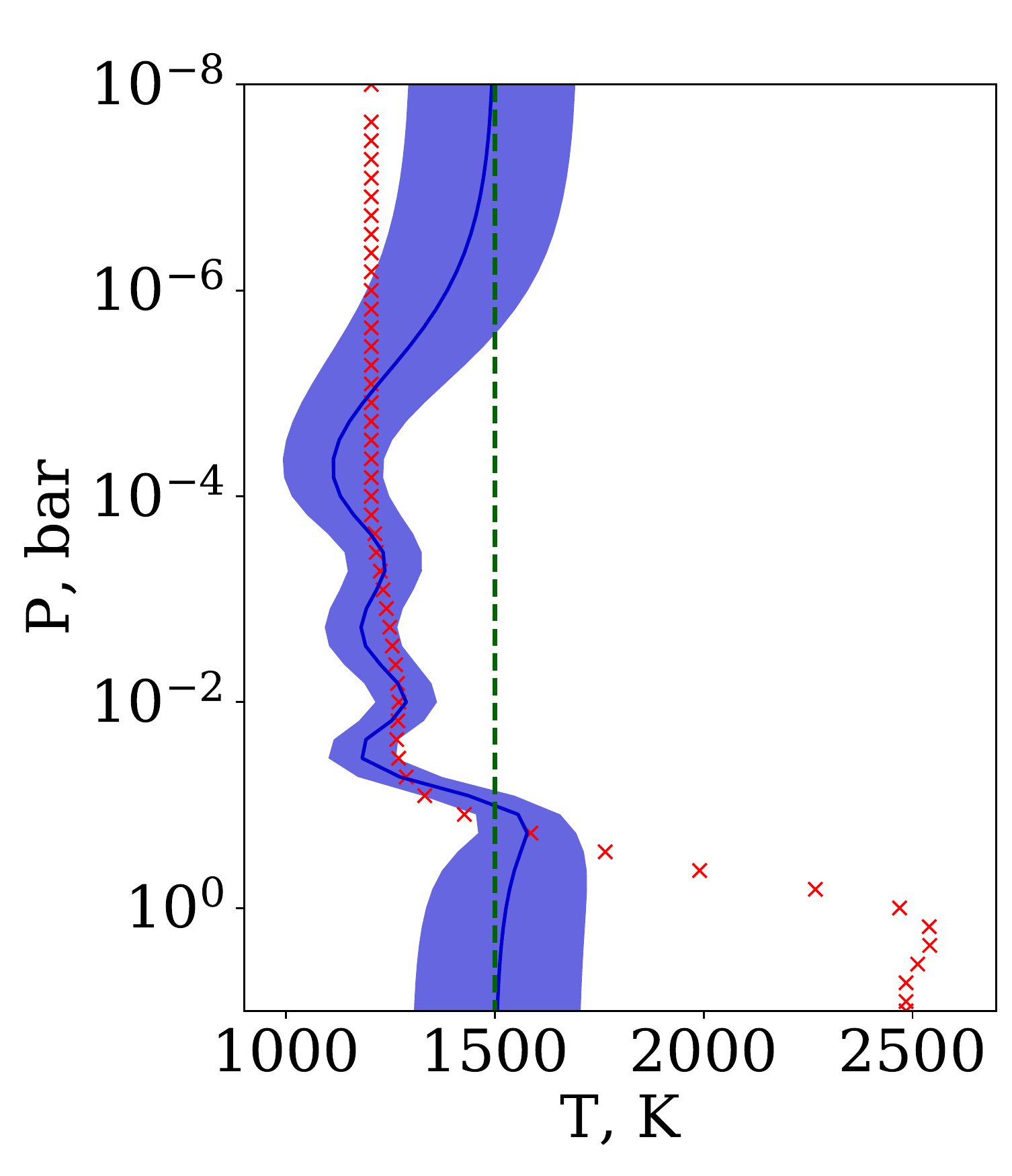}
\includegraphics[trim={0 0 25.0cm 0},clip,height=0.13\hsize]{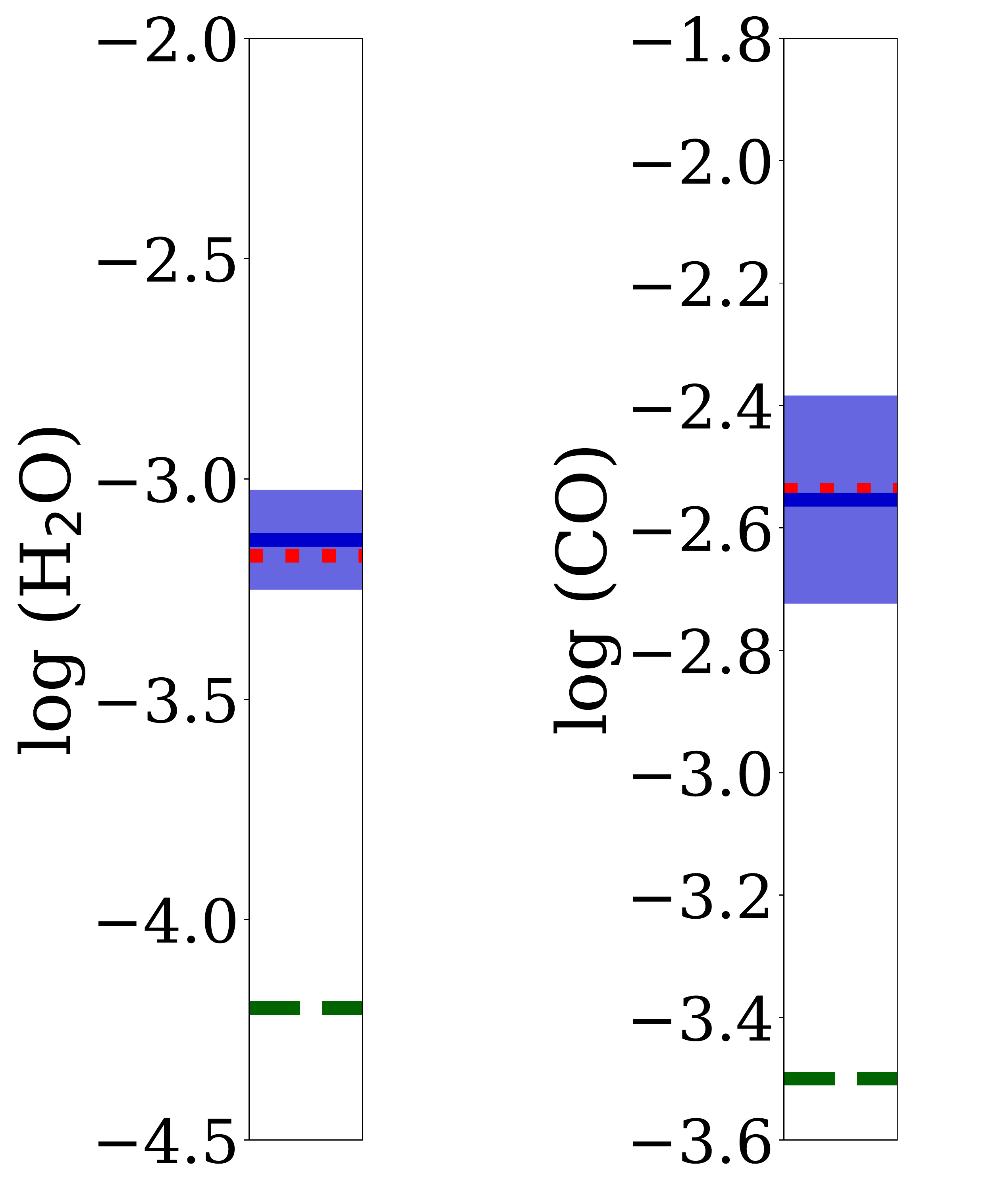}
\hspace{0.02\hsize}
\includegraphics[trim={0 0 35.5cm 0},clip,height=0.13\hsize]{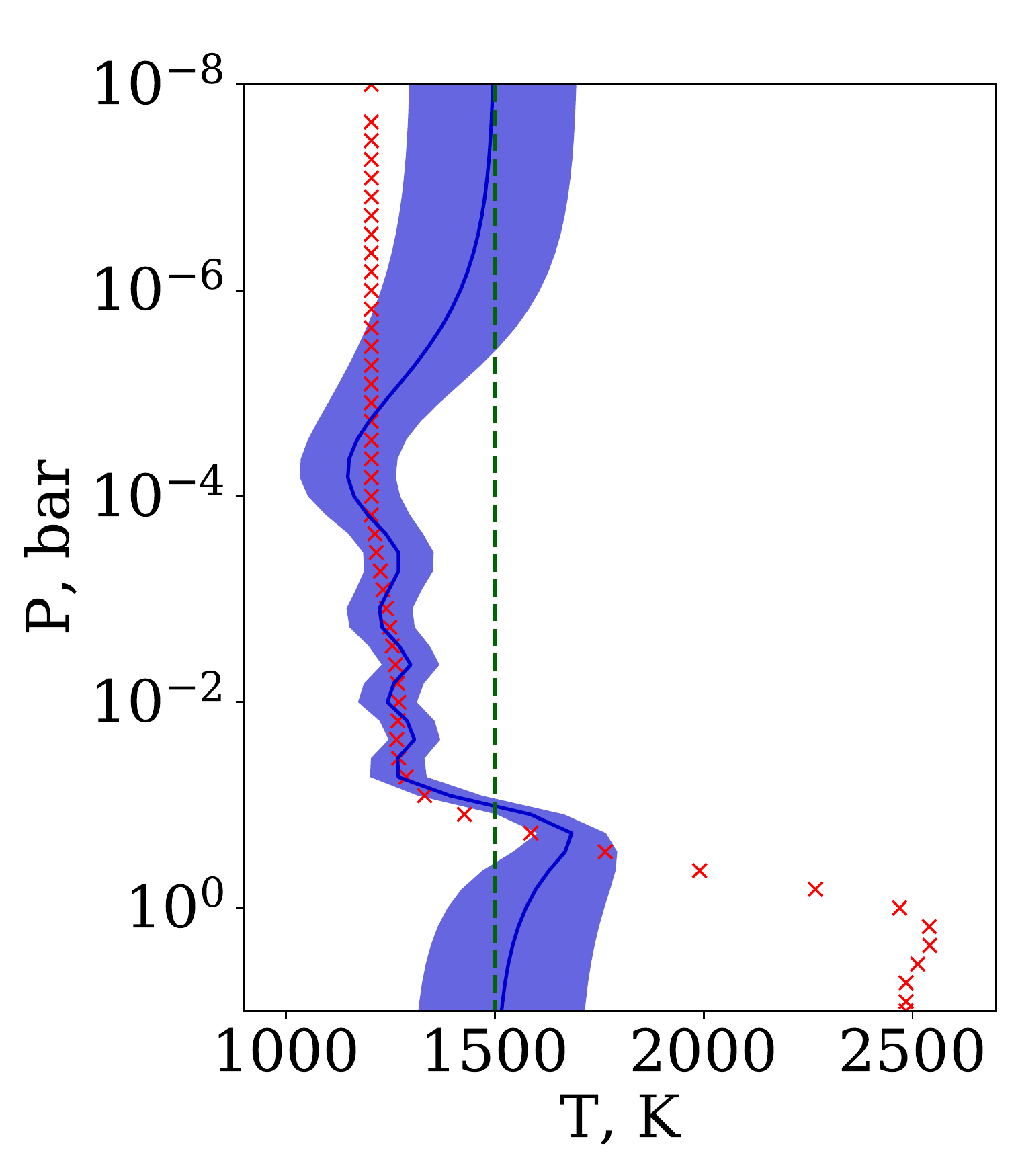}
\includegraphics[trim={0 0 25.0cm 0},clip,height=0.13\hsize]{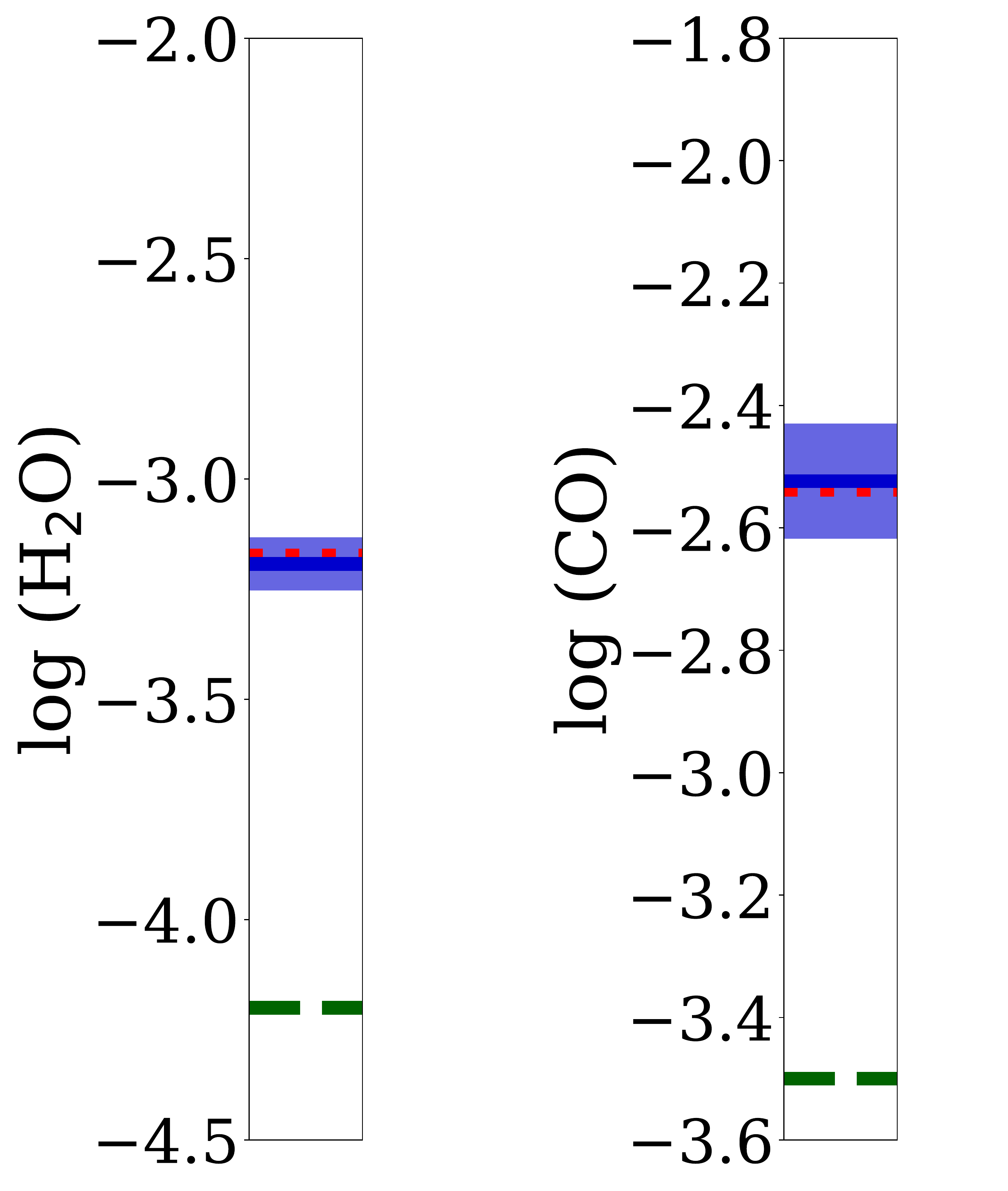}
\hspace{0.02\hsize}
\includegraphics[trim={0 0 35.5cm 0},clip,height=0.13\hsize]{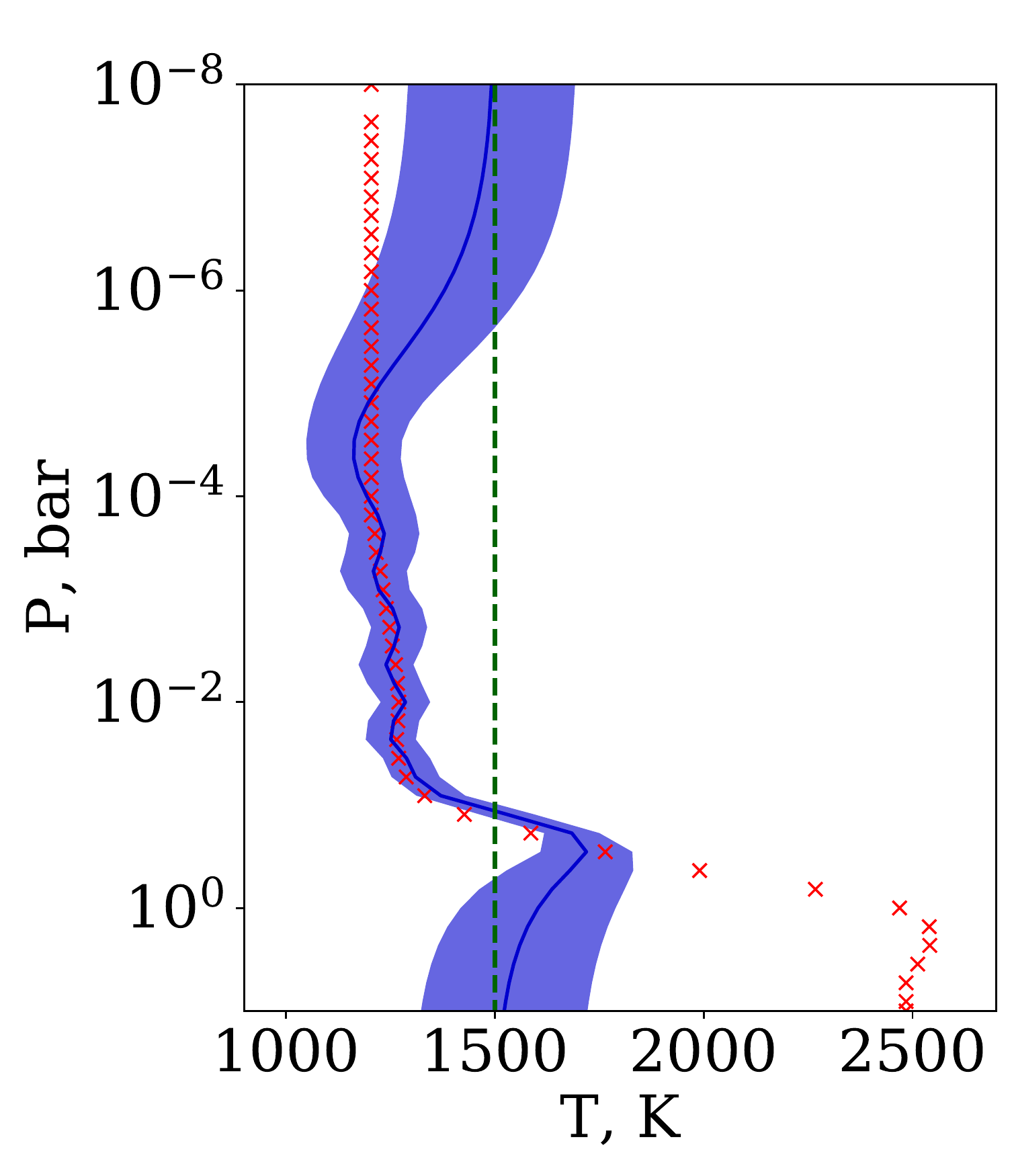}
\includegraphics[trim={0 0 25.0cm 0},clip,height=0.13\hsize]{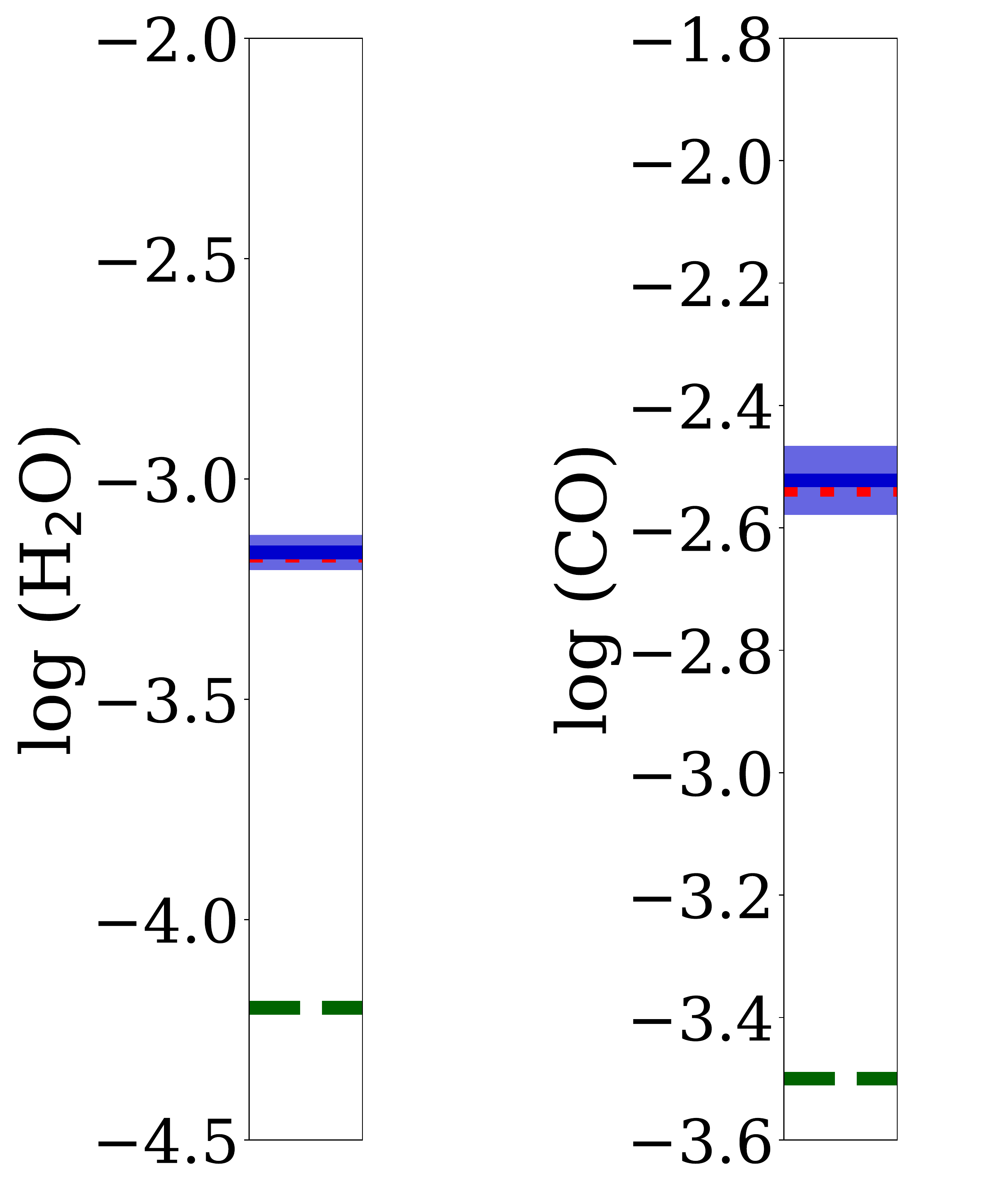}
}
\centerline{
\includegraphics[trim={0 0 35.5cm 0},clip,height=0.13\hsize]{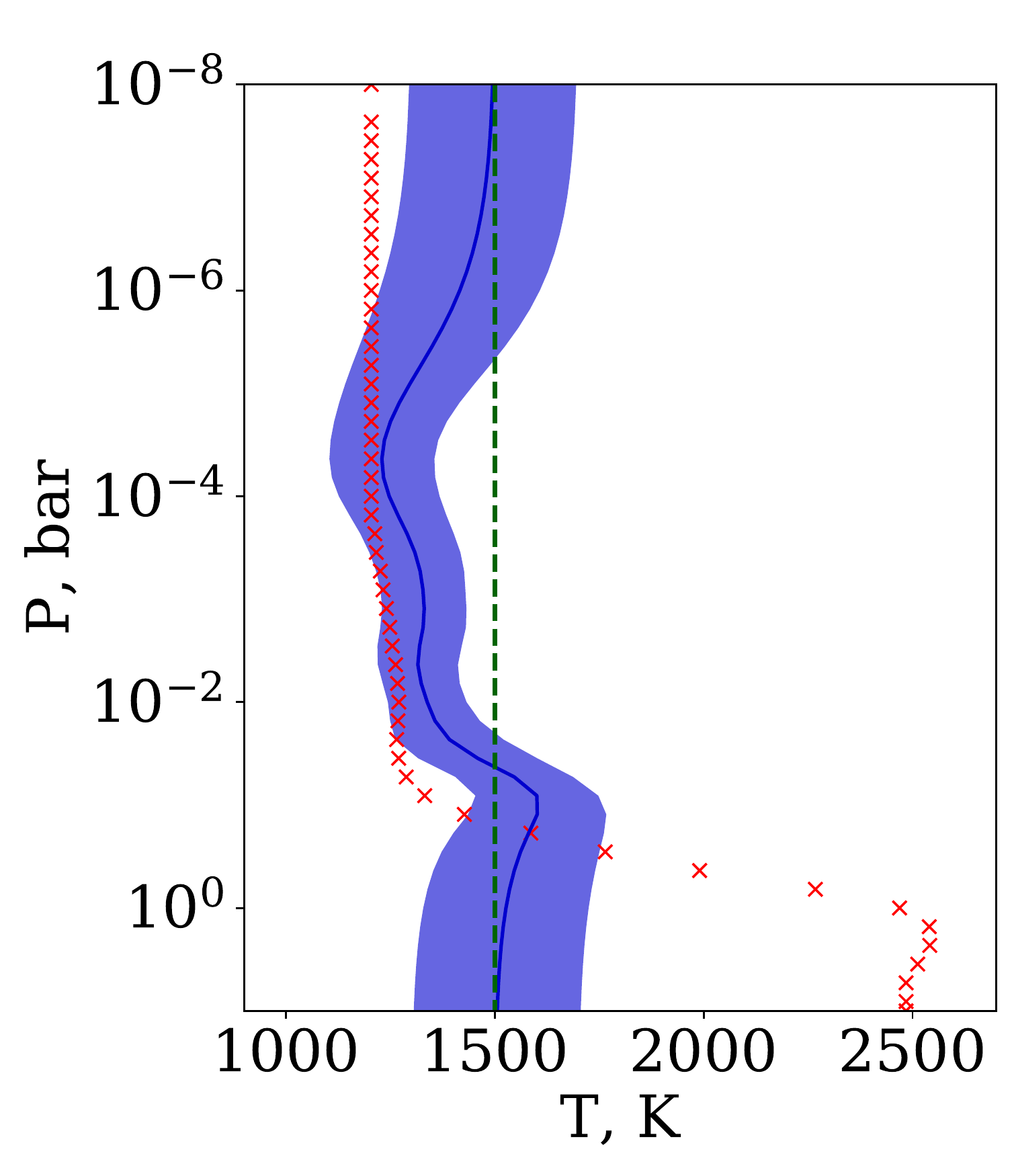}
\includegraphics[trim={0 0 25.0cm 0},clip,height=0.13\hsize]{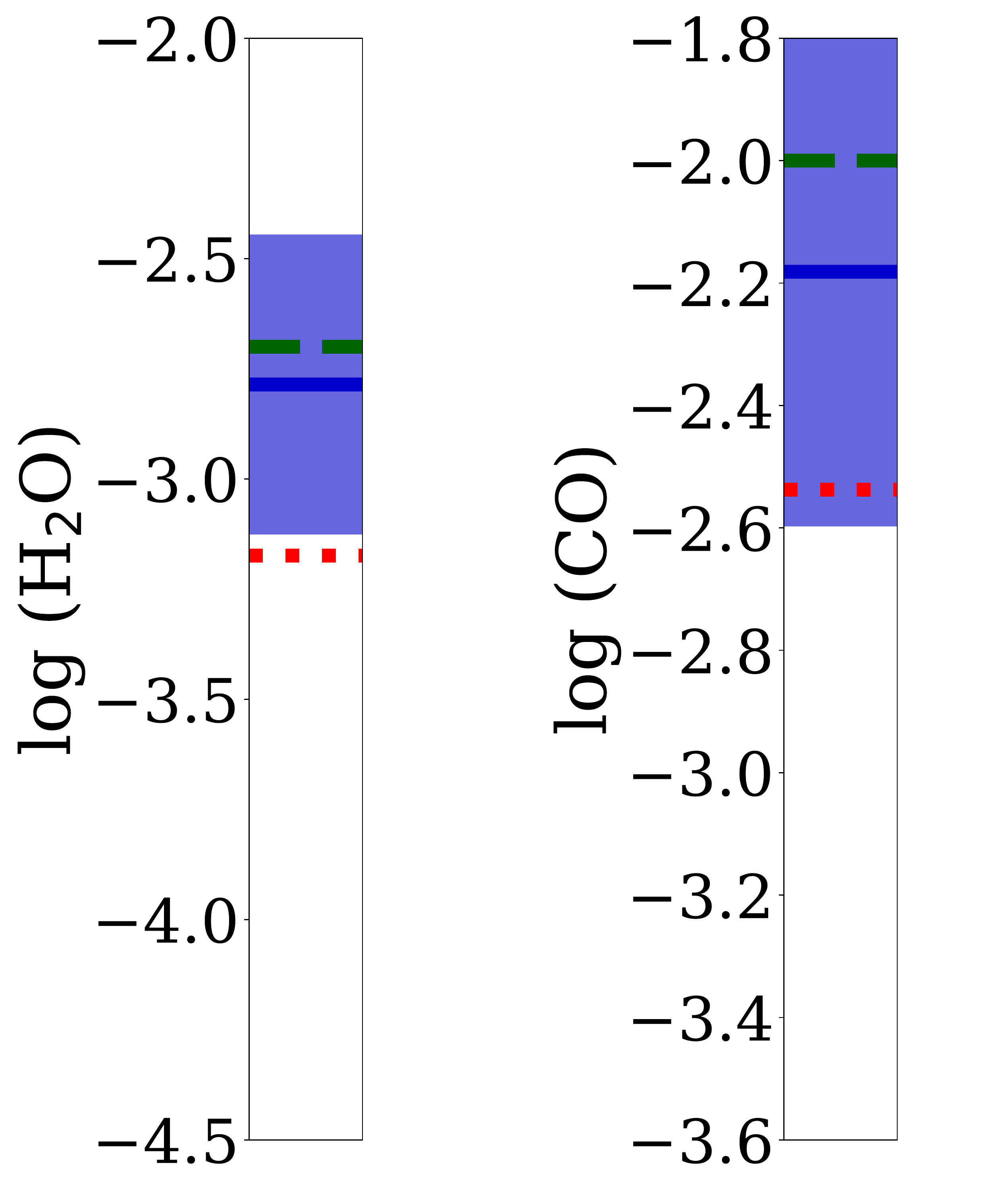}
\hspace{0.02\hsize}
\includegraphics[trim={0 0 35.5cm 0},clip,height=0.13\hsize]{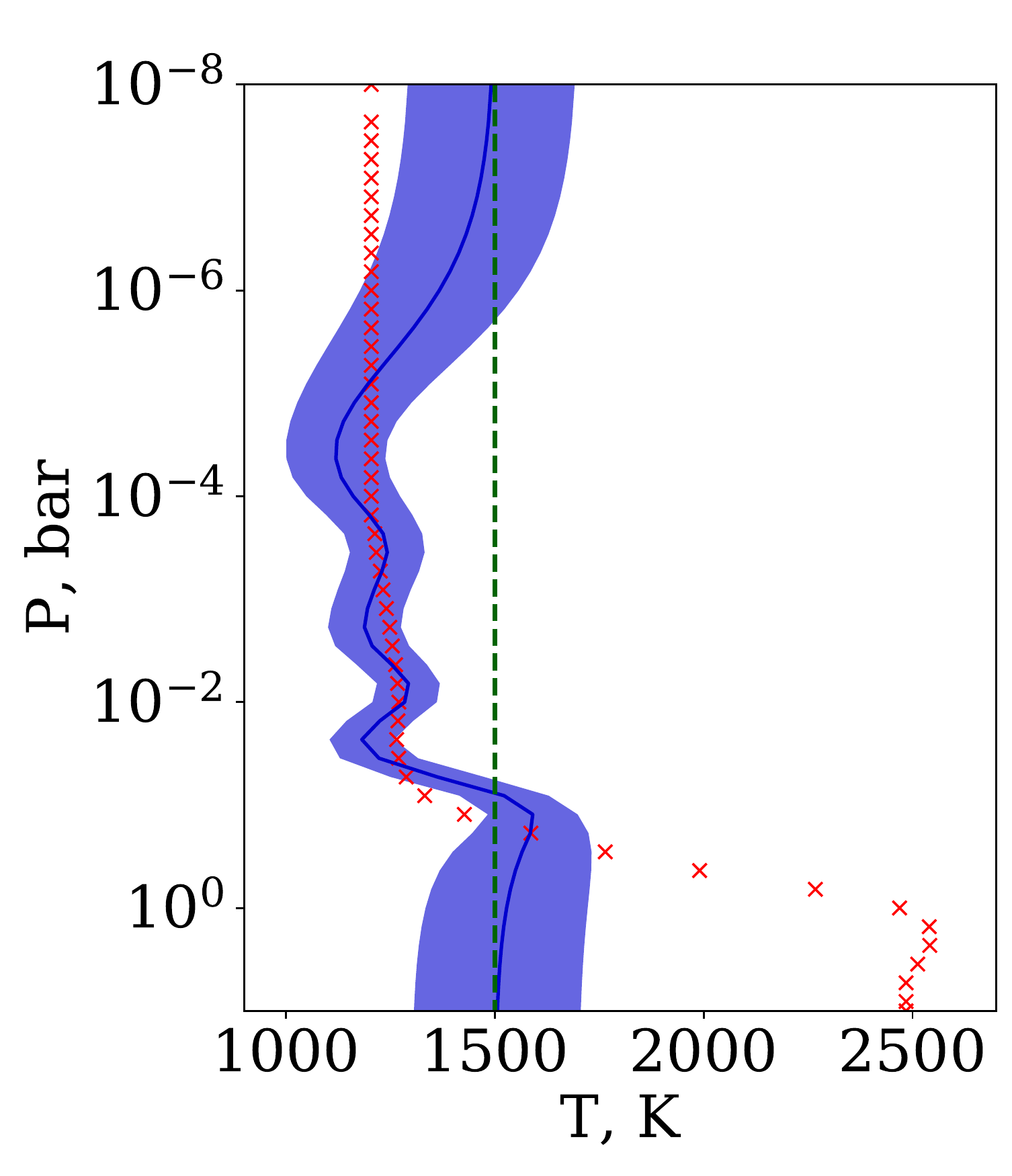}
\includegraphics[trim={0 0 25.0cm 0},clip,height=0.13\hsize]{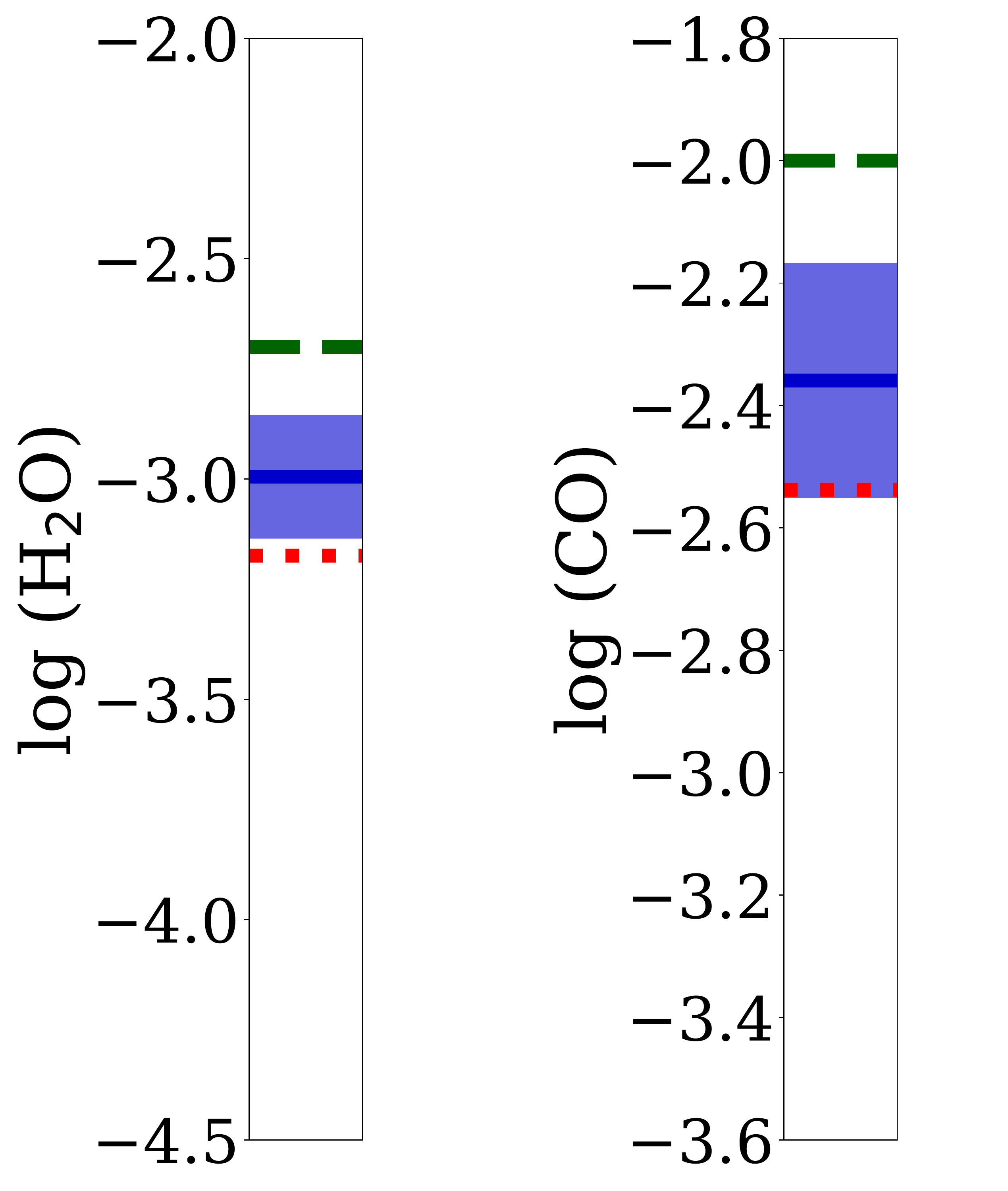}
\hspace{0.02\hsize}
\includegraphics[trim={0 0 35.5cm 0},clip,height=0.13\hsize]{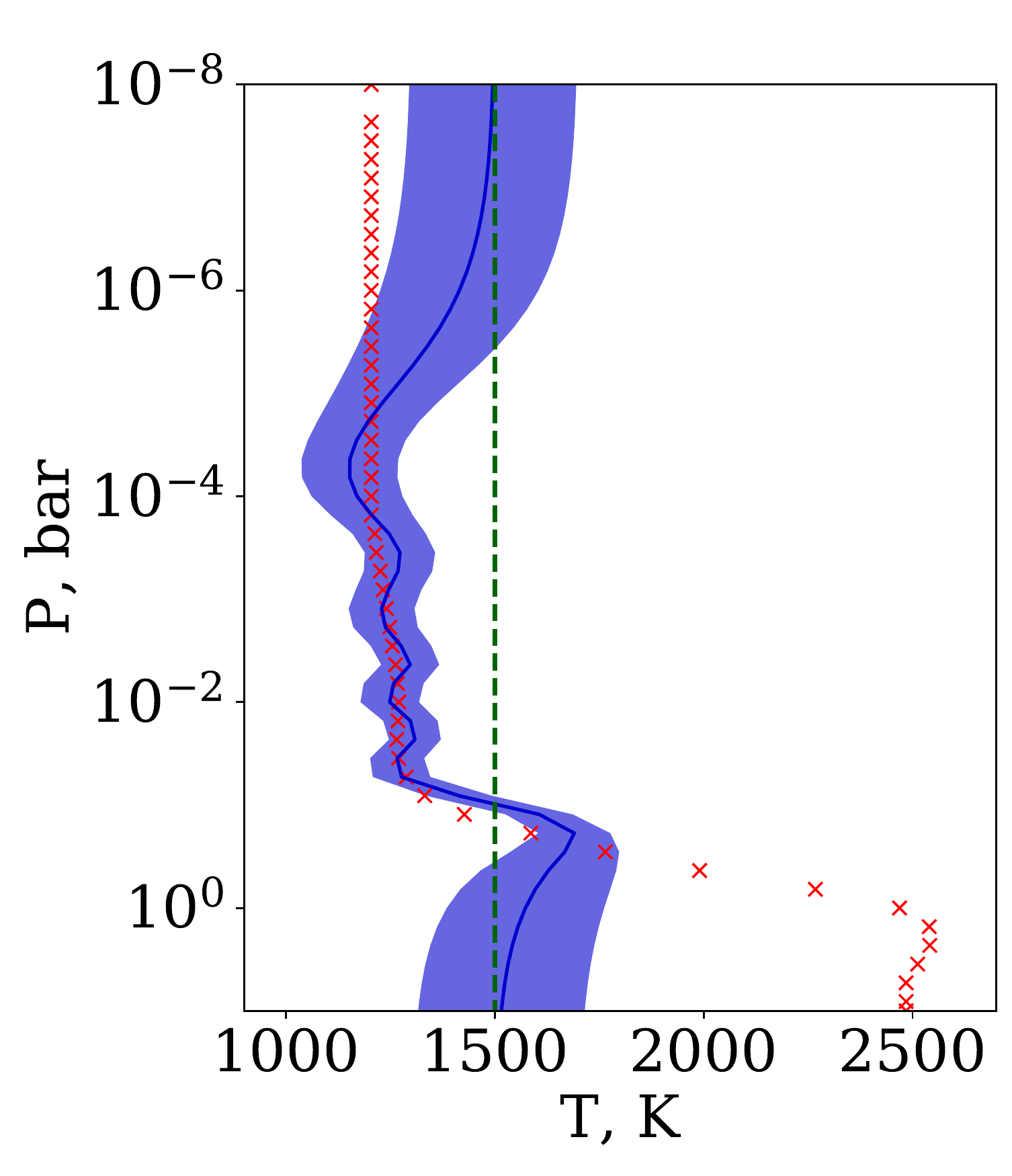}
\includegraphics[trim={0 0 25.0cm 0},clip,height=0.13\hsize]{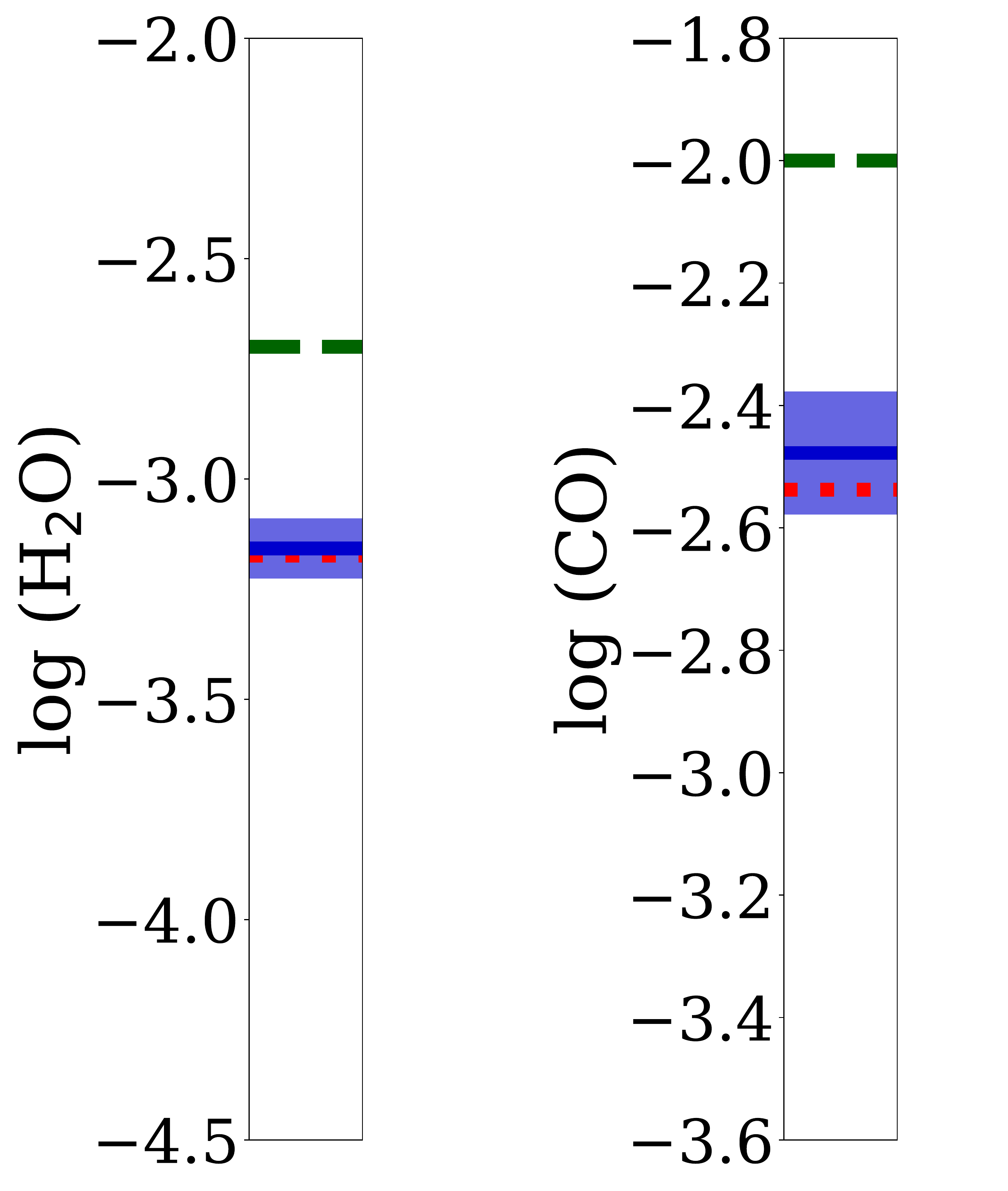}
\hspace{0.02\hsize}
\includegraphics[trim={0 0 35.5cm 0},clip,height=0.13\hsize]{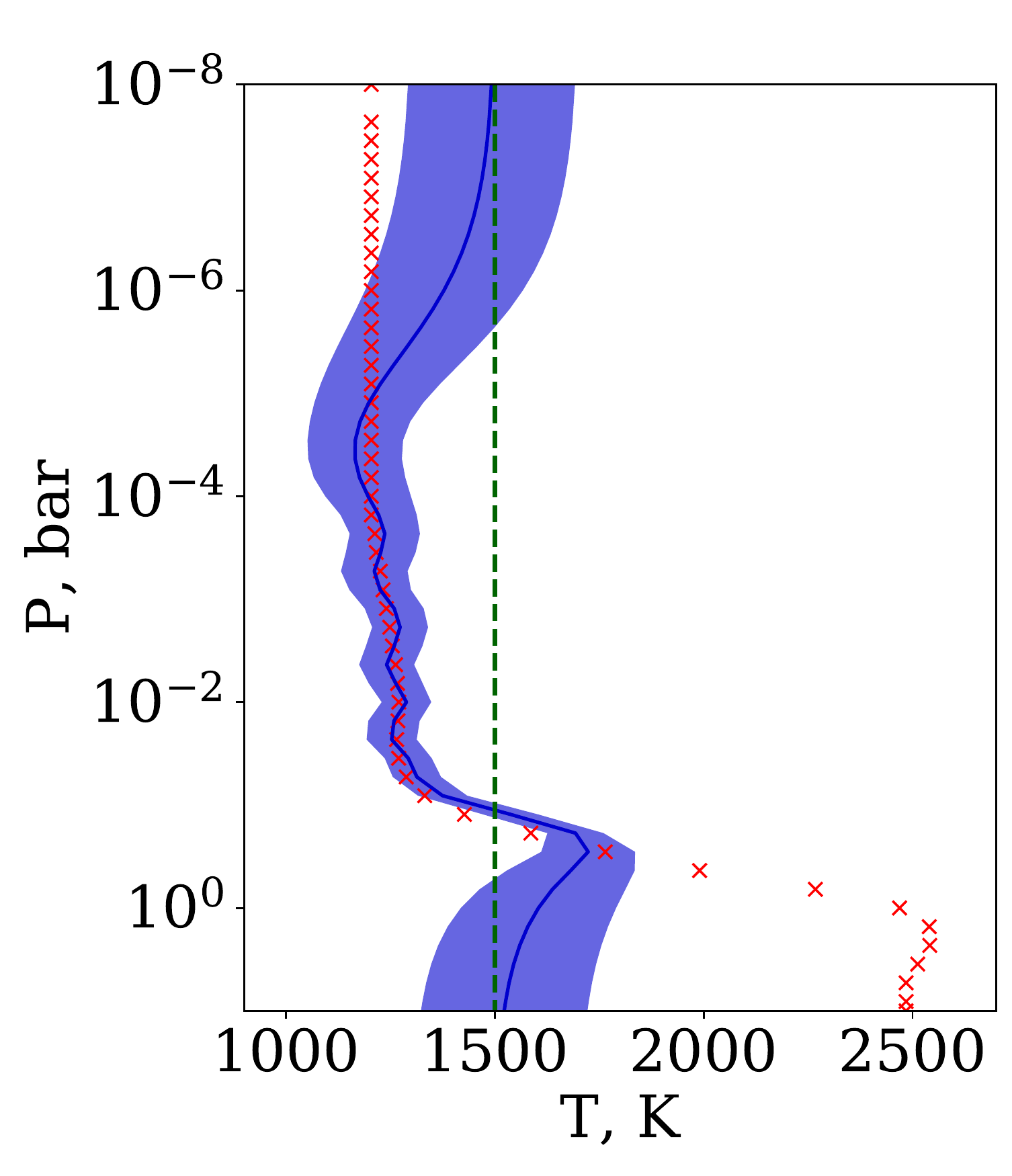}
\includegraphics[trim={0 0 25.0cm 0},clip,height=0.13\hsize]{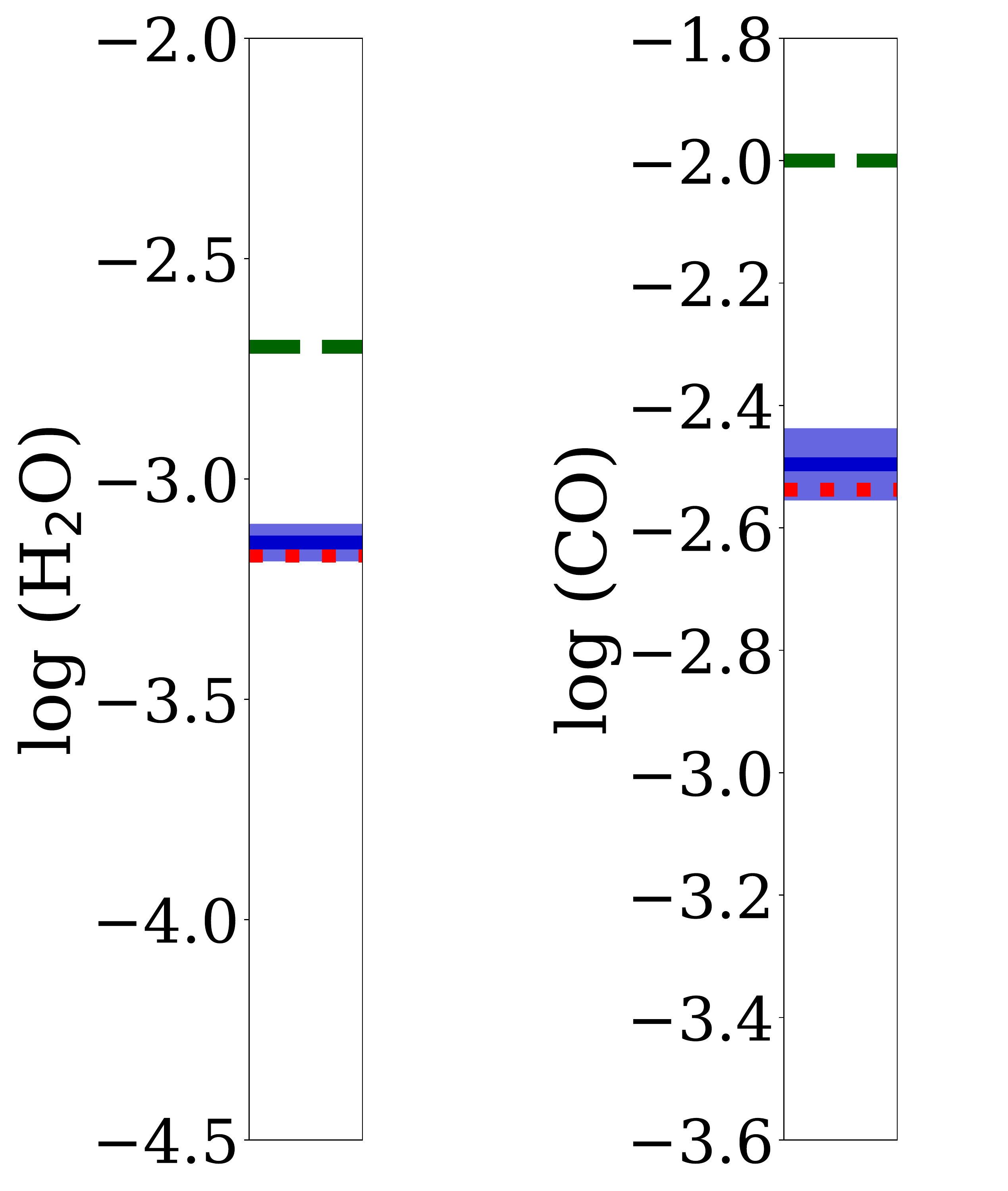}
}
\caption{\label{fig:init-guess-vmr}
Retrievals of the \hho\ and \co\ mixing ratios from the ($2.28-2.38$)~\mum\ region assuming three different
initial guesses (from top to bottom) and four S/N values (from left to right). The green dashed and red dotted lines
are the initial guess and true solutions, respectively. The solid blue  line and the shaded area are the retrieved
value and its 1$\sigma$ uncertainty, respectively.
}
\end{figure*}

In Fig.~\ref{fig:snr10-r100k} we show the results of retrievals from five spectral regions
and S/N=$10$ (the results for other S/N values are provided in Figs.~\ref{fig:snr5-r100k}, \ref{fig:snr25-r100k}, and \ref{fig:snr50-r100k} 
in the online material). The data was simulated using temperature structure and mixing ratios of HD~189733~b as derived in \citet{2012MNRAS.420..170L} 
(red dashed lines in Fig.~\ref{fig:snr10-r100k}).
In all cases we start from an isothermal temperature distribution with T=$1500$~K.
In this way we can better study the ability of the EO method to recover the true temperature
distribution at different altitudes.
The initial values for mixing ratios were taken to be $-0.5$~dex from the true ones with the uncertainty of $1$~dex.
It is seen that the spectra of HJs in the wavelength range covered by \criresplus\, 
is sensitive to a wide range of atmospheric pressures between $1$~bar and $10^{-4}$~bar. 
At these altitudes we recover temperatures very close to the true
solution. However, everything that is outside this altitude range remains unconstrained due to the lack
of information provided by the observed spectra.
The typical temperature errors in the pressure range between $1$~bar and $10^{-4}$~bar is $120$~K for S/N=$5$,
$100$~K for S/N=$10$, $90$~K for S/N=$25$, and $80$~K for S/N=50.
All spectral windows except $4.80-5.00$~\mum\, appear to be relevant to
study temperature stratification in atmospheres of HJs, but only two of them, ($1.50-1.70$)~\mum\ and  ($2.28-2.38$)~\mum,
allow us to constrain VMRs of at least some molecules.

As was mentioned in \citet{2012MNRAS.420..170L}, the degeneracy between fit parameters
may lead to incorrectly derived mixing ratios. We find that this is also the case for high-resolution low S/N data.
As an example calculation, our Fig.~\ref{fig:init-guess-vmr} shows retrievals of \hho\ and \co\ from the ($2.28-2.38$)~\mum\ region
using three different initial guesses for the molecular VMRs. For S/N=5 the retrievals converge to nearly the same
solution if the initial values of VMRs are $0.5$~dex and $1.0$~dex below the true one.
However, when the initial guess is $0.5$~dex higher than the true value then the algorithm
prefers to increase the local temperature in the atmosphere keeping the VMRs at their high values.
The increase in local temperature makes lines of molecules weaker, thus compensating for the high initial VMR.
The molecular VMRs and temperature are thus degenerated as all solutions result in the same
final normalized cost function ($\phi=0.73$).
As the S/N of the data increases, the solutions for  temperature and VMRs both tend to converge to the true values, 
while for S/N$\leqslant$10 the final values may deviate from the true ones (but are still  within the $3\sigma$ error bars)  depending on the initial guess.

\begin{figure*}
\centerline{
\includegraphics[height=0.14\hsize]{figures/r100k/snr10/h2o+co+ch4+co2/ll-2.28-2.38/out-vmr_init_m0.5dex-vmr_err_1.0dex.bwd_fit.pdf}
\hspace{0.02\hsize}
\includegraphics[trim={0 0 20.0cm 0},clip,height=0.14\hsize]{figures/r100k/snr10/h2o+co+ch4+co2/ll-2.28-2.38/out-vmr_init_m0.5dex-vmr_err_1.0dex.bwd_param1.pdf}
\hspace{0.02\hsize}
\includegraphics[height=0.14\hsize]{figures/r100k/snr10/h2o+co+ch4+co2/ll-2.28-2.38/out-vmr_init_m0.5dex-vmr_err_1.0dex.bwd_param_vmr.pdf}
}
\centerline{
\includegraphics[height=0.14\hsize]{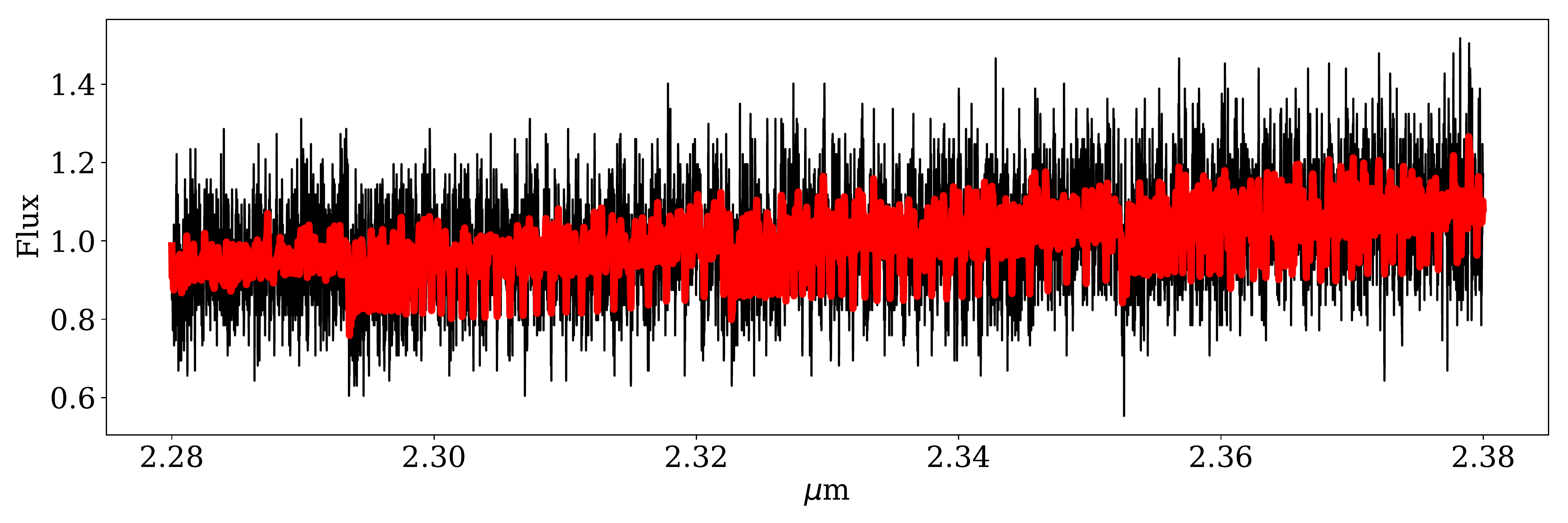}
\hspace{0.02\hsize}
\includegraphics[trim={0 0 20.0cm 0},clip,height=0.14\hsize]{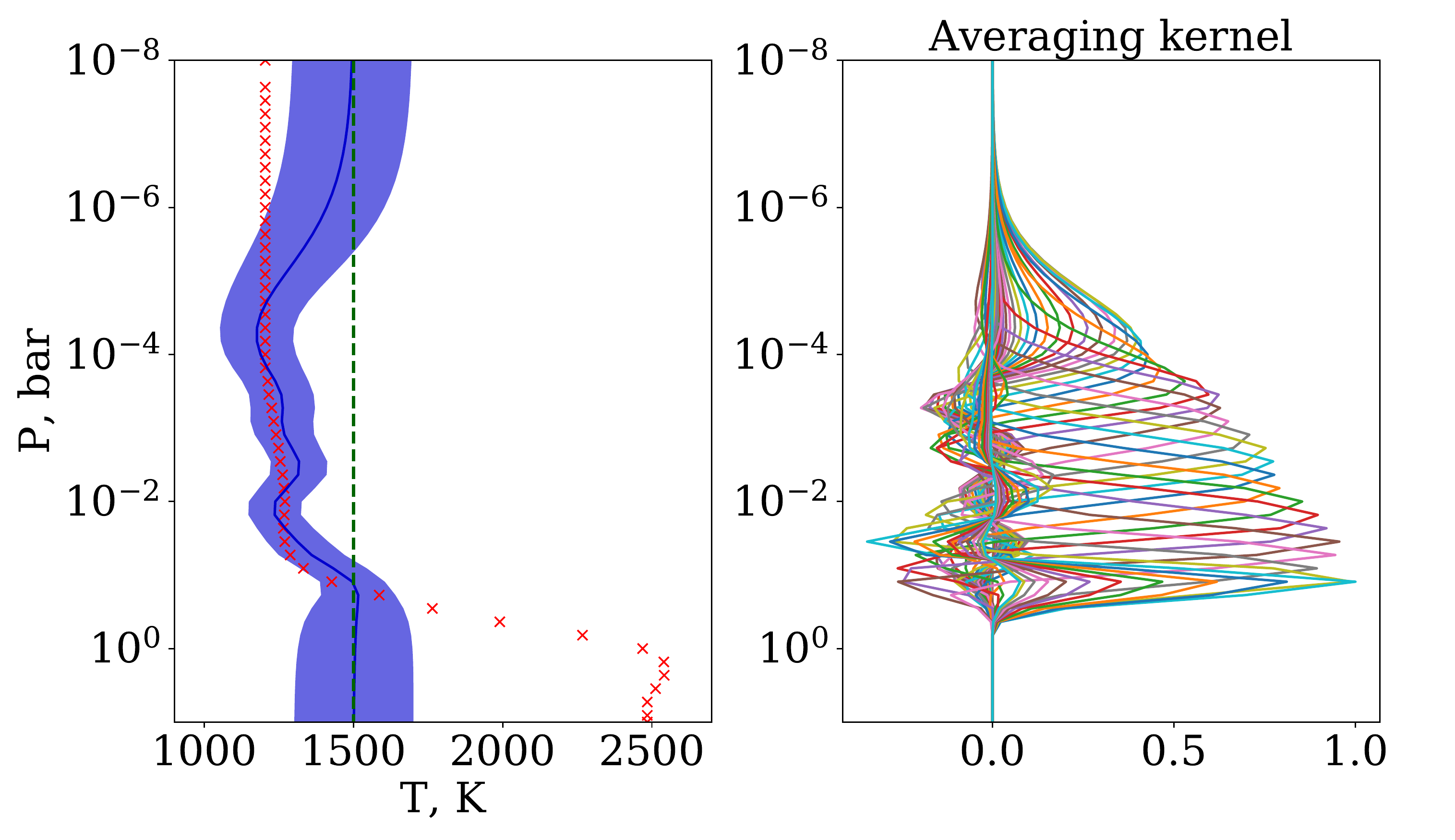}
\hspace{0.02\hsize}
\includegraphics[height=0.14\hsize]{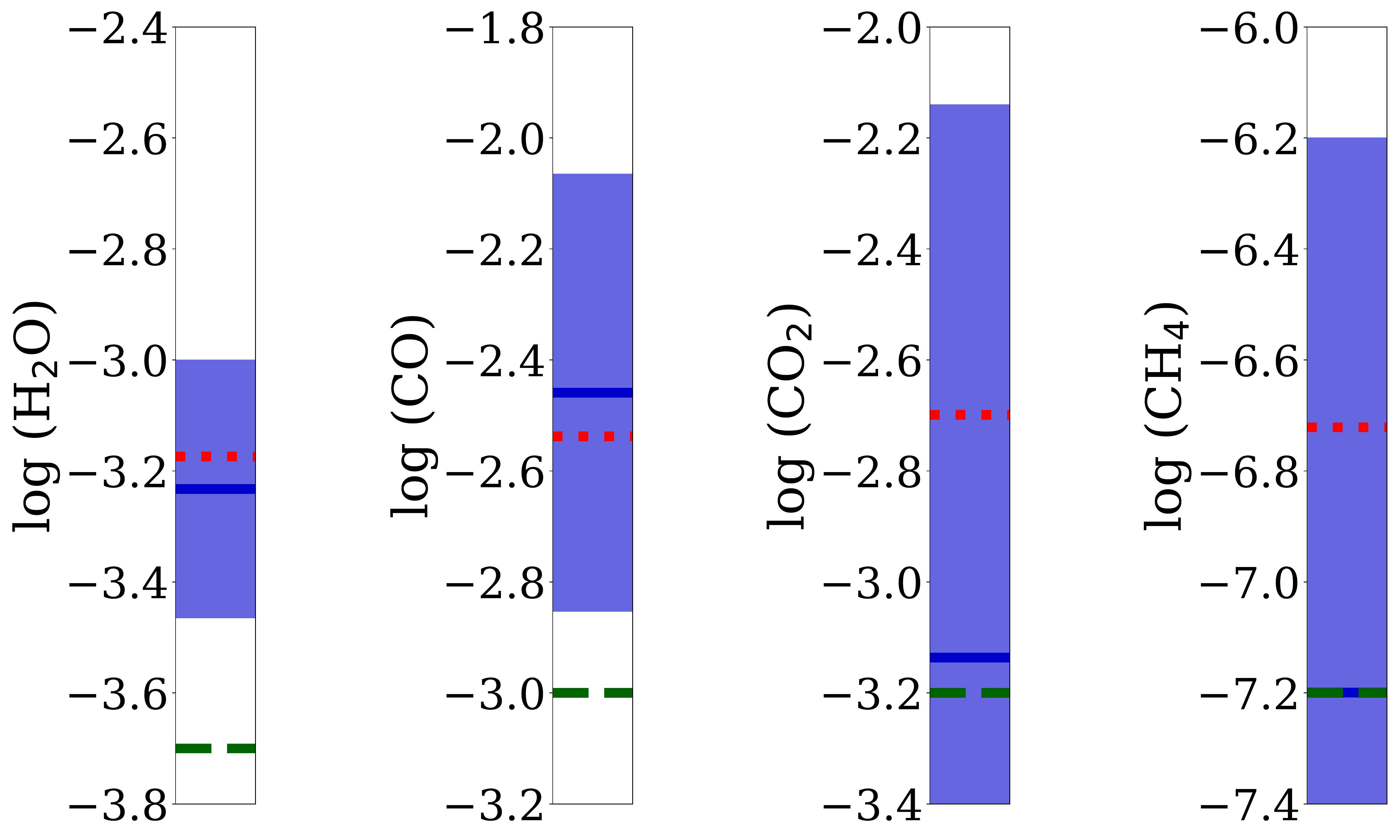}
}
\caption{\label{fig:snr10-r50k}
Comparison of retrieved temperature and mixing ratios of four molecules assuming S/N=$10$ and two spectral resolutions
of R=$100\,000$ and R=$50\,000$ for the top and bottom panels, respectively.
The color-coding is the same as in Fig.~\ref{fig:snr10-r100k}.
}
\end{figure*}

Finally, degrading spectral resolution by a factor of two from R=$100\,000$ to R=$50\,000$
does not significantly affect  the results of our temperature retrievals, while the accurate retrievals of molecular VMRs
become problematic because the obtained uncertainties are large.
An example is shown in Fig.~\ref{fig:snr10-r50k} for the region
$2.28-2.38$~\mum\ and S/N=$10$.  With the degrading of spectral resolution,
the amount of information that we can learn from the observed spectra mostly depends on the number
of spectral features resolved. Decreasing  the resolving power weakens the depth of spectral
lines, and thus directly affects retrieval of mixing ratios. 
At the same time, the profiles of strong molecular lines 
are still satisfactorily resolved, which provides enough information for the temperature retrieval.
It is thus possible to rebin the obtained high-resolution spectrum to a smaller resolution 
and boost the S/N without greatly affecting our ability to retrieve temperature stratification. 
Alternatively, our Fig.~\ref{fig:snr10-r50k}  reflects the case of two observations with different slit widths,
but after reaching the same S/N. This would require less telescope time for a wider slit width to reach the same S/N,
which could be an affordable choice   to reduce the integration time for a fixed S/N.

\subsection{Retrievals from multiple spectral regions}

\begin{figure*}
\centerline{\textbf{$\mathbf{(1.50-1.70)+(2.28+2.38)}$~$\mathbf{\mu}$m}}
\centerline{
\includegraphics[trim={0 0 20.0cm 0},clip,height=0.14\hsize]{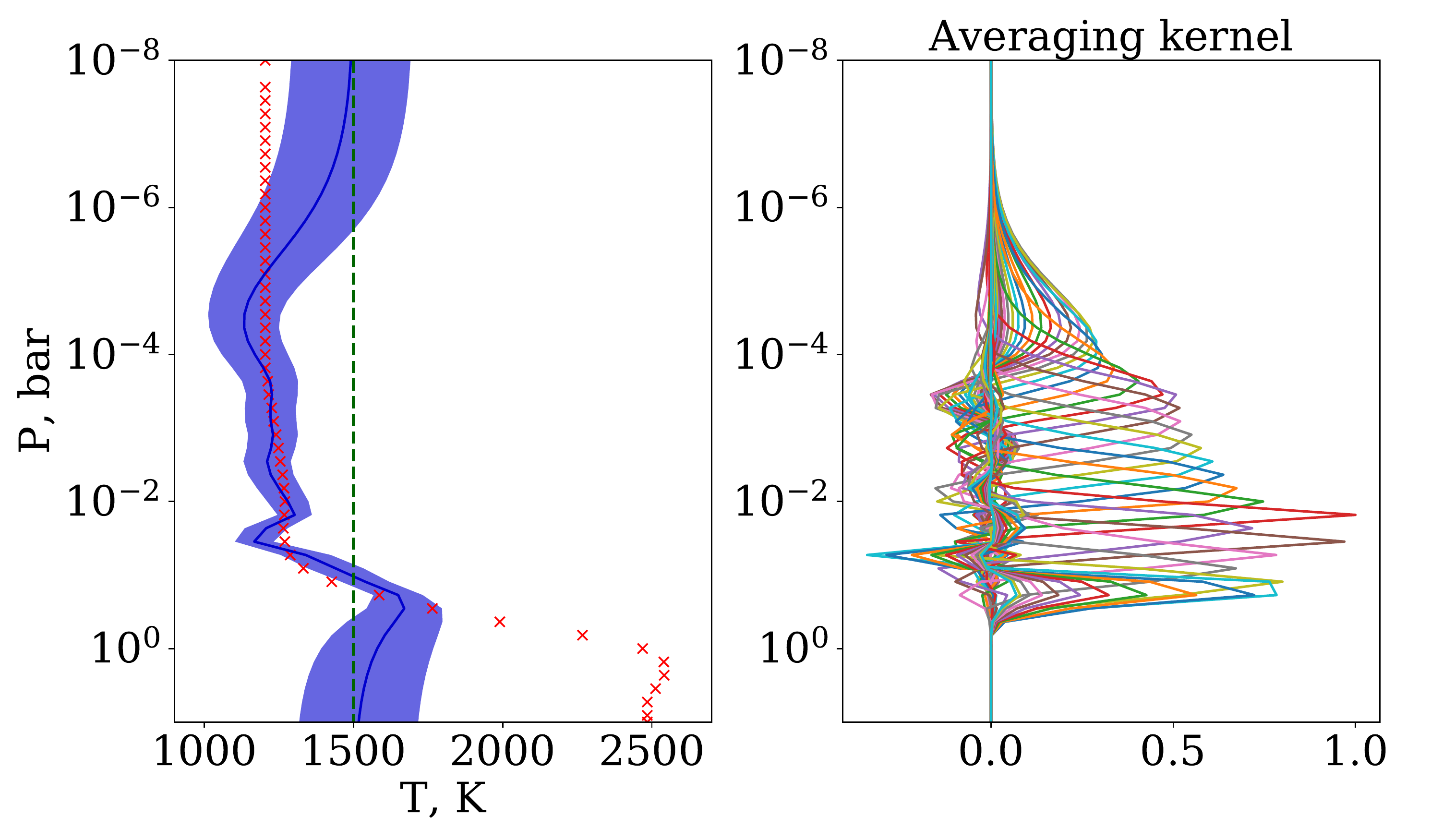}
\includegraphics[height=0.14\hsize]{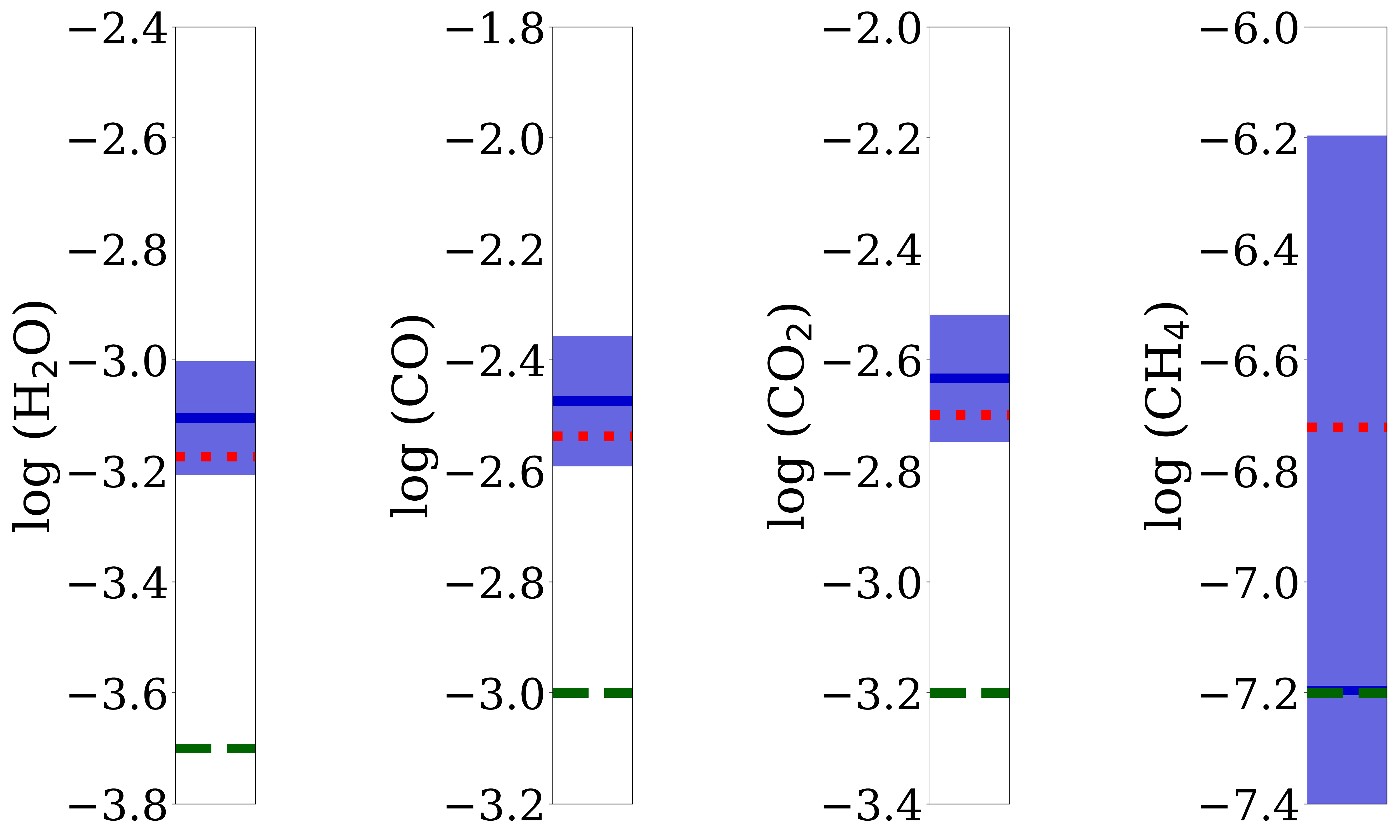}
}
\centerline{\textbf{$\mathbf{(2.10+2.28)+(2.28-2.38)}$~$\mathbf{\mu}$m}}
\centerline{
\includegraphics[trim={0 0 20.0cm 0},clip,height=0.14\hsize]{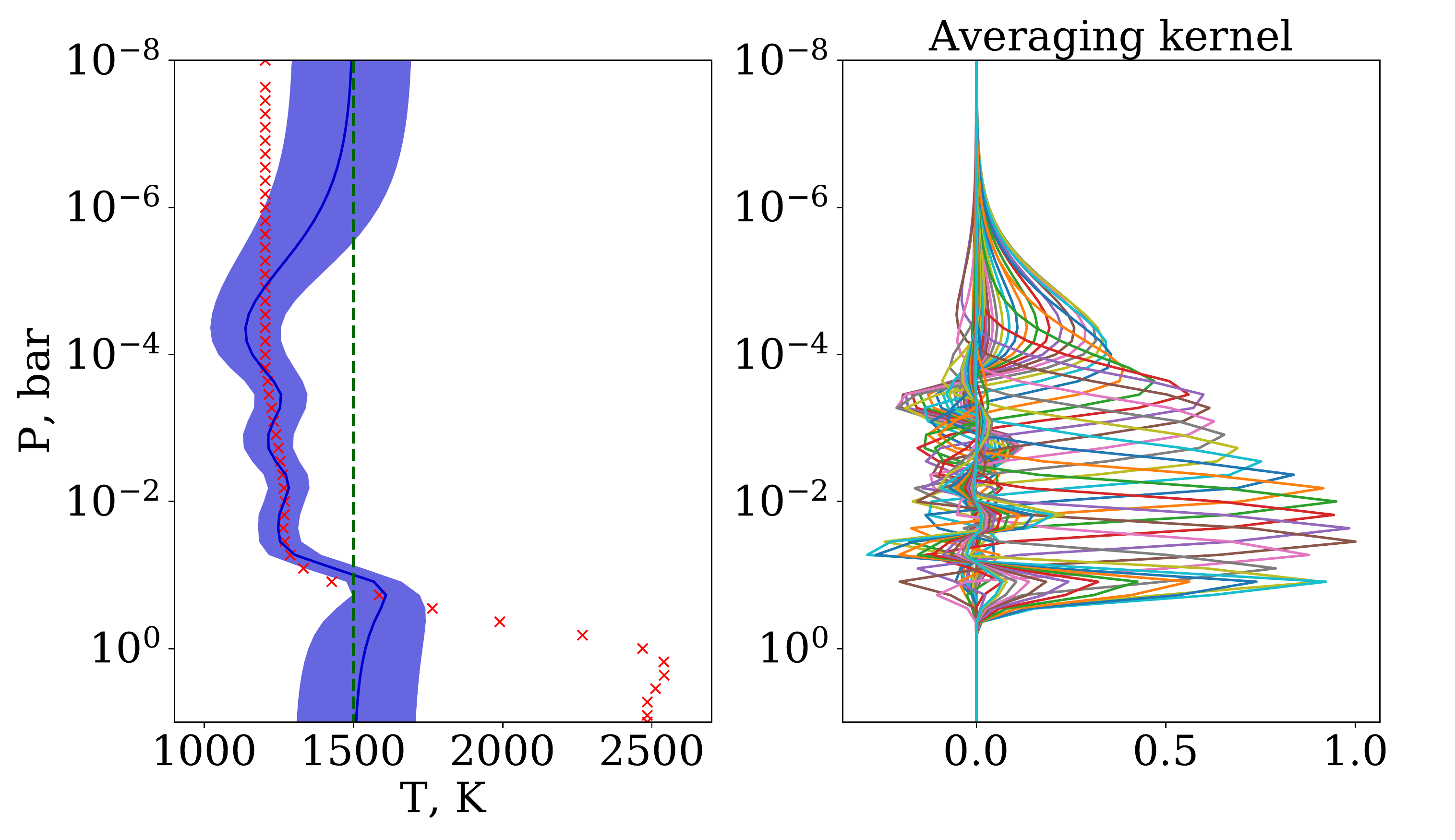}
\includegraphics[height=0.14\hsize]{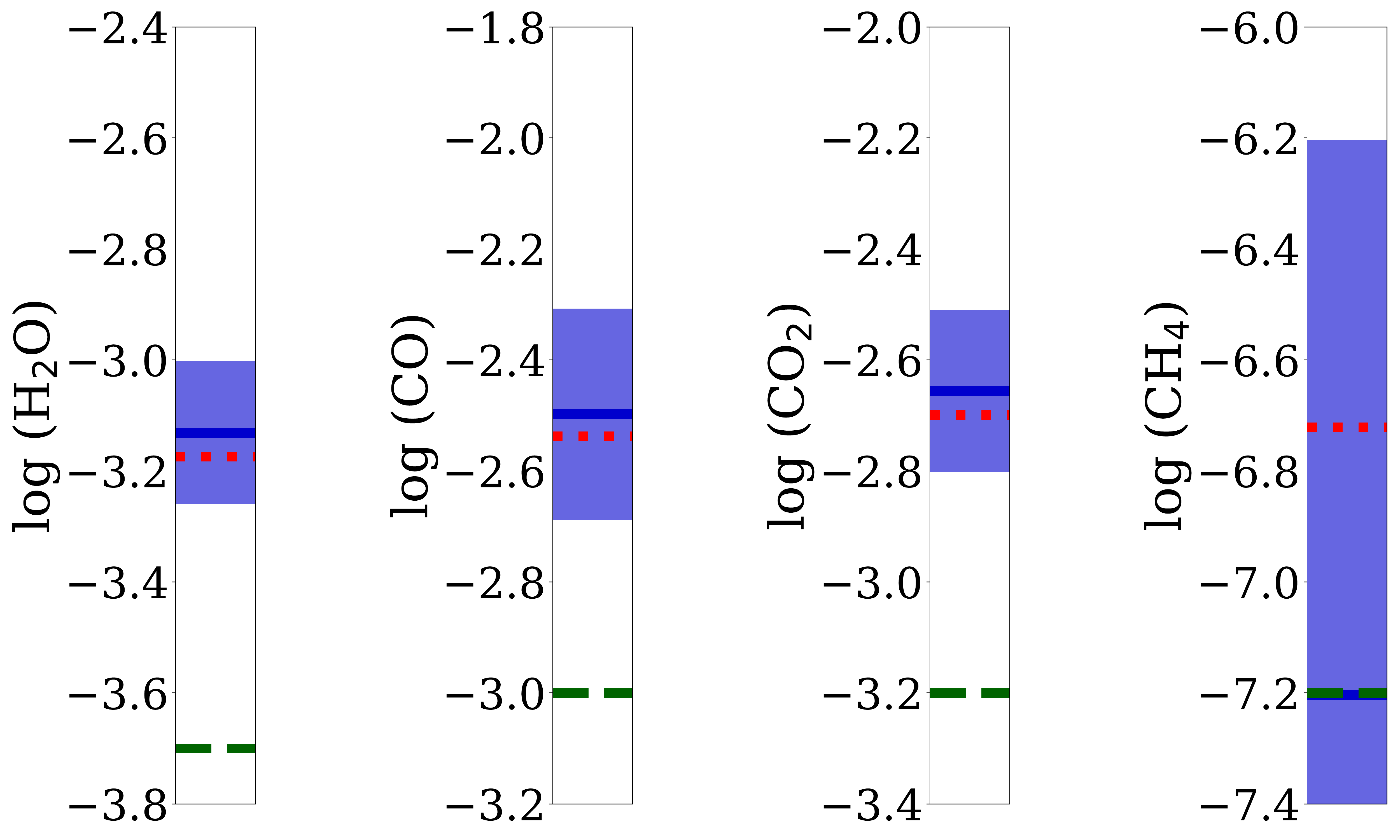}
}
\centerline{\textbf{$\mathbf{(2.28-2.38)+(3.80-4.00)}$~$\mathbf{\mu}$m}}
\centerline{
\includegraphics[trim={0 0 20.0cm 0},clip,height=0.14\hsize]{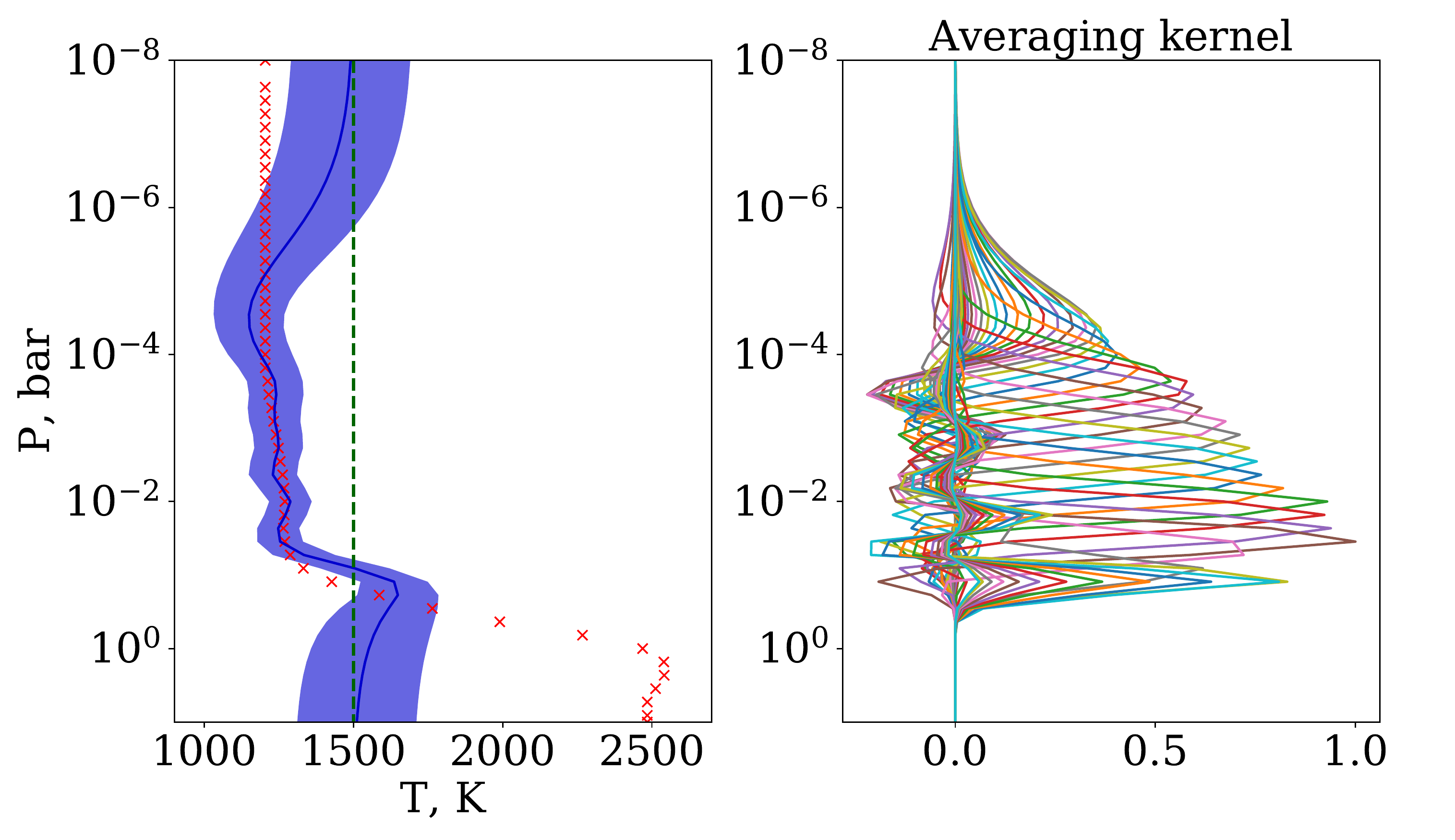}
\includegraphics[height=0.14\hsize]{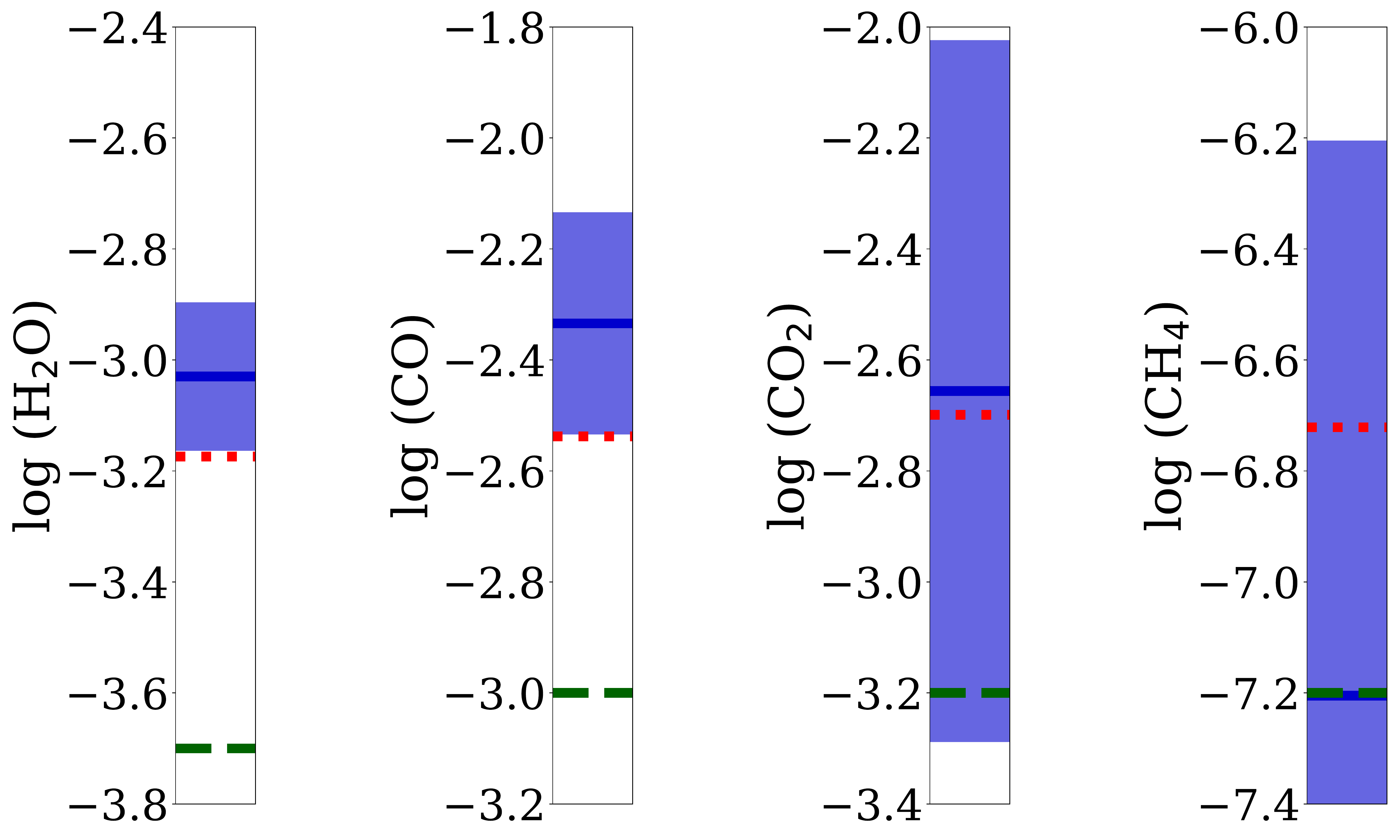}
}
\centerline{\textbf{$\mathbf{(2.28-2.38)+(4.80-5.00)}$~$\mathbf{\mu}$m}}
\centerline{
\includegraphics[trim={0 0 20.0cm 0},clip,height=0.14\hsize]{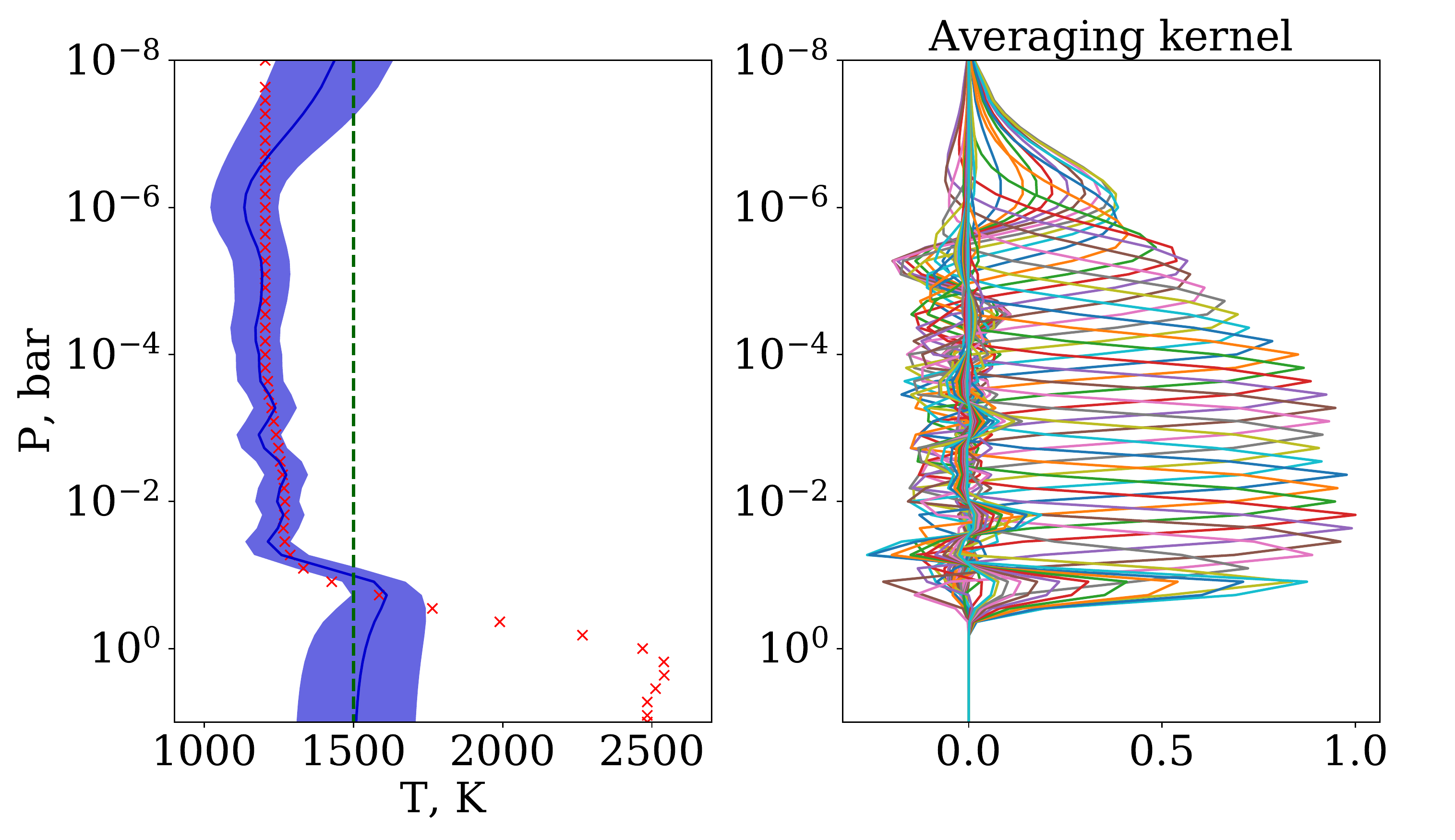}
\includegraphics[height=0.14\hsize]{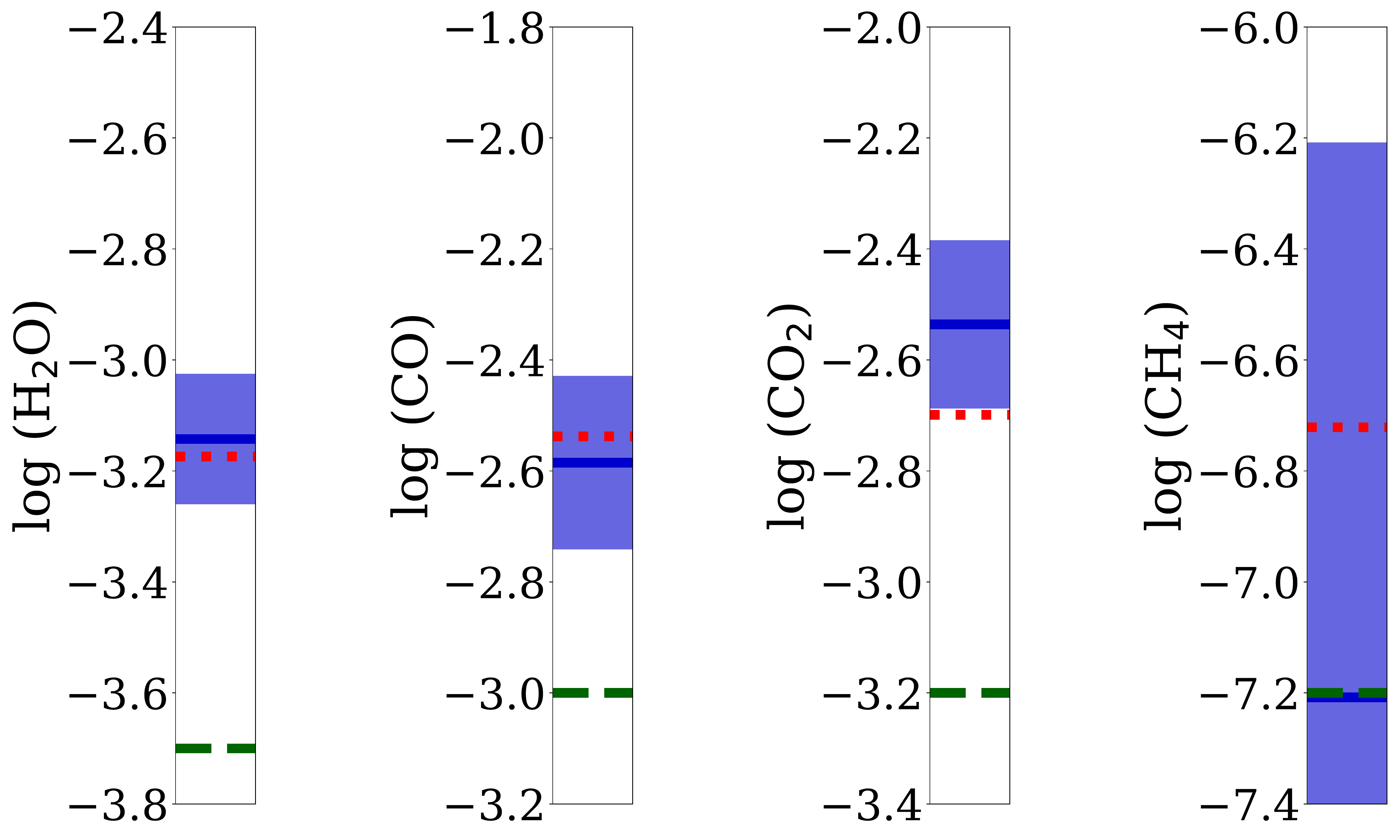}
}
\caption{\label{fig:snr10-r100k-2reg}
Retrieved temperature and mixing ratios of four molecular species from a combination of different spectral regions with a reference region $2.28-2.38$~\mum.
The retrievals are shown for the case of S/N=$10$. The color-coding is the same as in Fig.~\ref{fig:snr10-r100k}.
}
\end{figure*}

As a next step we studied the scenario in which the retrievals are done from a combination of different
spectral regions. In particular, we investigated combinations of the $(2.28-2.38)$~\mum\, region with the others
because, as   noted in the previous section, this region is often used for the detection of molecular lines in atmospheres of HJs, 
and it also provides rich information about the altitude structure of their atmospheres. 
Figure~\ref{fig:snr10-r100k-2reg} compares the results of retrievals considering S/N=$10$ (the results for other S/N values are provided
in the online material, see Figs.~\ref{fig:snr5-r100k-2reg}, \ref{fig:snr25-r100k-2reg}, and \ref{fig:snr50-r100k-2reg}).
As expected, we find that a combination
of $(1.50-1.70)$+$(2.28-2.38)$~\mum\, regions is one of the best in terms of retrieved information. 
When analyzed separately both these regions already constrain accurate temperatures, and combining them helps to improve our retrievals even further.
On the other hand, the combination of $(2.28-2.38)$+$(4.80-5.00)$~\mum\, gives us the possibility to probe even
higher altitudes up to $P=10^{-6}$~bar, but the concentration of \co\, is then derived less accurately compared to the previous case
with S/N=$10$, and noticeably worse with S/N=$5$ (Fig.~\ref{fig:snr5-r100k-2reg}).
In general, retrievals from two different spectral regions help to constrain molecular number densities better. 
For the same case of $(1.50-1.70)$+$(2.28-2.38)$~\mum\ we can now pin down
the mixing ratio of \co\ and \coo\ by a factor of about two compared to the case when either of these regions is used
(compare the VMR uncertainties in the top panel of Fig.~\ref{fig:snr10-r100k} and Fig.~\ref{fig:snr10-r100k-2reg}).
The typical temperature errors in the pressure range between $1$~bar and $10^{-4}$~bar is $100$~K for S/N=$5$,
$90$~K for S/N=$10$, $80$~K for S/N=$25$, and $70$~K for S/N=50.
We conclude that retrievals from two different spectral regions
should be preferred in order to constrain molecular number densities as accurately as possible compared to single-band retrievals,
but such observations would require two times longer integration times for the same noise levels.

Finally, simultaneous retrievals from three and four spectral windows (with a maximum wavelength coverage of about $1$~\mum)
did not significantly improve our results for the temperature and mixing ratios compared to the retrievals from two spectral ranges.
Thus, the only benefit of observing at many different spectral regions is to measure molecular number densities of as many molecules as possible.
However, enough lines of at least three molecules that we considered (\hho, \co, \coo) can be captured already 
with two cross-disperser settings of \criresplus. Thus, including additional spectral windows is not justified
for the case of atmospheric retrievals.

\section{Suggested observation strategies and potential targets}

Our analysis shows that with a single setting, \criresplus\,  covers a spectral range
containing enough features for  exoplanet atmospheric studies. However, not all the spectral regions that we considered in this study 
are equally good for the retrieval of temperature and molecular concentrations. In the case of a single observation setting, 
accurately determining the temperature and mixing ratios depends only on the S/N of the spectrum of an exoplanetary atmosphere. Our simulations showed that
under the same observing conditions the retrievals of accurate mixing ratios will generally require higher S/N compared to the case when
only temperature stratification is investigated. For instance, with S/N=$5$ it is possible to derive accurate temperatures
in the line forming region of an HJ atmosphere ($P>1$~bar), while realistic values of  the  
\hho\ mixing ratio, for example,  can only be obtained with two times higher S/N (see the results for the $(2.28-2.38)$~\mum\ region 
in Figs.~\ref{fig:snr10-r100k} and~\ref{fig:snr5-r100k}). 

In order to retrieve accurate temperature stratification and mixing ratios of \hho, \co, and \coo, 
the simultaneous observations at two spectral windows $(1.50-1.70)$+$(2.28-2.38)$~\mum\ and S/N=$10$ should be used.
As a next choice the $(2.10-2.28)$+$(2.28-2.38)$~\mum\ or $(2.28-2.38)$+$(4.80-5.00)$~\mum\ regions can also be used for robust retrievals.
We note that the $M$ band contains many more telluric lines and the performance of the old CRIRES in that region
was noticeably worse compared to the performance in  $K$ band.

When only S/N=$5$ could be reached, the choice of a single observation setting is not obvious, and there are two options.
On the one hand, using $(1.50-1.70)$~\mum\ region could provide VMRs of \hho, \co, and \coo, but accurate temperatures
could only be retrieved up to $10^{-3}$~bar. On the other hand, the $(2.28-2.38)$~\mum\ region would still provide
accurate measurements of \hho\ and temperatures up to $10^{-4}$~bar, but less accurate \co\ and no constraints for \coo\ (see Fig.~\ref{fig:snr5-r100k}).
Furthermore, the planet-to-star flux contrast in the $(1.50-1.70)$~\mum\ wavelengths is expected to be lower in HJs
compared to the $(2.28-2.38)$~\mum\ region, thus requiring longer integration times for the same S/N.

Observing in two spectral regions with S/N=$5$ in each of them, we still favor the combination of $(1.50-1.70)$+$(2.28-2.38)$~\mum.
However, for the same integration time, we suggest   observing only the $(2.28-2.38)$~\mum\ region obtaining higher S/N=$\sqrt{2}\cdot5=7$, 
which  allows more precise measurements of \hho\ and \co.

In order to provide estimates for the exposure times needed to study atmospheres of HJs we used predictions
of our models and the ESO Exposure Time Calculator (ETC) for CRIRES\footnote{\tt http://etimecalret-p95-2.eso.org/observing/etc/bin/gen/form?INS.NAME=CRIRES+INS.MODE=swspectr}. 
Because there is no ETC for \criresplus\ available yet, 
we note that the listed exposure times are upper limits due to the expected improved performance of the new instrument compared to the old one.
From the list of known HJs we selected several of the best targets that can be  observed with high-resolution ground-based spectroscopy
similar to the capabilities of \criresplus. We additionally restricted the list
of objects by including only those that require less than 100~h of integration time.
We list these targets in Table~\ref{tab:obs}. 
The integration times were computed using stellar and planetary parameters listed in the current version 
of the Extrasolar Planets Encyclopaedia\footnote{\tt http://exoplanet.eu/}
and The Exoplanet Orbit Database\footnote{\tt http://exoplanets.org/}. 
The stellar magnitudes in the $V$ and $K$ bands were extracted from
the SIMBAD database\footnote{\tt http://simbad.u-strasbg.fr/simbad/sim-fid}.
The planetary equilibrium temperatures
were computed assuming atmospheric albedo $a=0.03$ expected for HJs \citep{2000ApJ...538..885S}.
We calculated planet-to-star flux ratios as an average between $1$~\mum\ and $5$~\mum\ assuming
a blackbody approximation for stellar and planetary flux. This values will be larger in the red part of the spectrum
(typically by a factor of two), and thus the corresponding exposure times that we list in the table will be shorter for the same S/N
at the red end of \criresplus\ spectra.
However, the performance of old CRIRES detectors in the $M$ band was noticeably lower compared to  the $K$ band
so that the estimation of realistic exposure times for \criresplus\ at different spectral bands is currently not possible.
This is why we provide exposure times for the mean flux contrast and use one of the most efficient instrument settings 
at the echelle order No.25 ($\lambda_{\rm eff}=2.267$~\mum). We note that the S/N for the planetary spectrum
is defined by the planet-to-star flux ratio S/N(planet)=F$_{\rm p}$/F$_{\rm s}$$\cdot$S/N(star).

Due to the demand for very high-quality exoplanetary spectra
the number of potential targets is very small. Among them, it will be possible to observe only
two HJs from the ESO Paranal observatory and with affordable integration times.
These exoplanets are 51~Peg~b and $\tau$~Boo~b. 
For 51~Peg~b, a little more than one full night (e.g., 12~h) will be needed to reach S/N=$5$ using two instrument settings
or S/N=$7$ in one setting. This will be sufficient for accurate atmospheric inversions. 
For our preferable configuration of S/N=$10$ and two settings a 48~h  integration time will be needed.
In case of $\tau$~Boo~b two full nights will be needed to reach S/N=$5$ in two settings, or eight nights
for  S/N=$10$ and two settings. The choice of the corresponding instrument settings
will depend on the performance of the \criresplus\ at different infrared bands. 
Our simulations favors the ($1.50-1.70$)+($2.28-2.38$)~\mum\ region because the VMR of three molecules
can be accurately constrained, as can the  temperature distribution, until $P<10^{-4}$~bar.
If one wants to look at higher altitudes then the ($2.28-2.38$)+($4.80-5.00$)~\mum\ region should be used,
but the uncertainties on   \co\ will be slightly larger.
In the case of a single setting we suggest the ($2.28-2.38$)~\mum\ region; however, the ($1.50-1.70$)~\mum\ region
can also be an option if one wants to constrain the VMRs of as many molecules as possible.
Should the sensitivity of \criresplus\ be considerably improved, then the integration times
listed in Table~\ref{tab:obs} would be smaller, thus favoring a detailed analysis 
of the atmosphere of 51~Peg~b with two setting observations.

\input{tables/obs.tex}

\section{Discussion}

In our analysis of high-resolution simulated observations we used the OE method
for the retrieval of atmospheric parameters, such as temperature and molecular mixing ratios.
We did not attempt to utilize other methods (e.g., MCMC) primarily
because the number of free parameters in our model is high. In this particular case the OE method
has the advantage of being relatively fast but still robust. 
It should be noted that retrieving a detailed structure of planetary atmospheres
relies primarily on the data quality and the presence of lines that probe different 
atmospheric altitudes. In cases when the temperature distribution is expected to be 
very smooth, or when the quality of the spectrum is poor so that the altitude structure
could not be satisfactorily resolved, it is possible to parameterize altitude-dependent 
quantities with simple functions containing only few free parameters.
For example, we could have parameterized the temperature distribution using
a commonly used profile after,  \citet{2014A&A...562A.133P}, among others, which requires only four free parameters
instead of retrieving a temperature value at each atmospheric layer. This would allow us
to use a standard Bayesian parameter search. However, in this work we decided to carry out
assumption-free retrievals.
Hence, our application of the OE method should not be considered as the only suggested approach to the atmospheric
retrievals at high spectral resolution, but rather as one possibility among others, with the ability to retrieve
accurate parameters with a proper application.

Near-future instruments will deliver data of much better quality and wider spectral range coverage compared to what is available now.
That means that at high spectral resolution the shapes of spectroscopic lines originating from the planetary atmospheres
could be studied directly. This was not possible to achieve before, and the   high-resolution spectroscopy of exoplanets
was pushed forward mainly by the detection of molecular species via the cross-correlation technique. This was proven
to be very efficient in disentangling the stellar and planetary fluxes from the noisy observations that are additionally
contaminated by telluric absorption. Recently, \citet{2019AJ....157..114B} made another step forward and suggested
the attractive approach of using cross-correlation technique for the retrieval of atmospheric properties.
This is done by the mapping of the cross-correlation coefficients to log-likelihood. After this mapping was performed, 
the standard Bayesian methods could be used to derive the best fit parameters and their errors. The choice of the function that maps
cross-correlation coefficients to the log-likelihood space is arbitrary, but should satisfy some basic
criteria to make retrievals unambiguous \citep[see][for more details]{2019AJ....157..114B}.
The new mapping proposed by \citet{2019AJ....157..114B} has certain benefits. 
For instance, it can be applied to already existing data when the planetary lines 
are drawn in the  noise-dominated spectra.
%At the same time, the degeneracy between fit parameters can not be avoided, too, simply because
%the qiality of the present data is low (e.g., the degeneracy between parameters that describe
%temperature stratification). 

Mapping of the cross-correlation coefficient and the OE method
both rely on the presence of hundreds or even thousands of atmospheric lines, 
although the former method works with the residual spectra, while our
implementation of the OE method works with spectra in arbitrary units and the corresponding flux scaling
is a free parameter(s) of the model that could be optimized simultaneously with other (atmospheric) parameters.
In our particular case the application of the new cross-correlation mapping 
was not strongly justified because the number of free parameters in our model is too high, making
Bayesian post-processing very time consuming.
In addition, we note that with the sufficiently high  S/N values of the planetary spectra that are expected to be obtained
with future instruments, the profiles of individual planetary lines could be satisfactorily resolved
and directly analyzed. This will likely make the use of any cross-correlation mapping unnecessary favoring
other approaches. We note that cross-correlation itself will   still be used as a powerful tool 
for disentangling stellar and planetary spectra.

In this work we analyzed five different spectral regions between $1$~\mum\ and $5$~\mum. In these regions 
the telluric absorption is expected to be relatively weak compared to the rest of the spectrum.
In addition, these regions contain strong bands of molecular species that we investigated (\hho, \co, \coo, \chhhh).
Another  region where it is possible to study planetary atmospheres that we did not consider in this study is the one around $3.5$~\mum, which
is dominated by numerous lines of \chhhh\ and \hho.
We carried out several additional retrievals using the $3.50-3.75$~\mum\ spectral region, but
retrieved much less information compared to the other four regions. We conclude that although this region contains numerous
molecular lines that can be easily detected from the usual cross-correlation analysis as shown in   \citet{2014A&A...561A.150D}, among others, 
all these lines are strongly blended with each other and are not favorable for retrieving accurate molecular number densities.

In this work we did not account for the instrumental effects such as  flat-fielding, order merging, or wavelength calibration.
Some of these effects may influence the shape of the observed spectrum (which is a common problem for every echelle spectrograph) 
and therefore have direct impact on the retrieved parameters.
Unfortunately, at this moment we cannot quantify these effects for \criresplus\ because many parameters of the instrument are still
under investigation and the actual values will only be known once the instrument has been tested against real observations.
Obtaining the best quality planetary spectrum  requires a very accurate removal of instrument effects
corresponding to the ratio between stellar and planetary continuum fluxes, which is about $\approx10^{-3}$ for known HJs. 
Obviously, this accuracy will be very hard to achieve. 
On the other hand, the global shape of individual molecular bands will likely be preserved. In this case only arbitrary flux normalization
to each spectral region under investigation will be needed (as we did in this study) in order to carry out 
accurate retrievals. Such a normalization ideally keeps the ratio of the planetary (pseudo)continuum-to-line
depth unchanged even if the information regarding the true stellar continuum was lost during the data processing stage.
For instance, in the wavelength domain covered by \criresplus, the stellar continuum is very smooth.
Therefore, instead of using simple continuum scaling, we also attempted to fit low-order polynomials to the simulated observations 
in each spectral region and did the same to the predicted spectra at each iteration in our retrievals. 
By doing this exercise we found  that the choice of normalization function does not
affect the final result very much. The temperature and mixing ratios varied from retrieval to retrieval, but always within the error bars.
This exercise is very simplistic and cannot reflect all the  details of the data reduction process
planned for \criresplus\ and should not be considered as an approximation for  reality.
Nevertheless, it tells us that information about the true stellar continuum is not  likely  to be  a major obstacle  toward
atmospheric retrievals as long as the relative shape of spectral lines is preserved.
We note that in our approach the pseudo-continuum could be fit using low-order polynomials to be found during the retrieval process.
This leaves the S/N of the planetary spectrum as the main limitation for atmospheric retrievals. 
Thus, accurate retrievals can still be performed provided that the order merging within a single \criresplus\ setting
is accurately done by the pipeline of the instrument. We plan to improve our simulations by considering
realistic instrumental effects in our next work.

Even if all the instrumental effects are properly taken into account, the telluric absorption represents
the next obvious problem. Telluric lines are present in all near-infrared bands covered by \criresplus. 
Even in the five carefully chosen regions that we investigated in this paper, many of the telluric lines can be strong depending on
local weather conditions.
Normally, moderately strong telluric lines can be successfully removed 
\citep[e.g.,][]{2018NatAs...2..714Y,2016ApJ...817..106B}, 
while spectral regions containing the strongest telluric lines are excluded from the analysis. 
However, in cases when telluric lines need to be removed, the accuracy of this procedure should  again be very high,  about $\approx10^{-3}$ or better.
More importantly, there might be numerous very weak telluric lines, with intensities of just a fraction of a percent relative to the stellar continuum,
that are unaccounted for in existing telluric absorption models and that could potentially distort the shape of planetary lines.
We note that this is of minor importance if only a detection of exoplanetary lines using cross-correlation technique is needed.
In that case, the amplitude of the cross-correlation function depends on  matching the position of exoplanet lines
against a chosen template spectrum. Thus, even if the depth of exoplanetary lines is affected by unremoved telluric absorption,
it will not hide the detection of the molecular species. However, it could still be an issue when making retrievals from those distorted
planetary lines. The overall impact of these very weak telluric lines should be tested with real observations.
One obvious advantage of \criresplus\ is the ability to observe many molecular lines simultaneously, which will surely help to reduce
the impact of telluric absorption in atmospheric retrievals.

In our simulated observations we used mixing ratios of four molecules (\hho, \co, \coo, \chhhh) as derived in \citet{2012MNRAS.420..170L}.
Our retrievals showed that we can derive number densities of \hho, \co, and \coo\ by using different
spectral regions. At the same time, we could not constrain a mixing ratio of \chhhh\ because the assumed mixing ratio
is very low ($1.9\times10^{-7}$) and the lines of \chhhh\ are too weak compared to the lines of other three molecules. 
This is in agreement with the predictions of equilibrium chemical calculations \citep[e.g.,][]{2017MNRAS.472.2334G,2017AJ....153...56M,2015ApJ...813...47M}.
We note that \chhhh\ has a very rich spectrum in the near-infrared with many bands present in the wavelength range of \criresplus.
In atmospheres of HJs with equilibrium temperatures below $1000$~K or with C/O>1, the lines of \chhhh\ should become strong
enough for the purpose of abundance analysis. As a test case we simulated an additional set of observations with
the increased methane mixing ratio of \chhhh=$5\times10^{-4}$ and performed retrievals from the
($3.80-4.00$)~\mum\ region and from the combined ($2.28-2.38$)+($3.80-4.00$)~\mum\ region. 
In both cases we were able to obtain definite constraints on the methane mixing ratio, with slightly more accurate values when derived
from the the combined ($2.28-2.38$)+($3.80-4.00$)~\mum\ region, just as expected.
Thus, using \criresplus\ it will also be possible to constrain methane concentration in planets with cooler
atmospheric temperatures than those considered here.

\section{Summary}

High-resolution spectroscopy in the near-infrared is a powerful tool for the  study of absorption and emission
lines to estimate the state of extra-solar atmospheres. Until now, it was mainly used for the detection of different molecular species
via   transit spectroscopy. In this work we made another step forward and investigated the potential of
high-resolution spectroscopy to study molecular number densities and temperature stratification in exoplanet atmospheres
by applying the optimal estimation method to the simulated observations at different near-infrared bands.
We used a modified \taurex\, forward model to simulate emission (i.e., out-of-transit) 
spectra for different molecular mixing ratios, spectral resolutions,
and S/N values. The results of our investigation are summarized below.

\begin{itemize}
\item
The spectra of HJs in the wavelength range of \criresplus\, is sensitive to a very wide range 
of atmospheric pressures between $1$~bar and $10^{-6}$~bar.
\item
Avoiding regions with strong telluric contamination, the best region for deriving mixing ratios of \hho, \co\ and \coo\ is around $1.6$~\mum, 
and the best region for studying temperature stratification and mixing ratios of \hho\ and \co\ is around $2.3$~\mum.
The mixing ratio of \chhhh\, is impossible to constrain accurately in any wavelength range and at the S/N that we considered,
due to the very low number density that we assumed in our synthetic observations. It can be retrieved though, for example 
in  cooler exoplanets that may have a \chhhh\ mixing ratio of $\approx10^{-4}-10^{-3}$.
\item
Retrieving mixing ratios simultaneously from two separate spectral regions helps to obtain accurate results for \hho, \co, and \coo. 
In this regard the combination of the $1.6$~\mum\ and $2.3$~\mum\ regions looks promising at any S/N, and the combination
of $2.3$~\mum\ with any other region except $3.9$\mum\ when S/N$\geqslant$10.
However, the retrieval of accurate number densities
always requires higher S/N compared to the S/N required for the temperature retrievals. 
The latter can be accurately retrieved even with the lowest S/N$=5$ that we assumed in our simulations.
\item
In the case of single setting retrievals, observing at the $2.3$~\mum\ region
provides accurate VMRs of \hho\ and \co\ only if S/N$\geqslant$10.
When S/N$\geqslant$25 the $1.6$~\mum\ region provides very accurate VMRs of \hho, \co, and  \coo.
However, to achieve such high S/N values would exceed typical allocated telescope observational times. 
The best strategy for obtaining reliable mixing ratios of as many molecules as possible 
is thus to observe at two different infrared bands, at $2.3$~\mum\ and at any of the others except the one at $3.9$~\mum.
\item
Degrading spectral resolution results in decreasing the sensitivity of our retrievals of molecular concentrations,
but the temperature distribution can still be accurately derived. Rebinning the spectra to a lower resolution could
help to increase S/N and to obtain estimates on the molecular number densities and atmospheric temperature structure
in cases when the S/N of the original data is not sufficient (e.g., S/N$<$5).
%\item
%The spectra in the chosen ranges does not allow to measure abundance of \chhhh\, with any SNR tried. We attribute this to a very low
%\chhhh\, mixing ratio assumed in our simulations and degeneracy between temperature and abundance retriavals
%which was already mentioned in the study by \citet{2012MNRAS.420..170L}.
\item
With \criresplus\, it will be possible to carry out detailed atmospheric retrievals in two HJs known so far.
Planet-hunting missions such as TESS\footnote{\tt https://heasarc.gsfc.nasa.gov/docs/tess/}, 
PLATO\footnote{\tt http://sci.esa.int/plato/}, and CHEOPS\footnote{\tt http://sci.esa.int/cheops/} 
are expected to significantly increase the number of exoplanets accessible
for current and future ground-based spectroscopic studies.
\end{itemize}

Our study suggests that even though the spectra of exoplanets will likely be obtained with very high noise levels,
the atmospheric retrievals can still benefit from numerous molecular lines observed thanks to the wide wavelength coverage
of \criresplus. Moreover, additional important constraints can be provided by utilizing low-resolution data \citep{2017ApJ...839L...2B}.
This data is already available for a number of HJs and could be used to break the degeneracy between the retrieved parameters.
Thus, in our future work we plan to make another step forward and create a suite for the simultaneous retrieval of atmospheric parameters
from low- and high-resolution data.

\begin{acknowledgement}
We wish to thank L.~Fletcher for providing us with the data of HD~189733~b.
We acknowledge the support of the DFG priority program SPP-1992 “Exploring the Diversity of ExtrasolarPlanets” (DFG PR 36 24602/41).
This work made use of the GWDG IT infrastructure. 
The GWDG is a shared corporate facility of the Georg-August-University G\"ottingen and the nonprofit Max-Planck Society 
(Max-Planck Gesellschaft zur F\"orderung der Wissenschaften e. V.).
We also acknowledge the use of electronic databases SIMBAD and NASA's ADS.
\end{acknowledgement}

\bibliographystyle{aa}
\bibliography{biblio}

\begin{appendix}

\section{Retrievals of temperature and mixing ratios from high-resolution spectroscopy}

\begin{figure*}
\centerline{
\includegraphics[height=0.11\vsize]{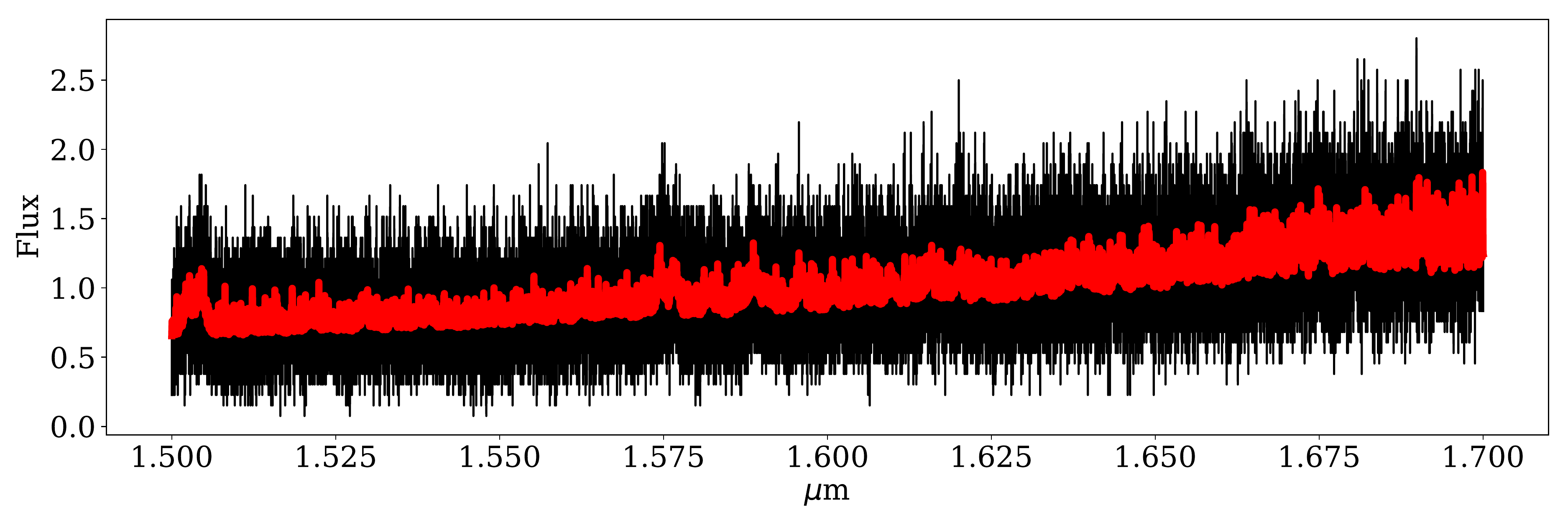}
\hspace{0.02\hsize}
\includegraphics[trim={0 0 20.0cm 0},clip,height=0.11\vsize]{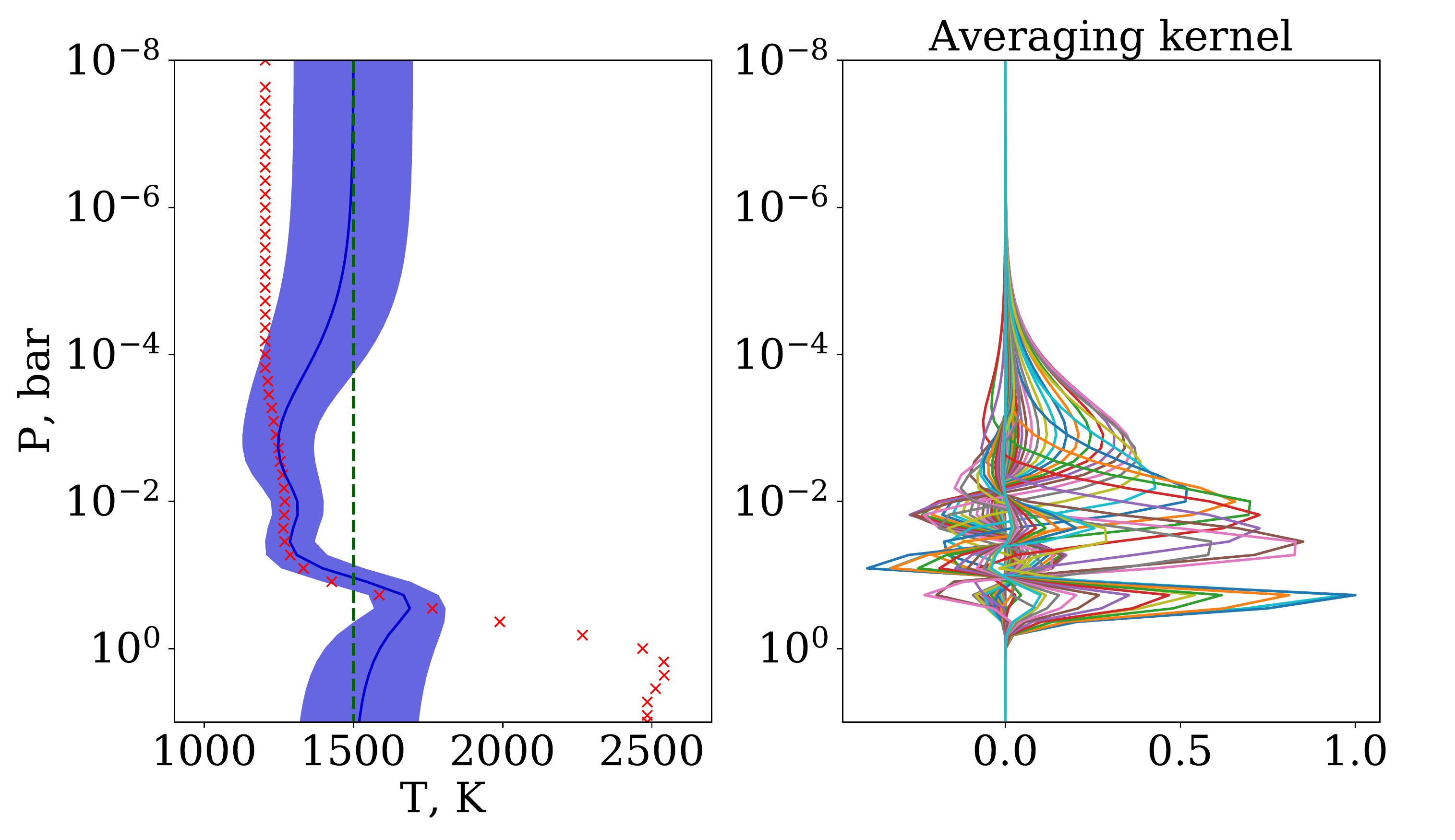}
\hspace{0.02\hsize}
\includegraphics[height=0.11\vsize]{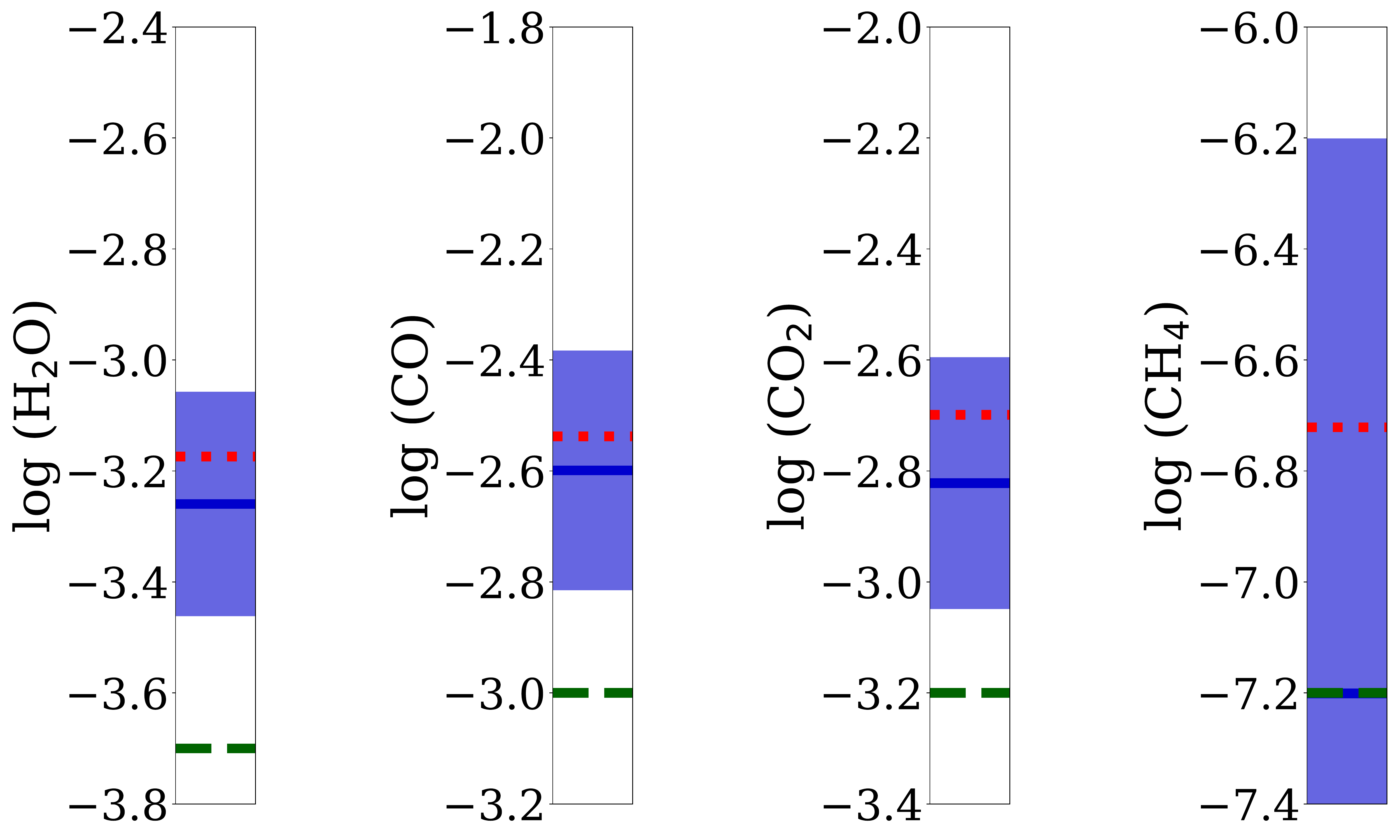}
}
\centerline{
\includegraphics[height=0.11\vsize]{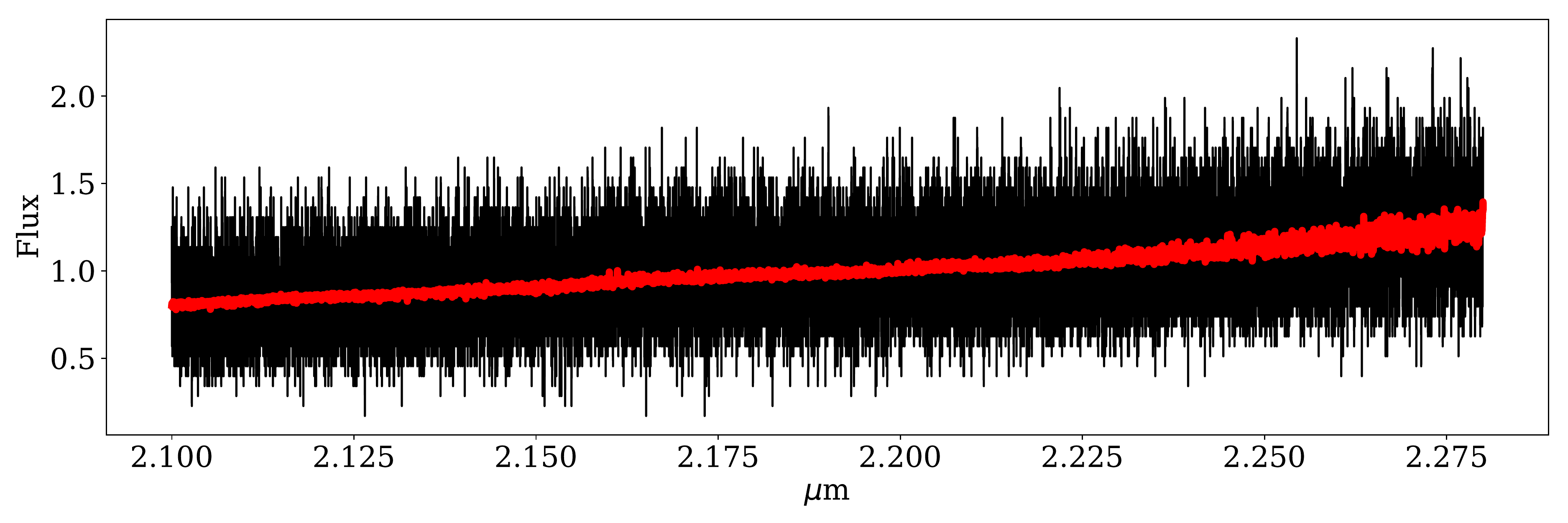}
\hspace{0.02\hsize}
\includegraphics[trim={0 0 20.0cm 0},clip,height=0.11\vsize]{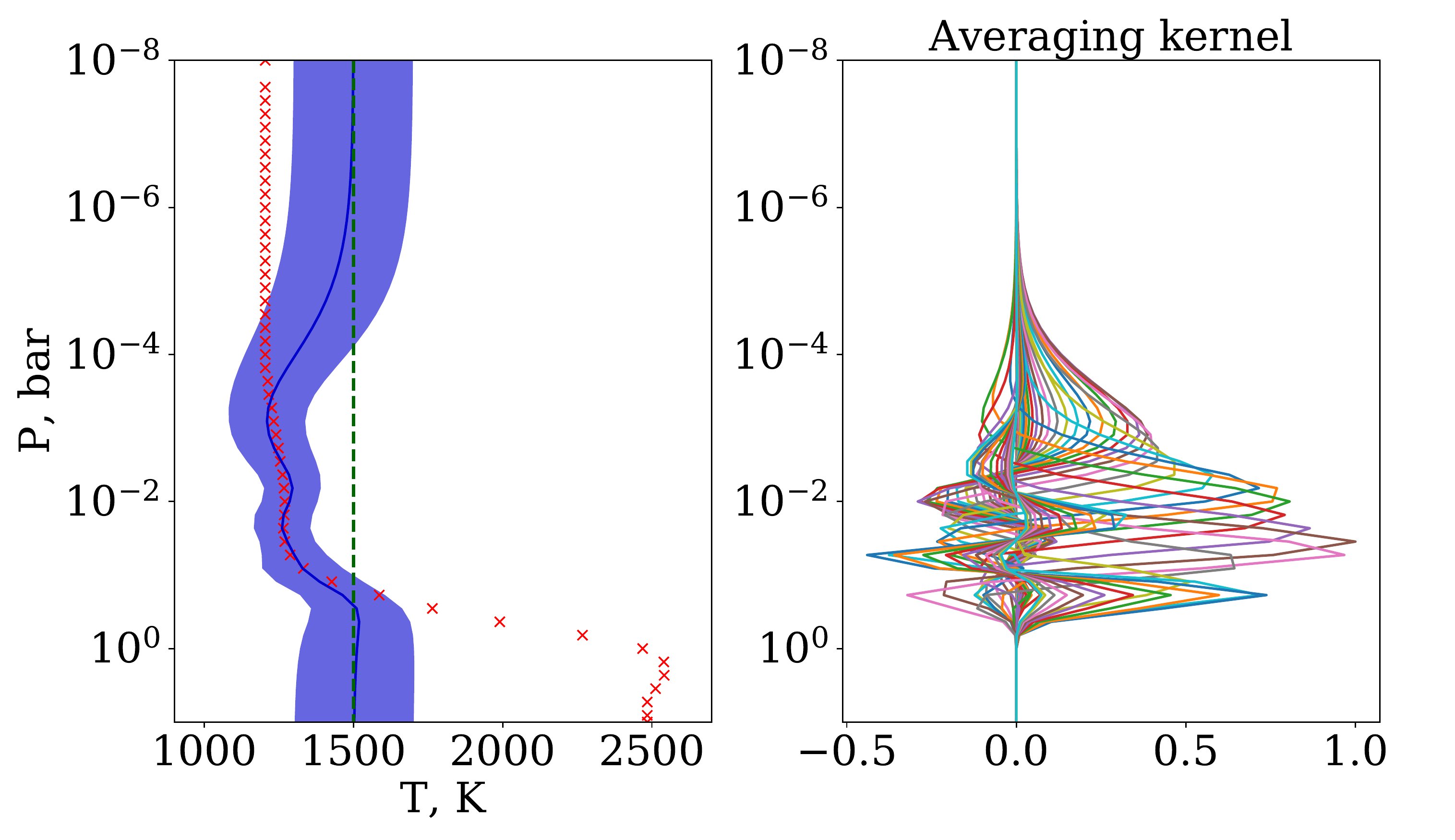}
\hspace{0.02\hsize}
\includegraphics[height=0.11\vsize]{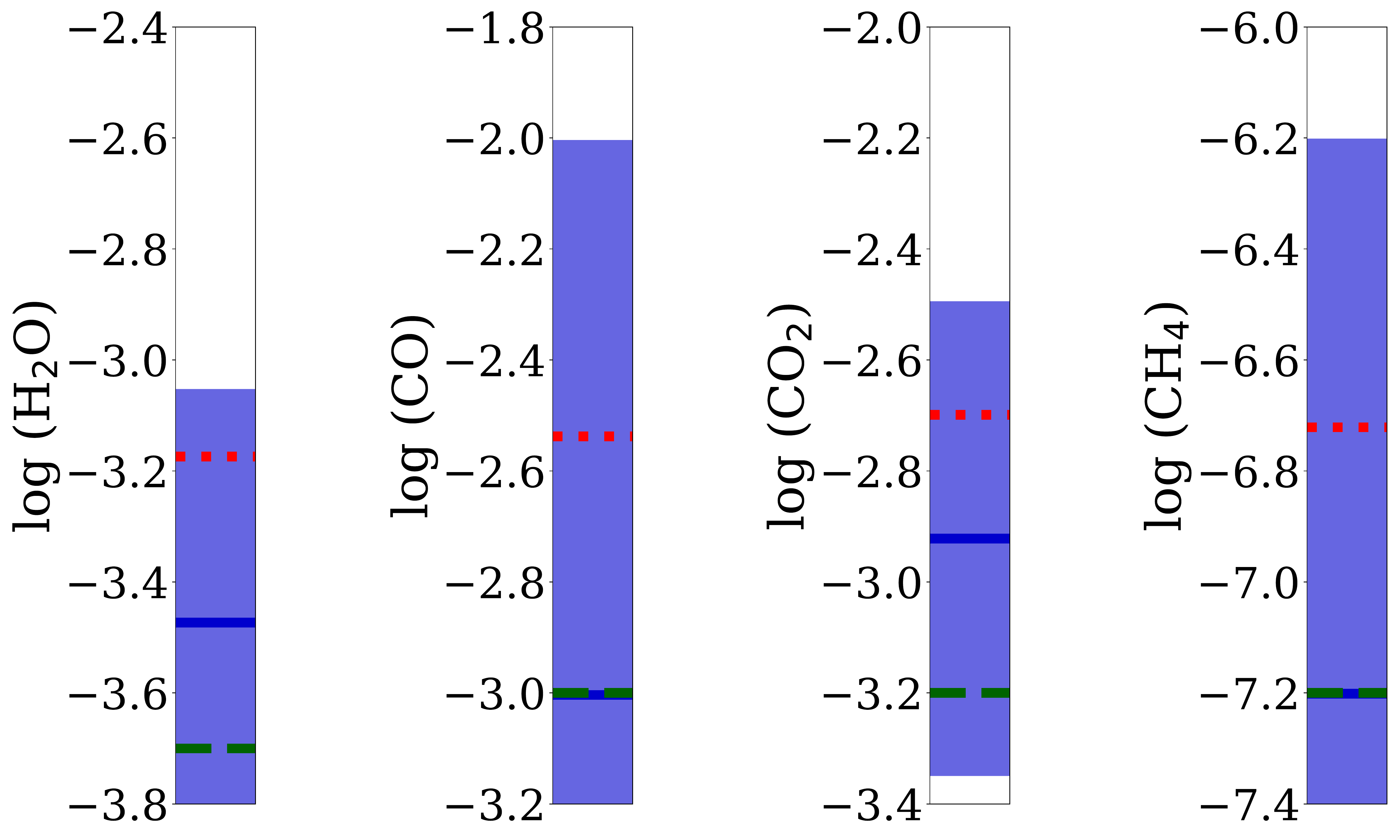}
}
\centerline{
\includegraphics[height=0.11\vsize]{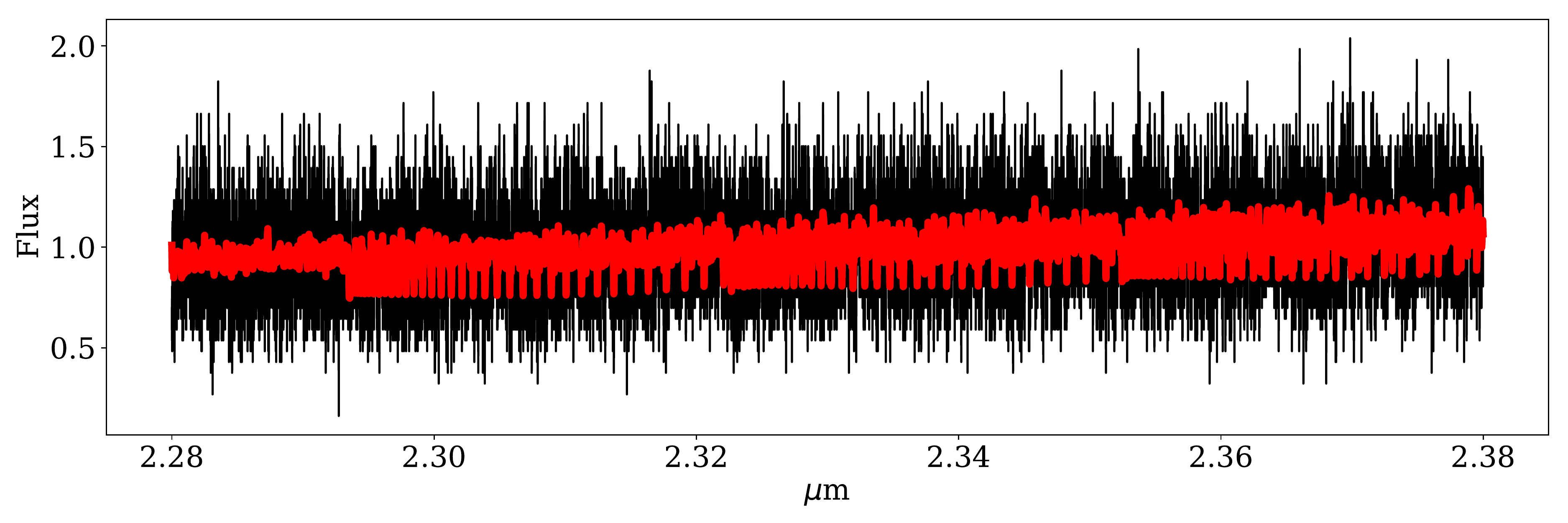}
\hspace{0.02\hsize}
\includegraphics[trim={0 0 20.0cm 0},clip,height=0.11\vsize]{figures/r100k/snr5/h2o+co+ch4+co2/ll-2.28-2.38/out-vmr_init_m0.5dex-vmr_err_1.0dex.bwd_param1.pdf}
\hspace{0.02\hsize}
\includegraphics[height=0.11\vsize]{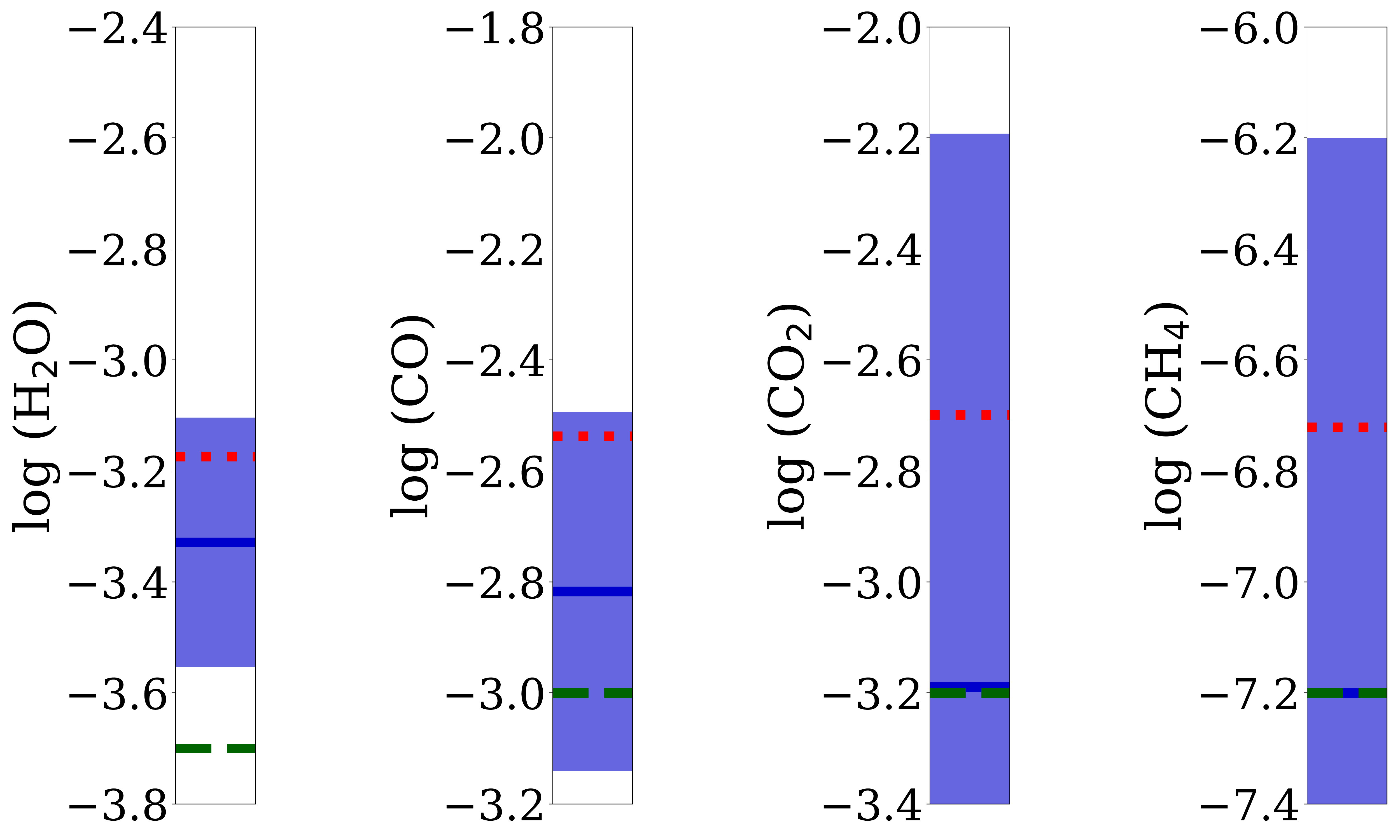}
}
\centerline{
\includegraphics[height=0.11\vsize]{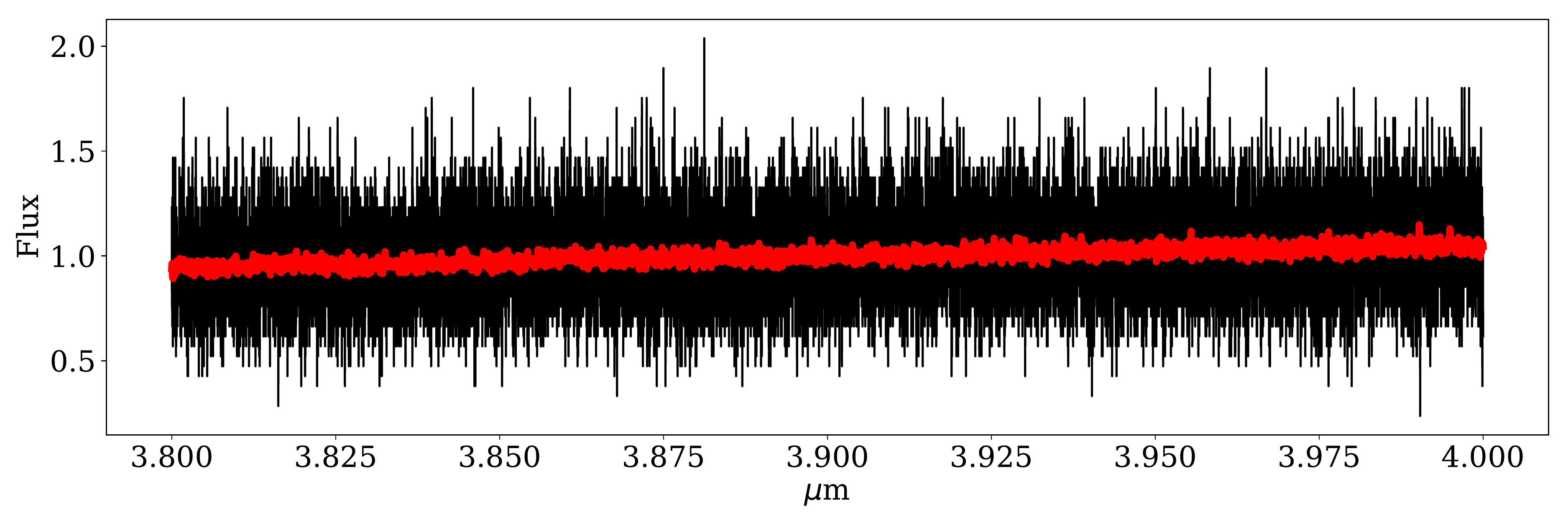}
\hspace{0.02\hsize}
\includegraphics[trim={0 0 20.0cm 0},clip,height=0.11\vsize]{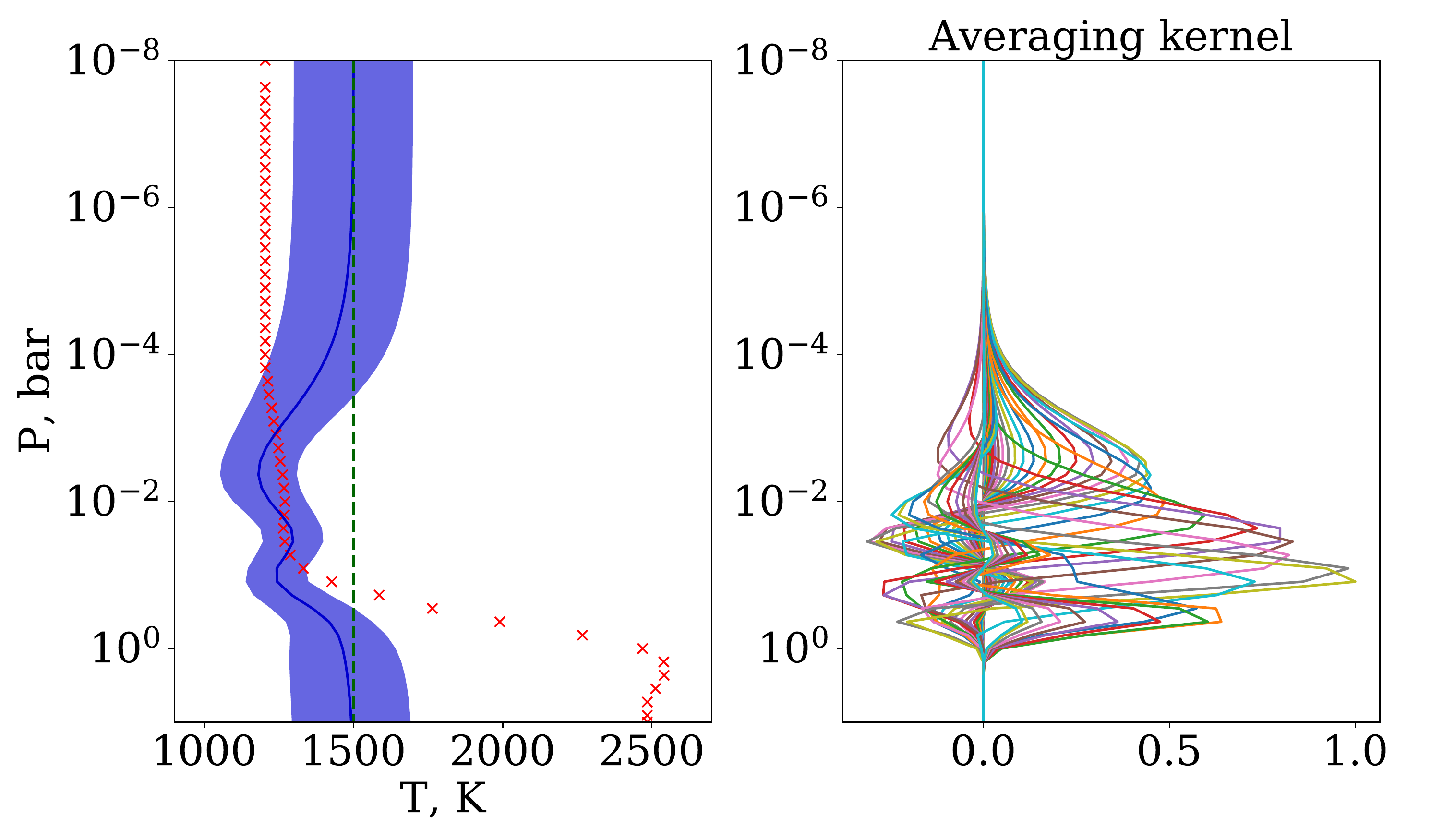}
\hspace{0.02\hsize}
\includegraphics[height=0.11\vsize]{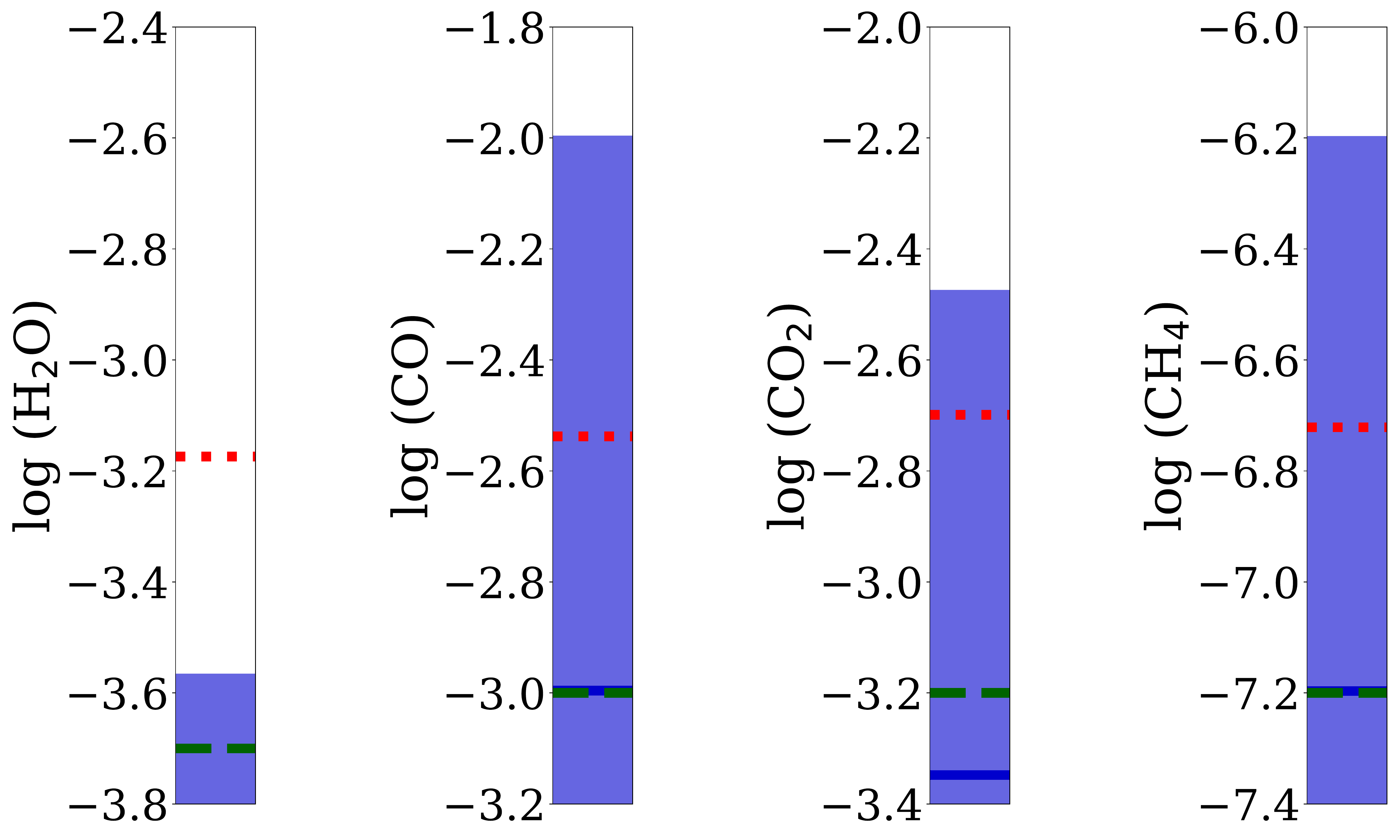}
}
\centerline{
\includegraphics[height=0.11\vsize]{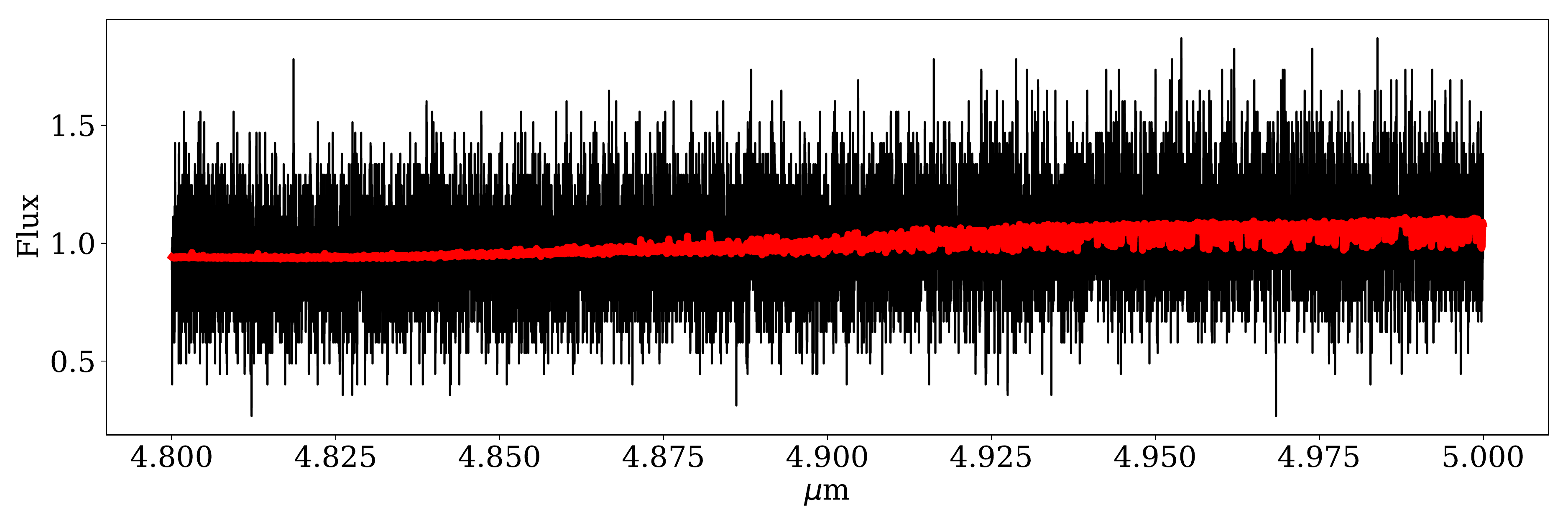}
\hspace{0.02\hsize}
\includegraphics[trim={0 0 20.0cm 0},clip,height=0.11\vsize]{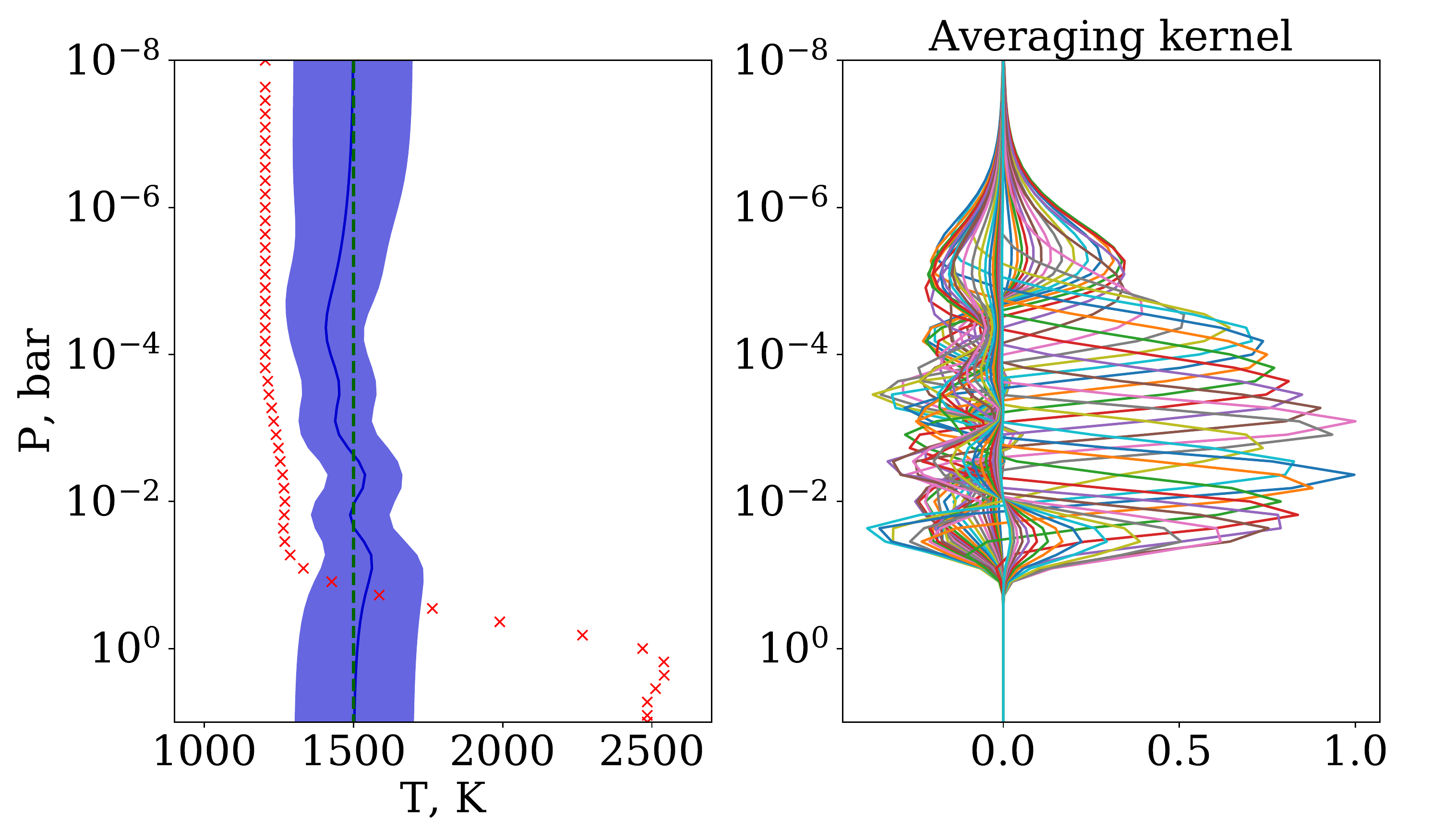}
\hspace{0.02\hsize}
\includegraphics[height=0.11\vsize]{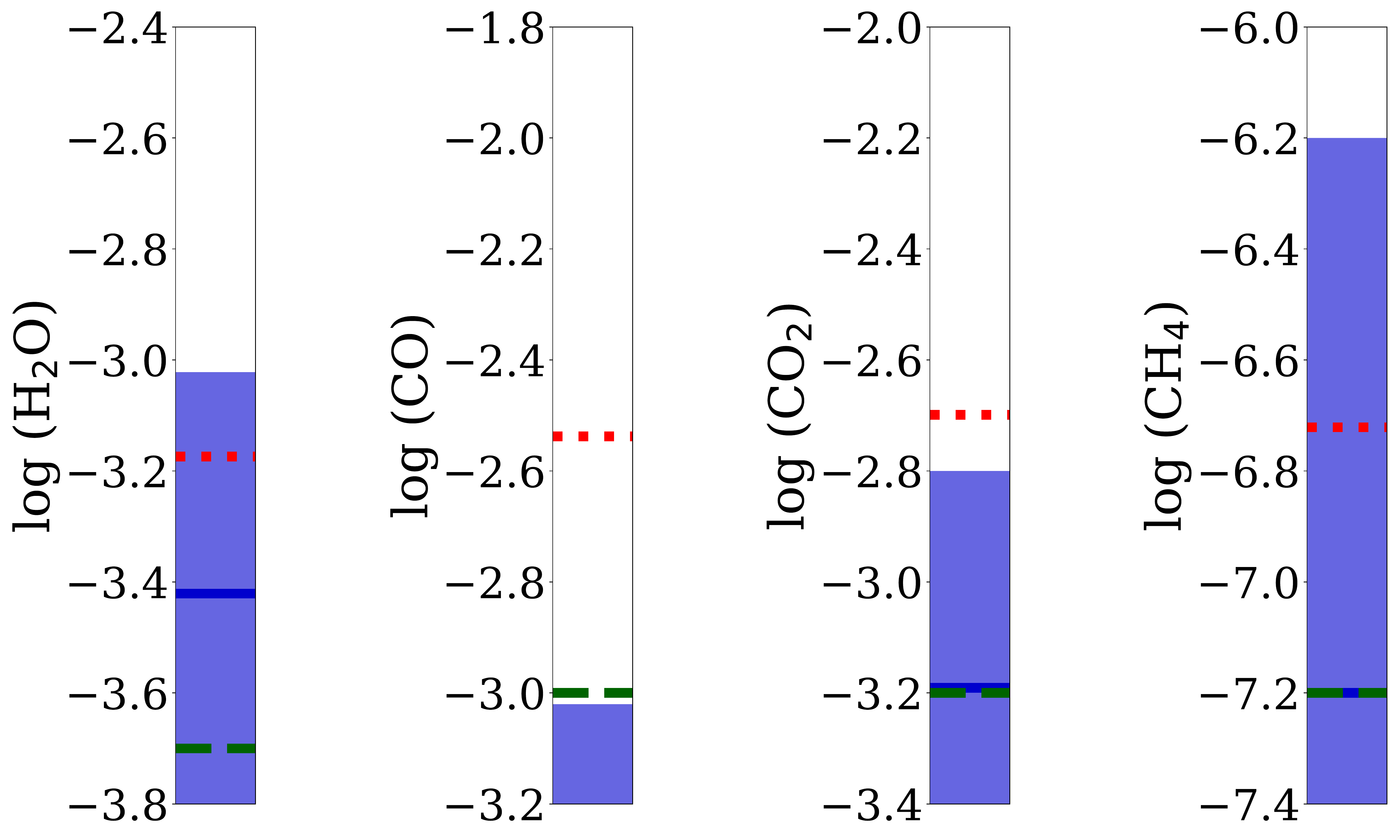}
}
\caption{\label{fig:snr5-r100k}
Retrieved temperature and mixing ratios from five spectral regions at spectral resolution R=$100\,000$ and S/N=$5$. In each panel the first plot is the 
temperature distribution as a function of atmospheric pressure (solid blue line) and with error bars shown as shaded area.
The red crosses and green dashed line are the true solution and initial guess, respectively. The second plot shows
averaging kernels derived for the temperature distribution. Here averaging kernels for different atmospheric depths are randomly
color-coded for better representation. The next four plots are the values of the retrieved mixing ratios of four
molecular species (color-coded   as in the first plot). The mixing ratios were assumed to be constant with 
atmospheric depth, and we show their values on the $y$-axis. The last plot compares the best fit predicted spectrum (solid red  line) with
the observed one (black circles).
}
\end{figure*}

\begin{figure*}
\centerline{
\includegraphics[height=0.11\vsize]{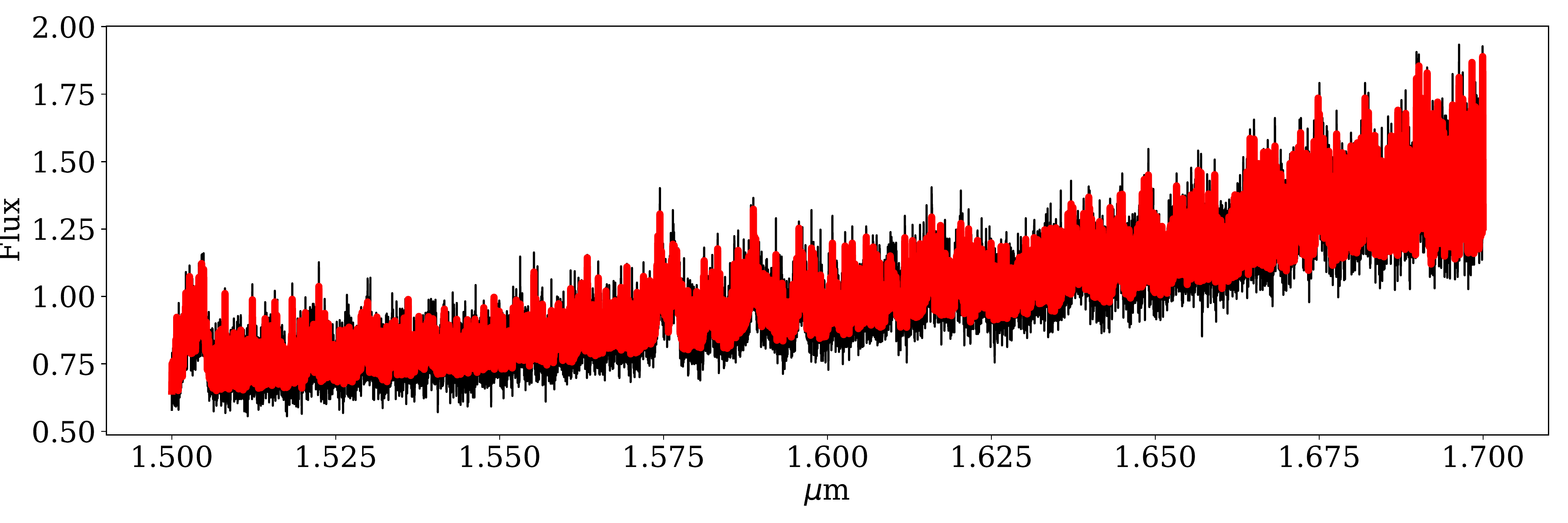}
\hspace{0.02\hsize}
\includegraphics[trim={0 0 20.0cm 0},clip,height=0.11\vsize]{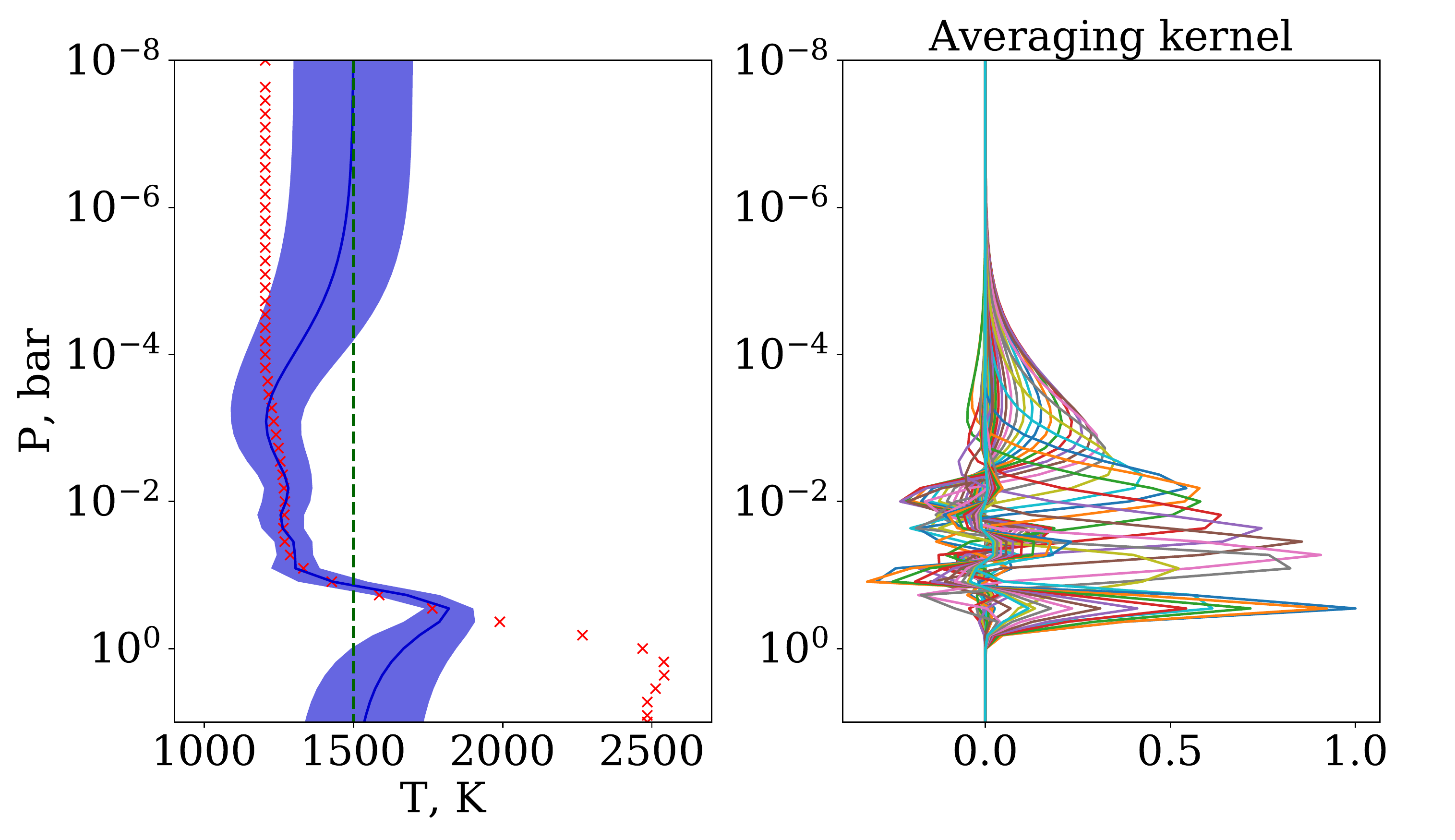}
\hspace{0.02\hsize}
\includegraphics[height=0.11\vsize]{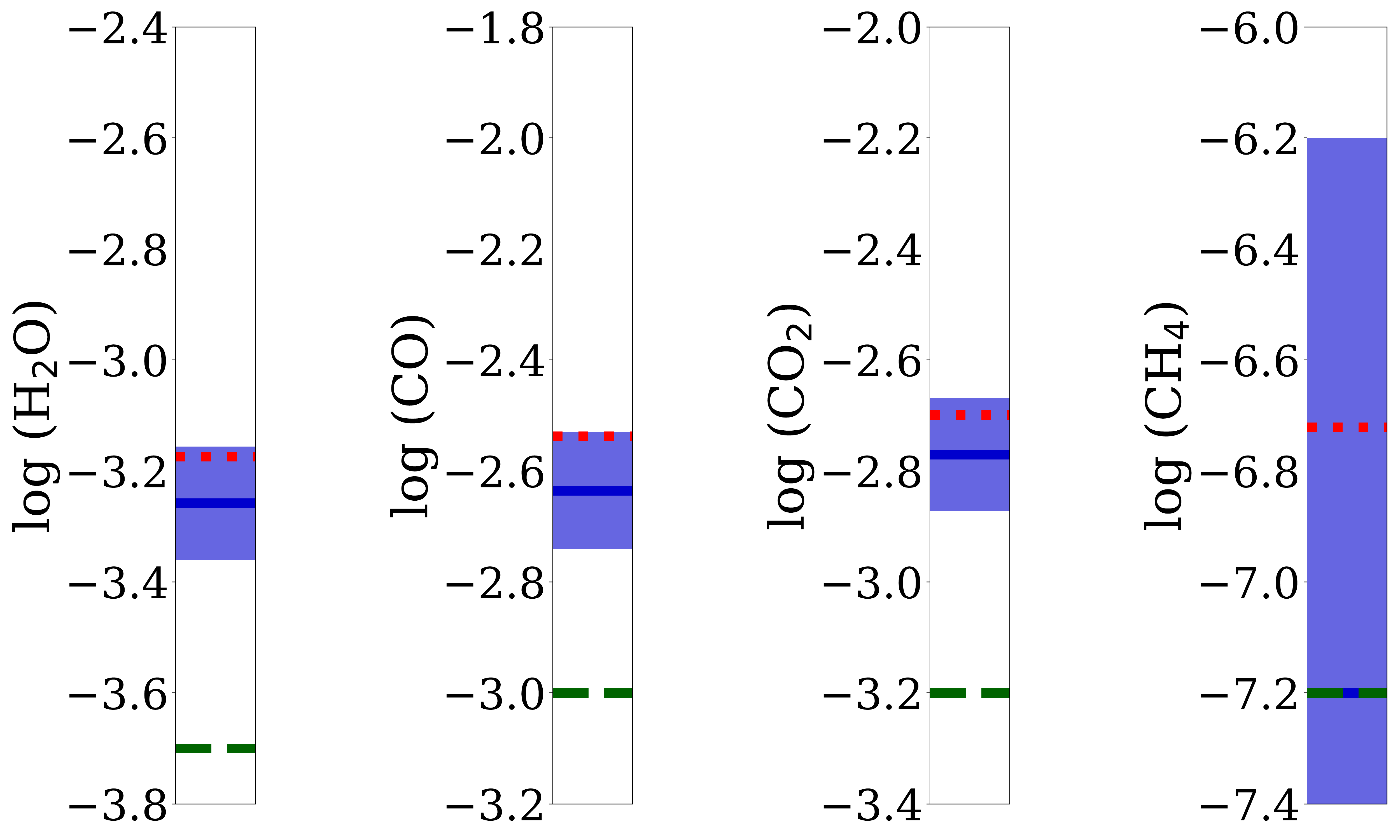}
}
\centerline{
\includegraphics[height=0.11\vsize]{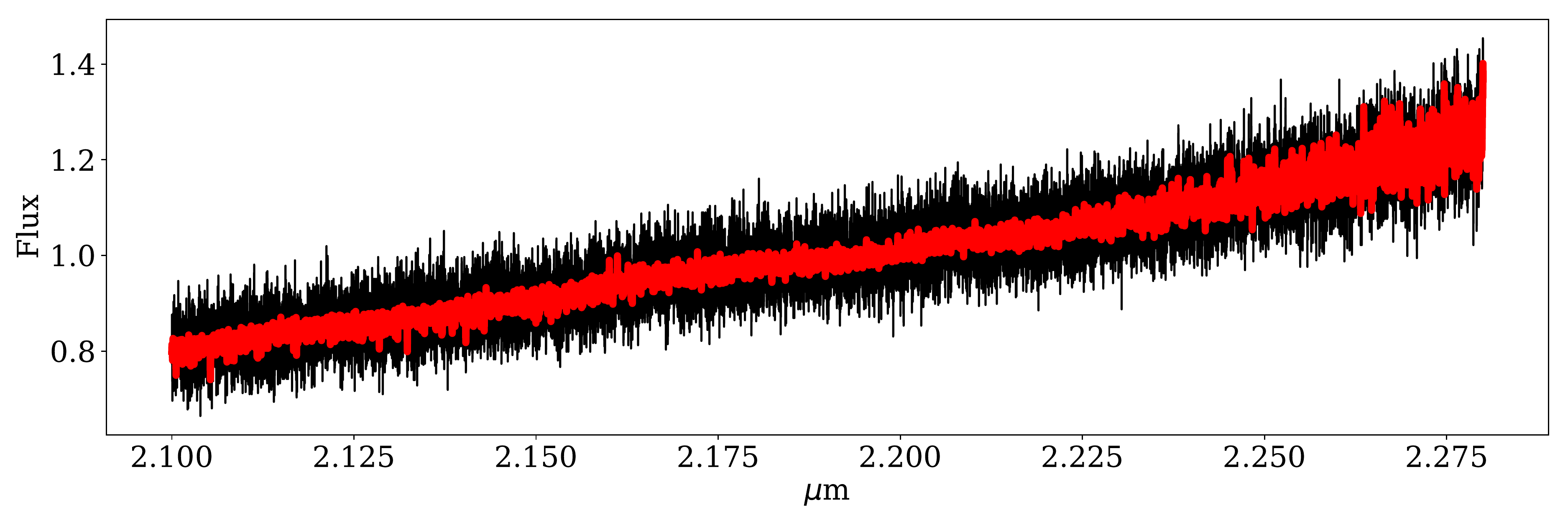}
\hspace{0.02\hsize}
\includegraphics[trim={0 0 20.0cm 0},clip,height=0.11\vsize]{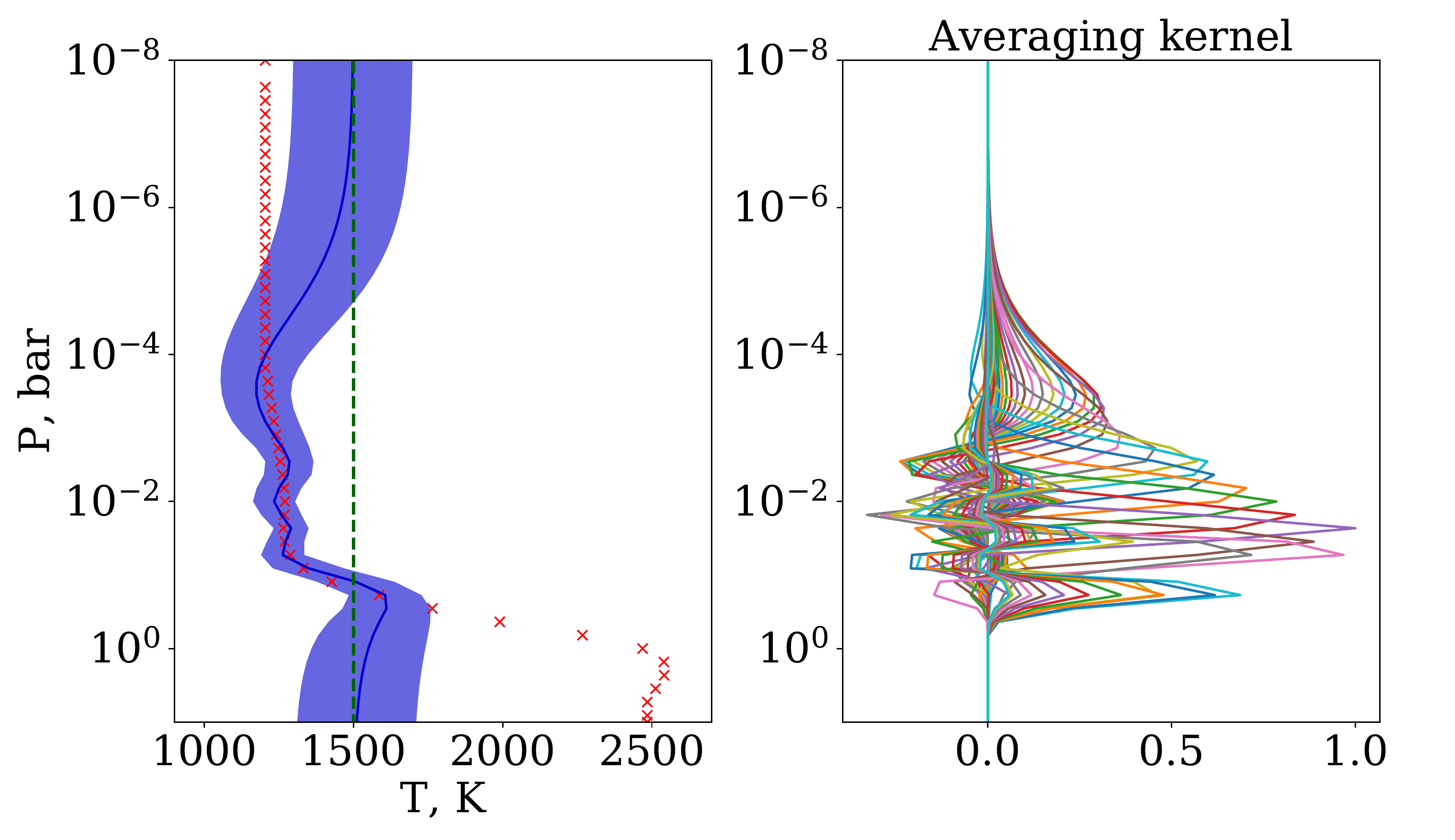}
\hspace{0.02\hsize}
\includegraphics[height=0.11\vsize]{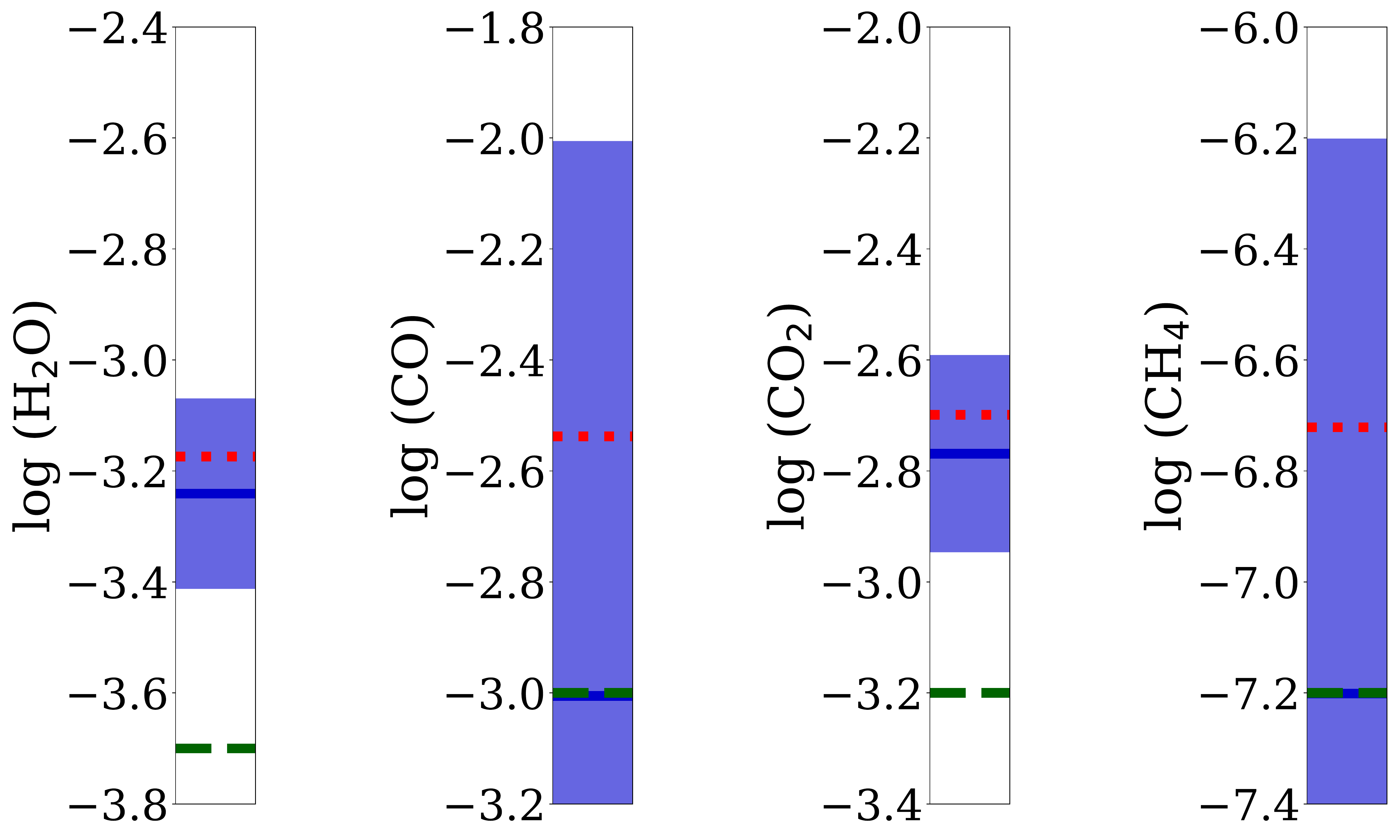}
}
\centerline{
\includegraphics[height=0.11\vsize]{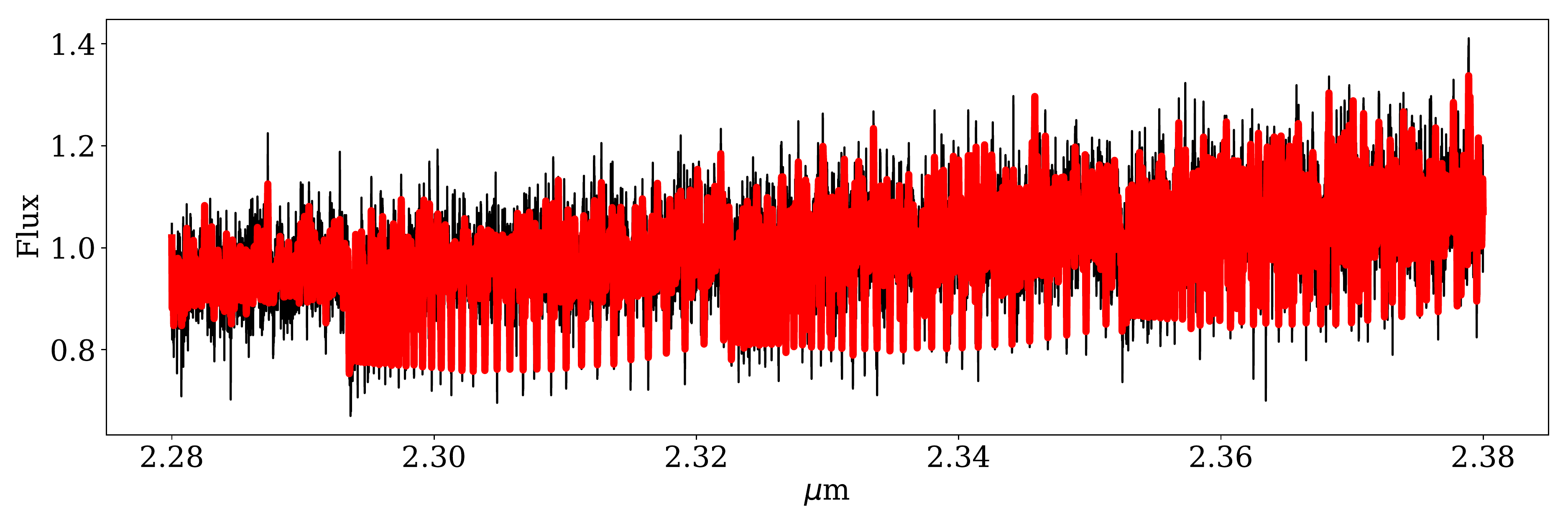}
\hspace{0.02\hsize}
\includegraphics[trim={0 0 20.0cm 0},clip,height=0.11\vsize]{figures/r100k/snr25/h2o+co+ch4+co2/ll-2.28-2.38/out-vmr_init_m0.5dex-vmr_err_1.0dex.bwd_param1.pdf}
\hspace{0.02\hsize}
\includegraphics[height=0.11\vsize]{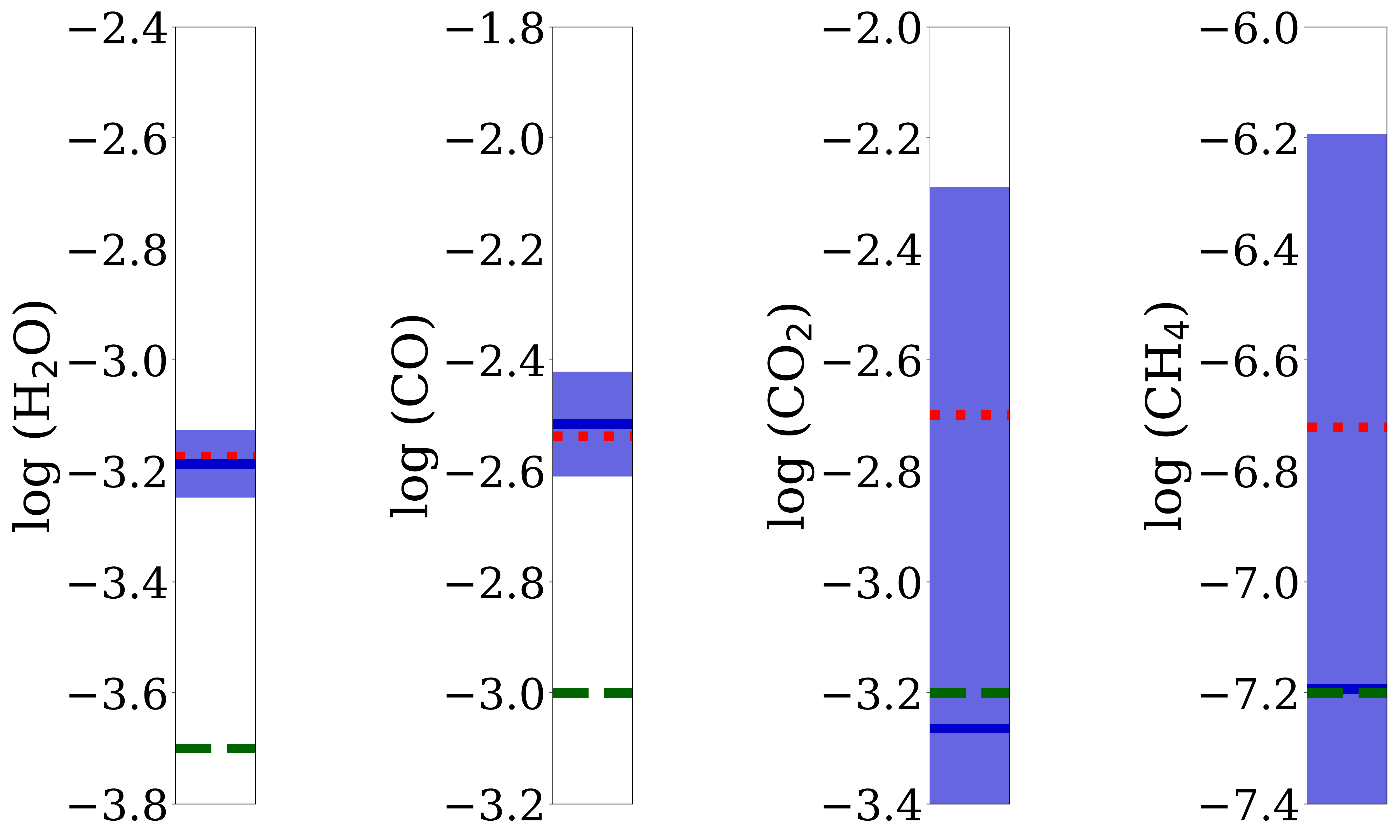}
}
\centerline{
\includegraphics[height=0.11\vsize]{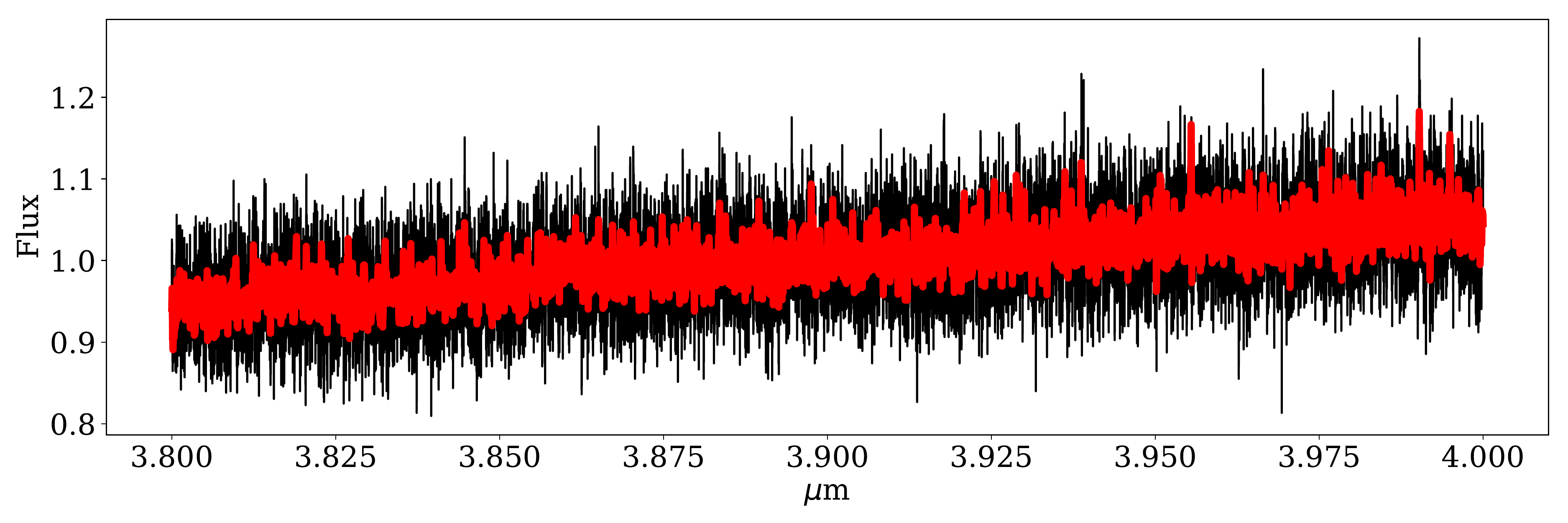}
\hspace{0.02\hsize}
\includegraphics[trim={0 0 20.0cm 0},clip,height=0.11\vsize]{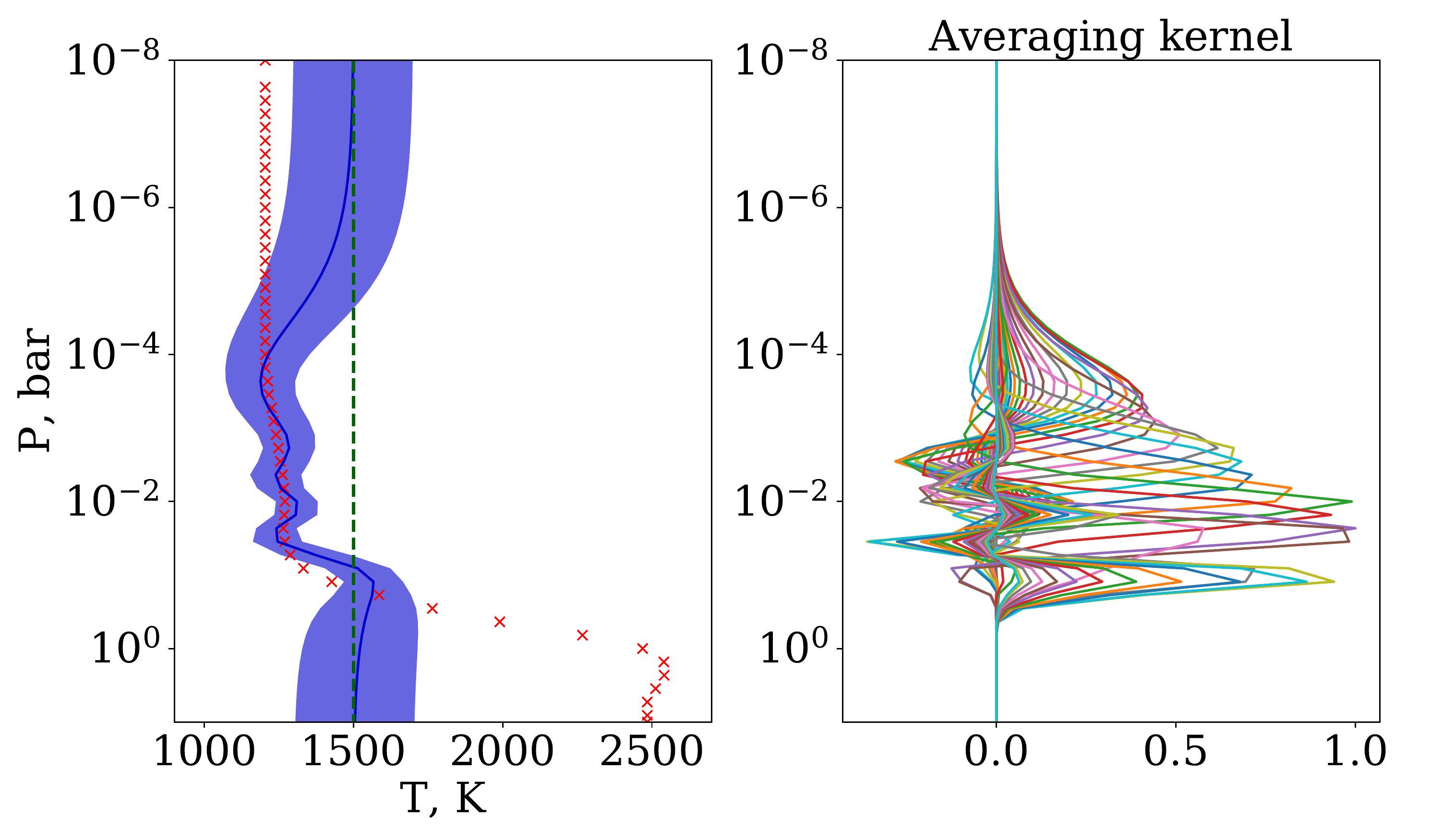}
\hspace{0.02\hsize}
\includegraphics[height=0.11\vsize]{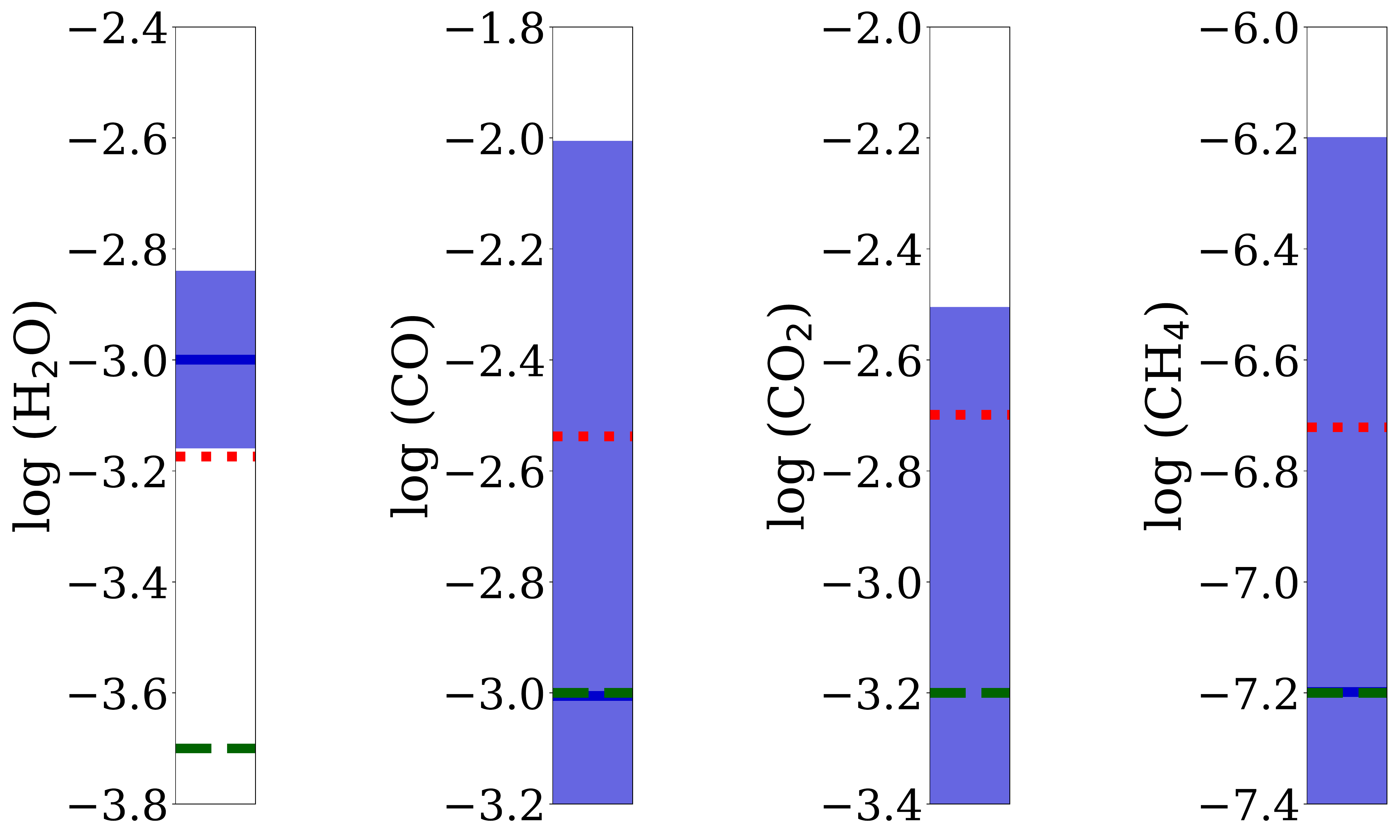}
}
\centerline{
\includegraphics[height=0.11\vsize]{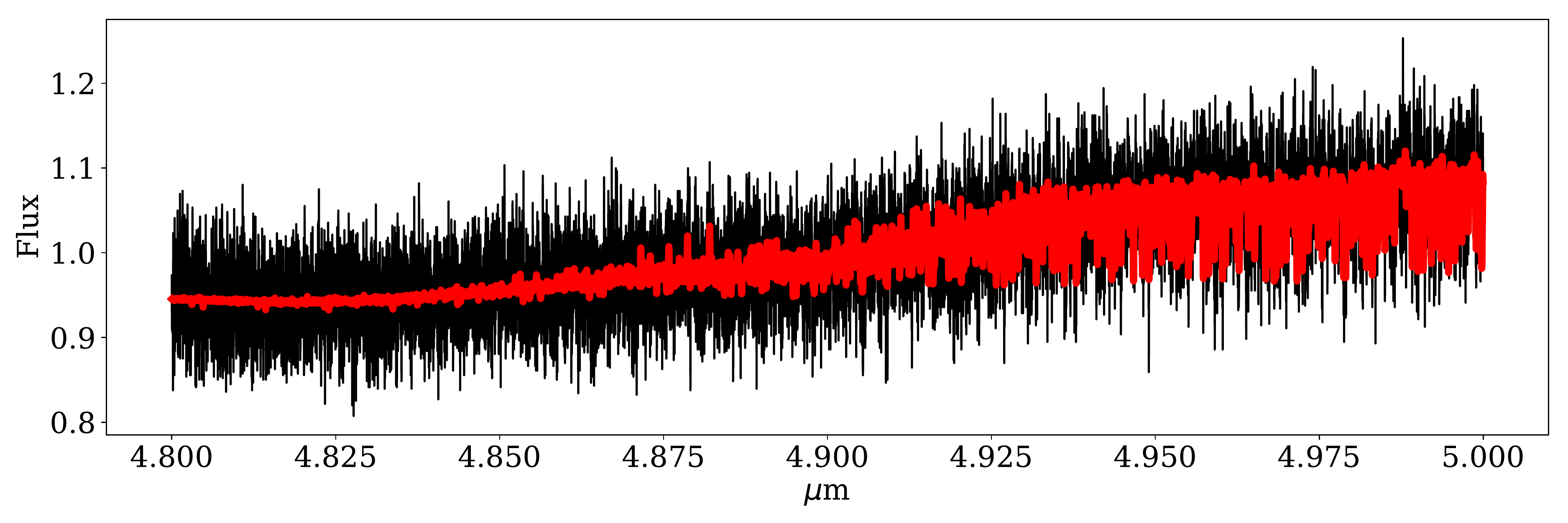}
\hspace{0.02\hsize}
\includegraphics[trim={0 0 20.0cm 0},clip,height=0.11\vsize]{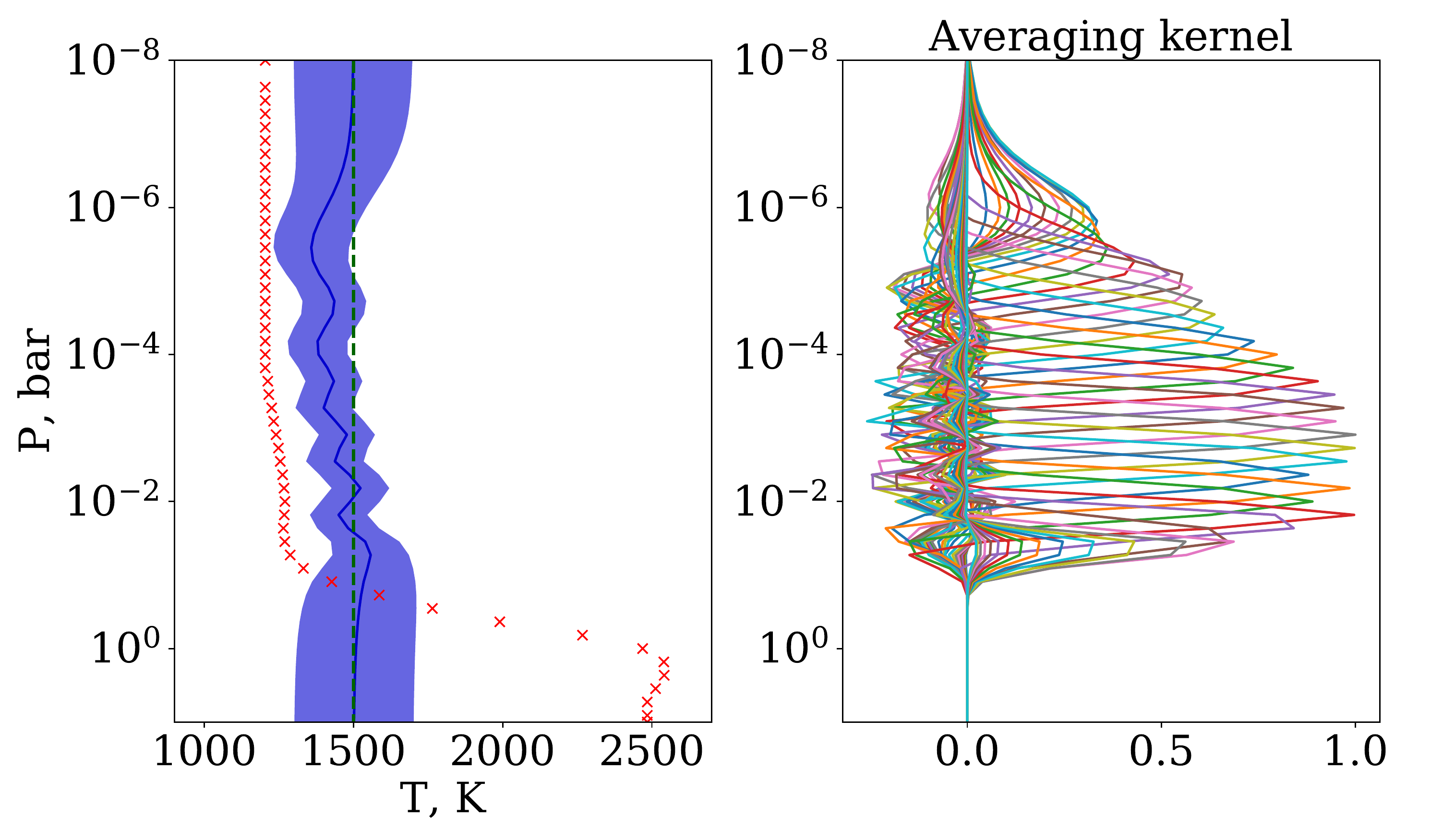}
\hspace{0.02\hsize}
\includegraphics[height=0.11\vsize]{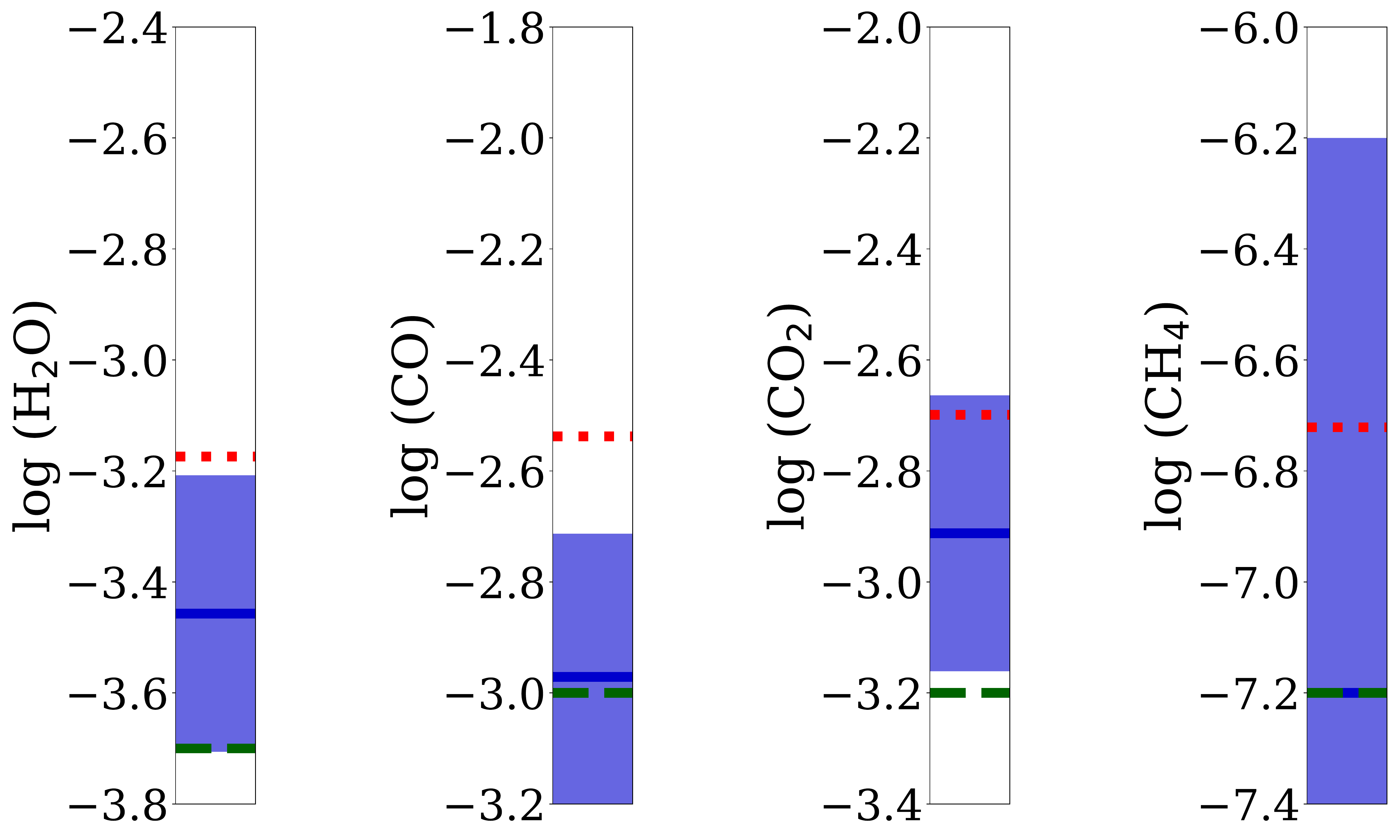}
}
\caption{\label{fig:snr25-r100k}
Same as   Fig.~\ref{fig:snr5-r100k}, but for S/N=$25$.
}
\end{figure*}

\begin{figure*}
\centerline{
\includegraphics[height=0.11\vsize]{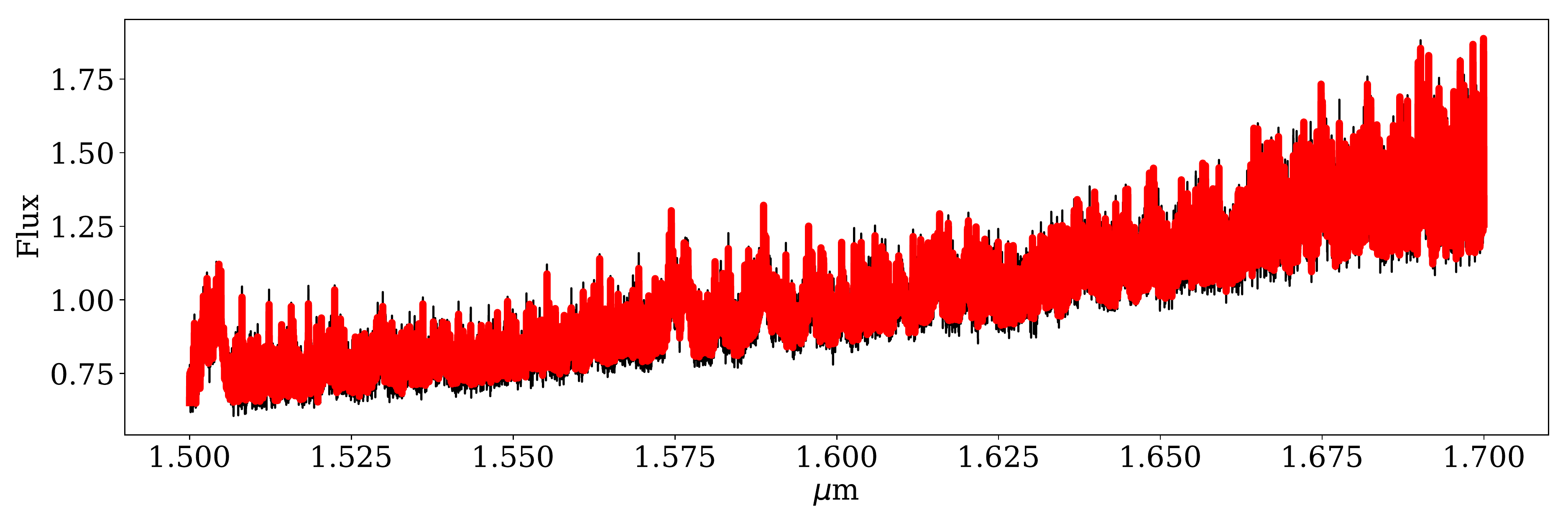}
\hspace{0.02\hsize}
\includegraphics[trim={0 0 20.0cm 0},clip,height=0.11\vsize]{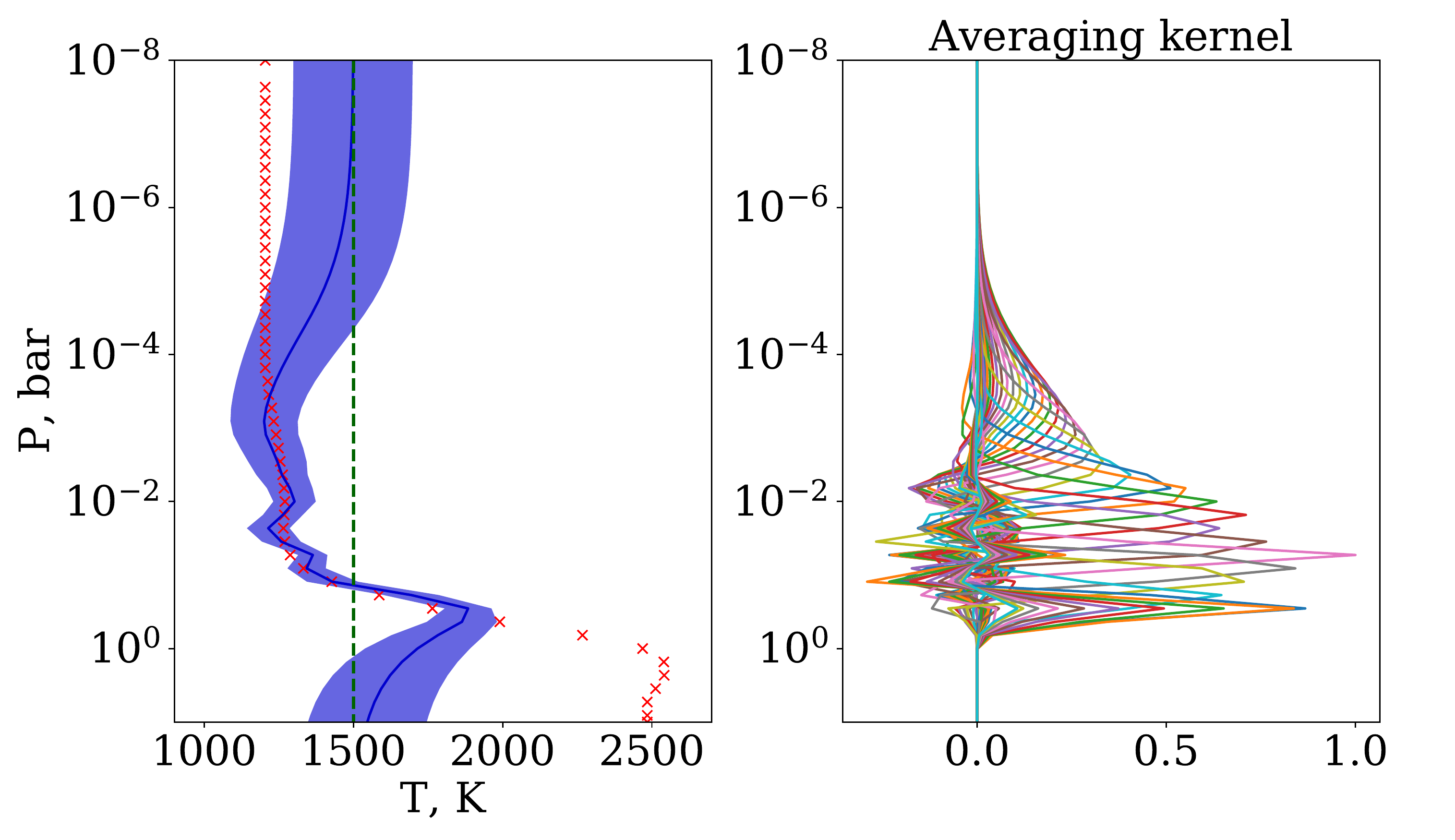}
\hspace{0.02\hsize}
\includegraphics[height=0.11\vsize]{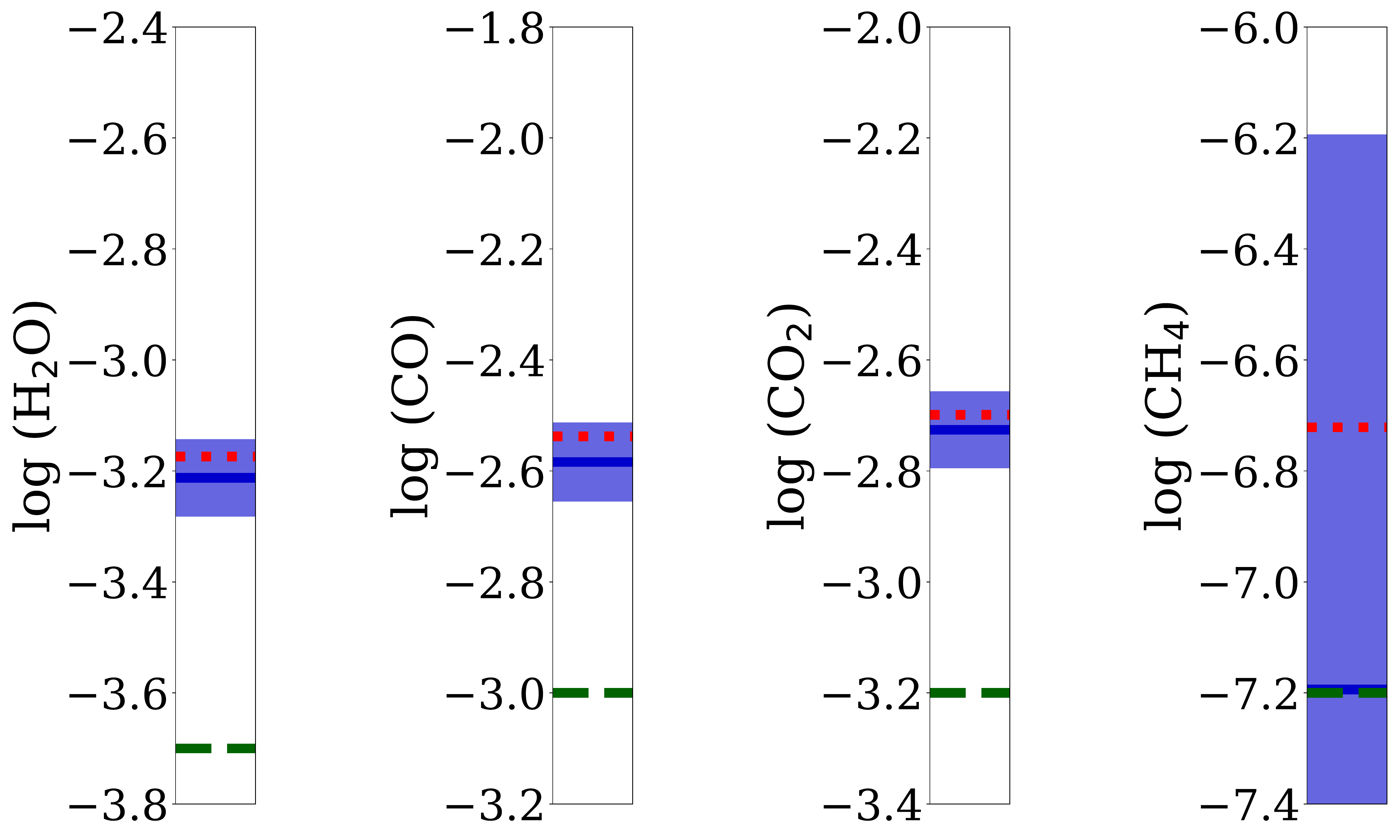}
}
\centerline{
\includegraphics[height=0.11\vsize]{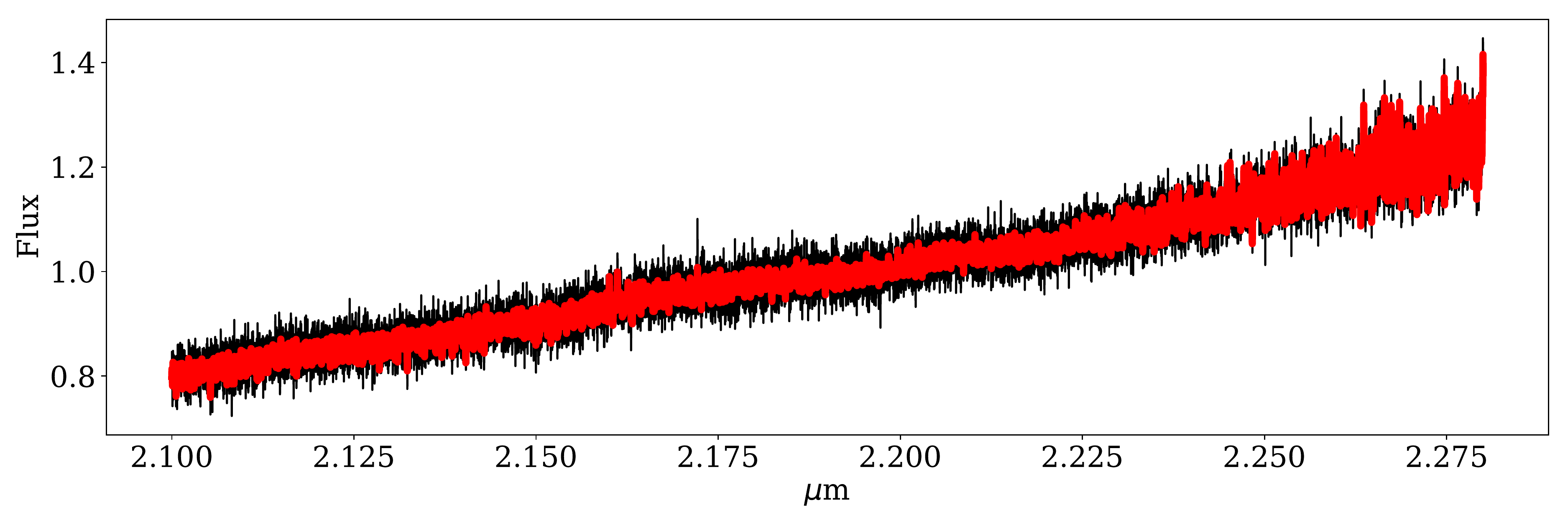}
\hspace{0.02\hsize}
\includegraphics[trim={0 0 20.0cm 0},clip,height=0.11\vsize]{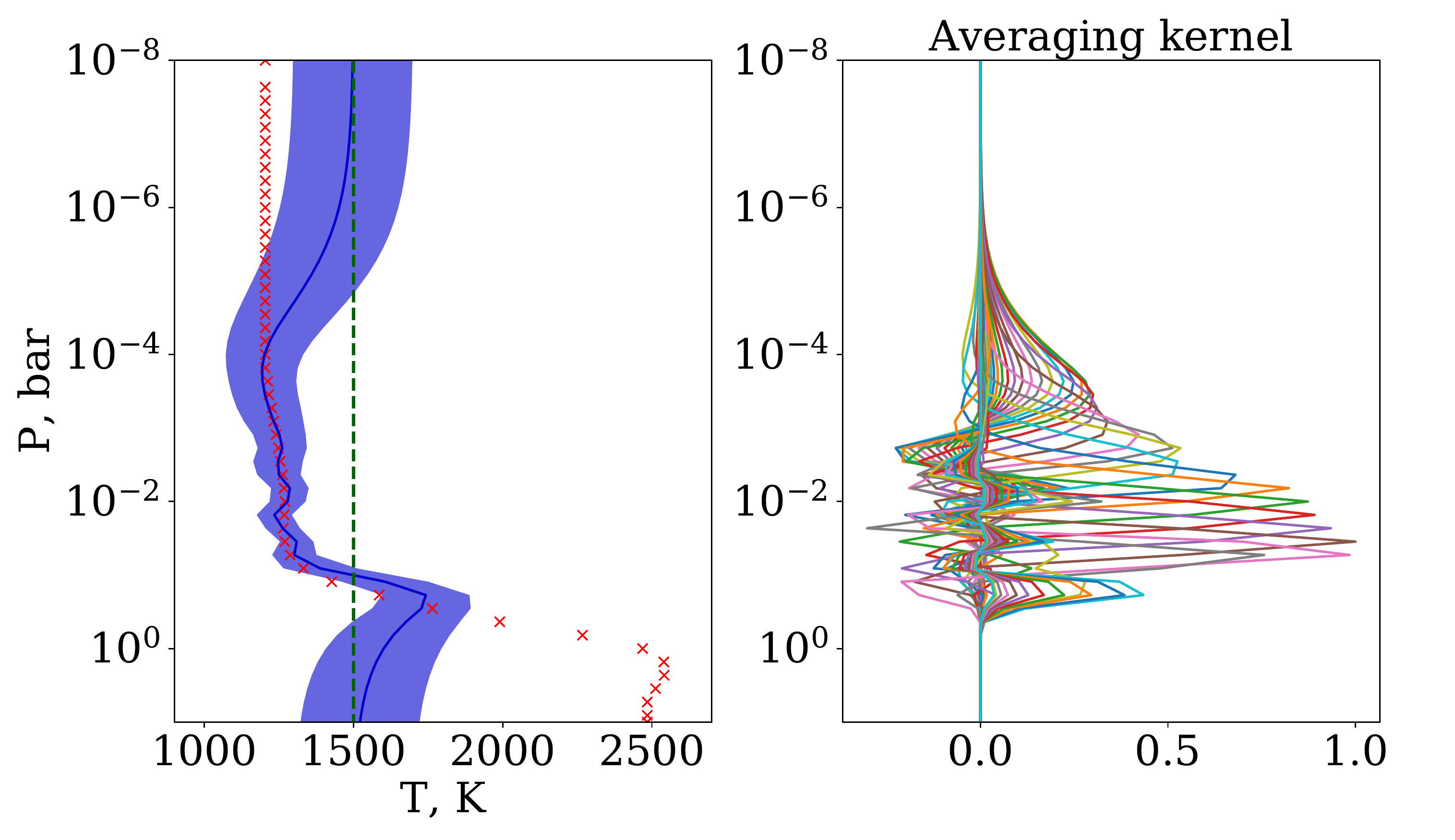}
\hspace{0.02\hsize}
\includegraphics[height=0.11\vsize]{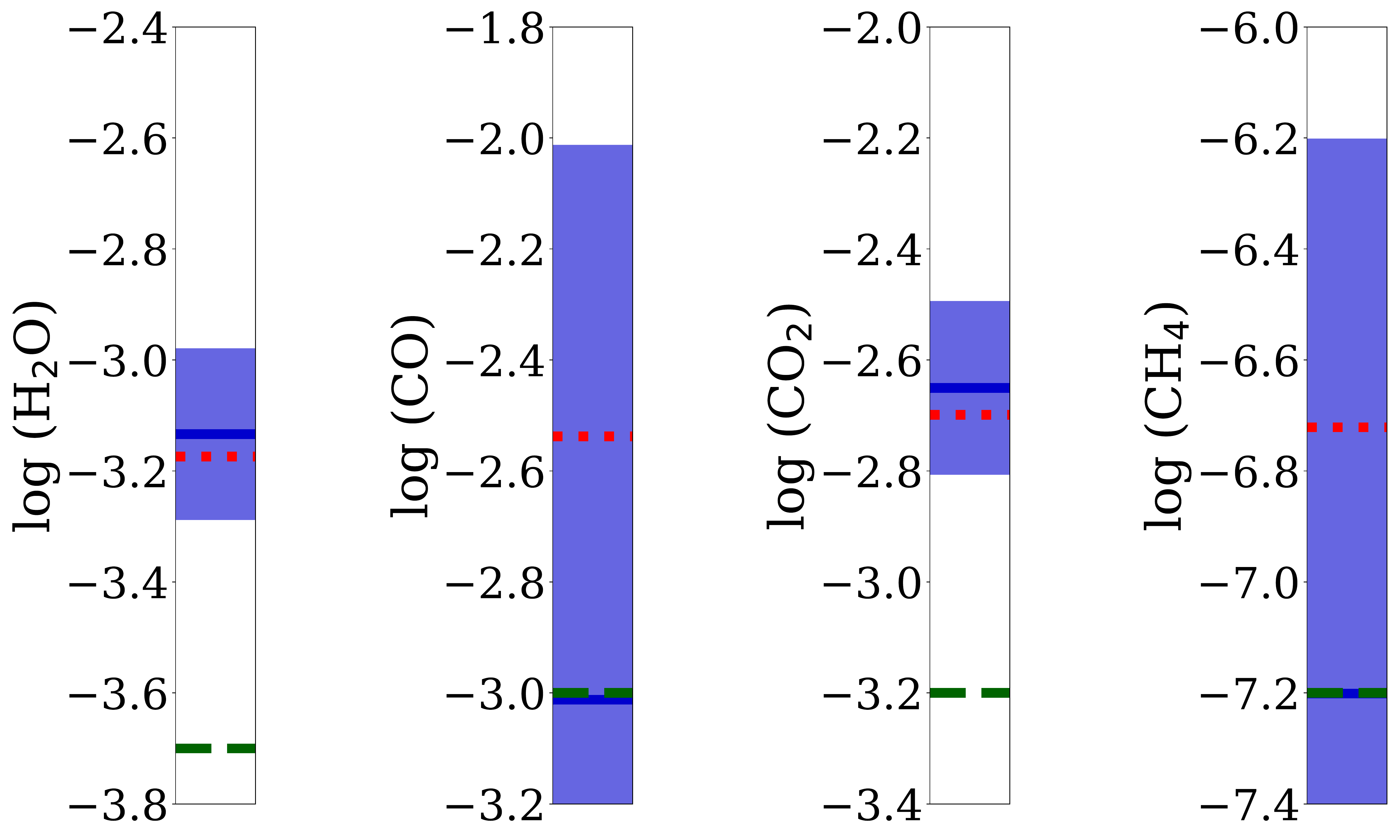}
}
\centerline{
\includegraphics[height=0.11\vsize]{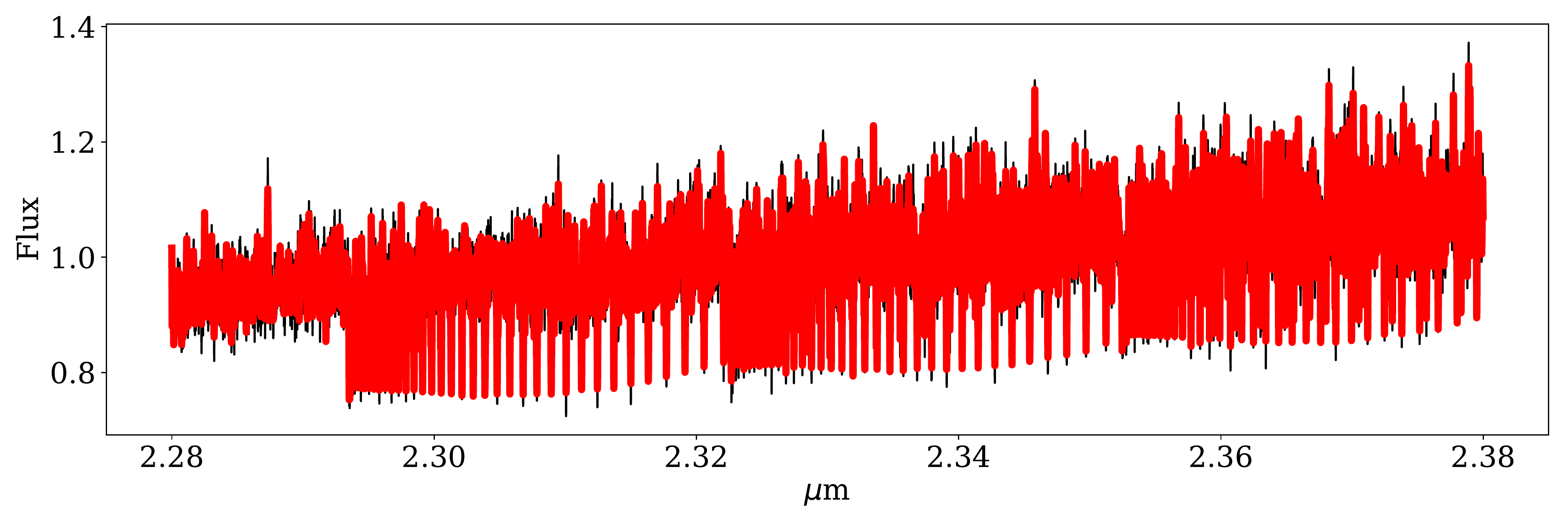}
\hspace{0.02\hsize}
\includegraphics[trim={0 0 20.0cm 0},clip,height=0.11\vsize]{figures/r100k/snr50/h2o+co+ch4+co2/ll-2.28-2.38/out-vmr_init_m0.5dex-vmr_err_1.0dex.bwd_param1.pdf}
\hspace{0.02\hsize}
\includegraphics[height=0.11\vsize]{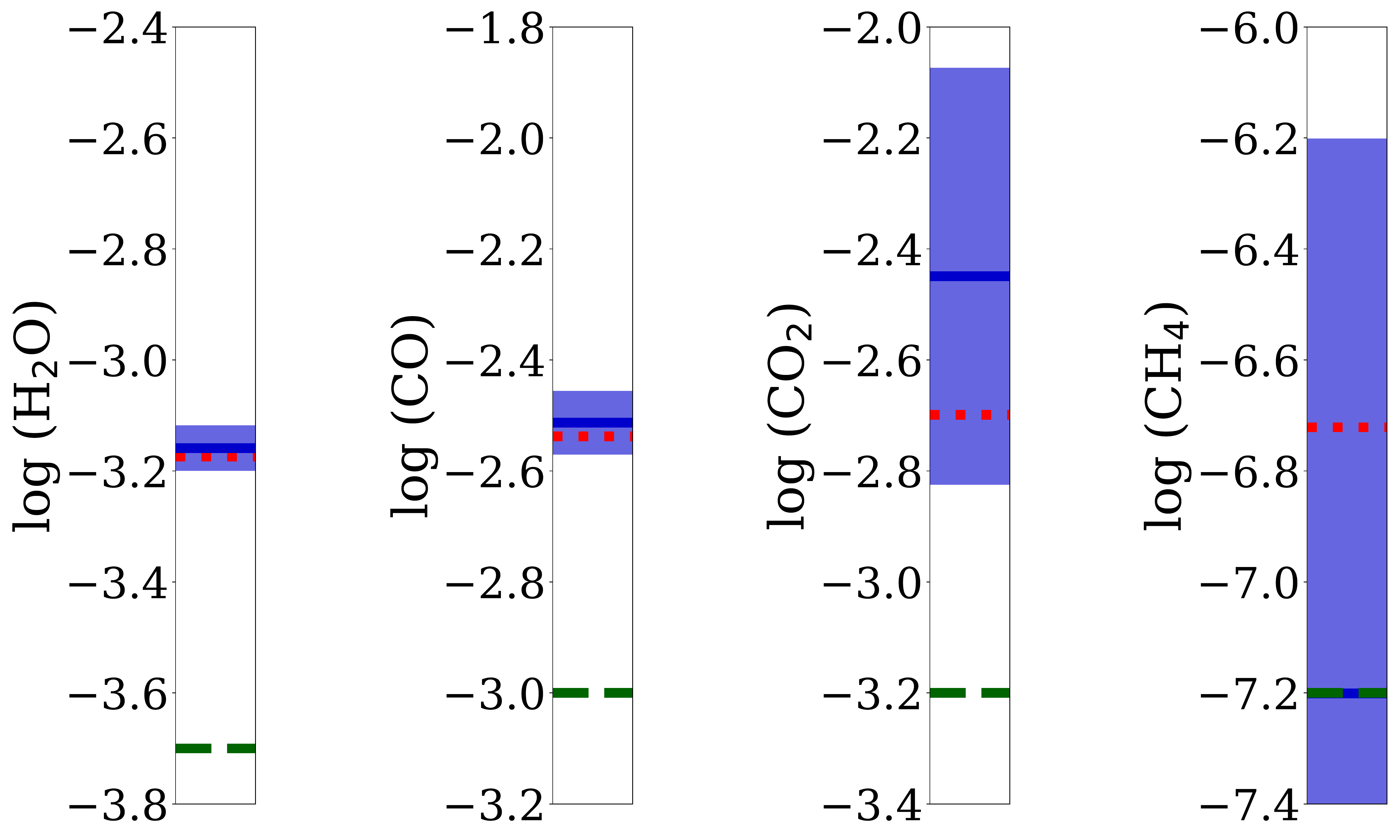}
}
\centerline{
\includegraphics[height=0.11\vsize]{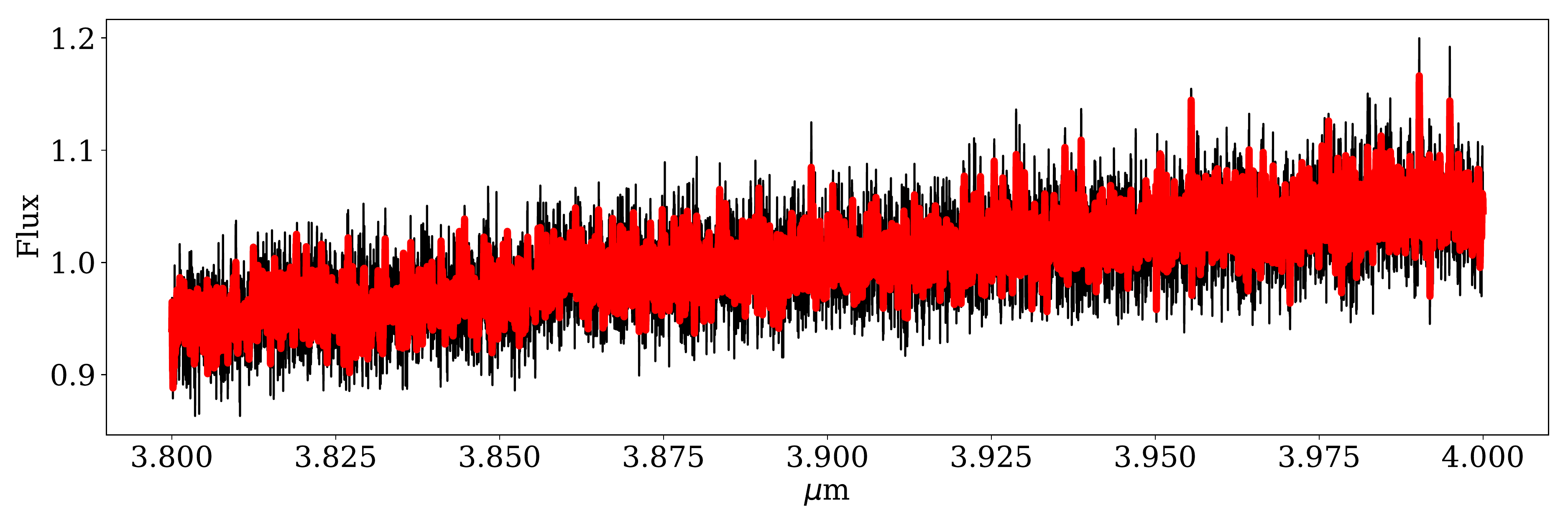}
\hspace{0.02\hsize}
\includegraphics[trim={0 0 20.0cm 0},clip,height=0.11\vsize]{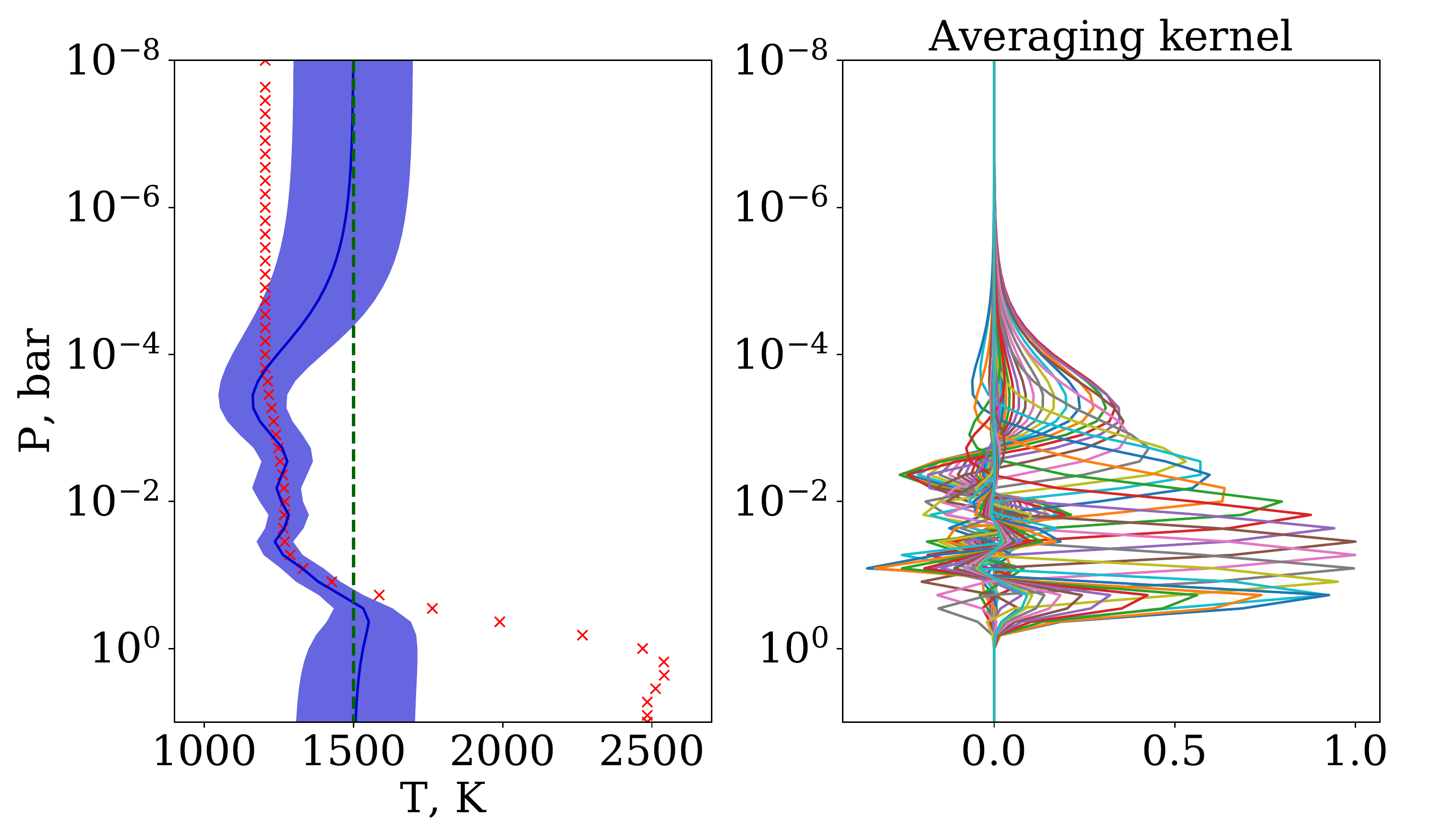}
\hspace{0.02\hsize}
\includegraphics[height=0.11\vsize]{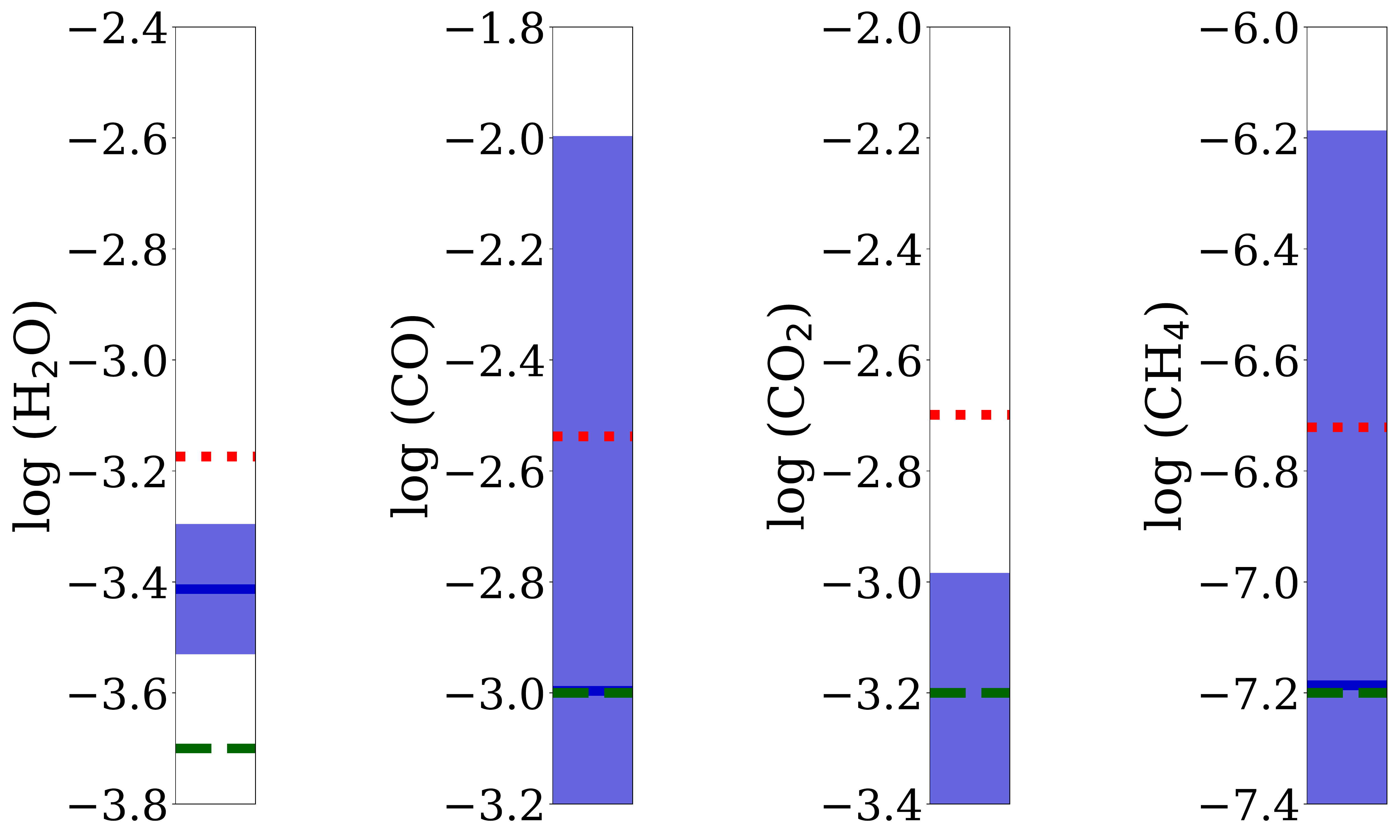}
}
\centerline{
\includegraphics[height=0.11\vsize]{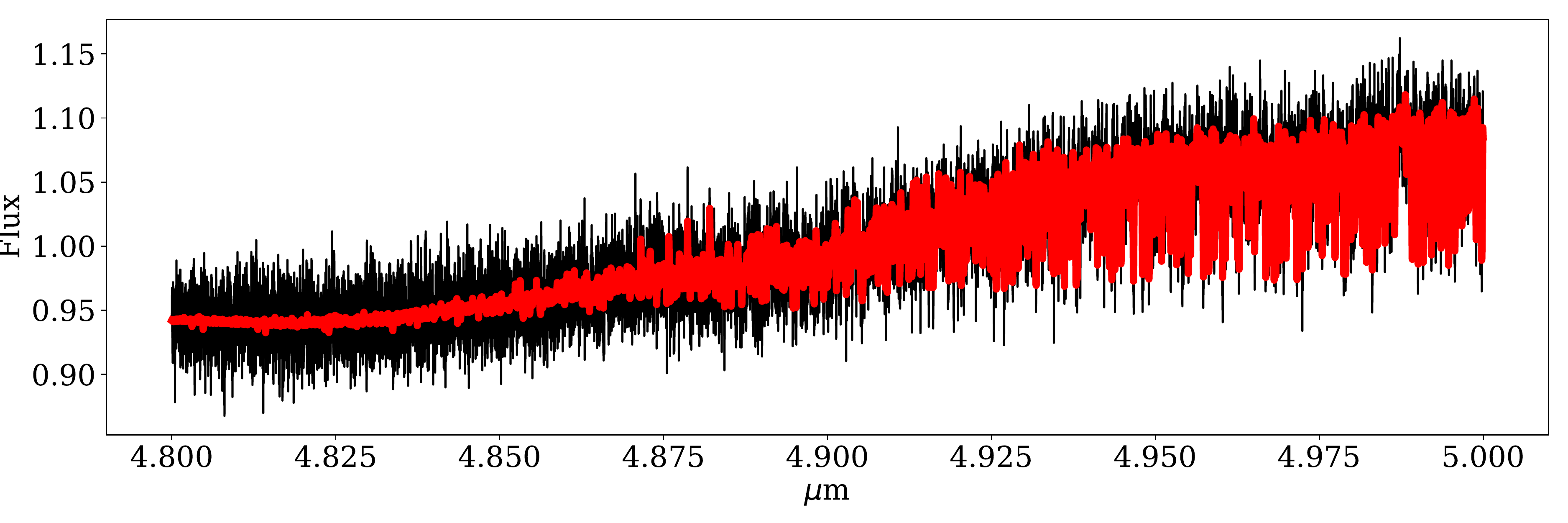}
\hspace{0.02\hsize}
\includegraphics[trim={0 0 20.0cm 0},clip,height=0.11\vsize]{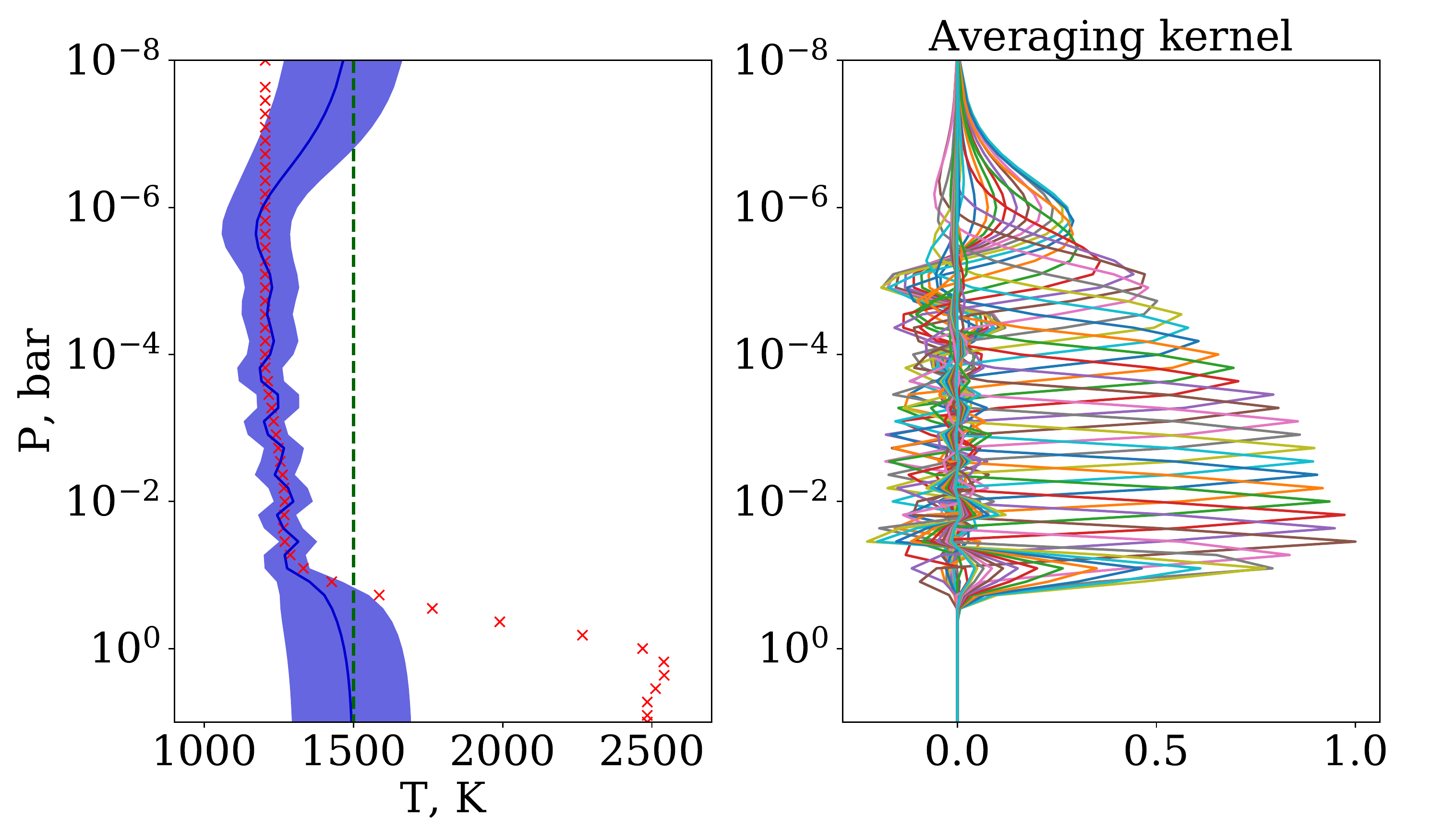}
\hspace{0.02\hsize}
\includegraphics[height=0.11\vsize]{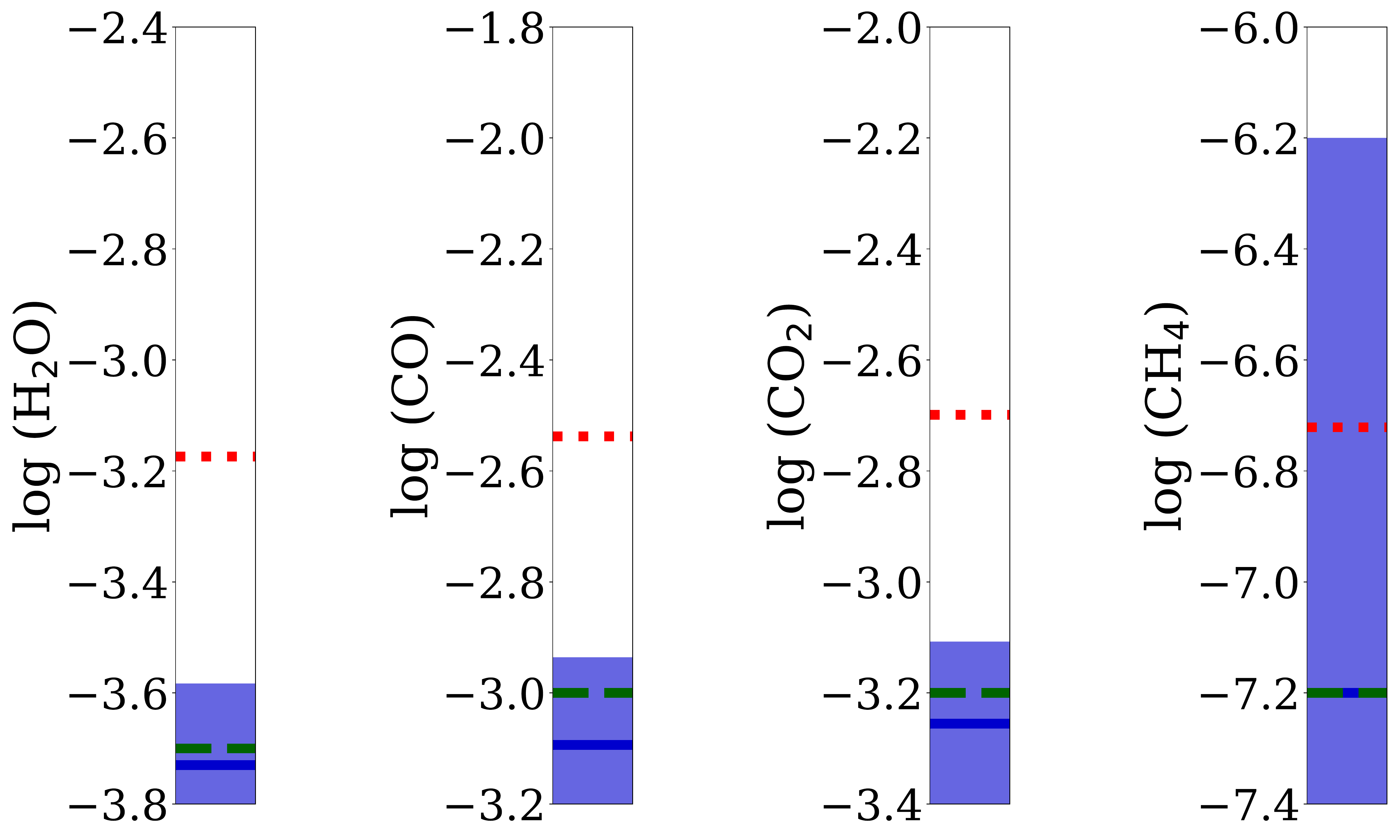}
}
\caption{\label{fig:snr50-r100k}
Same as   Fig.~\ref{fig:snr5-r100k}, but for S/N=$50$.
}
\end{figure*}

\begin{figure*}
\centerline{\textbf{$\mathbf{(1.50-1.70)+(2.28+2.38)}$~$\mathbf{\mu}$m}}
\centerline{
\includegraphics[trim={0 0 20.0cm 0},clip,height=0.14\vsize]{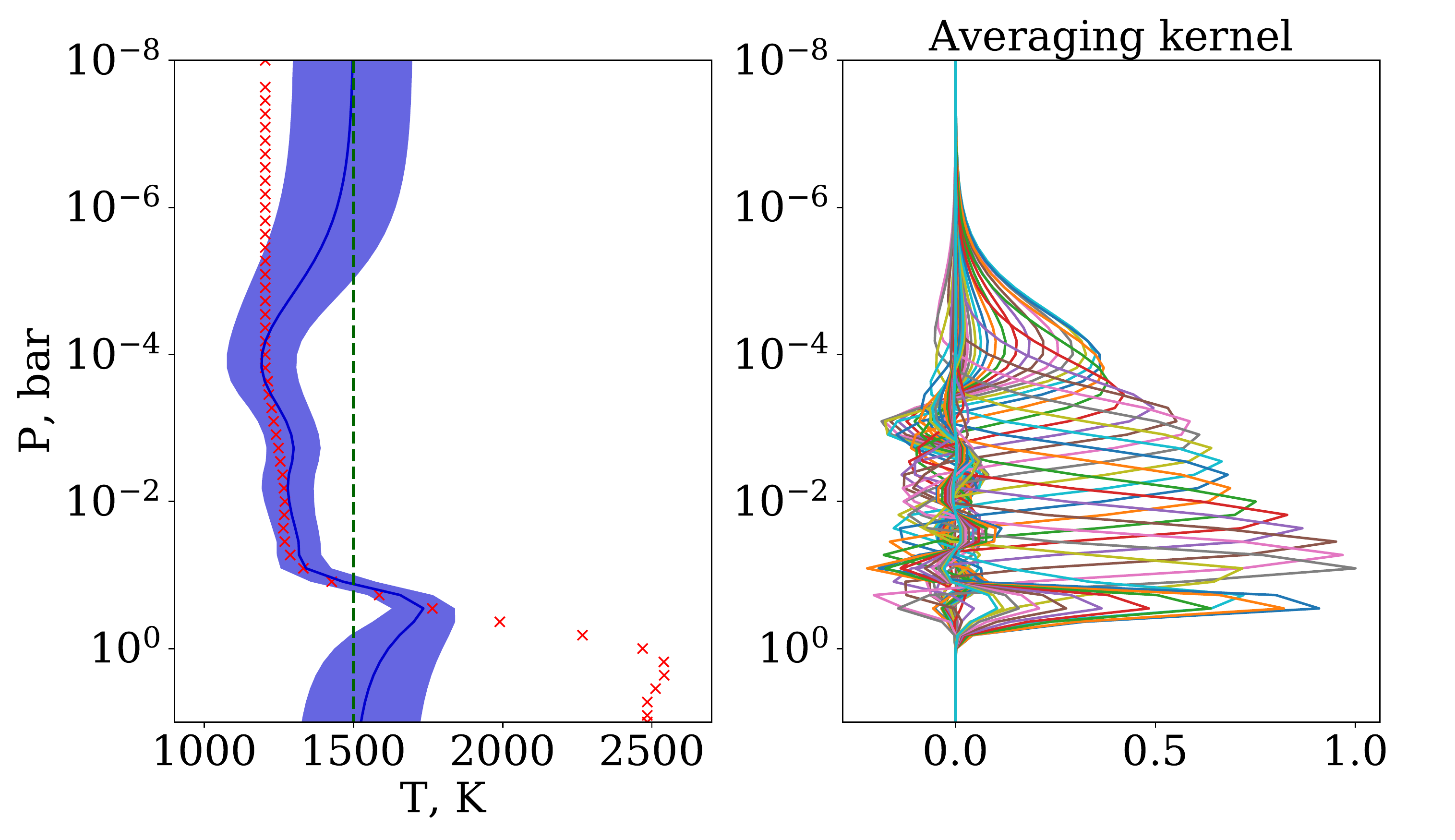}
\includegraphics[height=0.14\vsize]{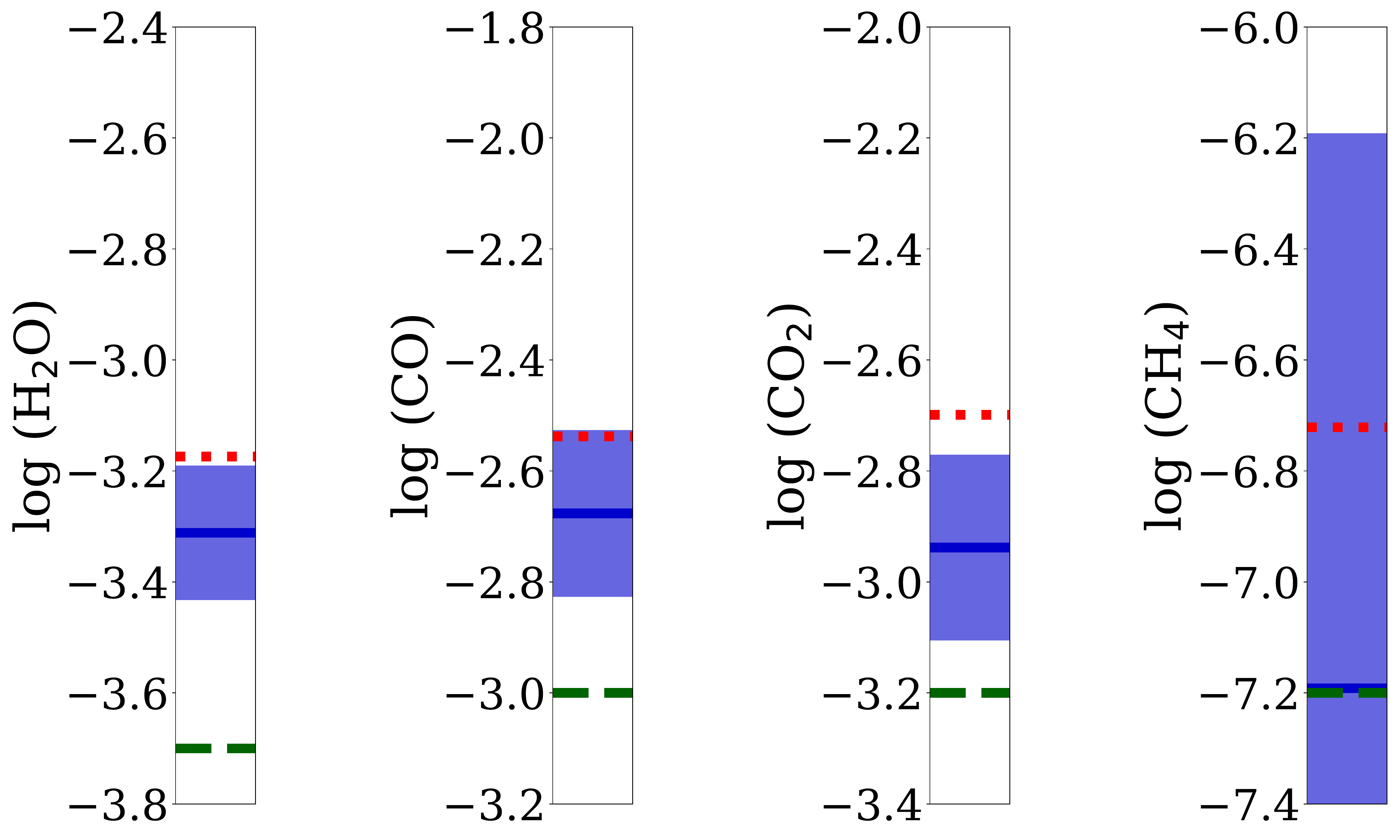}
}
\centerline{\textbf{$\mathbf{(2.10+2.28)+(2.28-2.38)}$~$\mathbf{\mu}$m}}
\centerline{
\includegraphics[trim={0 0 20.0cm 0},clip,height=0.14\vsize]{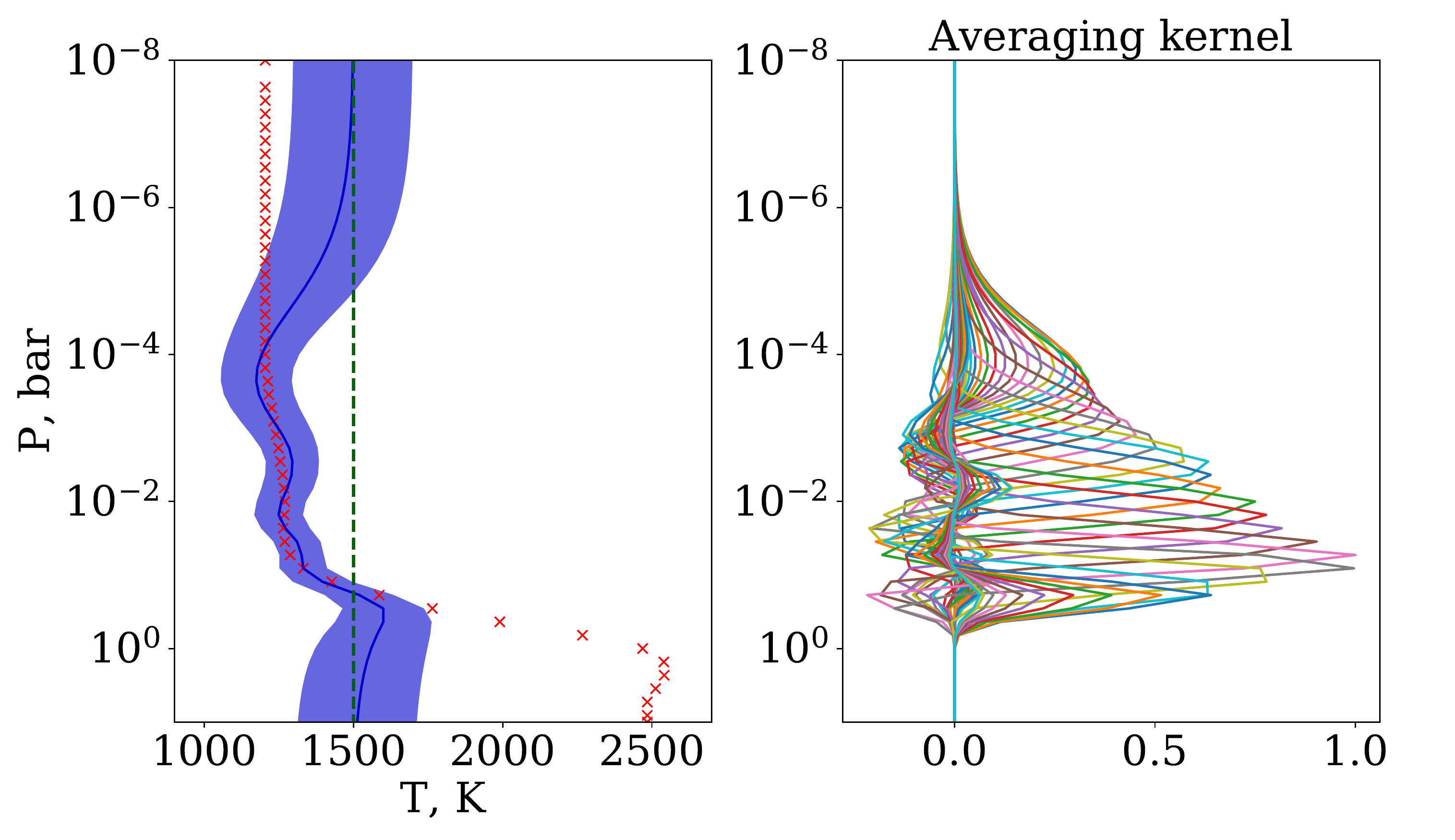}
\includegraphics[height=0.14\vsize]{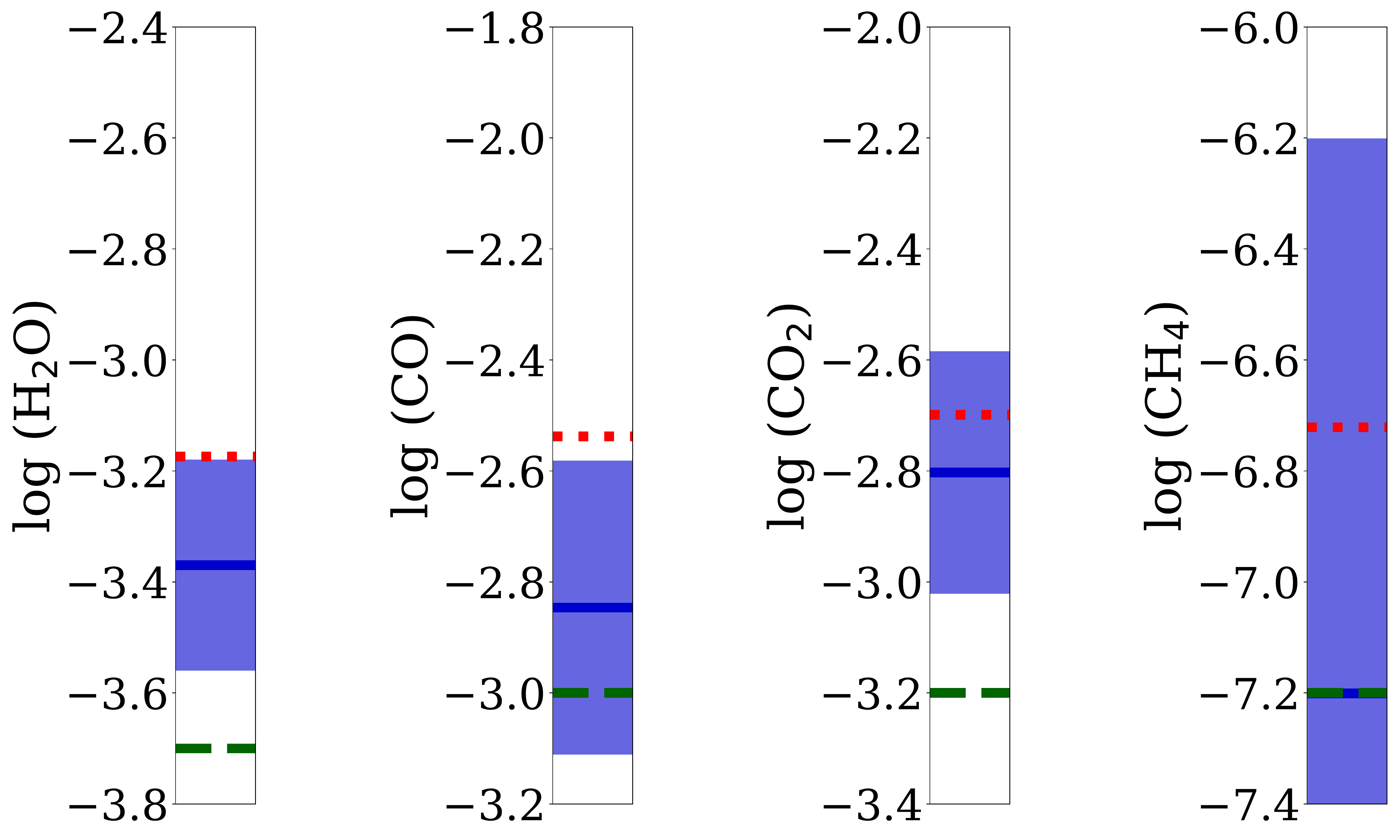}
}
\centerline{\textbf{$\mathbf{(2.28-2.38)+(3.80-4.00)}$~$\mathbf{\mu}$m}}
\centerline{
\includegraphics[trim={0 0 20.0cm 0},clip,height=0.14\vsize]{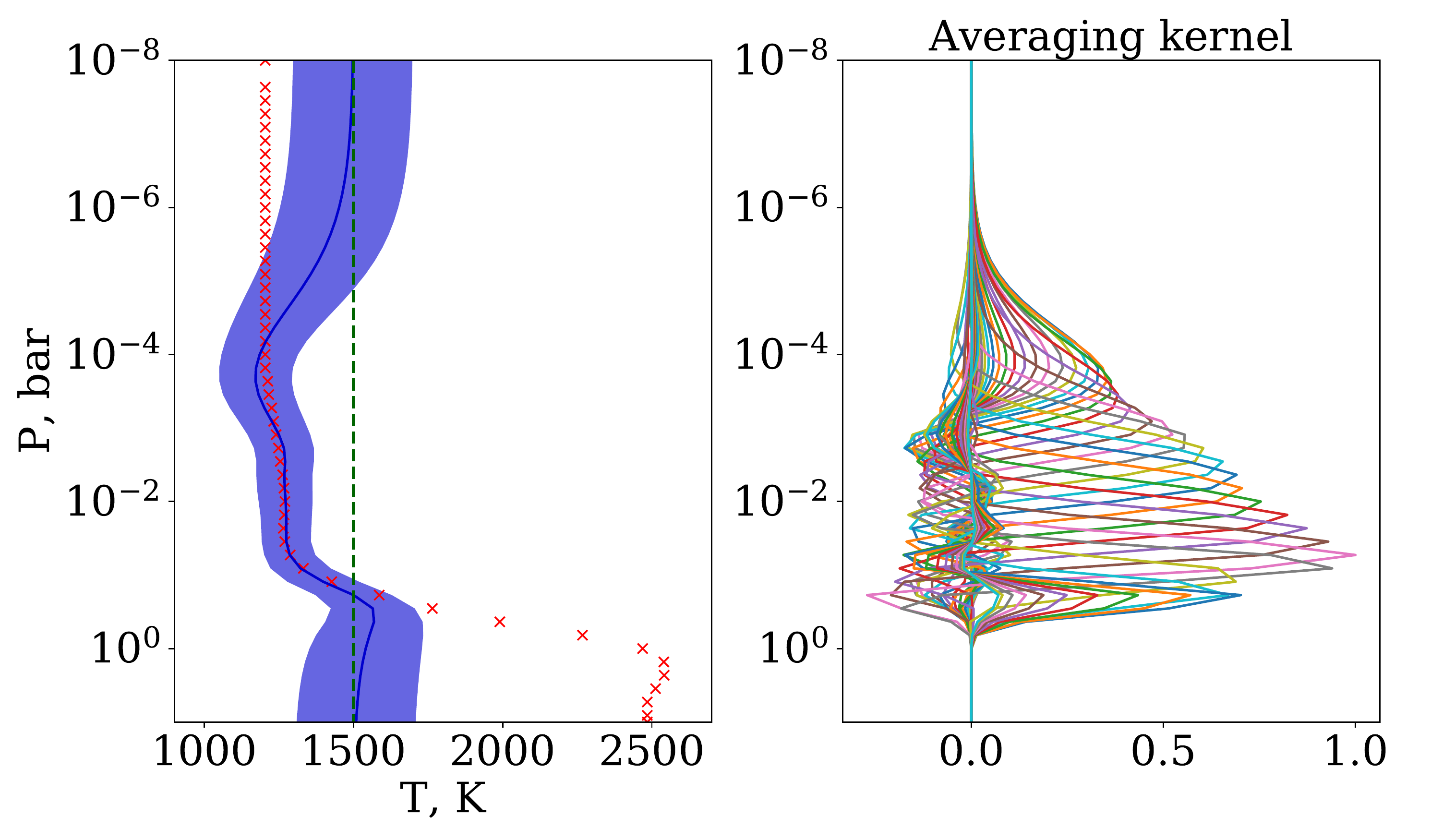}
\includegraphics[height=0.14\vsize]{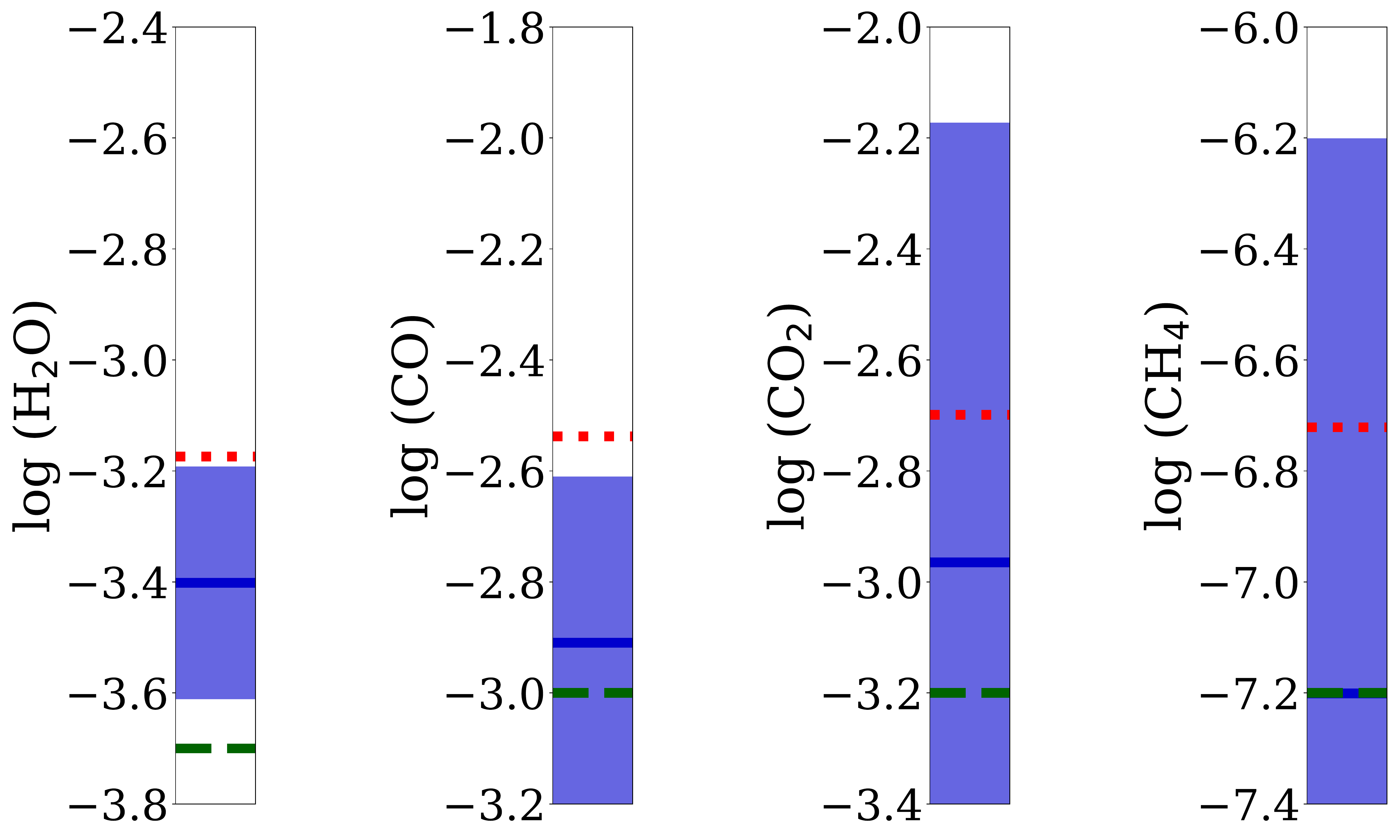}
}
\centerline{\textbf{$\mathbf{(2.28-2.38)+(4.80-5.00)}$~$\mathbf{\mu}$m}}
\centerline{
\includegraphics[trim={0 0 20.0cm 0},clip,height=0.14\vsize]{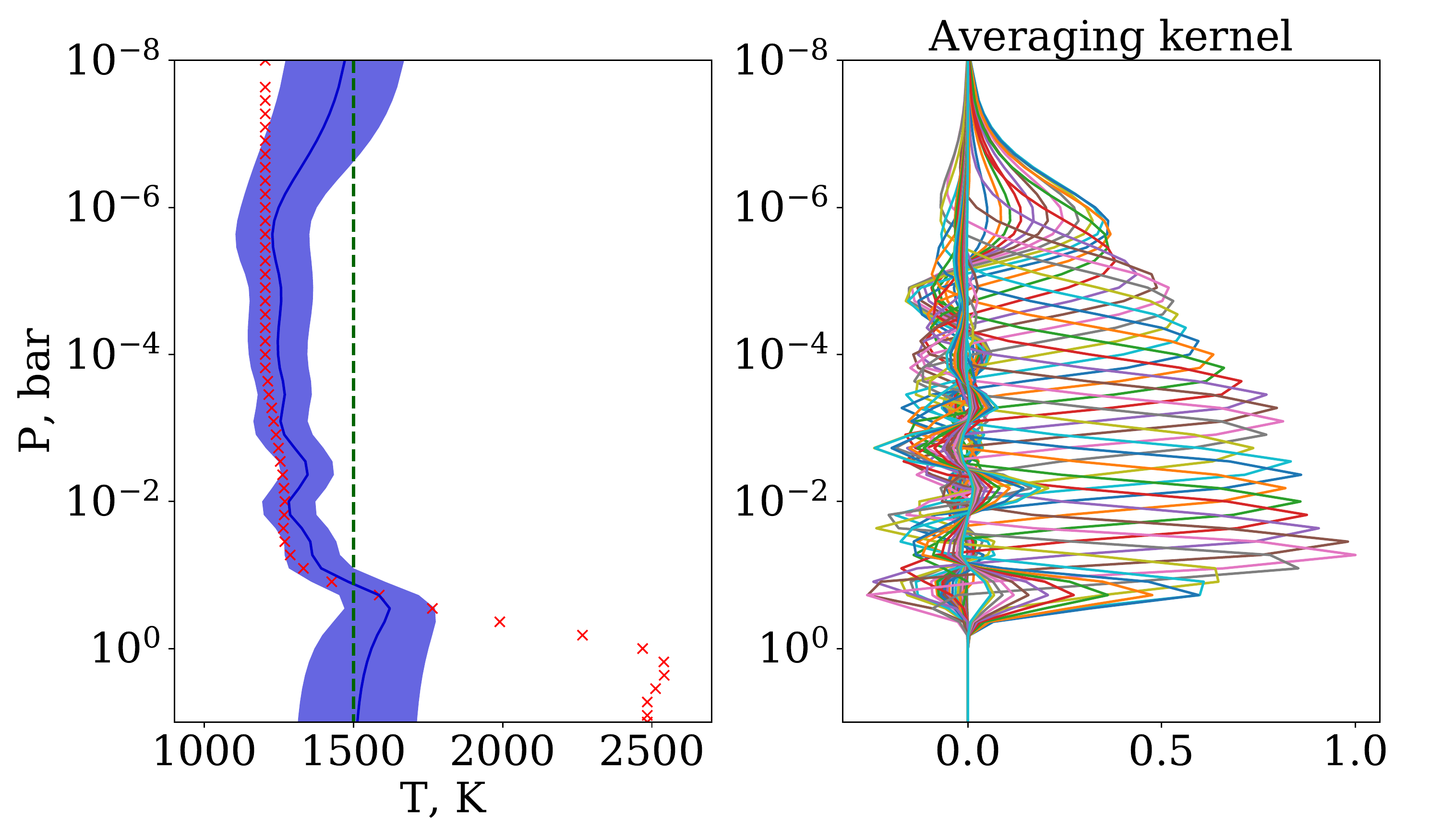}
\includegraphics[height=0.14\vsize]{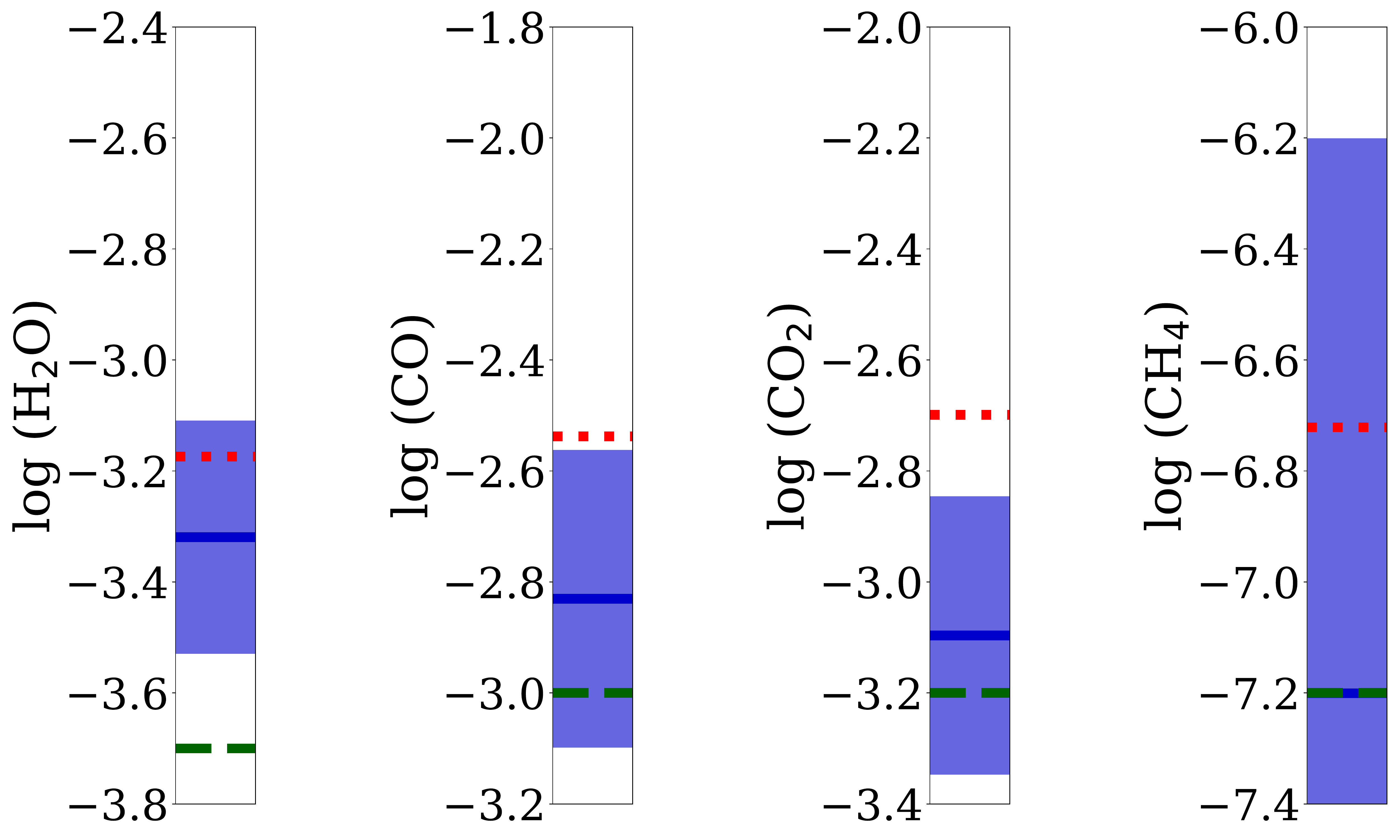}
}
\caption{\label{fig:snr5-r100k-2reg}
Retrieved temperature and mixing ratios of four molecular species from a combination of different spectral regions with a reference region $2.28-2.38$~\mum.
The retrievals are shown for the case of S/N=$5$ and spectral resolution R=$100\,000$.
The color-coding is the same as in Fig.~\ref{fig:snr10-r100k}.
}
\end{figure*}

\begin{figure*}
\centerline{\textbf{$\mathbf{(1.50-1.70)+(2.28+2.38)}$~$\mathbf{\mu}$m}}
\centerline{
\includegraphics[trim={0 0 20.0cm 0},clip,height=0.14\vsize]{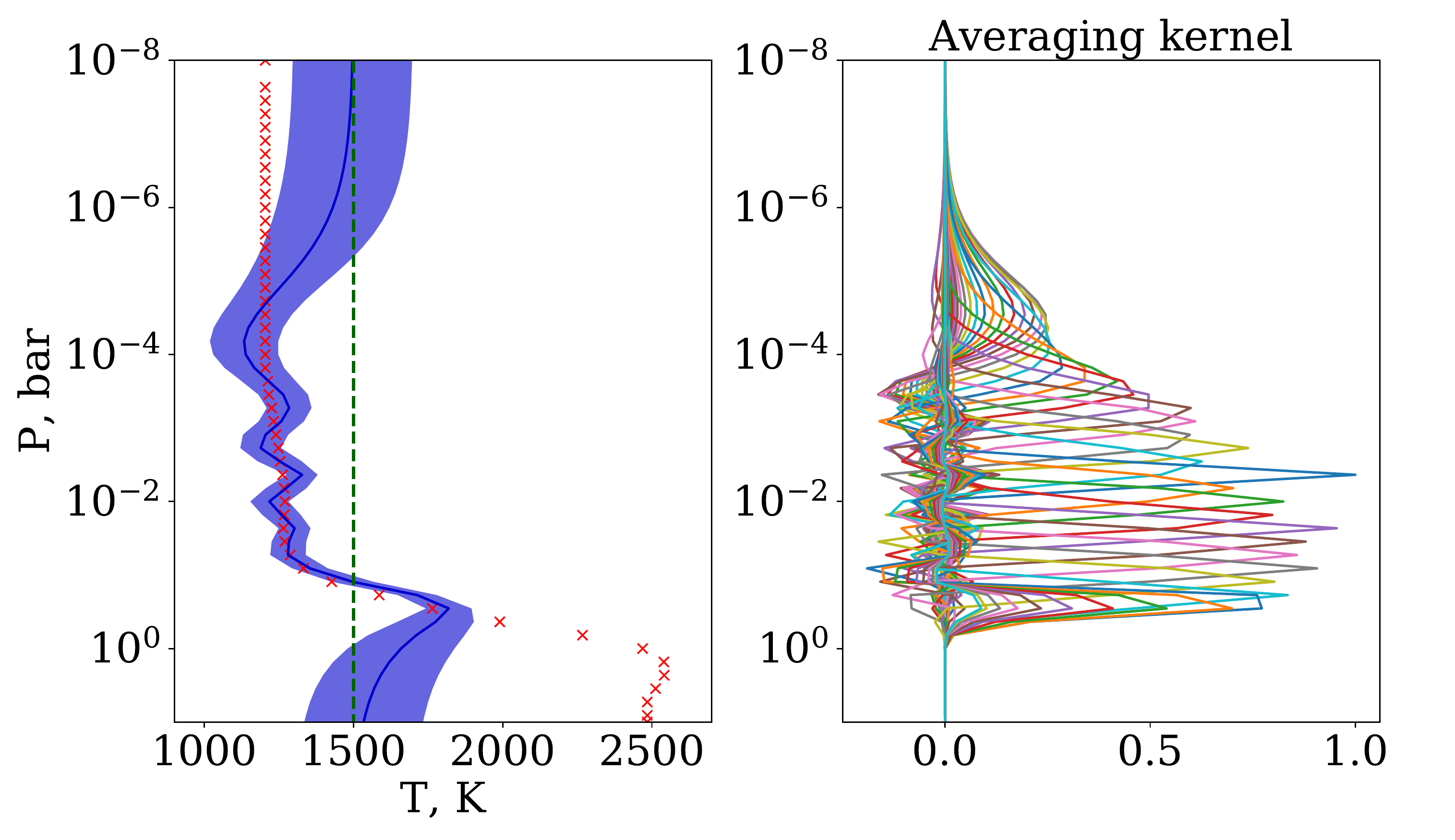}
\includegraphics[height=0.14\vsize]{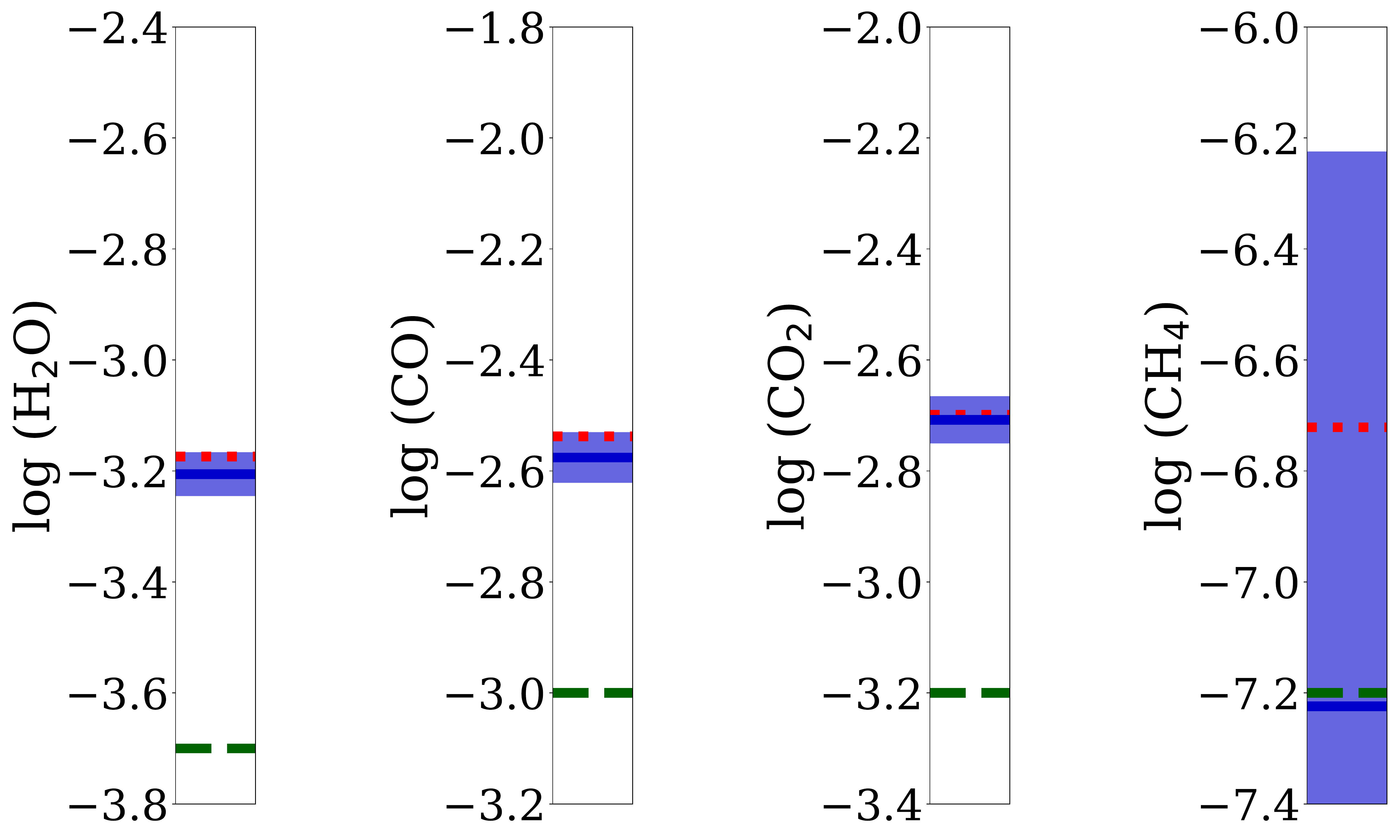}
}
\centerline{\textbf{$\mathbf{(2.10+2.28)+(2.28-2.38)}$~$\mathbf{\mu}$m}}
\centerline{
\includegraphics[trim={0 0 20.0cm 0},clip,height=0.14\vsize]{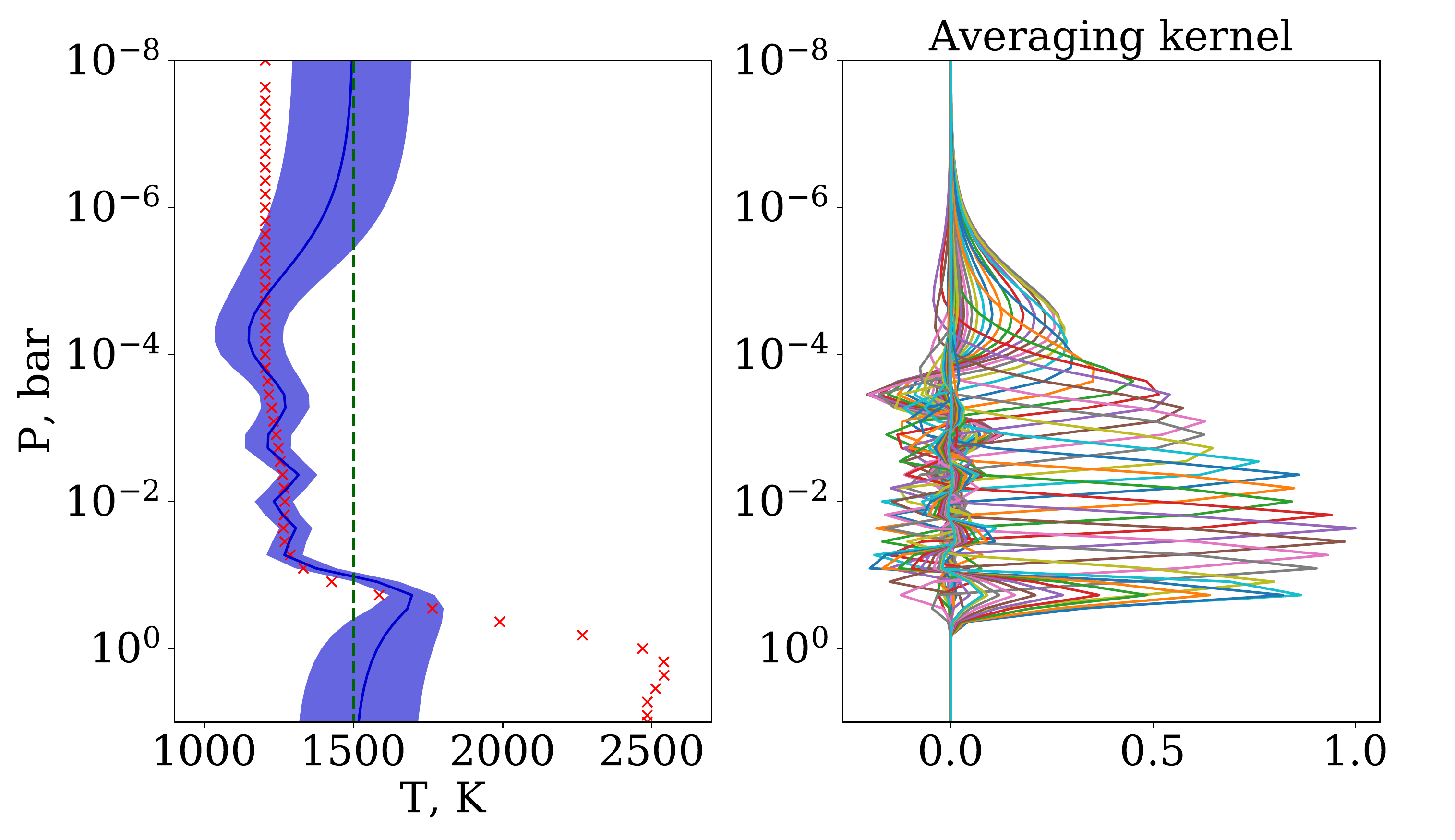}
\includegraphics[height=0.14\vsize]{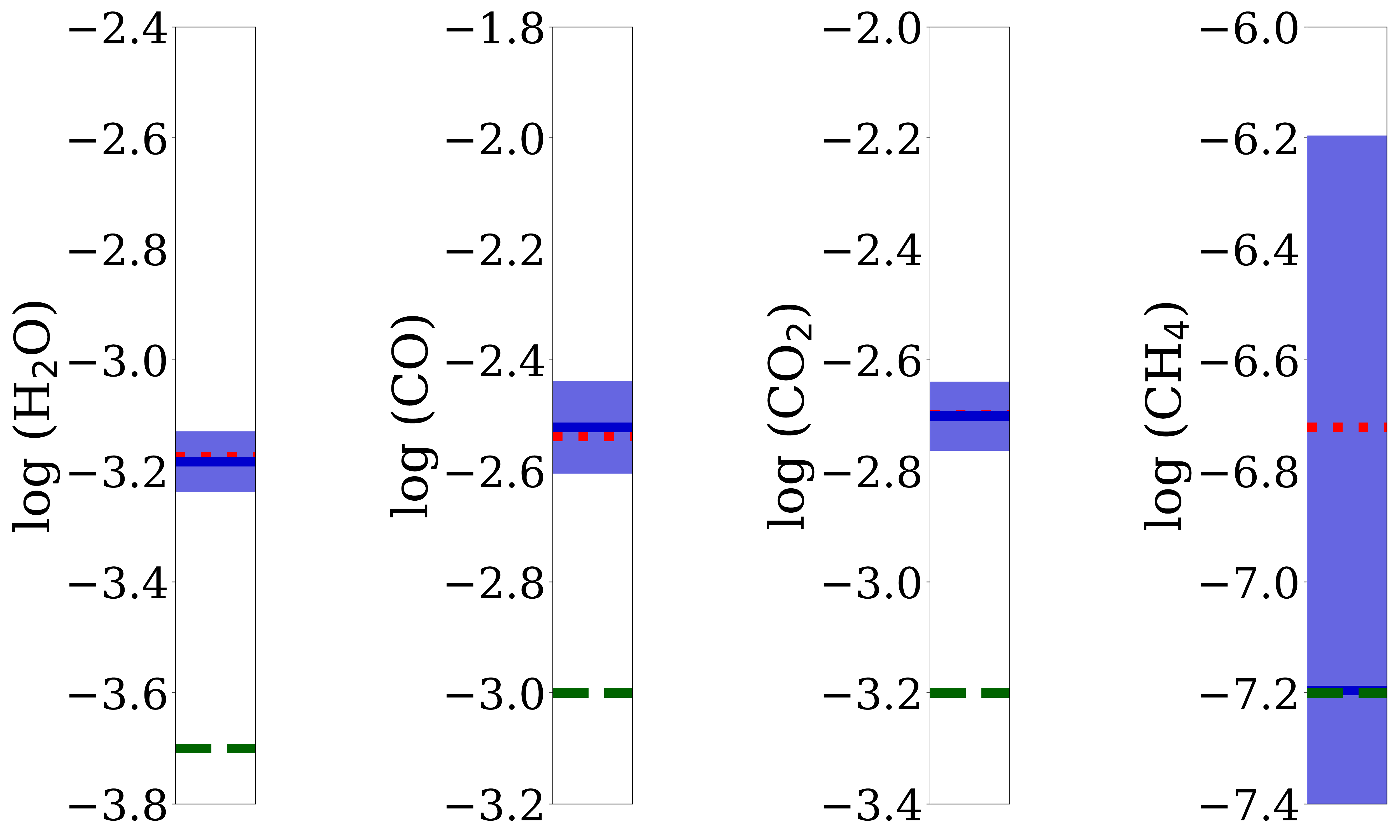}
}
\centerline{\textbf{$\mathbf{(2.28-2.38)+(3.80-4.00)}$~$\mathbf{\mu}$m}}
\centerline{
\includegraphics[trim={0 0 20.0cm 0},clip,height=0.14\vsize]{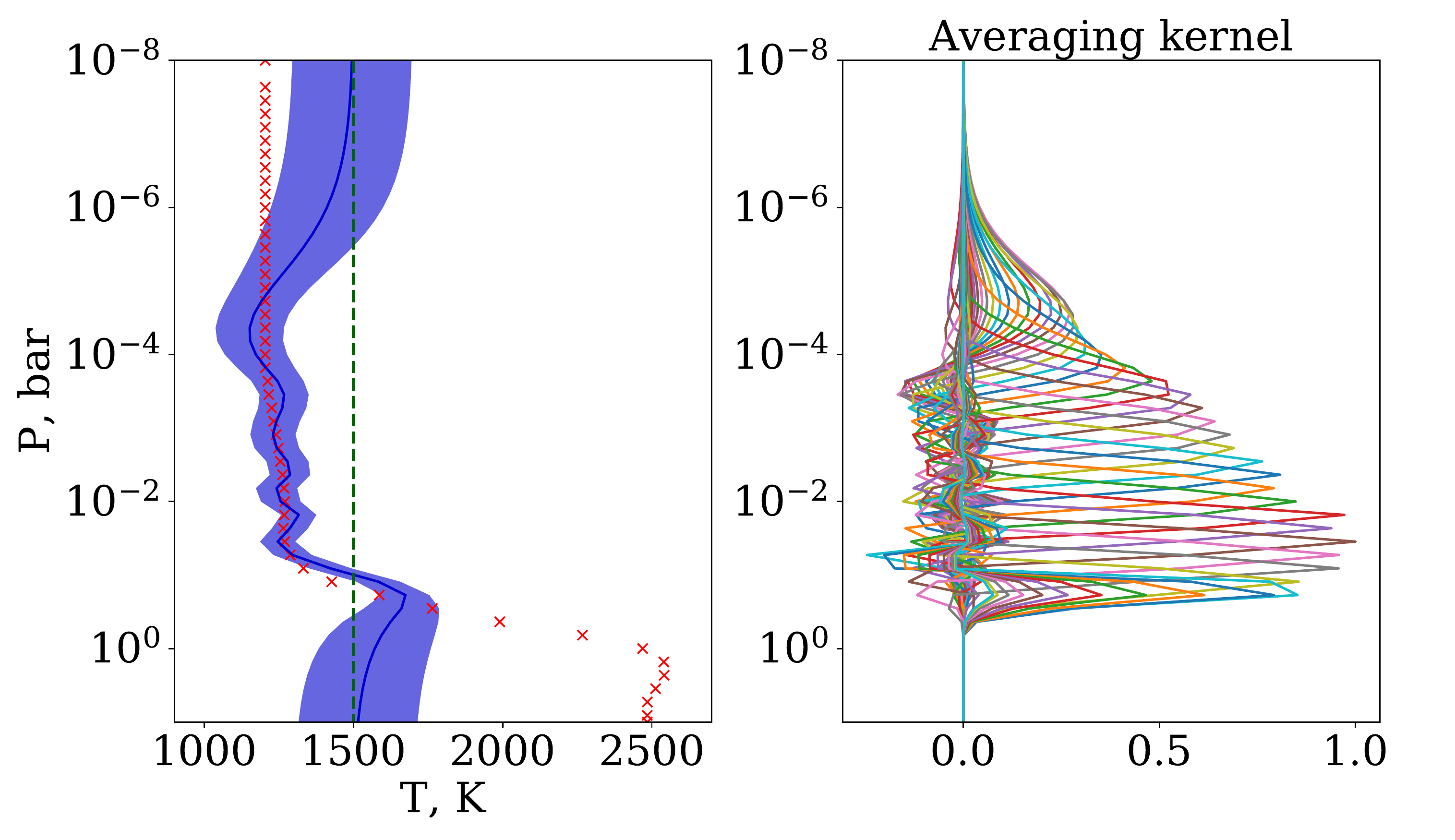}
\includegraphics[height=0.14\vsize]{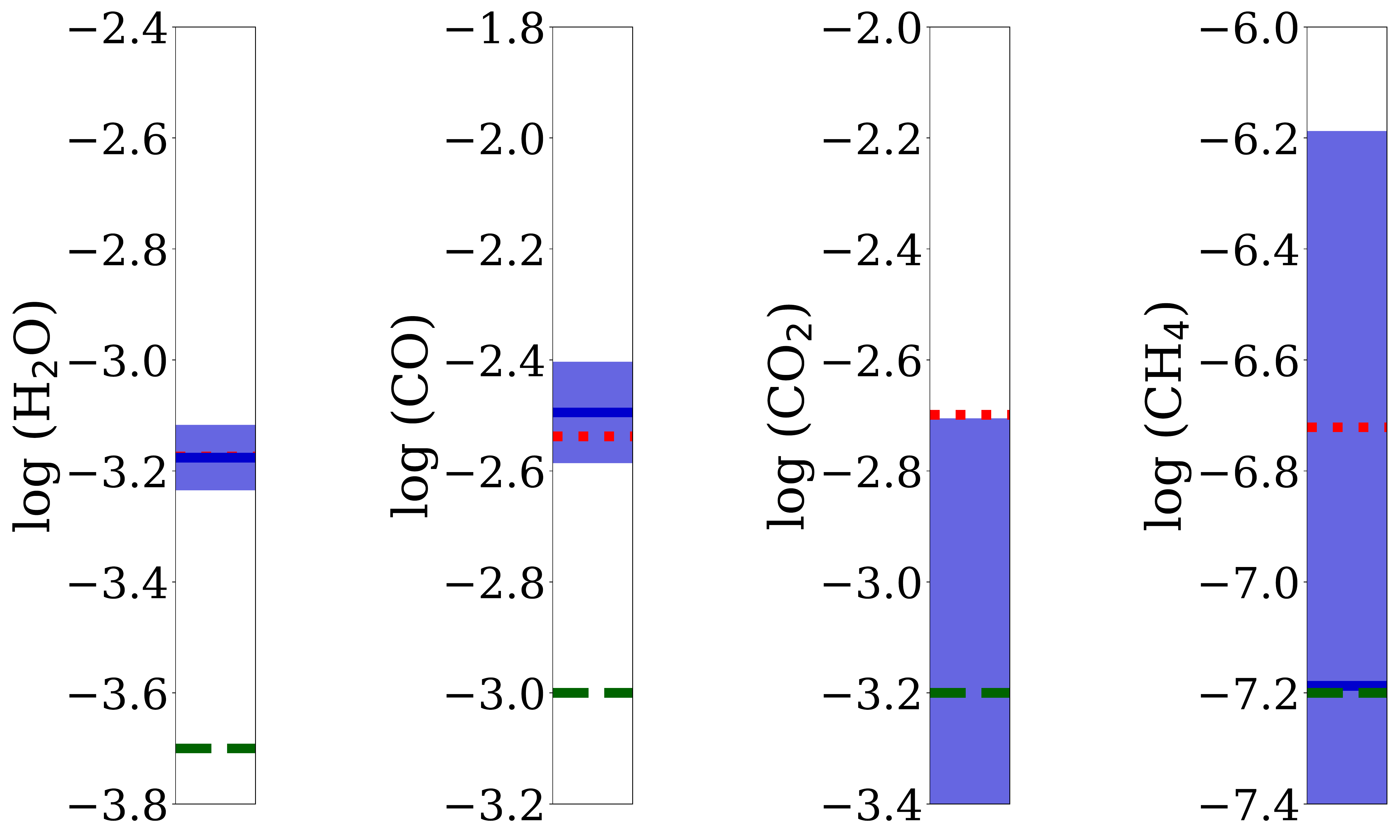}
}
\centerline{\textbf{$\mathbf{(2.28-2.38)+(4.80-5.00)}$~$\mathbf{\mu}$m}}
\centerline{
\includegraphics[trim={0 0 20.0cm 0},clip,height=0.14\vsize]{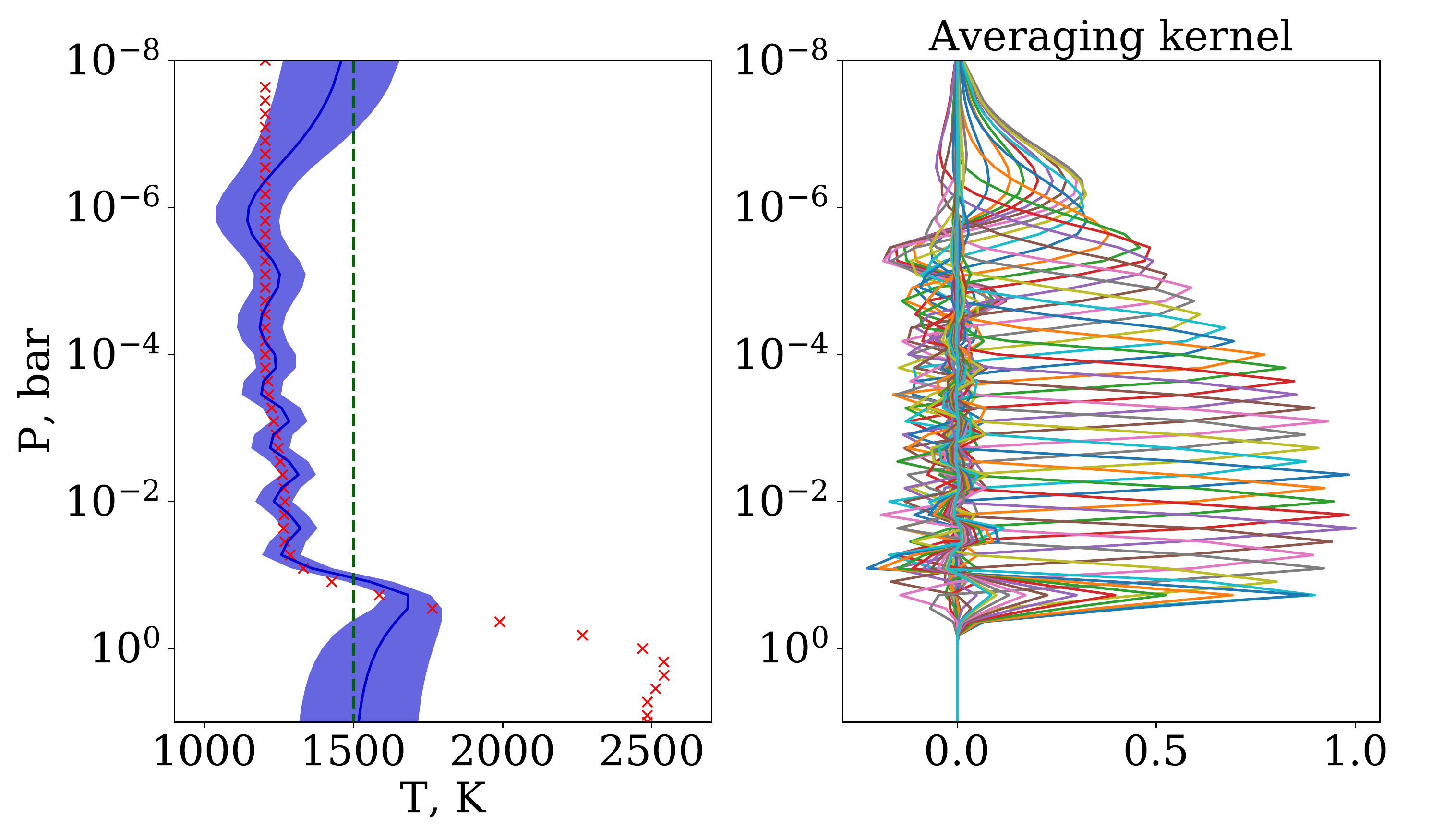}
\includegraphics[height=0.14\vsize]{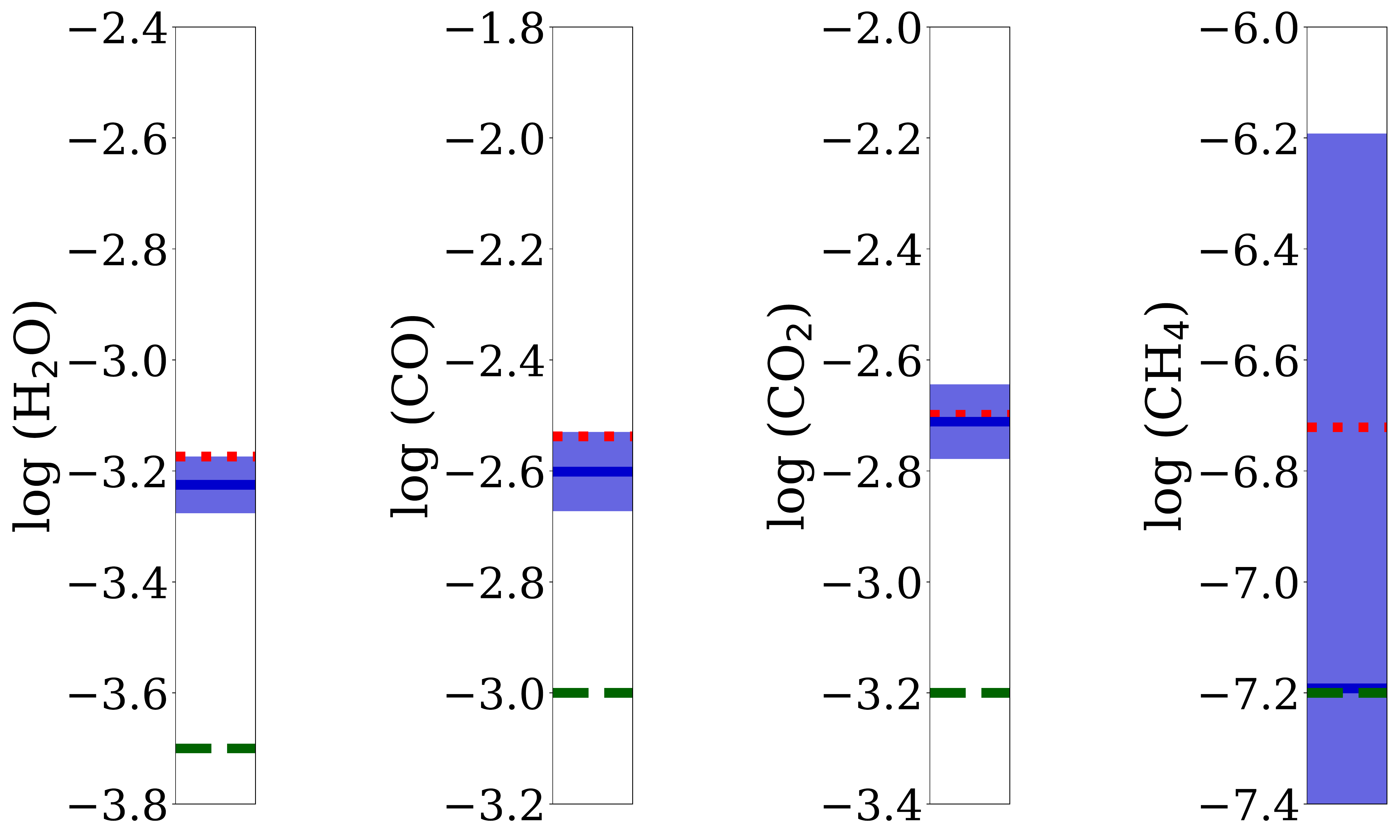}
}
\caption{\label{fig:snr25-r100k-2reg}
Same as  Fig.~\ref{fig:snr5-r100k-2reg}, but for S/N=$25$.
}
\end{figure*}

\begin{figure*}
\centerline{\textbf{$\mathbf{(1.50-1.70)+(2.28+2.38)}$~$\mathbf{\mu}$m}}
\centerline{
\includegraphics[trim={0 0 20.0cm 0},clip,height=0.14\vsize]{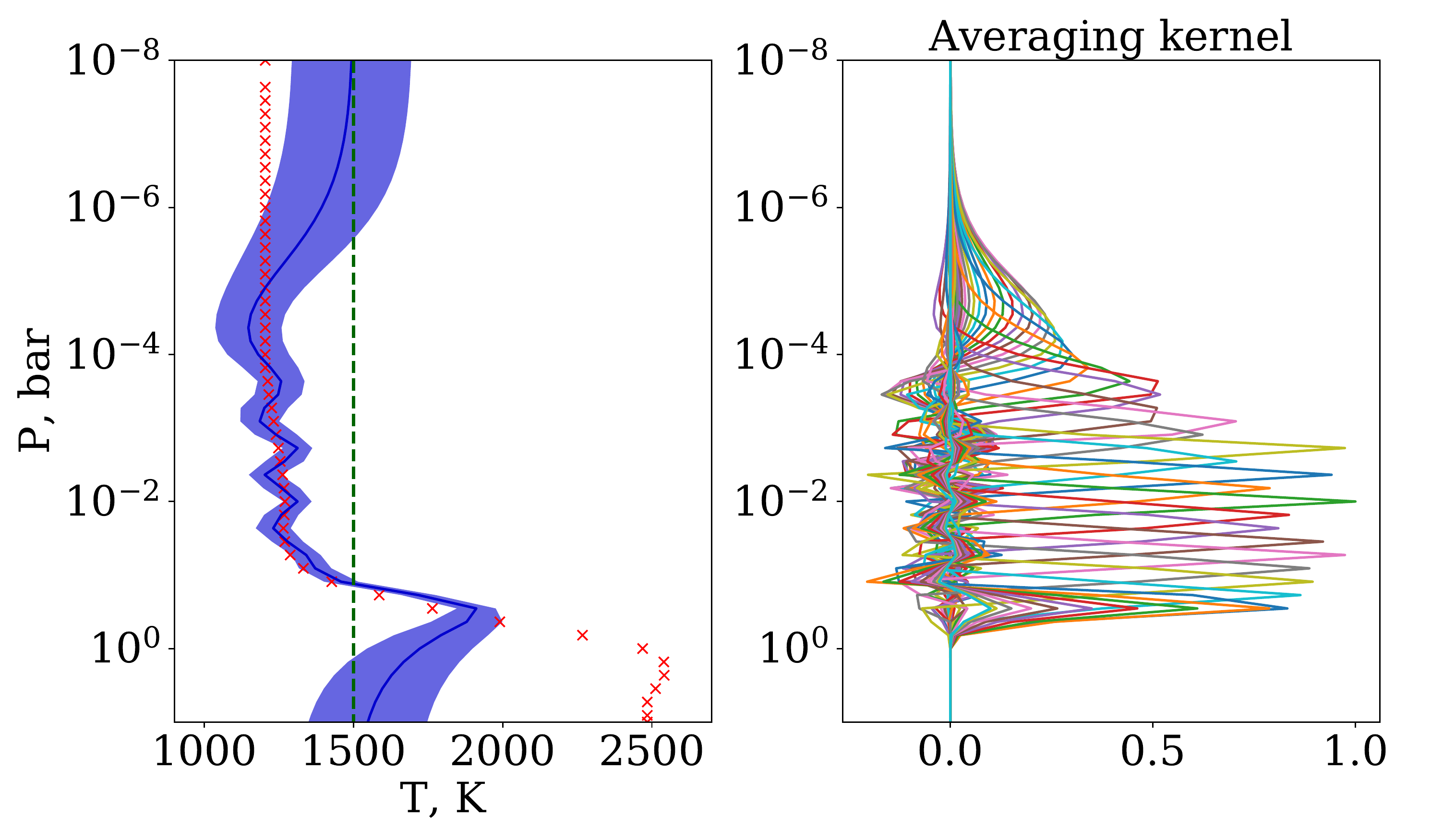}
\includegraphics[height=0.14\vsize]{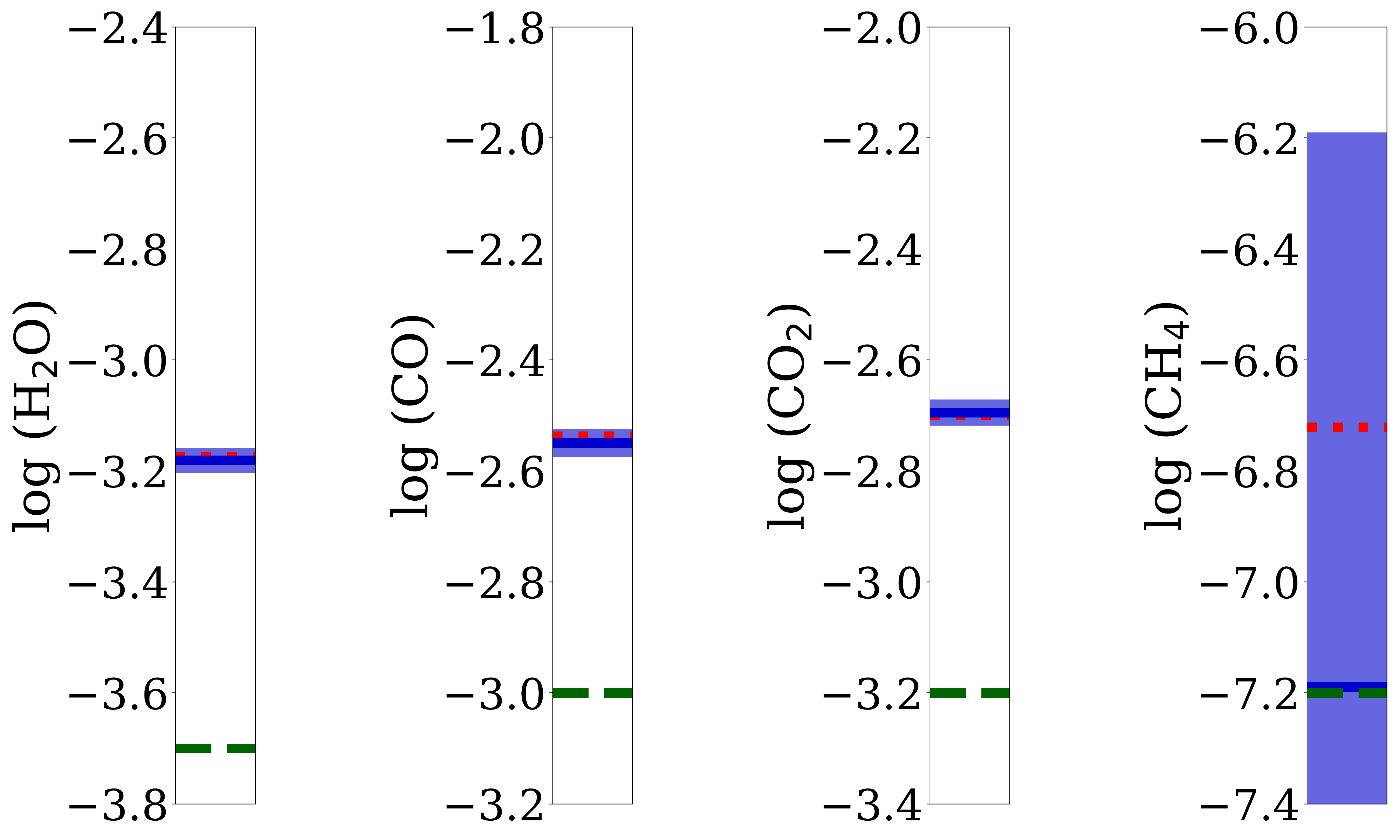}
}
\centerline{\textbf{$\mathbf{(2.10+2.28)+(2.28-2.38)}$~$\mathbf{\mu}$m}}
\centerline{
\includegraphics[trim={0 0 20.0cm 0},clip,height=0.14\vsize]{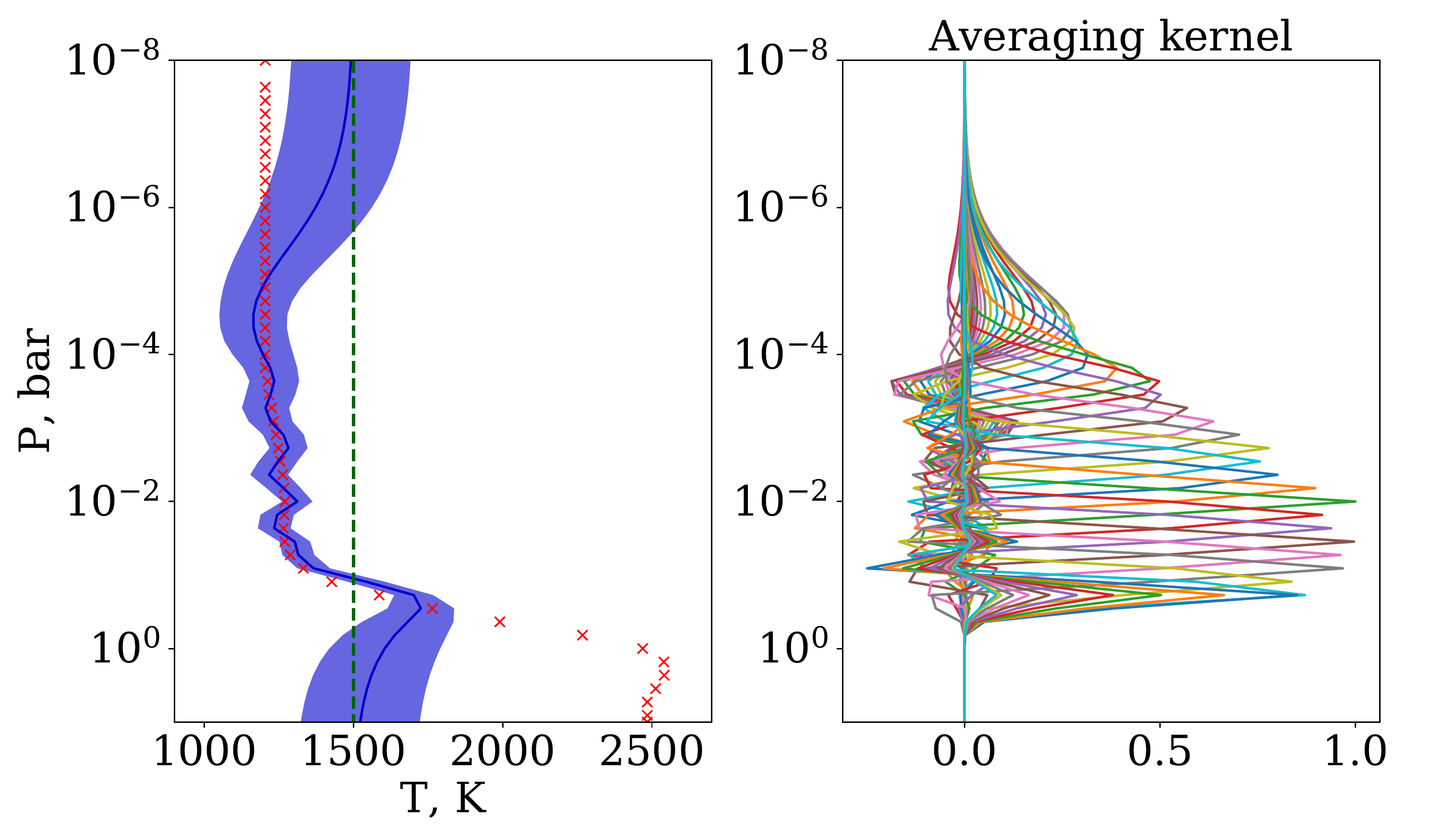}
\includegraphics[height=0.14\vsize]{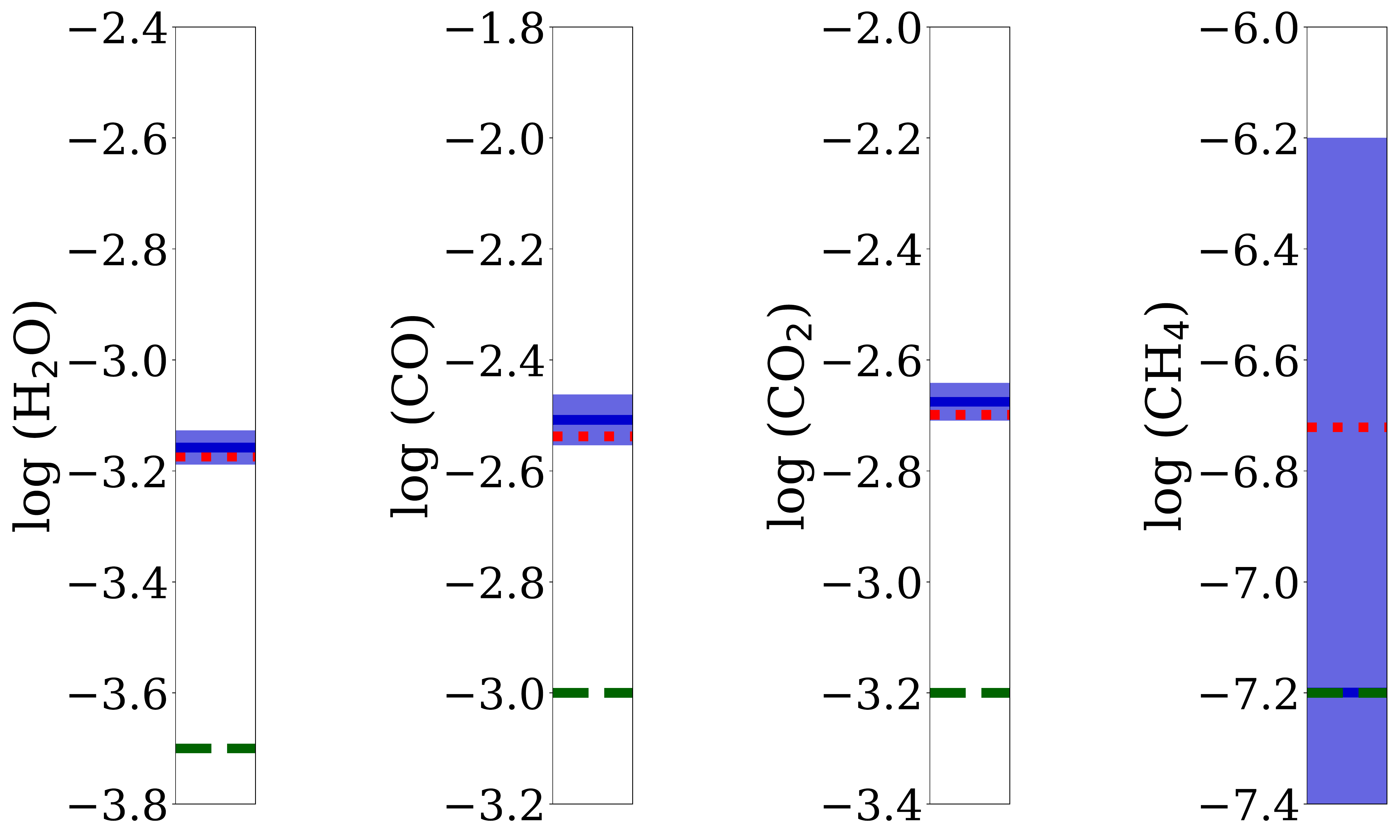}
}
\centerline{\textbf{$\mathbf{(2.28-2.38)+(3.80-4.00)}$~$\mathbf{\mu}$m}}
\centerline{
\includegraphics[trim={0 0 20.0cm 0},clip,height=0.14\vsize]{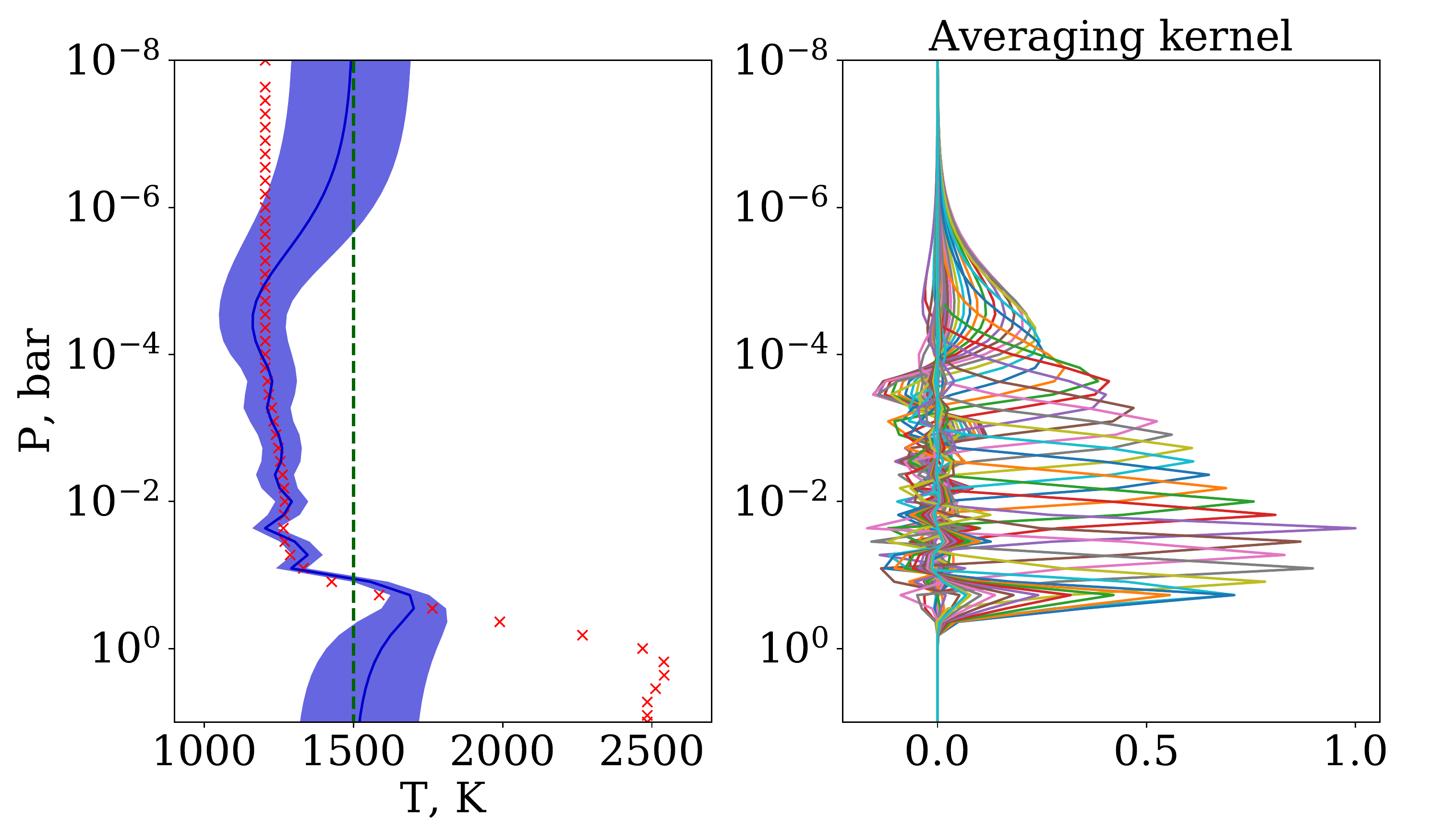}
\includegraphics[height=0.14\vsize]{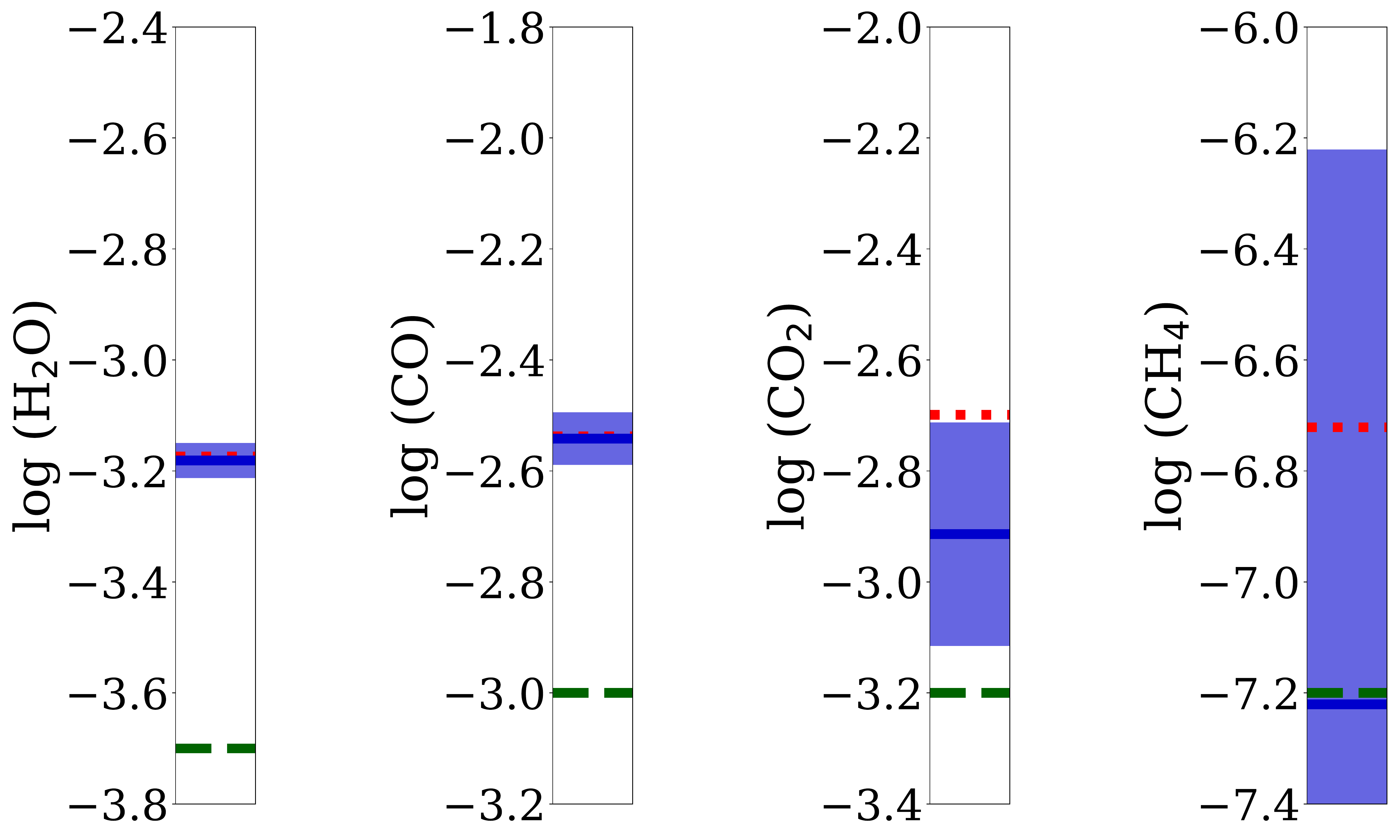}
}
\centerline{\textbf{$\mathbf{(2.28-2.38)+(4.80-5.00)}$~$\mathbf{\mu}$m}}
\centerline{
\includegraphics[trim={0 0 20.0cm 0},clip,height=0.14\vsize]{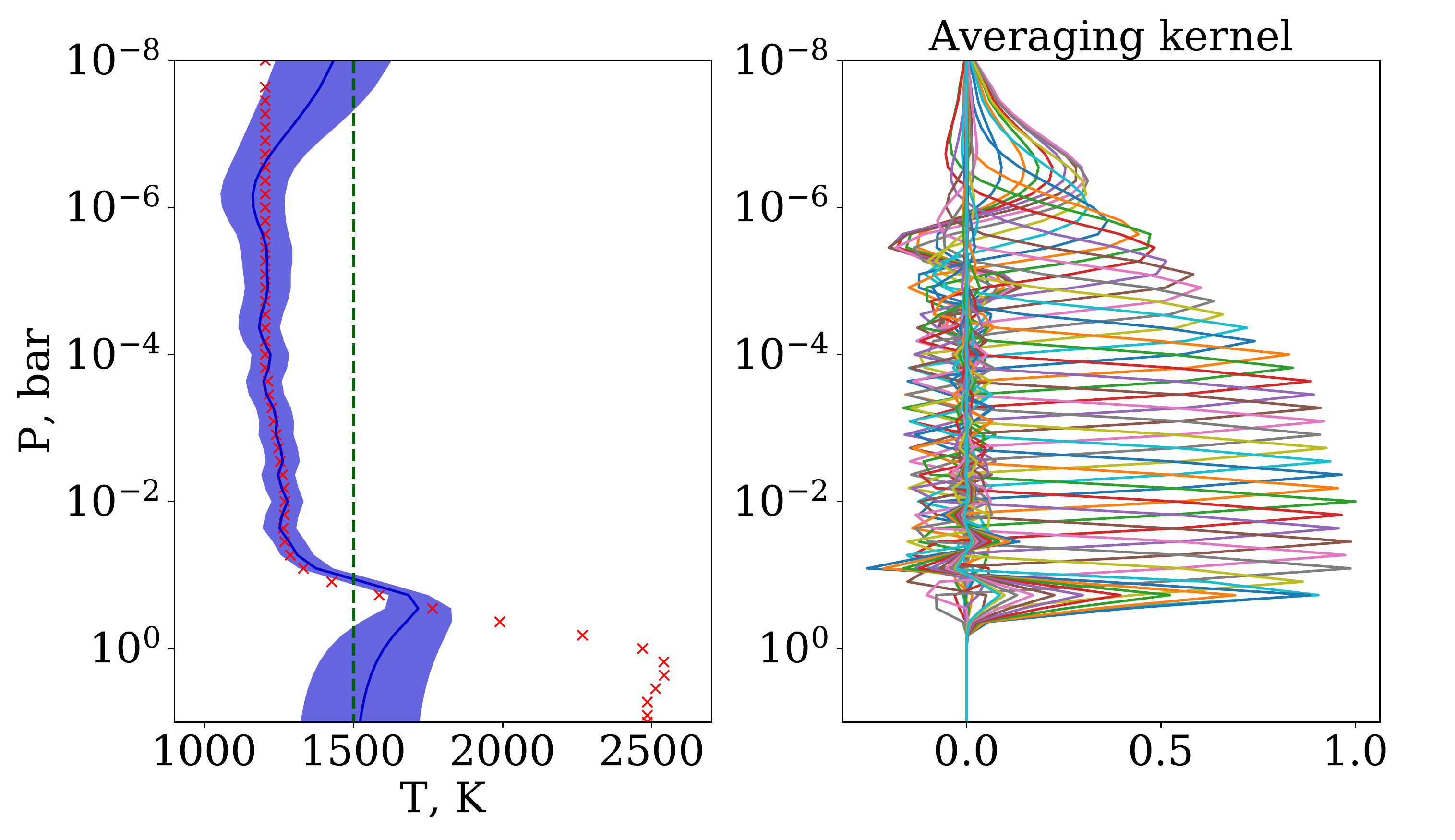}
\includegraphics[height=0.14\vsize]{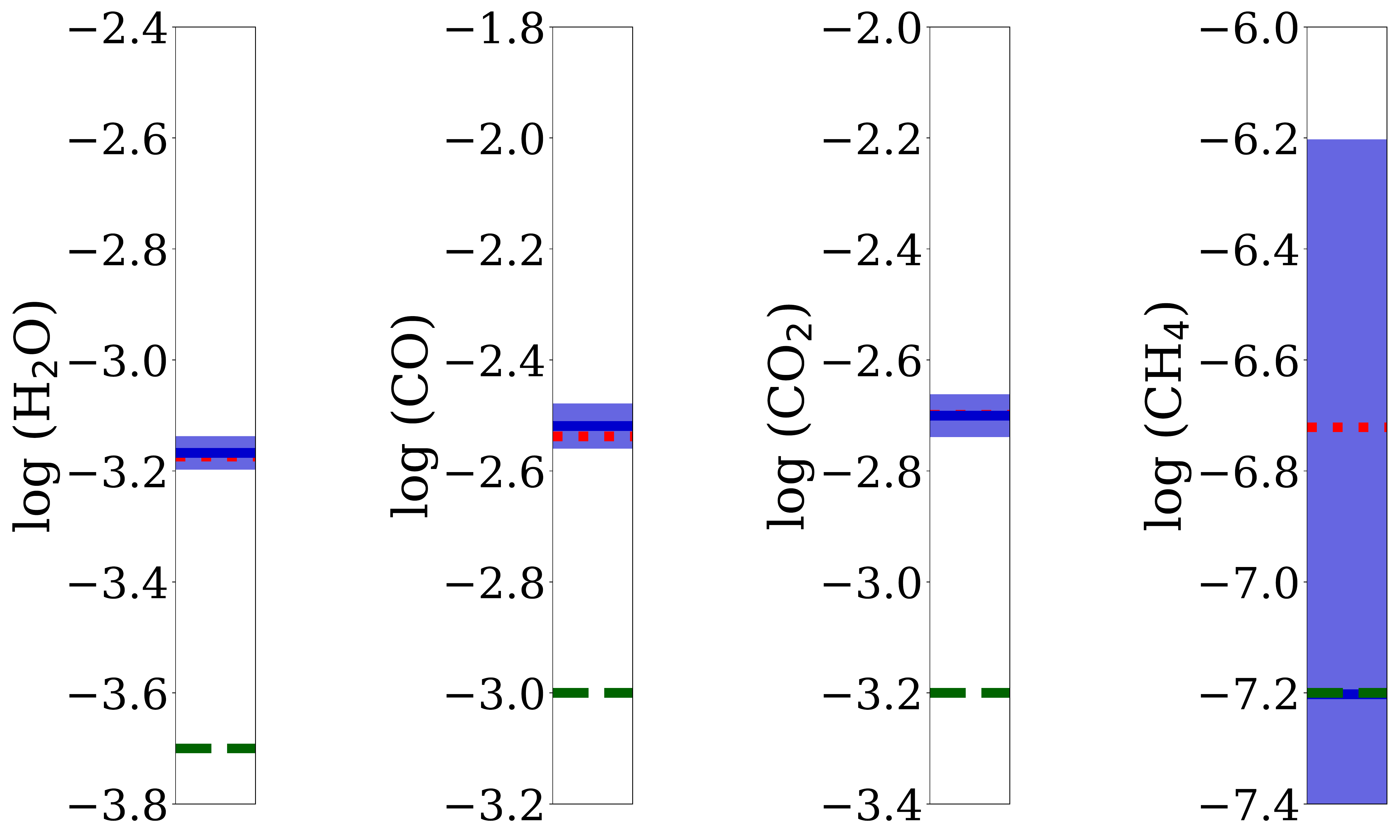}
}
\caption{\label{fig:snr50-r100k-2reg}
Same as  Fig.~\ref{fig:snr5-r100k-2reg}, but for S/N=$50$.
}
\end{figure*}

\end{appendix}

\listofobjects

\end{document}

%% file: common-def.tex
\def\synth3{\textsc{Synth3}}

\def\Mjup{M$_{\rm J}$}

\def\mum{$\mu$m}

\ifdefined\ion
\renewcommand{\ion}[2]{\textup{#1\,\textsc{\uppercase{#2}}}}
\else
\newcommand{\ion}[2]{\textup{#1\,\textsc{\uppercase{#2}}}}
\fi

%% file: tables/obs.tex
\begin{table*}[!ht]
\caption{\label{tab:obs}
Observability of favorable Hot Jupiters for atmospheric retrieval studies with ground-based high resolution spectroscopy.}
\begin{center}
\begin{tabular}{lcccccccccc}
\hline
\hline
  Name                &      d       &  R$_{\rm s}$ & R$_{\rm p}$ &  T$_{\rm s}$   & T$_{\rm p}$  & V$^m$ star  & K$^m$ star &  $\langle$F$_{\rm p}$/F$_{\rm s}$$\rangle$  & t$_{\rm exp}$ & Visible from ESO?\\
                      &   (a.u.)     & (R$_{\odot}$)&(R$_{\rm J}$)&    (K)         &    (K)       &  (mag)      &  (mag)     &                                             & (h)           & (Yes/No)     \\
                      &              &              &             &                &              &             &            &                                             & (R=100k/50k)  &              \\
\hline                                                                                                                                                                     
\textbf{51 Peg b}     &   0.052      & 1.27         & 1.90        & 5800           & 1370         &   5.46      & 3.91       & 1e-3                                        &   6 /   5     & Yes         \\
HD 46375 b            &   0.041      & 1.00         & 1.02        & 5200           & 1229         &   7.84      & 6.00       & 4e-4                                        &  71 /  61     & Yes         \\
%HD 63454 b            &   0.036      & 0.80         & 1.06        & 4841           & 1092         &   9.36      & 7.00       & 5e-4                                        & 109 /  95     & No (too low)\\
HD 73256 b            &   0.037      & 1.22         & 1.00$^{a}$  & 5636           & 1545         &   8.06      & 6.26       & 4e-4                                        &  91 /  79     & Yes         \\
HD 75289 b            &   0.046      & 1.25         & 1.03        & 6120           & 1527         &   6.36      & 5.06       & 4e-4                                        &  29 /  26     & Yes         \\
%HD 86081 b            &   0.039      & 1.22         & 1.08        & 6030           & 1614         &   8.70      & 7.30       & 5e-4                                        & 149 / 128     & Yes         \\
HD 143105 b           &   0.038      & 1.23         & 1.00$^{a}$  & 6380           & 1739         &   6.75      & 5.52       & 5e-4                                        &  27 /  23     & No          \\
%HD 149143 b           &   0.052      & 1.71         & 1.05        & 5730           & 1573         &   7.89      &            & 3e-4                                        & 164 / 143     & Yes         \\
HD 179949 b           &   0.045      & 1.19         & 1.05        & 6260           & 1541         &   6.24      & 4.94       & 4e-4                                        &  26 /  23     & Yes         \\
%HD 185269 b           &   0.077      & 1.88         & 1.00$^{a}$  & 5980           & 1414         &   7.00      & 5.26       & 1e-4                                        & 522 / 454     & No          \\
HD 187123 b           &   0.043      & 1.17         & 1.00$^{a}$  & 5714           & 1433         &   7.83      & 6.34       & 4e-4                                        &  98 /  85     & No (too low)\\
HD 189733 b           &   0.031      & 0.81         & 1.14        & 4880           & 1185         &   7.65      & 5.54       & 7e-4                                        &  78 /  68     & Yes         \\
HD 209458 b           &   0.048      & 1.20         & 1.38        & 6100           & 1467         &   7.63      & 6.31       & 7e-4                                        &  95 /  83     & Yes         \\
HD 212301 b           &   0.036      & 1.25         & 1.07        & 6000           & 1692         &   7.76      &            & 6e-4                                        &  37 /  32     & No (too low)\\
HD 217107 b           &   0.073      & 1.08         & 1.00$^{a}$  & 5666           & 1043         &   6.16      & 4.54       & 2e-4                                        &  67 /  58     & No          \\
%HD 330075 b           &   0.039      & 1.20         & 1.06        & 6300           & 1672         &   9.33      & 7.17       & 5e-4                                        & 131 / 113     & Yes         \\
KELT-20 b             &   0.054      & 1.56         & 1.74        & 8981           & 2307         &   7.58      & 7.42       & 1e-3                                        &  43 /  37     & No (too low)\\
%KOI-13 b              &   0.034      & 1.71         & 1.41        & 7650           & 2588         &   9.95      & 9.43       & 1e-3                                        & 426 / 352     & No (too low)\\
\textbf{$\tau$ Boo b} &   0.046      & 1.33         & 1.06        & 6300           & 1624         &   4.49      & 3.36       & 4e-4                                        &  10 /   9     & Yes         \\
$\upsilon$ And b      &   0.059      & 1.63         & 1.00$^{a}$  & 6200           & 1560         &   4.10      & 2.86       & 2e-4                                        &  14 /  12     & No (too low)\\
%WASP-14 b             &   0.036      & 1.31         & 1.28        & 6475           & 1866         &   9.75      & 8.62       & 9e-4                                        & 188 / 158     & Yes         \\
WASP-18 b             &   0.021      & 1.23         & 1.17        & 6400           & 2374         &   9.30      & 8.13       & 2e-3                                        &  22 /  19     & Yes         \\
WASP-33 b             &   0.026      & 1.44         & 1.60        & 7400           & 2661         &   8.14      & 7.47       & 2e-3                                        &  13 /  11     & No (too low)\\
%WASP-74 b             &   0.034      & 1.64         & 1.40        & 5970           & 1972         &   9.75      & 8.22       & 8e-4                                        & 141 / 120     & Yes         \\
WASP-76 b             &   0.033      & 1.73         & 1.83        & 6250           & 2166         &   9.52      & 8.24       & 2e-3                                        &  25 /  21     & Yes         \\
\hline
\end{tabular}
\end{center}
For each target the table lists its name, distance to the parent star, radii of the star and the planet, stellar effective temperature, planetary equilibrium temperature, 
stellar magnitudes in $V$ and $K$ bands, mean planet-to-star flux ratio $\langle$F$_{\rm p}$/F$_{\rm s}$$\rangle$ calculated between $1$~\mum\ and $5$~\mum, 
and exposure time needed to reach an SNR=$5$ for the finally obtained exoplanet spectrum using single \criresplus\ setting. The exposure times are given separately
for spectral resolutions R=$100$k and R=$50$k, respectively.
The last column contains a flag for the observability of the star at the ESO Paranal observatory (Chile).
The two primary targets for \criresplus\, are marked with bold.\\
\textbf{Note:} $^{a}$We assumed R$_{\rm p}$=1.0~R$_{\rm J}$ due to the lack of information.
\end{table*}

%slit = 0.2, SNR=1e4
%
%35    117960
%35     95790
%35     80040 
%35     86670
%35    235200
%35    381270
%      
%26    113070
%26    102030
%26     95040
%26     99570
%26    178530
%26    228990
%     
%25    123960
%25    103800
%25     97020
%25    101280
%25    176880
%25    225120
%     
%14   1150800
%14    940500
%14    613680
%14    614070
%14    881340
%14   1000590
%14   2619930
%14   3298980
%
%11  12666960
%11  11981820
%11   8377680
%11  11146440
%11  68427870
%11 156214800
%11 357698310
%11 258259050

%% file: oooeee.bbl
\begin{thebibliography}{41}
\expandafter\ifx\csname natexlab\endcsname\relax\def\natexlab#1{#1}\fi

\bibitem[{{Abel} {et~al.}(2011){Abel}, {Frommhold}, {Li}, \&
  {Hunt}}]{2011mss..confEFC07A}
{Abel}, M., {Frommhold}, L., {Li}, X., \& {Hunt}, K.~L.~C. 2011, in 66th
  International Symposium On Molecular Spectroscopy, EFC07

\bibitem[{{Abel} {et~al.}(2012){Abel}, {Frommhold}, {Li}, \&
  {Hunt}}]{2012JChPh.136d4319A}
{Abel}, M., {Frommhold}, L., {Li}, X., \& {Hunt}, K.~L.~C. 2012, \jcp, 136,
  044319

\bibitem[{{Barstow} {et~al.}(2013){Barstow}, {Aigrain}, {Irwin}, {Fletcher}, \&
  {Lee}}]{2013MNRAS.434.2616B}
{Barstow}, J.~K., {Aigrain}, S., {Irwin}, P.~G.~J., {Fletcher}, L.~N., \&
  {Lee}, J.~M. 2013, \mnras, 434, 2616

\bibitem[{{Barstow} {et~al.}(2014){Barstow}, {Aigrain}, {Irwin}, {Hackler},
  {Fletcher}, {Lee}, \& {Gibson}}]{2014ApJ...786..154B}
{Barstow}, J.~K., {Aigrain}, S., {Irwin}, P.~G.~J., {et~al.} 2014, \apj, 786,
  154

\bibitem[{{Barstow} {et~al.}(2017){Barstow}, {Aigrain}, {Irwin}, \&
  {Sing}}]{2017ApJ...834...50B}
{Barstow}, J.~K., {Aigrain}, S., {Irwin}, P.~G.~J., \& {Sing}, D.~K. 2017,
  \apj, 834, 50

\bibitem[{{Borysow}(2002)}]{2002A&A...390..779B}
{Borysow}, A. 2002, \aap, 390, 779

\bibitem[{{Borysow} \& {Frommhold}(1989)}]{1989ApJ...341..549B}
{Borysow}, A. \& {Frommhold}, L. 1989, \apj, 341, 549

\bibitem[{{Borysow} {et~al.}(2001){Borysow}, {Jorgensen}, \&
  {Fu}}]{2001JQSRT..68..235B}
{Borysow}, A., {Jorgensen}, U.~G., \& {Fu}, Y. 2001, \jqsrt, 68, 235

\bibitem[{{Bouchy} {et~al.}(2005){Bouchy}, {Udry}, {Mayor}, {Moutou}, {Pont},
  {Iribarne}, {da Silva}, {Ilovaisky}, {Queloz}, {Santos}, {S{\'e}gransan}, \&
  {Zucker}}]{2005A&A...444L..15B}
{Bouchy}, F., {Udry}, S., {Mayor}, M., {et~al.} 2005, \aap, 444, L15

\bibitem[{{Brogi} {et~al.}(2016){Brogi}, {de Kok}, {Albrecht}, {Snellen},
  {Birkby}, \& {Schwarz}}]{2016ApJ...817..106B}
{Brogi}, M., {de Kok}, R.~J., {Albrecht}, S., {et~al.} 2016, \apj, 817, 106

\bibitem[{{Brogi} {et~al.}(2014){Brogi}, {de Kok}, {Birkby}, {Schwarz}, \&
  {Snellen}}]{2014A&A...565A.124B}
{Brogi}, M., {de Kok}, R.~J., {Birkby}, J.~L., {Schwarz}, H., \& {Snellen},
  I.~A.~G. 2014, \aap, 565, A124

\bibitem[{{Brogi} {et~al.}(2017){Brogi}, {Line}, {Bean}, {D{\'e}sert}, \&
  {Schwarz}}]{2017ApJ...839L...2B}
{Brogi}, M., {Line}, M., {Bean}, J., {D{\'e}sert}, J.-M., \& {Schwarz}, H.
  2017, \apjl, 839, L2

\bibitem[{{Brogi} \& {Line}(2019)}]{2019AJ....157..114B}
{Brogi}, M. \& {Line}, M.~R. 2019, \aj, 157, 114

\bibitem[{{de Kok} {et~al.}(2014){de Kok}, {Birkby}, {Brogi}, {Schwarz},
  {Albrecht}, {de Mooij}, \& {Snellen}}]{2014A&A...561A.150D}
{de Kok}, R.~J., {Birkby}, J., {Brogi}, M., {et~al.} 2014, \aap, 561, A150

\bibitem[{{Dorn} {et~al.}(2014){Dorn}, {Anglada-Escude}, {Baade}, {Bristow},
  {Follert}, {Gojak}, {Grunhut}, {Hatzes}, {Heiter}, {Hilker}, {Ives}, {Jung},
  {K{\"a}ufl}, {Kerber}, {Klein}, {Lizon}, {Lockhart}, {L{\"o}winger},
  {Marquart}, {Oliva}, {Origlia}, {Pasquini}, {Paufique}, {Piskunov}, {Pozna},
  {Reiners}, {Smette}, {Smoker}, {Seemann}, {Stempels}, \&
  {Valenti}}]{2014Msngr.156....7D}
{Dorn}, R.~J., {Anglada-Escude}, G., {Baade}, D., {et~al.} 2014, The Messenger,
  156, 7

\bibitem[{{Follert} {et~al.}(2014){Follert}, {Dorn}, {Oliva}, {Lizon},
  {Hatzes}, {Piskunov}, {Reiners}, {Seemann}, {Stempels}, {Heiter}, {Marquart},
  {Lockhart}, {Anglada-Escude}, {L{\"o}winger}, {Baade}, {Grunhut}, {Bristow},
  {Klein}, {Jung}, {Ives}, {Kerber}, {Pozna}, {Paufique}, {Kaeufl}, {Origlia},
  {Valenti}, {Gojak}, {Hilker}, {Pasquini}, {Smette}, \&
  {Smoker}}]{2014SPIE.9147E..19F}
{Follert}, R., {Dorn}, R.~J., {Oliva}, E., {et~al.} 2014, in \procspie, Vol.
  9147, Ground-based and Airborne Instrumentation for Astronomy V, 914719

\bibitem[{{Gandhi} \& {Madhusudhan}(2017)}]{2017MNRAS.472.2334G}
{Gandhi}, S. \& {Madhusudhan}, N. 2017, \mnras, 472, 2334

\bibitem[{{Hartogh} {et~al.}(2010){Hartogh}, {Jarchow}, {Lellouch}, {de
  Val-Borro}, {Rengel}, {Moreno}, {Medvedev}, {Sagawa}, {Swinyard},
  {Cavali{\'e}}, {Lis}, {B{\l}{\c e}cka}, {Banaszkiewicz},
  {Bockel{\'e}e-Morvan}, {Crovisier}, {Encrenaz}, {K{\"u}ppers}, {Lara},
  {Szutowicz}, {Vandenbussche}, {Bensch}, {Bergin}, {Billebaud}, {Biver},
  {Blake}, {Blommaert}, {Cernicharo}, {Decin}, {Encrenaz}, {Feuchtgruber},
  {Fulton}, {de Graauw}, {Jehin}, {Kidger}, {Lorente}, {Naylor}, {Portyankina},
  {S{\'a}nchez-Portal}, {Schieder}, {Sidher}, {Thomas}, {Verdugo}, {Waelkens},
  {Whyborn}, {Teyssier}, {Helmich}, {Roelfsema}, {Stutzki}, {Leduc}, \&
  {Stern}}]{2010A&A...521L..49H}
{Hartogh}, P., {Jarchow}, C., {Lellouch}, E., {et~al.} 2010, \aap, 521, L49

\bibitem[{{Irwin} {et~al.}(2008){Irwin}, {Teanby}, {de Kok}, {Fletcher},
  {Howett}, {Tsang}, {Wilson}, {Calcutt}, {Nixon}, \&
  {Parrish}}]{2008JQSRT.109.1136I}
{Irwin}, P.~G.~J., {Teanby}, N.~A., {de Kok}, R., {et~al.} 2008, \jqsrt, 109,
  1136

\bibitem[{{Kreidberg} {et~al.}(2015){Kreidberg}, {Line}, {Bean}, {Stevenson},
  {D{\'e}sert}, {Madhusudhan}, {Fortney}, {Barstow}, {Henry}, {Williamson}, \&
  {Showman}}]{2015ApJ...814...66K}
{Kreidberg}, L., {Line}, M.~R., {Bean}, J.~L., {et~al.} 2015, \apj, 814, 66

\bibitem[{{Lee} {et~al.}(2012){Lee}, {Fletcher}, \&
  {Irwin}}]{2012MNRAS.420..170L}
{Lee}, J.-M., {Fletcher}, L.~N., \& {Irwin}, P.~G.~J. 2012, \mnras, 420, 170

\bibitem[{{Lee} {et~al.}(2014){Lee}, {Irwin}, {Fletcher}, {Heng}, \&
  {Barstow}}]{2014ApJ...789...14L}
{Lee}, J.-M., {Irwin}, P. G.~J., {Fletcher}, L.~N., {Heng}, K., \& {Barstow},
  J.~K. 2014, \apj, 789, 14

\bibitem[{{Line} {et~al.}(2013){Line}, {Wolf}, {Zhang}, {Knutson}, {Kammer},
  {Ellison}, {Deroo}, {Crisp}, \& {Yung}}]{2013ApJ...775..137L}
{Line}, M.~R., {Wolf}, A.~S., {Zhang}, X., {et~al.} 2013, \apj, 775, 137

\bibitem[{{Line} {et~al.}(2012){Line}, {Zhang}, {Vasisht}, {Natraj}, {Chen}, \&
  {Yung}}]{2012ApJ...749...93L}
{Line}, M.~R., {Zhang}, X., {Vasisht}, G., {et~al.} 2012, \apj, 749, 93

\bibitem[{{Malik} {et~al.}(2017){Malik}, {Grosheintz}, {Mendon{\c c}a},
  {Grimm}, {Lavie}, {Kitzmann}, {Tsai}, {Burrows}, {Kreidberg}, {Bedell},
  {Bean}, {Stevenson}, \& {Heng}}]{2017AJ....153...56M}
{Malik}, M., {Grosheintz}, L., {Mendon{\c c}a}, J.~M., {et~al.} 2017, \aj, 153,
  56

\bibitem[{{Molli{\`e}re} {et~al.}(2015){Molli{\`e}re}, {van Boekel},
  {Dullemond}, {Henning}, \& {Mordasini}}]{2015ApJ...813...47M}
{Molli{\`e}re}, P., {van Boekel}, R., {Dullemond}, C., {Henning}, T., \&
  {Mordasini}, C. 2015, \apj, 813, 47

\bibitem[{{Parmentier} \& {Guillot}(2014)}]{2014A&A...562A.133P}
{Parmentier}, V. \& {Guillot}, T. 2014, \aap, 562, A133

\bibitem[{{Rengel} {et~al.}(2008){Rengel}, {Hartogh}, \&
  {Jarchow}}]{2008P&SS...56.1368R}
{Rengel}, M., {Hartogh}, P., \& {Jarchow}, C. 2008, \planss, 56, 1368

\bibitem[{{Rodgers}(1976)}]{1976RvGSP..14..609R}
{Rodgers}, C.~D. 1976, Reviews of Geophysics and Space Physics, 14, 609

\bibitem[{{Rodgers}(2000)}]{2000imas.book.....R}
{Rodgers}, C.~D. 2000, {Inverse Methods for Atmospheric Sounding: Theory and
  Practice} (World Scientific Publishing Co)

\bibitem[{{Rothman} {et~al.}(2010){Rothman}, {Gordon}, {Barber}, {Dothe},
  {Gamache}, {Goldman}, {Perevalov}, {Tashkun}, \&
  {Tennyson}}]{2010JQSRT.111.2139R}
{Rothman}, L.~S., {Gordon}, I.~E., {Barber}, R.~J., {et~al.} 2010, \jqsrt, 111,
  2139

\bibitem[{{Schwarz} {et~al.}(2015){Schwarz}, {Brogi}, {de Kok}, {Birkby}, \&
  {Snellen}}]{2015A&A...576A.111S}
{Schwarz}, H., {Brogi}, M., {de Kok}, R., {Birkby}, J., \& {Snellen}, I. 2015,
  \aap, 576, A111

\bibitem[{{Sing} {et~al.}(2016){Sing}, {Fortney}, {Nikolov}, {Wakeford},
  {Kataria}, {Evans}, {Aigrain}, {Ballester}, {Burrows}, {Deming},
  {D{\'e}sert}, {Gibson}, {Henry}, {Huitson}, {Knutson}, {Lecavelier Des
  Etangs}, {Pont}, {Showman}, {Vidal-Madjar}, {Williamson}, \&
  {Wilson}}]{2016Natur.529...59S}
{Sing}, D.~K., {Fortney}, J.~J., {Nikolov}, N., {et~al.} 2016, \nat, 529, 59

\bibitem[{{Snellen} {et~al.}(2010){Snellen}, {de Kok}, {de Mooij}, \&
  {Albrecht}}]{2010Natur.465.1049S}
{Snellen}, I.~A.~G., {de Kok}, R.~J., {de Mooij}, E.~J.~W., \& {Albrecht}, S.
  2010, \nat, 465, 1049

\bibitem[{{Sudarsky} {et~al.}(2000){Sudarsky}, {Burrows}, \&
  {Pinto}}]{2000ApJ...538..885S}
{Sudarsky}, D., {Burrows}, A., \& {Pinto}, P. 2000, \apj, 538, 885

\bibitem[{{Tennyson} \& {Yurchenko}(2012)}]{2012MNRAS.425...21T}
{Tennyson}, J. \& {Yurchenko}, S.~N. 2012, \mnras, 425, 21

\bibitem[{{Waldmann} {et~al.}(2015{\natexlab{a}}){Waldmann}, {Rocchetto},
  {Tinetti}, {Barton}, {Yurchenko}, \& {Tennyson}}]{2015ApJ...813...13W}
{Waldmann}, I.~P., {Rocchetto}, M., {Tinetti}, G., {et~al.} 2015{\natexlab{a}},
  \apj, 813, 13

\bibitem[{{Waldmann} {et~al.}(2015{\natexlab{b}}){Waldmann}, {Tinetti},
  {Rocchetto}, {Barton}, {Yurchenko}, \& {Tennyson}}]{2015ApJ...802..107W}
{Waldmann}, I.~P., {Tinetti}, G., {Rocchetto}, M., {et~al.} 2015{\natexlab{b}},
  \apj, 802, 107

\bibitem[{{Wang} {et~al.}(2015){Wang}, {Fischer}, {Horch}, \&
  {Huang}}]{2015ApJ...799..229W}
{Wang}, J., {Fischer}, D.~A., {Horch}, E.~P., \& {Huang}, X. 2015, \apj, 799,
  229

\bibitem[{{Winn} {et~al.}(2010){Winn}, {Fabrycky}, {Albrecht}, \&
  {Johnson}}]{2010ApJ...718L.145W}
{Winn}, J.~N., {Fabrycky}, D., {Albrecht}, S., \& {Johnson}, J.~A. 2010, \apjl,
  718, L145

\bibitem[{{Yan} \& {Henning}(2018)}]{2018NatAs...2..714Y}
{Yan}, F. \& {Henning}, T. 2018, Nature Astronomy, 2, 714

\end{thebibliography}
